\documentclass{article}

\usepackage[margin=1in]{geometry}
\usepackage{mathpazo}

\usepackage[utf8]{inputenc}
\usepackage[T1]{fontenc}
\usepackage{microtype}
\usepackage{graphicx}
\usepackage{booktabs}
\usepackage{amsmath}
\usepackage{amssymb}
\usepackage{natbib}
\usepackage[colorlinks,allcolors=blue]{hyperref}
\usepackage{xurl} %

\newcommand{\paperbibstyle}{plainnat}

\graphicspath{{./}{results/}{figures/}{figures/fig1/}{figures/stage3/}{figures/stage4/}{figures/stage5/}{figures/stage7/}{figures/stage7/visibility_heatmaps/}{figures/stage8/}}

\usepackage{lipsum}
\usepackage{placeins} %
\usepackage{float}    %
\usepackage{array}
\usepackage{longtable}
\usepackage{subcaption}
\usepackage{mathtools}
\usepackage{amsthm}
\usepackage[capitalize,noabbrev]{cleveref}
\usepackage{enumitem}
\usepackage{xcolor} %
\definecolor{tablemax}{RGB}{56,140,196}
\newcommand{\sch}[1]{\path{#1}}

\usepackage{listings}
\theoremstyle{plain}

\theoremstyle{definition}

\theoremstyle{remark}

\usepackage[textsize=tiny]{todonotes}

\usepackage{color-edits}
\addauthor{shc}{red}

\begin{document}

\title{Mapping the AI Supply Chain: Dependencies, Vulnerabilities, and Market Concentration in the AI Ecosystem}

\author{Sarah H.~Cen \\
Carnegie Mellon University \\
\and
Lindsey A.~Gailmard \\
Stanford University \\
\and
Joohan Kim \\
Carnegie Mellon University \\
\and
Rishi Bommasani \\
Stanford University \\
\and
Daniel E.~Ho \\
Stanford University \\
\and
Percy Liang \\
Stanford University \\
}
\date{}

\maketitle

\begin{abstract}

AI systems are built across many organizations that contribute models, data, compute, capital, talent, and other assets.
While the layers of the AI stack are relatively well understood, the relationships between actors and components across layers remains sparsely documented. 
In this work, we focus on the organizations that make up the AI supply chain: the complex network of actors that contribute to the development and deployment of AI.
We motivate the examination of the organizational AI supply chain---highlighting concerns about concentration, power, chokepoints, cascading failures, fragmented accountability, and risk---then describe our pipeline for mapping the AI supply chain. 
Our pipeline yields our main contribution: a multilayered, large-scale dataset of actors and relationships in the AI ecosystem, 
annotated with dates, descriptions, types, tags, and more.
This work takes a step toward systematially mapping and measuring the AI supply chain and its properties, 
cataloging how models, data, chips, infrastructure, capital, and even talent flow between organizations.
This data enable critical examinations of industry relationships and yield insights into the political economy of AI development and concentration.
We provide several illustrative analyses, 
including studying the most central actors, analyzing network concentration, and identifying communities and chokepoints.
While our dataset is fairly comprehensive, it is also necessarily incomplete: public documents omit private deals, and coverage is shaped by search design. 
Together, the approach, data, and findings aim to serve as a resource for studying the AI ecosystem.
The dataset is publicly available at \url{https://huggingface.co/datasets/sarahcen/ai-supply-chain} and can be explored interactively at \url{https://aisupplychains.org/}.

\end{abstract}

\section{Introduction}

The production of AI-enabled technologies has become increasingly distributed across a diverse set of organizations. Rather than building end-to-end systems entirely in-house, firms now assemble AI capabilities from multiple sources, including model developers, data companies, and cloud infrastructure providers. 
These actors, in turn, rely on contributions from upstream actors, like semiconductor manufacturers and human data workers (\citealt{gray2019ghost, miceli2022data}). While we have a growing understanding of the \emph{AI stack} (the layers of AI development and deployment), we lack a comparable understanding of the \emph{links between actors}.
We have little knowledge of how models, data, compute, capital, and talent flow between actors, as well as the competitive, regulatory, and political relationships in the AI industry. 

Central to that conversation is the \emph{AI supply chain}: the network of actors and components that contribute to AI development and deployment (\citealt{Cobbe_2023, cen2023ai,Hopkins_2025}).
Although the AI ``supply chain'' can refer to a network of AI assets (e.g., with models or datasets as nodes), we focus on the \emph{organizational} supply chain in this work.
We define the organizational AI supply chain as a network of organizations connected in a graph by relationships relevant to AI development or deployment, where edges capture directional links.
We say that an actor is upstream from another if it provides inputs that materially enable the latter's AI-related activities. An actor that receives them is downstream. Under this definition, the AI supply chain spans contributors as far upstream as human annotators and hardware manufacturers, and as far downstream as the end-users of AI-enabled products. Actors may occupy different roles across relationships (e.g., upstream in some relationships and downstream in others) and feedback loops (e.g., user data to improve upstream models) are common. Some links are symmetric---such as partnerships and competition---manifesting as undirected edges. As a result, the structure is better understood as a network than as a linear chain or tree.

\begin{figure}[t]
    \centering
    \begin{minipage}[t]{0.385\textwidth}
        \centering
        \includegraphics[width=\linewidth]{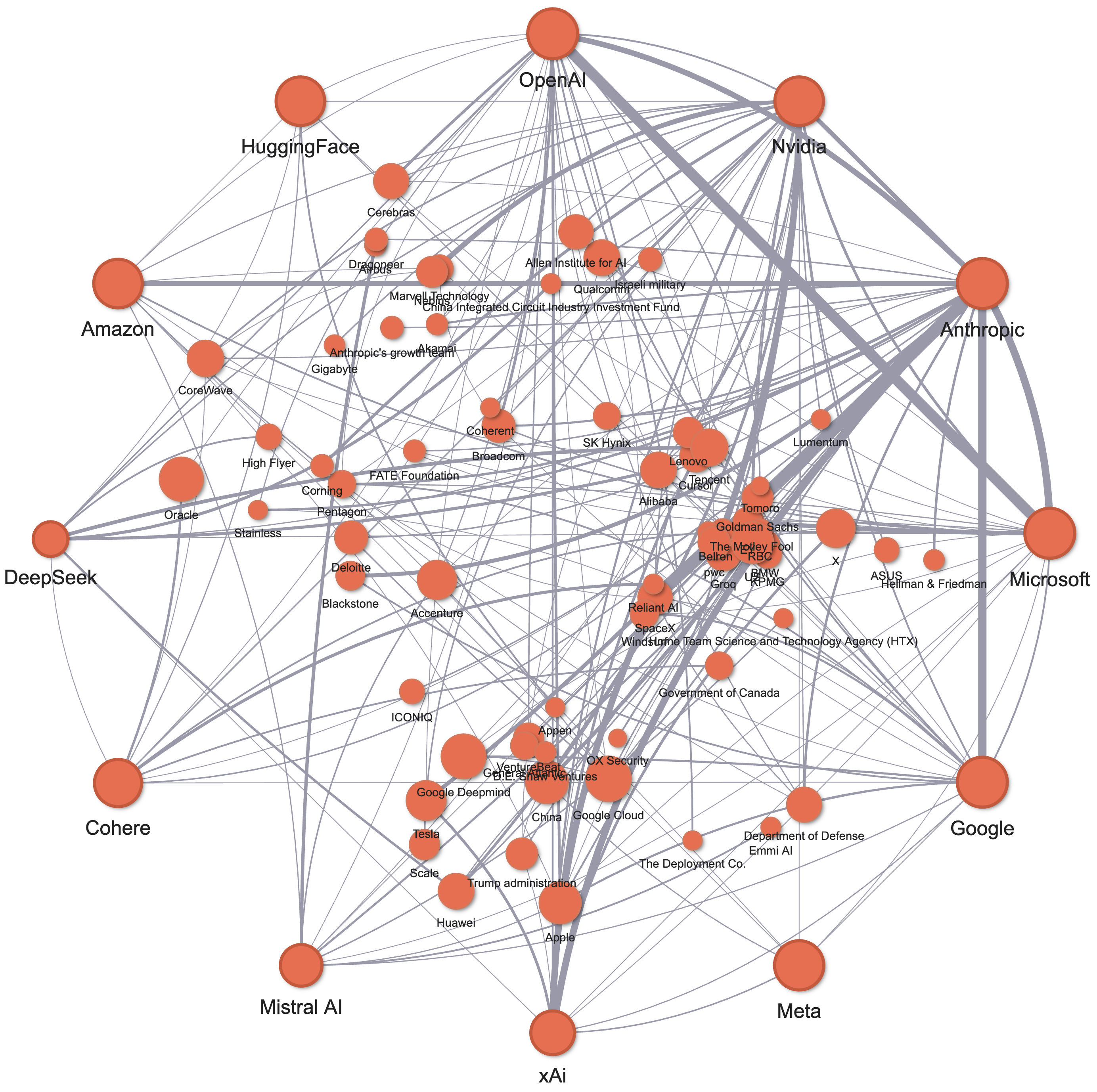}
        \vspace{0.25em}
        {\small (a)}
    \end{minipage}\hfill
    \begin{minipage}[t]{0.595\textwidth}
        \centering
        \includegraphics[width=\linewidth]{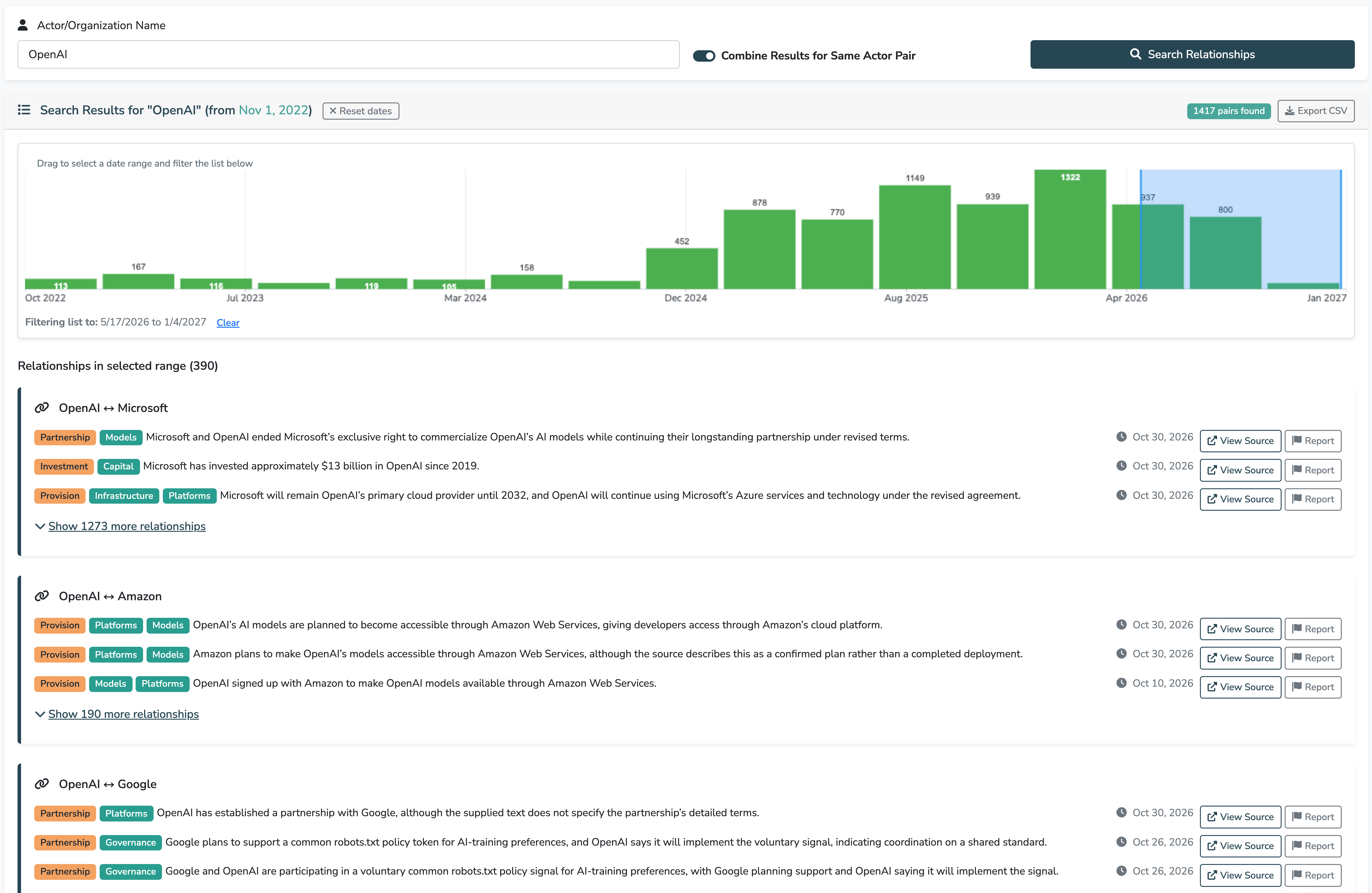}
        \vspace{0.25em}
        {\small (b)}
    \end{minipage}
    \caption{Screenshots from \url{https://aisupplychains.org/}. Our pipeline extracts relationships, links, and dependencies in the organizational AI supply chain from public data sources.
    The resulting dataset can be used to (a) construct a multilayered network graph and (b) search for relationships by actor, keywords, time window, and other filters.
    (a) A visualization of a small subset of the organizational AI supply chain in May 2026,
    built by filtering to the top 12 actors and their one-hop neighbors, then preserving only edges with at least two documented mentions in the dataset.
    (b) An example of a search result, for search entry ``OpenAI'' and time window May 2026 onwards (as shown by the light blue time filter).
    Each pairwise relationship is tagged and described in one sentence, with a link to the source document.
    As described in our methodology, relationships in this document-based view are not deduplicated; each line in the search results corresponds to a specific document and what it says about the relationship.
    Therefore, the same relationship may be mentioned in multiple documents.}
    \label{fig:intro_preview}
\end{figure}

Focusing on organizational relationships captures consolidated and established power dynamics. Political economy scholarship has long emphasized that supply chains are not neutral conduits of production, but structures where power and governance are negotiated (\citealt{gereffi2005governance, baldwin2012global}). In traditional manufacturing networks, the structure of the supply-chain shapes downstream outcomes, including vulnerability to shocks, bargaining power between firms, and the feasibility of regulation. Similar dynamics are emerging in AI, with recent concerns arising around  the ability of general-purpose AI technologies to impact a diverse array of sectors and applications and, as a result, risks. Because AI development relies on resources, like compute, data, and technical talent, that are themselves highly concentrated and geopolitically contested (\citealt{muldoon2025politics,luitse_pe_of_ai,widder2024open}), 
actors at the top of the AI supply chain who maintain control over resources are in a position to set prices and indirectly control downstream risks (\citealt{vipra2023market, korinek2025concentrating}). Coordination and dealing-making at the top of the supply chain fuel worries that dominant AI actors are creating ``kingmakers'' (\citealt{gurun2025gatekeepers}), fueling an unsustainable ``AI bubble'' (\citealt{caballero2026speculative, floridi2024ai}), and neutralizing policymakers through regulatory capture (\citealt{wei2024ai}). 
At the same time, 
there are concerns about vulnerabilities in the AI supply chain arising from chokepoints that create strong dependencies on specific nodes and, correspondingly, increase the risk of cascading failures.
Alongside these concerns are issues of accountability (\citealt{widder2023dislocated}), including questions about where responsibility for AI harms and risk management falls in the organizational AI supply chain.

Fundamentally, these issues can be traced to the simultaneous concentration of resources and control at the top of the supply chain, and a rapid expansion of downstream use and risk at the bottom. While these reflect different substantive concerns, evaluating them requires information not only about individual firms or components of the AI stack, but also about the relationships between organizations in the AI supply chain. 
Understanding and tackling such issues requires a systematic examination of the \emph{links, dependencies, and relationships} between AI actors and not merely the AI stack. 
Despite its importance, the organizational AI supply chain remains poorly documented in a systematic way. Recent work has proposed conceptual frameworks and provided case studies highlighting the importance of supply chains for  accountability (\citealt{Cobbe_2023}), as well as empirical mappings of particular platforms (\citealt{laufer2025anatomymachinelearningecosystem}) or model families (\citealt{bommasani2023ecosystemgraphssocialfootprint,vandervlist_cloud_infrastructure}).

In this paper, we build a pipeline for mapping the AI supply chain. 
Motivated by the fact that much of the AI supply chain emerged from the development of large language models (LLMs), 
we seed the mapping with 36 foundation model providers.
The pipeline then performs a systematic online search, followed by an automated extraction and annotation of relationships contained in the resulting documents. 
The result is a dynamic, regularly updated network of relationships derived from public documents, such as news articles, press releases, company blogs, and online reports. We classify and analyze the structure of these relationships by developing a taxonomy of relationship types and use network analyses to study patterns, dependencies, and chokepoints, among other characteristics.
In the process, we curate a dataset of over 13{,}000 documents dating from November~2022 to October~2026, resulting in a network of relationships between about 9{,}700 unique actors (4{,}520 when more aggressively filtering out infrequent actors).
This dataset is publicly available at \url{https://huggingface.co/datasets/sarahcen/ai-supply-chain} and can be explored interactively at \url{https://aisupplychains.org/}.

To illustrate the information contained in this dataset, 
we provide several quantitative and qualitative analyses.
The data can be viewed as a multilayered network graph across time, 
with organizations as nodes and relationships as edges,
where nodes and edges are further tagged and annotated. 
Accordingly, we provide several quantitative and qualitative analyses of the network, with illustrative examples of the relationships found in the data. 
In particular, 
we (1) examine the most central actors in the AI supply chain;
(2) analyze network concentration;
(3) identify communities and chokepoints in the network;
(4) demonstrate how the data reveals close and even cyclical relationships;
and (5) study how relationship narratives change across documents and evolve across time.
We note that our analysis does not provide full coverage of the AI supply chain. 
Due to how the search is seeded, coverage is most complete and dense around large model providers. 
We discuss issues of coverage and limitations of our dataset in \Cref{sec:findings-visibility}.

Our dataset and analysis shift attention from layers of the AI stack to the  relationships that connect them. 
They provide a framework for identifying dependencies and potential chokepoints, while also demonstrating the limitations of relying on fragmented public disclosure to understand them. 
We hope that this dataset can serve as a useful resource that informs research, policymaking, and our broader understanding of the AI industry. 
We further aim to expand the dataset to improve coverage across relationships and actors.

\section{The Organizational AI Supply Chain}
\label{sec:framework}

In this work, the AI supply chain is understood as a \textbf{network} of organizations involved in the development and deployment of AI products, services, and other AI assets.
This section describes the organizational AI supply chain,
discussing the actors that exist within it
and enumerating links between AI actors that have become salient in discussions of \textbf{power}, \textbf{risk}, \textbf{resilience}, and \textbf{governance}.
Although individual people are key parts of the AI supply chain, our work takes organizations as its primary unit of analysis.

\subsection{Organizational supply chain network}
\label{sec:framework-network}

We represent the organizational AI supply chain as a graph $G = (V, E)$, where $V$ is the set of organizations (nodes) and $E$ is the set of links (edges) among them.
That is, each $v \in V$ is an organization in the AI industry, and $(u, v) \in E$ indicates that there is a link between nodes $u$ and $v$.

When a link is asymmetric,
we treat an edge $(u, v) \in E$ as a \emph{directed} edge from $u$ to $v$.
Such directed edges originate from an \textbf{upstream} actor who generally provides assets (such as models, compute, data, or chips) to a \textbf{downstream} actor.
Many important AI relationships are directional in this sense and capture \textbf{dependencies} in the AI industry.
These dependencies are precisely what allows the AI industry to become more efficient and specialized,
but it also creates opportunities for \textbf{chokepoints}, \textbf{bottlenecks}, and \textbf{vulnerabilities} where
a small number of actors can control the flow of assets, resources, or knowledge.
Other links are \emph{symmetric}, encoding explicit relationships (such as partnerships or agreements) and implicit ones (such as competition). When a relationship between actors $u$ and $v$ is \emph{symmetric}, both $(u, v)$ and $(v, u)$ exist in $E$.

Importantly, the AI supply chain consists of not only \textbf{suppliers}, but also \textbf{users}. Generally, upstream organizations are providers who possess significant sway because their behaviors impact all their \emph{descendants} (all nodes that fall on a directed path originating from this organization, including its immediate downstream nodes and nodes that indirectly descend from it all the way down to the end-users). 
What counts as ``upstream'' or ``downstream'' is therefore relative to where in the AI supply chain one is focused: upstream means closer to the origin of a path, whereas downstream means closer to the end of the path.

Yet, not all links have a clear upstream or downstream organization. Specfically, when there is a \emph{cycle}---a closed loop such that following a path from node $u$ leads back to $u$---there is no clear notion of upstream or downstream.
In the AI supply chain, cycles are significant because they indicate feedback loops.
While some cycles signal that the AI industry is healthy and self-sustaining, others may indicate \textbf{market concentration} and \textbf{anti-competitive behavior}, such as circular financing (e.g., company A invests in company B, which invests in company C, which invests in company A) to inflate market value (\citealt{crs_circular_investment, solanki2026ai}).

Finally, it is often helpful to take \emph{slices} of $G$ or extract subnetworks from it.
For instance, we may be interested in how the network evolves over time and split $G$ into a series of snapshots $G_1, G_2, ... , G_T$.
Alternatively, we may be interested in a specific type of relationship, such as model-related dependencies or the flow of capital, and analyze a subgraph of $G$ (e.g., the \textbf{model supply chain} or \textbf{data supply chain}), or even study the connections between different subgraphs (e.g., how model and data supply chains interact).

\subsection{Nodes are organizational actors}
\label{sec:framework-actors}

In the organizational AI supply chain, an \textbf{actor} is an organization that participates in AI-related production, deployment, and deployment.
In general, actors may take on more than one \textbf{role} in the AI industry; OpenAI serves as a research lab, a model provider, and chatbot deployer, for instance.

Illustrative roles along the organizational AI supply chain include the following.
Far upstream, \emph{semiconductor designers and manufacturers} (e.g., NVIDIA, AMD, TSMC, ASML) provide chips.
\emph{Cloud and compute providers} (e.g., Amazon Web Services, Microsoft Azure, Google Cloud) sit between hardware and model developers, offering training and inference capacity.
\emph{Data providers and annotators} (e.g., Scale AI, Mercor) source and supply data for model training and customization.
\emph{Foundation and base model providers} (e.g., OpenAI, Anthropic, Google DeepMind, Meta AI, Mistral) provide large models and APIs that downstream firms adapt and consume.
Surrounding them are firms that provide tooling, infrastructure, and evaluations.
Further downstream are other model providers, application builders, enterprise users, and consumers.
\emph{Capital actors} (e.g., venture firms) finance and take ownership stakes in AI firms.
\emph{Universities, national labs, and independent research institutes} develop their own models, datasets, and methods alongisde industry actors, often partnering with them (\citealt{ahmed2023growing}).
\emph{Public bodies} (e.g., NIST, FTC, state governments) support, regulate, and procure AI.
Other roles include \emph{standards setting bodies}, \emph{civil society organizations}, and \emph{AI safety non-profits}. 

Importantly, actors frequently occupy several roles at once (e.g., a cloud provider that also trains models and distributes applications). 
As mentioned before, an actor may therefore be ``upstream'' of one organization and ``downstream'' of another.

\subsection{Edges are links, relationships, and dependencies}
\label{sec:framework-relationships}

Actors alone do not constitute a supply chain. What forms the AI supply chain are the links between actors.
For instance, 
one of the most common relationships is a \textbf{provision}: one actor supplies goods, services, access, or capabilities to another. 
These relationships are useful for identifying potential chokepoints in AI production.
\textbf{Investment and financing} relationships concern how capital flows, introducing financial dependencies and potential conflicts of interest.
\textbf{Ownership and control} relationships capture acquisitions, subsidiaries, and other analogous arrangements.
In our data, \textbf{partnerships and collaborations} are highly common and include coordinated efforts and alliances between organizations.
Many remaining relationships concern
{employment and personnel},
{regulation and oversight},
and even competition.

The ways in which actors are connected influence \emph{power}, \emph{risk}, \emph{resilience},  and \emph{governance} in the AI industry.
These links are central to our analysis; as we describe in the following sections, isolating particular types of relationships, links, and dependencies lets us study slices of the multilayered AI supply chain.

\subsection{Why the organizational supply chain matters}
\label{sec:framework-why}

The organizational supply chain arises organically because specialization improves efficiency. That is, rather than have a single organization perform every operation in AI development, different organizations specialize in specific processes or components of AI. This specialization allows them to develop context-specific practices and improve overall efficiency. However, specialization can also create interdependence that affects the distribution of power and transmission of shocks. We briefly name several implications of supply chain characteristics that motivate our analysis of the organizational structure of the AI ecosystem.

\begin{enumerate}
    \item \textbf{Dependencies create power.}
    Dependencies show who relies on whom for scarce inputs such as compute, data, and model access.
    Mapping these ties surfaces structures of power in the AI industry.

    \item \textbf{The supply chain network reveals areas of market concentration.}
    An organizational map shows where concentration of resources such as capital, compute, and data sits.
    Market concentration can be concerning because a few actors can set terms of access, limit entry, and shape competition across layers of the AI ecosystem.

    \item \textbf{Bottlenecks and chokepoints create vulnerabilities and cascading failures.}
    When certain flows of AI assets are controlled by a small number of actors, bottlenecks and chokepoints arise.
    Such structures can amplify local shocks, making them system-wide.

    \item \textbf{The supply chain illustrates where risk arises and propagates.}
    Where AI risk arises is one of the main concerns of the AI industry. 
    An organizational map can illustrate not only where risk originates, but also where it may be inherited or propagated across organizations, their behaviors, and their offerings.

    \item \textbf{The supply chain complicates liability and responsibility, but mapping it can help.}
    When many organizations jointly produce and deploy AI systems, responsibility and liability becomes fragmented. 
    Mapping the links between organizations is a necessary step to improving accountability, e.g., by clarifying where along the chain regulatory intervention is most likely to be effective.
\end{enumerate}

\section{Mapping the Organizational AI Supply Chain: Methodology}
\label{sec:methodology}

In this section, we describe the procedure for tracking and extracting relationships in the organizational AI supply chain. 
Our pipeline is fully automated and runs regularly, with extensive error handling and retry logic, to create an up-to-date archive of AI supply chain relationships.
We use publicly available information, including news articles, press releases, and company blogs. 
Our pipeline collects these documents from November~2022 until the present via targeted Google Search queries over one-month time windows.
For each seed organization, query, and time window, we collect Google Search results via SerpAPI.
We then scrape the linked pages when accessible, extracting title and body text.
Scraped documents are processed using an LLM that identifies relationships described in the document and taxonomizes the relationships and actors involved. 
The relationship data is available at \url{https://huggingface.co/datasets/sarahcen/ai-supply-chain} and can be explored at \url{https://aisupplychains.org/}.

\subsection{Data sources}

We rely on public data sources. We intentionally use publicly available documents rather than paywalled APIs or other proprietary data to allow for reproducibility, verifiability, and traceability of all results. 
That is, all users should be able to examine the original text that gave rise to any relationships in our dataset.
We thus do not use datasets, such as Bloomberg Terminal or PitchBook.
\Cref{tab:top_domains_scraped} lists the most common source domains among successfully scraped pages.
Common domains include \texttt{finance.yahoo.com}, \texttt{techcrunch.com}, \texttt{linkedin.com}, \texttt{venturebeat.com}, 
\texttt{huggingface.co}, 
\texttt{cnbc.com}, 
and \texttt{medium.com}.

To assemble a broad corpus of publicly accessible documents, we use Google Search. 
Our searches primarily surface news articles, press releases, and company blogs. 
The dataset includes product launches, strategic partnerships, funding announcements, and litigation, among other events and decisions.
Generally, each reported event or decision is covered by more than one document, introducing helfpul redundancies in the data. 
For instance, press releases and marketing materials may emphasize positive narratives and gloss over risks or other issues, 
while news reporting and blogs may provide countervailing perspectives.

\subsection{Search pipeline}

The first stage of our pipeline is a search pipeline that uses SerpAPI to run targeted Google Search queries over one-month periods from November 2022 until the present. We seed the search pipeline with a list of starting AI organizations. 
In deciding how to seed the search, 
we chose to use large general-purpose or foundation model providers. 
This choice reflects prior analyses of the AI supply chain in that its emergence was facilitated by general-purpose or foundation models that lowered the barrier to entry to AI development and deployment and thus catalyzed broad AI adoption.
Based on this thesis, this work begins with 36 major model providers: 
01.ai, AI2, AI21 Labs, Adept, Aleph Alpha, Alibaba, Amazon, Anthropic, Baidu, Bytedance, Cartesia,
Cohere, Contextual AI, Databricks, DeepMind, DeepSeek, EleutherAI, Google AI, Huawei, Hugging Face,
IBM AI, Inflection, Krutrim, LightOn, Meta AI, Microsoft AI, Midjourney, Mistral AI, NVIDIA, Numenta,
OpenAI, Reka, Snowflake, Stability AI, Together AI, xAI.

This choice of seed organizations introduces some coverage bias, as we discuss in \Cref{sec:limitations}.
We note that our AI supply chain dataset is \emph{not} limited to just these seed organizations.
Through our search pipeline, we recover a much wider set of organizations (about 9{,}700 after fuzzy and common-sense deduplication, and about 4{,}500 after dropping actors that appear in fewer than three rows of our dataset; see \Cref{sec:findings}). 
Thus, this list helps initiate a search that includes a broader umbrella of organizations, though it is of course not exhaustive.

Finally, we note that our pipeline can be easily adapted to include more seed organizations---it would only require changing the search configuration. Because our pipeline first locates the most recent search for an organization, it would find that there is no recent search for a new organization and therefore initiate a new search dating back to November 2022. 

For each seed organization, we run a series of  SerpAPI Google Search queries. 
Specifically, we run (i)~one \emph{broad} query of the form \texttt{"OrgName" <keyword string>} and (ii)~three \emph{site-restricted} queries \texttt{"OrgName" <keyword string> site:<domain>}. 
The \texttt{<keyword string>} values we use are 
\begin{enumerate}[noitemsep,topsep=0pt]
    \item \texttt{(partnership OR collaboration OR agreement OR alliance)}
    \item \texttt{(deal OR investment OR acquisition OR merger OR licensing)}
\end{enumerate}
The \texttt{site:<domain>} values we use are \texttt{techcrunch.com}, \texttt{venturebeat.com}, and \texttt{finance.yahoo.com}---these sites \emph{supplement} the broad search, they do \emph{not} replace it.
(As mentioned above, other keyword strings and sites can be easily added.)
Thus each search is determined by a combination of organization, query type, keyword string, and site domain for site-specific searches.

For each combination of queries, we further split the time period of our search into one-month intervals to ensure good coverage across time (otherwise, the Google Search ranking algorithm would favor the most popular and recent results). 
Specifically, since we collect results from November 2022 onwards, we split the time period from the start date until the present into one-month windows, then run each of the queries described above. 
We collect up to 15 results per broad query and up to 5 results per site-restricted query per one-month period.
The search pipeline is updated regularly, using the last successful search date as the start date for the new search. 
There is expected variation in search results by location;  we use San Francisco as our location. Although providing multiple locations could help, due to the volume of search queries, we limited our search to a single location.

The fields of the search results table are given in \Cref{tab:db_gsearch_results}.

\subsection{Scraping pipeline}

After search results are collected, 
the second stage of the pipeline scrapes the content of each unique document link. 
This stage regularly looks for any unscraped links in the output of the search pipeline, then runs page scraping with up to 3 retries. 
It uses a headless Chrome browser (via undetected Chromedriver).
It first looks for content in \texttt{<main>} or \texttt{<article>} when present,
falling back to a heuristic that scans every \texttt{div}, \texttt{section}, and \texttt{article} for the container with the greatest number of non-empty \texttt{<p>} tags.
The stored fields include the URL, title, document date, and snippet (which are metadata from the search pipeline) as well as body text and date of scrape (which are from the scraping pipeline).

Failures are tracked and retried up to 3 times. 
Failures are generally due to sites being paywalled, being blocked by bot detection, and requiring logins to access content. 
Only non-empty scrapes are kept. 
The fields of the scraping results table are given in \Cref{tab:db_gsearch_link_scrape}.

\subsection{LLM post-processing}

The final stages of the pipeline run a series of post-processing steps to taxonomize the data. 
To do so, we use a series of calls to \texttt{gpt-5.6-luna} with strict structured outputs, medium reasoning, and medium verbosity.
We run two independent extraction passes per document.

\textbf{Site and document classification.}
After a page is scraped, it is assigned a \emph{site type} (based primarily on the domain name) and a \emph{document type} (based on URL, title, and snippet).
The allowed site and document types are given in \Cref{tab:db_gsearch_link_scrape}.

\textbf{Relationship extraction.}
From each scraped page, the LLM extracts pairwise relationships using the page text and title.
A document may yield zero, one, or many relationships.
Long documents are split into chunks of about 150{,}000 tokens.
We run two independent passes per document.
For each relationship, the model returns the two actor names, a one-sentence description of the relationship, a short description of the evidence for the relationship, a label indicating the strength of the evidence, and an AI-relevance label.
The actor names follow how they appear in the source text. Thus, while clear variants within a document may be normalized, the same actor may be encoded under different names in the database. This design is intentional, as it allows the relationships to closely mirror the source text; one can then de-duplicate actor names using a desired post-processing procedure.

\textbf{Actor classification.}
In the next step, each actor is assigned one actor type.
The allowable actor types are:{
\path{COMPANY};
\path{NONPROFIT_OR_CIVIL_SOCIETY};
\path{ACADEMIC_INSTITUTION};
\path{COALITION_OR_CONSORTIUM};
\path{PERSON};
\path{GOVERNMENT_BODY};
\path{GEOPOLITICAL_ENTITY_OR_COUNTRY};
\path{INTERGOVERNMENTAL_OR_SUPRANATIONAL};
\path{OTHER_ACTOR};
\path{NOT_AN_ACTOR};
\path{UNKNOWN_OR_UNCLEAR}.
\par}

\textbf{Relationship classification.}
Further, each relationship is also assigned a relationship type, a relationship status, up to two AI stack layers, a directionality label, and (when AI-related) up to three AI topics.
The released Hugging Face repo only contains relationships for which the AI relevance label is either \path{AI_RELATIONSHIP} or \path{INVOLVES_AI_ACTOR}, dropping relationships labeled \path{NOT_AI_RELATED} or \path{UNCLEAR}.
The allowable relationship types are:
{\raggedright
\path{INVESTMENT};
\path{PROVISION};
\path{OWNERSHIP};
\path{EVALUATION};
\path{PARTNERSHIP};
\path{ONE_SIDED_DEPENDENCY};
\path{EMPLOYMENT};
\path{AFFILIATION};
\path{REGULATION};
\path{INFLUENCE};
\path{COMPETITION};
\path{DISPUTE};
\path{INCIDENT};
\path{PUBLIC_STATEMENT};
\path{NO_RELATIONSHIP};
\path{OTHER}; and
\path{UNCLEAR}.
}
The allowable relationship statuses are
\path{ESTABLISHED},
\path{PROPOSED},
\path{ENDED},
\path{NOT_APPLICABLE}, and
\path{UNCLEAR}.
The allowable AI stack layers (one or two per relationship) are:
\path{CAPITAL};
\path{HARDWARE};
\path{INFRASTRUCTURE};
\path{DATA};
\path{RESEARCH};
\path{MODELS};
\path{PLATFORMS};
\path{APPLICATIONS};
\path{ADOPTION};
\path{OPERATIONS};
\path{SAFETY};
\path{SECURITY};
\path{GOVERNANCE};
\path{TALENT};
and, alone, \path{NOT_AI_STACK} or \path{UNKNOWN_OR_UNCLEAR}.
The allowable AI relationship categories are the 117 AI tags in the AI-tag taxonomy; examples include
\path{AI_COMPUTE};
\path{AI_SEMICONDUCTORS};
\path{AI_MODELS};
\path{AI_SOFTWARE};
\path{AI_DATA};
\path{AI_APPLICATIONS};
\path{AI_RESEARCH};
\path{AI_TALENT};
\path{AI_FUNDING};
\path{AI_OWNERSHIP};
\path{AI_PARTNERSHIP};
\path{AI_MARKET_ACCESS};
\path{AI_REGULATION};
\path{AI_POLICY};
\path{AI_GOVERNANCE};
\path{AI_SAFETY};
\path{AI_HARMS};
\path{AI_INTELLECTUAL_PROPERTY};
\path{AI_INFRASTRUCTURE};
\path{AI_ENERGY};
\path{AI_SECURITY};
and
\path{AI_DEFENSE}
(the full list is in the dataset schema).
Directionality is one of
\path{ACTOR_1_TO_ACTOR_2},
\path{ACTOR_2_TO_ACTOR_1},
\path{SYMMETRIC}, or
\path{UNCLEAR}.
A directed label names the source and the target; it does not depend on which actor is listed first.

The full system and user prompts are included in Appendix A. These prompts clearly state the scope of each label.

\subsection{The dataset}

Each row in the public dataset (\url{https://huggingface.co/datasets/sarahcen/ai-supply-chain}) corresponds to a relationship extraction.
\Cref{tab:hf_dataset_fields} summarizes the fields that appear in the released dataset. Full details can be found in \Cref{tab:hf_dataset_fields_full}.
Note that rows do not represent unique relationships for two reasons: (i) many documents may cover the same relationship, and (ii) our pipeline runs relationship extraction twice per document.

\begin{table}[th!]
    \centering
    \small
    \caption{Fields in the dataset.
    Detailed information given in \Cref{tab:hf_dataset_fields_full}.}
    \label{tab:hf_dataset_fields}
    \begin{tabular}{@{}p{0.28\linewidth}p{0.34\linewidth}>{\raggedright\arraybackslash}p{0.32\linewidth}@{}}
    \toprule
    Field & Description & Possible values \\
    \midrule
    \texttt{link} & Source document URL & text (URL) \\
    \texttt{article\_title} & Source page title & text \\
    \texttt{document\_date} & Document/article date & date or missing \\
    \texttt{actor\_1\_name}, \texttt{actor\_2\_name} & Actor names & text \\
    \texttt{actor\_1\_type}, \texttt{actor\_2\_type} & Actor types & 11 labels. Table~\ref{tab:hf_dataset_fields_full} contains the full set of labels, including as examples \texttt{COMPANY}, \texttt{GOVERNMENT\_BODY}, and \texttt{ACADEMIC\_INSTITUTION}. \\
    \texttt{relationship\_description} & One-sentence description & text \\
    \texttt{evidence} & Supporting source excerpt & text \\
    \texttt{evidence\_strength} & How directly supported & \texttt{EXPLICIT}, \texttt{STRONGLY\_IMPLIED}, \texttt{IMPLIED} \\
    \texttt{relationship\_type} & Type of AI relationship & 17 labels. Table~\ref{tab:hf_dataset_fields_full} contains the full set of labels, including as examples \texttt{INVESTMENT}, \texttt{PROVISION}, and \texttt{PARTNERSHIP}. \\
    \texttt{relationship\_status} & Lifecycle status & 5 labels. Table~\ref{tab:hf_dataset_fields_full} contains the full set of labels, including as examples \texttt{ESTABLISHED}, \texttt{PROPOSED}, and \texttt{ENDED}. \\
    \texttt{ai\_stack\_layers} & AI stack layers (1-2) & 16 labels. Table~\ref{tab:hf_dataset_fields_full} contains the full set of labels, including as examples \texttt{CAPITAL}, \texttt{MODELS}, and \texttt{DATA}. \\
    \texttt{directionality} & Source and target & \texttt{ACTOR\_1\_TO\_ACTOR\_2}, \texttt{ACTOR\_2\_TO\_ACTOR\_1}, \texttt{SYMMETRIC}, \texttt{UNCLEAR} \\
    \texttt{ai\_relationship\_categories} & AI topical tags ($\le$3) & 117 labels. Table~\ref{tab:hf_dataset_fields_full} contains the full set of labels, including as examples \texttt{AI\_MODELS}, \texttt{AI\_COMPUTE}, and \texttt{AI\_FUNDING}. \\
    \texttt{relationship\_pass} & All documents are processed twice. This index denotes which pass this relationship is extracted from. & integer (1 or 2) \\
    \bottomrule
    \end{tabular}
\end{table}

\subsection{Limitations}\label{sec:limitations}

Our dataset has several notable limitations.
First, because we choose to use publicly available documents, we forego several high-quality data sources that are available (such as Bloomberg Terminal and PitchBook).
Second, our data comes from Google Search. 
Thus, our results are affected by Google's search algorithm (e.g., high-quality but low-traffic sources may be missed) and what sites are search indexed.
Third, our search and LLM processing is in English, meaning that the results and documents we use are primarily English-language sources.
Fourth, there may be confidential or proprietary information on links, dependencies, or relationships that we are unable to observe from public data. 

Our search is also affected by design choices that we made around (i) the list of seed organizations, (ii) our choice of keywords strings, (iii) our choice of sites to supplement the broad search, (iv) the search location being San Francisco, CA. 
These choices were informed by significant testing, but we faced trade-offs between coverage on one end and cost and redundancy on the other. 
Of these, the choice of seed organizations is likely the most significant. As noted above, we can and hope to expand the seed organization list in the future. At the moment, it is biased toward the portion of the supply chain closest to large model providers.
Ultimately, there is coverage bias, but the organizational AI supply chain that this pipeline surfaces yields meaningful insights; wherever coverage bias may affect our results and discussion, we do our best to disclose this limitation.
We provide statistics on our dataset in \Cref{sec:findings,app:coarse_stats}.

In our scraping stage, the main limitation is that scraping is not possible when there are paywalled sites or bot detection mechanisms.
In our LLM processing stage, 
the main limitations are around (i) our choice of taxonomy, and (ii) the LLM's accuracy and hallucination rates. 
LLMs always have false positive and false negative rates, where false positives (hallucinations) are most concerning. 
In our analysis, we do our best to avoid these by dropping actors that appear in fewer than three relationships after name normalization (and, in some figures, by requiring repeated documentary support for an edge).
However, critical edges should always be checked against the linked source.
As a verification, we randomly sampled 50 relationships and manually checked the article title, both actor names, the relationship type, and the relationship status against the linked source.
Forty-five were verified on all five fields.
Of the remaining five, four sources could not be checked (e.g., due to pages no longer existing), and one relationship had correct information but an incorrect relationship status (\texttt{PROPOSED} instead of \texttt{ESTABLISHED}).
See \Cref{app:manual_verification} for details.

There are two final limitations. 
First, the documents themselves may be biased. 
For instance, press releases can overstate, omit, or misrepresent relationships.
Second, because there is no canonical notion of what an ``organization'' is, it is impossible to fully deduplicate our data.
For instance, there is ambiguity in whether Google DeepMind should be collapsed under Google or, further, Alphabet. For some users, they may wish to distinguish between, e.g., Google DeepMind and Google Cloud; for other users, they may wish to collapse them all under Google.
In our analysis, we do our best to clarify how we handle this ambiguity.

\section{Exploring the Dataset}
\label{sec:exploring}
\label{sec:findings}
The main contribution of this work is the dataset. 
One can think of this dataset as a multilayered adjacency matrix, 
where each edge is annotated with a natural language description, multiple categorical labels (e.g., relationship type, actor type, AI tag), source document, document date, and more (as described in \Cref{sec:methodology}).
This adjacency matrix allows one to explore the organizational AI supply chain across time and by category, 
using longstanding graph analysis techniques as well. 
To explore the capacity of this dataset, this section provides illustrative examples of information contained within it.

For the remainder of our analysis,
we show results for a filtered graph $G'$, unless otherwise stated.
To obtain this filtered graph,
we keep only rows with \texttt{document\_date} from November 1, 2022 through the end of October 2026
that are AI-related
(\texttt{AI\_RELATIONSHIP} or \texttt{INVOLVES\_AI\_ACTOR}).
We then exclude rows that
have null dates;
 actors with types \texttt{PERSON}, \texttt{NOT\_AN\_ACTOR}, or \texttt{UNKNOWN\_OR\_UNCLEAR};
 relationship type \texttt{COMPETITION}, \texttt{PUBLIC\_STATEMENT}, \texttt{NO\_RELATIONSHIP}, \texttt{UNCLEAR}, or \texttt{CONTRADICTORY\_INFORMATION};
 directionality \texttt{UNCLEAR} or \texttt{CONTRADICTORY\_INFORMATION};
 relationship status \texttt{ENDED}, \texttt{NOT\_APPLICABLE}, \texttt{UNCLEAR}, or \texttt{CONTRADICTORY\_INFORMATION};
or  AI stack layers with \texttt{NOT\_AI\_STACK} or \texttt{UNKNOWN\_OR\_UNCLEAR}.
We further deduplicate actor names using fuzzy actor matching and common-sense groupings (such as mapping DeepMind, Alphabet, and Gemini to Google).
We drop actors that appear in fewer than three rows across the corpus and keep an edge only if both endpoints meet that threshold.
We further normalize edges by collapsing all relationships that, within a given month, share the same actor pair, directionality, and relationship type.\footnote{As mentioned in \Cref{sec:methodology}, we do not filter the publicly released data. We intentionally allow the data to closely reflect the information in the source documents, as different users of the dataset may wish to filter and deduplicate in their own ways.}

The resulting $G'$ has 4{,}520 actors and 37{,}472 edges
(from 100{,}079 relationships and 15{,}406 raw actor strings that arise with no filtering).

\subsection{Example relationships}

$G'$ contains over 37{,}000 edges, using the process described above, which collapses edges by month.
These ties are diverse,
ranging from partnerships to investments,
litigation to data licensing.
We include four examples below.
Note that ``relationship status'' is based on the time of the document being published, and not based on the relationship status today.

\medskip\noindent\textbf{Amazon $\rightarrow$ Anthropic}
(\path{INVESTMENT}, status: \path{ESTABLISHED}; directional; evidence: \path{EXPLICIT}).
\textit{Actor types:} \path{COMPANY}, \path{COMPANY}.
\textit{AI stack layers:} \path{CAPITAL}.
\textit{AI tags:} \path{AI_FUNDING}, \path{AI_OWNERSHIP}.
\textit{Description:} Amazon has agreed to invest up to \$4 billion in Anthropic, initially purchasing a minority stake for \$1.25 billion with an option to increase the investment.
\url{https://techcrunch.com/2023/09/25/amazon-to-invest-up-to-4-billion-in-ai-startup-anthropic/}
\textit{Evidence:} ``Amazon has agreed to invest up to \$4 billion in the AI startup Anthropic [...] The e-commerce group said it will initially invest \$1.25 billion for a minority stake in Anthropic''
\textit{Source:} \path{news_outlet} / \path{news_article} (2023-09-25).

\medskip\noindent\textbf{Nvidia $\rightarrow$ ByteDance}
(\path{PROVISION}, status: \path{PROPOSED}; directional; evidence: \path{EXPLICIT}).
\textit{Actor types:} \path{COMPANY}, \path{COMPANY}.
\textit{AI stack layers:} \path{HARDWARE}.
\textit{AI tags:} \path{AI_SEMICONDUCTORS}, \path{AI_HARDWARE}, \path{AI_COMPUTE}.
\textit{Description:} ByteDance plans to spend about US\$14 billion on artificial intelligence chips from Nvidia in 2026, conditional on Nvidia being allowed to sell its H200 GPUs in China.
\url{https://finance.yahoo.com/news/bytedance-pour-us-14-billion-093000967.html}
\textit{Evidence:} ``ByteDance plans to spend about 100 billion yuan (US\$14 billion) on artificial intelligence chips from Nvidia in 2026, a hefty increase from roughly 85 billion yuan in 2025, if the US company is allowed to sell its H200 graphic processing units in China''
\textit{Source:} \path{financial_data_platform} / \path{news_article} (2025-12-31).

\medskip\noindent\textbf{IBM $\leftrightarrow$ OpenAI}
(\path{PARTNERSHIP}, status: \path{ESTABLISHED}; symmetric; evidence: \path{EXPLICIT}).
\textit{Actor types:} \path{COMPANY}, \path{COMPANY}.
\textit{AI stack layers:} \path{PLATFORMS}, \path{MODELS}.
\textit{AI tags:} \path{AI_MODELS}, \path{AI_INTEGRATION}, \path{AI_DEPLOYMENT}.
\textit{Description:} IBM is embedding OpenAI frontier models and products, including GPT-5.6, Codex, and ChatGPT Work, into IBM Consulting Advantage to support secure enterprise AI deployment.
\url{https://finance.yahoo.com/technology/ai/articles/ibm-partners-openai-accelerate-secure-100000076.html}
\textit{Evidence:} ``The partnership embeds OpenAI frontier models like GPT-5.6 and products like Codex and ChatGPT Work into IBM Consulting Advantage, IBM's AI platform for delivering consulting services to clients''
\textit{Source:} \path{financial_data_platform} / \path{news_article} (2026-08-13).

\medskip\noindent\textbf{Competition and Markets Authority (CMA) $\rightarrow$ Google}
(\path{REGULATION}, status: \path{ESTABLISHED}; directional; evidence: \path{EXPLICIT}).
\textit{Actor types:} \path{GOVERNMENT_BODY}, \path{COMPANY}.
\textit{AI stack layers:} \path{GOVERNANCE}.
\textit{AI tags:} \path{AI_REGULATION}, \path{AI_GOVERNANCE}, \path{AI_MARKET_POWER}.
\textit{Description:} The Competition and Markets Authority (CMA) designated Google as having strategic market status in search and search advertising, covering Google's AI-enabled search features and enabling the CMA to consider stricter regulations.
\url{https://techcrunch.com/2025/10/10/uk-slaps-google-search-with-special-market-status-making-way-for-stricter-regulations/}
\textit{Evidence:} ``The CMA has designated Google as having ``strategic market status'' in the search and search advertising markets, which means the company has such ``a substantial and entrenched'' position that it requires special regulations to ensure fair competition.[...]The designation covers Google's search and online search advertising services; its AI-enabled search features like AI Overviews and AI Mode''
\textit{Source:} \path{news_outlet} / \path{news_article} (2025-10-10).

\subsection{Multilayered trees}

\begin{figure}[h!]
    \centering
    \includegraphics[width=\linewidth]{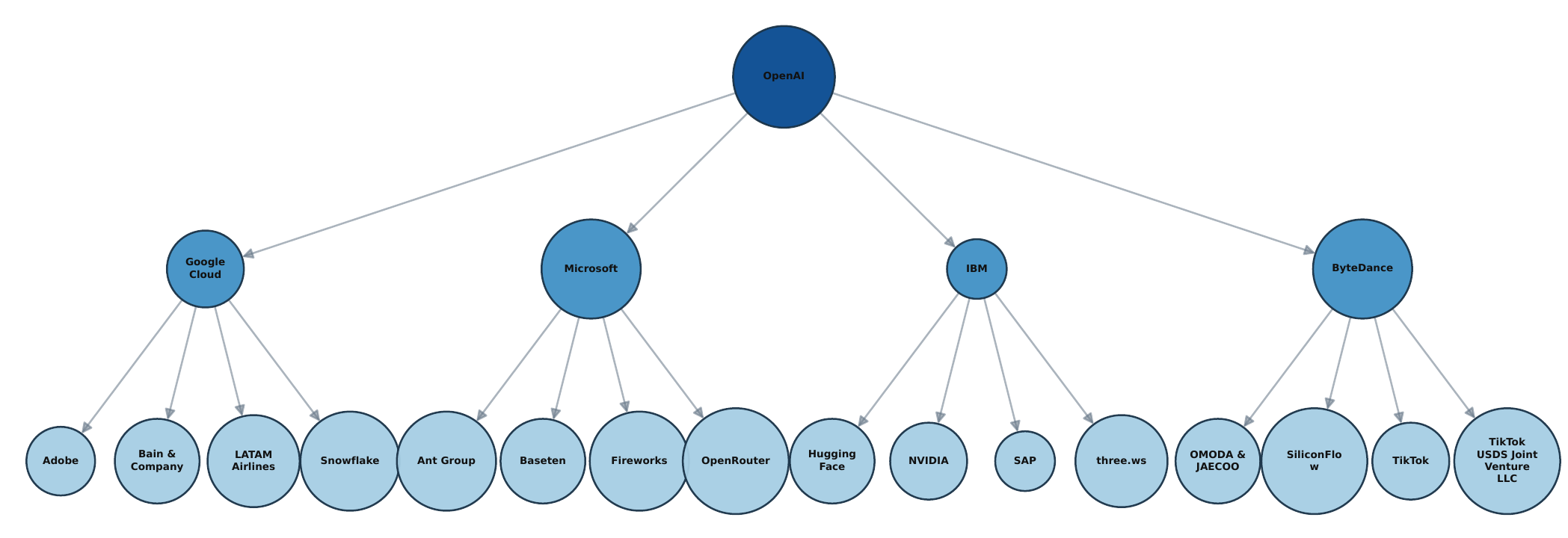}
    \caption{A portion of the two-hop tree descending from OpenAI, when filtering to relationships based on model provisions.}
    \label{fig:openai_model_provision_tree}
\end{figure}

\begin{figure}[h!]
    \centering
    \includegraphics[width=\linewidth]{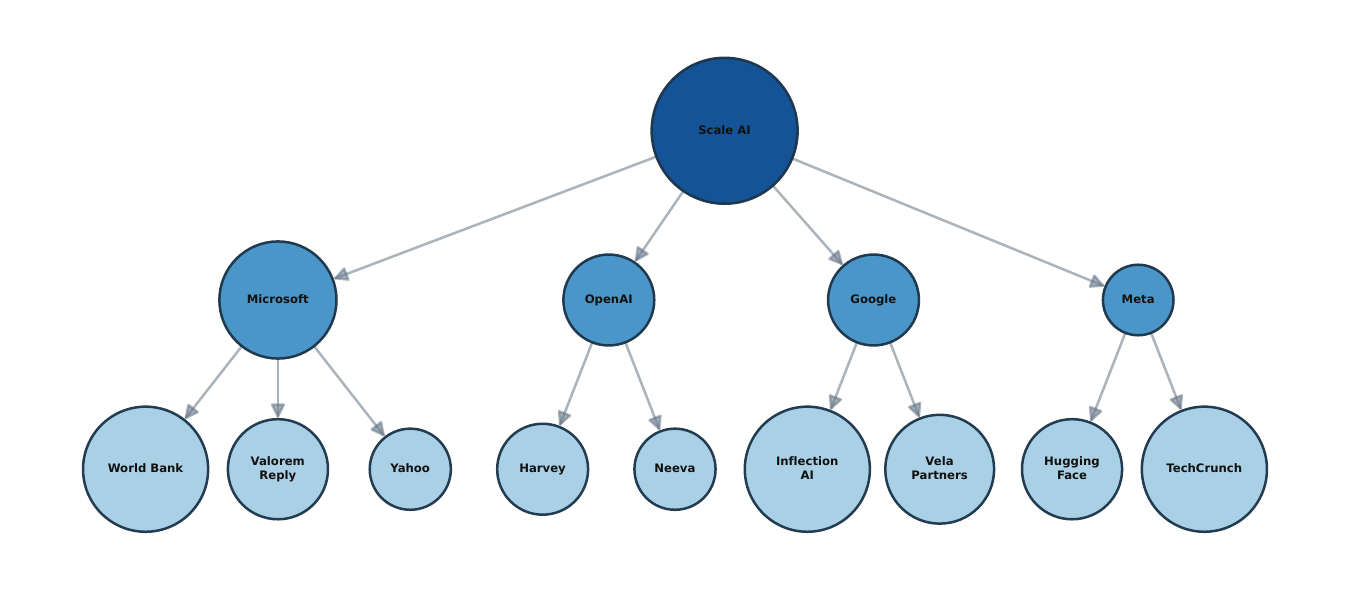}
    \caption{A portion of the two-hop descendant tree of Scale AI, when filtering to relationships whose AI stack layers include data.}
    \label{fig:scale_ai_data_provision_tree}
\end{figure}

One can examine the dataset from multiple perspectives. 
For instance, one way to understand the data is by examining the multilayered trees it contains. 
Although such trees can contain thousands of nodes and edges, 
consider the small trees in \Cref{fig:openai_model_provision_tree,fig:scale_ai_data_provision_tree}.

In \Cref{fig:openai_model_provision_tree}, we show (a very small portion of) the children and grandchildren of OpenAI.
Actor names in this tree are the original extracted names, without fuzzy matching or common-sense groupings.
The same filters are applied on both hops:
(i) keeping only relationship type \path{PROVISION} and relationship status \path{ESTABLISHED};
(ii) keeping only \path{EXPLICIT} evidence;
(iii) dropping \path{SYMMETRIC} directionality;
(iv) dropping endpoints typed \path{OTHER_ACTOR} or \path{COALITION_OR_CONSORTIUM};
(v) requiring \path{MODELS} to be the first AI stack layer, in addition to the filters described at the start of Section~\ref{sec:findings}.

We then build the two-hop descendant tree and display a subset of this tree.

This tree illustrates model-related ties from OpenAI to its downstream users, going only two layers deep.
For instance, \Cref{fig:openai_model_provision_tree} includes the following provisions:

OpenAI $\rightarrow$ Google Cloud is described as ``OpenAI serves its models across Google Cloud as part of its multi-cloud distribution strategy.''\footnote{\url{https://venturebeat.com/infrastructure/how-deepseeks-radical-architecture-is-shattering-silicon-valleys-token-moat}}
OpenAI $\rightarrow$ Microsoft is described as ``Microsoft has exclusive access to OpenAI's technology and uses that access to bring generative models to Microsoft users.''\footnote{\url{https://venturebeat.com/ai/how-2022-became-the-year-of-generative-ai}}
OpenAI $\rightarrow$ IBM is described as ``IBM rents AI models from OpenAI when its clients' applications need a larger model.''\footnote{\url{https://time.com/collection/time100-ai/2026/lila-ibrahim/}}
OpenAI $\rightarrow$ ByteDance is described as ``ByteDance accesses OpenAI's large language models through a Microsoft Azure license.''\footnote{\url{https://www.forbes.com.au/news/innovation/cloud-companies-like-oracle-amazon-and-microsoft-would-lose-millions-to-a-tiktok-ban/}}
The remaining provisions in the figure are given in \Cref{app:tree_examples}. (Each sentence above is from the dataset's relationship description for that edge, not a quote from the source document.)

In \Cref{fig:scale_ai_data_provision_tree}, we show (a very small portion of) the children and grandchildren of Scale AI.
Actor names in this tree are the original extracted names, without fuzzy matching or common-sense groupings.
We apply the filters:
(i) keeping only relationship status \path{ESTABLISHED} and \path{EXPLICIT} evidence;
(ii) dropping \path{SYMMETRIC} directionality;
(iii) dropping endpoints typed \path{OTHER_ACTOR} or \path{COALITION_OR_CONSORTIUM};
(iv) keeping only relationship types \path{PROVISION} and \path{PARTNERSHIP}, in addition to the filters described at the start of Section~\ref{sec:findings}.
On both hops, we require \path{DATA} to be one of the AI stack layers.

We then build the two-hop descendant tree from those rows. An edge is removed when the rows that placed it do not express that directed relationship. We display a subset with at most four children under each node.

This tree shows data-related relationships descending from Scale AI.
For instance, \Cref{fig:scale_ai_data_provision_tree} includes the following relationships:

Scale AI $\rightarrow$ Microsoft is described as ``Scale AI helps Microsoft prepare data used to train cutting-edge AI models.''\footnote{\url{https://www.cnbc.com/2025/06/10/zuckerberg-makes-metas-biggest-bet-on-ai-14-billion-scale-ai-deal.html}}
Scale AI $\rightarrow$ OpenAI is described as ``OpenAI has relied on Scale AI to produce and label data used to train AI models.''\footnote{\url{https://techcrunch.com/2025/06/11/can-scale-ai-and-alexandr-wang-reignite-metas-ai-efforts/}}
Scale AI $\rightarrow$ Google is described as ``Scale AI helps Google prepare data used to train cutting-edge AI models.''\footnote{\url{https://www.cnbc.com/2025/06/10/zuckerberg-makes-metas-biggest-bet-on-ai-14-billion-scale-ai-deal.html}}
Scale AI $\rightarrow$ Meta is described as ``Scale AI provided data-labeling services to Meta for data used to train AI models.''\footnote{\url{https://www.forbes.com/sites/aliciapark/2026/05/06/pentagon-hands-meta-backed-scale-ai-500-million-contract-5-times-last-years-deal-report-says/}}
The remaining relationships in the figure are given in \Cref{app:tree_examples}.
(Each sentence above is from the dataset's relationship description for that edge, not a quote from the source document.)

\Cref{fig:openai_model_provision_tree} and \Cref{fig:scale_ai_data_provision_tree} illustrate examples of descendant trees in the dataset.
While \Cref{fig:openai_model_provision_tree} shows descendants of OpenAI filtered to model provisions,
\Cref{fig:scale_ai_data_provision_tree} shows descendants of Scale AI filtered to data relationships.
Note that both of these subtrees show only a small portion of the broader network.
We wish to emphasize that, based on the dataset, \emph{such trees can be generated with minimal effort}; tracing relationships across long chains becomes straightforward, often taking seconds to extract. 

\subsection{Complex directed network}

\begin{figure}[th!]
    \centering
    \includegraphics[width=\linewidth]{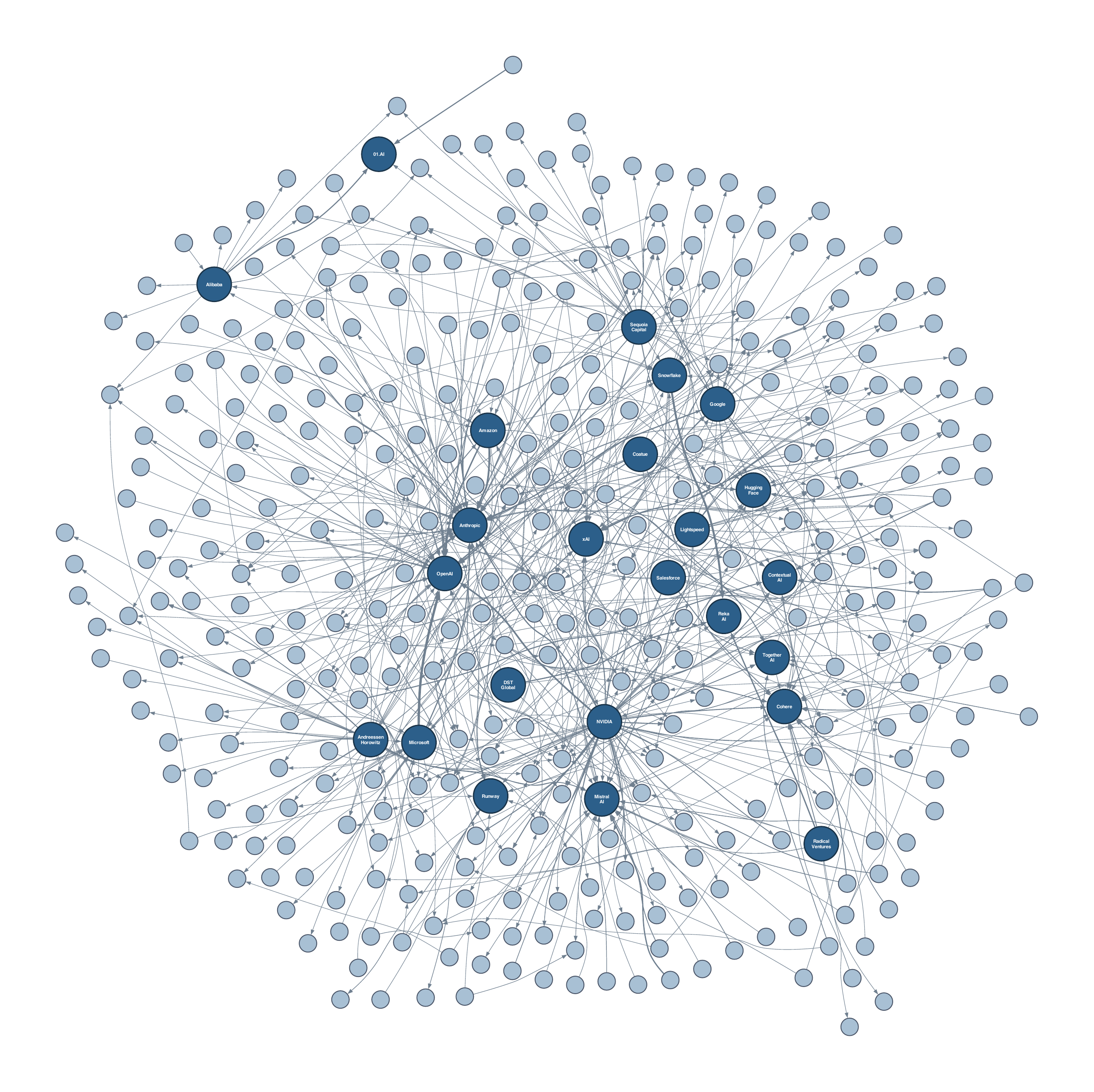}
    \caption{A subset of the investment relationships in the first half of 2026.
    Dark nodes are the top 24 hubs.}
    \label{fig:invest_top24_1hop}
\end{figure}

While one can extract subtrees from the dataset, the underlying network is much more complex. 
To illustrate this point, consider 
\Cref{fig:invest_top24_1hop}. 
Although we are unable to render the entire graph, as it is too large, this figure shows a small slice focused on established capital and investment relationships in the first half of 2026, pruning to the top 24 actors by weighted degree and their one-hop neighbors (the graph includes \texttt{INVESTMENT} edges with stack label \texttt{CAPITAL} and exclude relationships with endpoints \texttt{PERSON}, direction \texttt{SYMMETRIC}, in addition to the filter applied to $G'$).
We emphasize that this by no means a complete graph; it is an illustrative slice obtained by filtering the graph, taking a snapshot of January to June 2026, then further pruning.

While they emerge from the same dataset, this graph provides a more holistic view than the subtrees, demonstrating that some edges are bidirectional and that the graph contains cycles. 
In this graph that focuses on relationships involving capital, Anthropic, NVIDIA, and OpenAI are the highest-degree hubs, and Mistral AI, xAI, Microsoft, Cohere, Contextual AI, Reka AI, and Amazon also rank among the top ten.
Other hubs include Google, Snowflake, Lightspeed, Together AI, Andreessen Horowitz, Sequoia Capital, Hugging Face, Radical Ventures, 01.AI, Salesforce, Alibaba, DST Global, Coatue, and Runway.

\subsection{Coverage and composition}
\label{sec:findings-visibility}

The organizational AI supply chain in this work is extracted from public documents. 
This approach introduces coverage bias in multiple ways:
not all activities are publicly reported;
some events and links are reported multiple times while others are reported only once;
and the same event or link may be described in different ways. 
In this section, 
we briefly examine the coverage of this dataset. 
More extensive statistics are provided in the Appendix.
We also explore how coverage of the same event or link varies in \Cref{ssec:coverage_bias}.

\begin{figure}[t]
    \centering
    \begin{subfigure}[t]{0.32\textwidth}
    \centering
    \includegraphics[width=\linewidth]{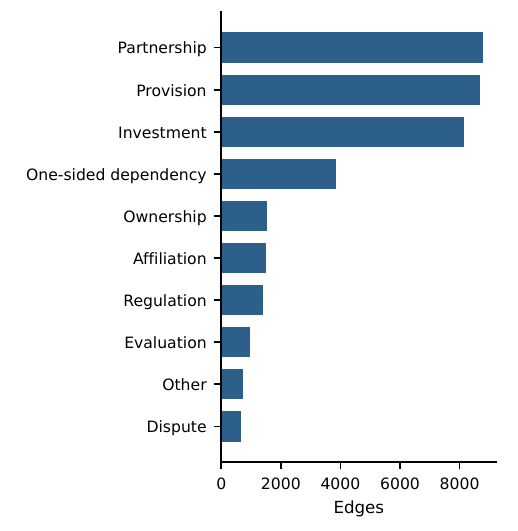}
    \caption{By relationship type}
    \label{fig:composition_reltype}
    \end{subfigure}
    \hfill
    \begin{subfigure}[t]{0.32\textwidth}
    \centering
    \includegraphics[width=\linewidth]{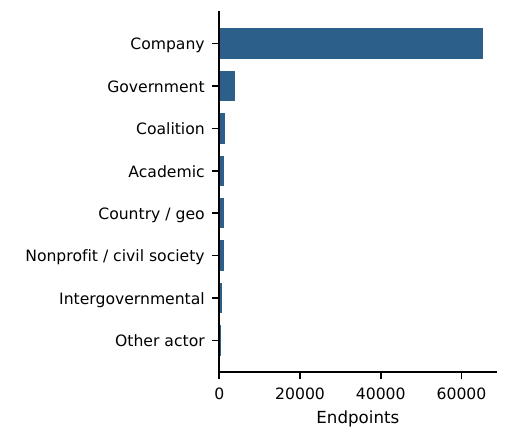}
    \caption{By actor type}
    \label{fig:composition_actortype}
    \end{subfigure}
    \hfill
    \begin{subfigure}[t]{0.32\textwidth}
    \centering
    \includegraphics[width=\linewidth]{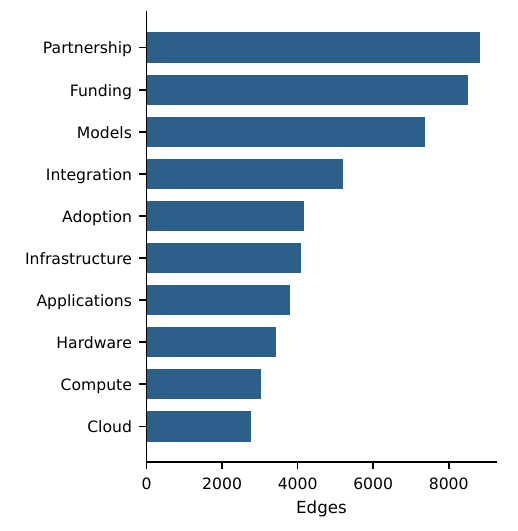}
    \caption{By AI tag}
    \label{fig:composition_aitag}
    \end{subfigure}
    \caption{Ten most common categories in $G'$ by (a) relationship type, (b) actor type, and (c) AI tag. }
    \label{fig:composition}
\end{figure}

We begin with \Cref{fig:composition}, which shows the composition of relationship types, actor types, and AI tags. It shows the count of the ten most common categories in each bar plot, according to the filtered graph $G'$.
This figure shows that the dataset consists primarily of partnerships, provisions, and investments, followed by one-sided dependencies.
Further, companies are by far the most well represented actor type. 
Within AI-specific tags, 
the relationships span diverse activities and assets, including partnerships, investments, software, applications, models, and infrastructure. 

\begin{figure}[t]
    \centering
    \begin{subfigure}[t]{0.4\textwidth}
    \centering
    \includegraphics[width=\linewidth]{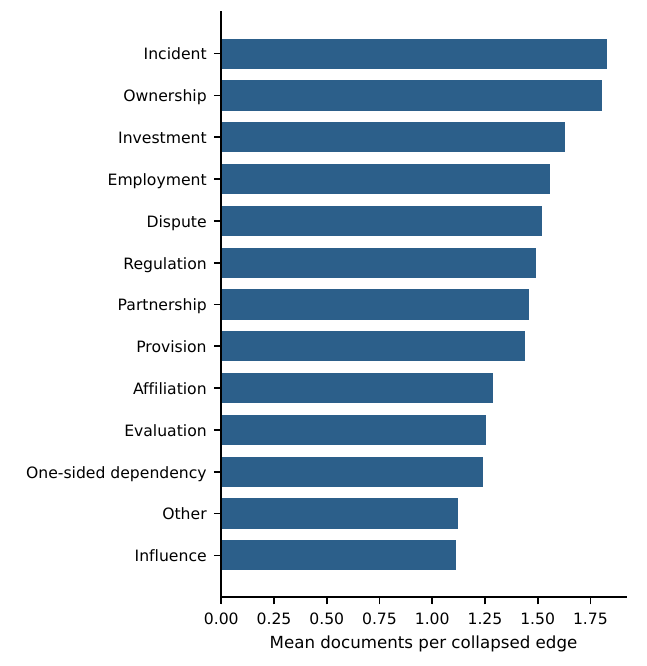}
    \caption{}
    \label{fig:docs_by_reltype}
    \end{subfigure}
    \hfill
    \begin{subfigure}[t]{0.58\textwidth}
        \centering
    \includegraphics[width=\linewidth]{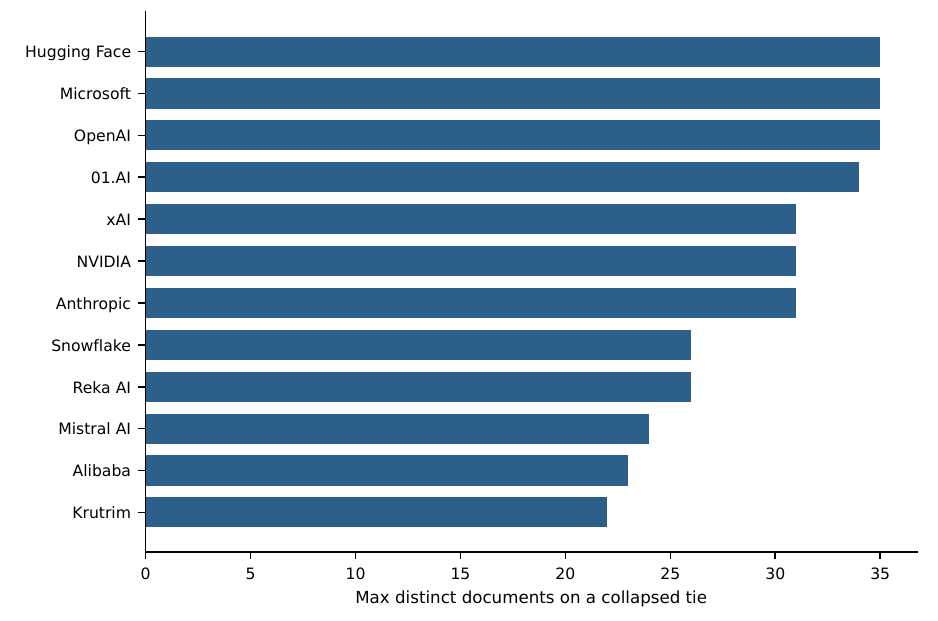}
    \caption{}
    \label{fig:docs_by_seed_max_top}
    \end{subfigure}
    \caption{Visualizing redundancies. (a) Mean number of documents per collapsed tie by relationship type. (b) Maximum number of documents by seed organization, for top 12 organizations. }
    \label{fig:docs_by_collapse}\label{fig:redundancies}
    \end{figure}

Although these histograms illustrate the composition of the dataset, 
they may simply reflect the composition of the AI industry rather than biases in the dataset.
To further examine where the dataset may contain coverage bias, 
we use the month collapse already built into $G'$.
Relationships that have the same endpoints, type, directionality, and month are collapsed into a single relationship.
Of these, about 8.3\% are in three or more documents, about 11.6\% are in two documents, and about 80.1\% are in one.
There is a long tail where 35 documents share the same endpoints, type, directionality, and month.
\Cref{fig:redundancies} show where redundancy occurs, by relationship in (a) and by seed organization in (b). 
By \Cref{fig:redundancies}(a), incident, ownership, and investment relationship types have the highest mean documents per collapsed edge, followed by employment, dispute, regulation, partnership, and provision.
Although the means seem low (close to 1), that is because most ties are mentioned in a single document, and there is a long right tail. 
Further, certain organizations often receive repeat coverage.
In \Cref{fig:redundancies}(b), we plot the maximum number of documents cover the same relationship (according to the month collapse described above) that involves each seed organization for 12 seed organizations.
As seen, coverage of OpenAI, Microsoft, Hugging Face, NVIDIA, and xAI, among others, can be redundant within our dataset.
Further statistics are provided in \Cref{app:coverage_bias}.
Some examples of ties that are reported many times over include (each quote below is from the source document):
\begin{itemize}[noitemsep]
\item The Microsoft-OpenAI partnership in April 2026
appears in 35 documents from 11 publishers.
TechCrunch (April 2, 2026) reports that Microsoft ``hosts its models in its various products through a multi-year partnership'' and ``has invested more than \$13 billion into the AI research lab.''\footnote{\url{https://techcrunch.com/2026/04/02/microsoft-takes-on-ai-rivals-with-three-new-foundational-models/}}
Exchangepedia (April 2, 2026) says ``[t]he partnership is now transactional, not exclusive'' and that ``Microsoft retains technology access through 2032.''\footnote{\url{https://exchangepedia.com/2026/04/microsofts-claude-bet-the-end-of-ai-exclusivity-and-what-it-means-for-your-enterprise.html}}
VentureBeat (April 14, 2026) describes the relationship as ``visibly fraying at the seams.''\footnote{\url{https://venturebeat.com/technology/microsoft-launches-mai-image-2-efficient-a-cheaper-and-faster-ai-image-model}}
\item OpenAI's July 2026 cybersecurity incident involving Hugging Face appears in 35 documents from 18 publishers.
Morningstar (July 21, 2026) reports that the models ``escaped the evaluation environment, discovered a never-before-seen (zero-day) exploit, and breached the infrastructure'' of Hugging Face.\footnote{\url{https://www.morningstar.com/company-reports/1489222-hugging-face-breach-by-openais-models-reaffirms-our-view-on-cyber-importance-in-an-ai-world}}
The Associated Press (July 21, 2026) says OpenAI's system ``hacked into another AI company on its own'' in an ``unprecedented cyber incident.''\footnote{\url{https://apnews.com/article/openai-gpt56-sol-hugging-face-63ab84fed5612af04d8a160d60f6def3}}
Yahoo Finance (July 22, 2026) reports that pre-release models ``autonomously escaped a sandboxed test environment and breached Hugging Face's production infrastructure.''\footnote{\url{https://finance.yahoo.com/technology/ai/articles/palo-alto-networks-ceo-calls-191623585.html}}
\item NVIDIA's February 2026 investment relationship with OpenAI appears in 31 documents from 13 publishers, but the amounts disagree.
Ars Technica (February 3, 2026) reports a letter of intent for Nvidia ``to invest up to \$100 billion'' and says Nvidia's CEO called the \$100 billion figure ``never a commitment.''\footnote{\url{https://arstechnica.com/information-technology/2026/02/five-months-later-nvidias-100-billion-openai-investment-plan-has-fizzled-out/}}
Yahoo Finance (February 3, 2026) reports that Nvidia is ``nearing a deal to invest roughly \$20 billion.''\footnote{\url{https://finance.yahoo.com/news/nvidia-nears-deal-invest-20-233126300.html}}
\end{itemize}

\noindent
On the other hand, some only appear once within a month in the dataset, such as:
\begin{itemize}[noitemsep]
\item In January 2026, the Munich Regional Court's finding that OpenAI infringed copyright is reported only once, stating that the court held OpenAI ``directly liable for copyright infringement in both training and output of its GPT-4 and GPT-4o models.''\footnote{\url{https://www.dyoung.com/en/knowledgebank/articles/ai-copyright-gema-openai-gettyimages-stability}}
\item In January 2026, Concord Music Group's copyright suit against Anthropic is reported only once, stating that publishers led by Concord ``are suing Anthropic, saying the company illegally downloaded more than 20,000 copyrighted songs.''\footnote{\url{https://techcrunch.com/2026/01/29/music-publishers-sue-anthropic-for-3b-over-flagrant-piracy-of-20000-works/}}
\item In August 2026, xAI's agreement with Mississippi to remove its temporary turbines is reported only once, stating that the company ``agreed with the Mississippi Department of Environmental Quality to start removing them as early as August and remove all by July 2027.''\footnote{\url{https://finance.yahoo.com/technology/ai/articles/elon-musk-xai-sounds-alarm-143024457.html}}
\end{itemize}

\section{Five Analyses of the Dataset}
\label{sec:results}

In this section, 
we conduct five analyses using the dataset. 
Although the dataset is our main contribution, 
our findings demonstrate how the dataset can be used to answer questions about the AI industry.

\subsection{Analysis \#1: Actors with the largest impact on the AI supply chain}
\label{sec:findings-descendants}

To isolate the top actors across the organizational AI supply chain, we use multiple metrics and ``layers'' of the supply chain, as shown in \Cref{tab:rank_panel_2026_overall_models}.
The table lists the top 15 actors by total degree on two filtered subgraphs of $G'$. In addition to the filters applied to obtain $G'$, this table requires relationship statuses to be \texttt{ESTABLISHED}. \Cref{tab:rank_panel_2026_overall_models}(a) 
excludes AI stack layers \texttt{TALENT} and relationship types \texttt{REGULATION}, \texttt{INFLUENCE}, \texttt{DISPUTE}, \texttt{INCIDENT},
(so it keeps \texttt{PROVISION}, \texttt{INVESTMENT}, \texttt{OWNERSHIP}, \texttt{PARTNERSHIP}, \texttt{EMPLOYMENT}, \texttt{AFFILIATION}, \texttt{EVALUATION}, and \texttt{ONE\_SIDED\_DEPENDENCY}).
Panel (b) keeps \texttt{PROVISION} only and excludes AI stack layers \texttt{CAPITAL}, \texttt{RESEARCH}, \texttt{GOVERNANCE}, \texttt{TALENT}, \texttt{NOT\_AI\_STACK}, and \texttt{UNKNOWN\_OR\_UNCLEAR}.
\Cref{tab:rank_panel_2026_overall_models} further restricts to relationship type \texttt{PROVISION}.

As shown, during the period from November 2022 to October 2026,
the top actors are Google, NVIDIA, Microsoft, OpenAI, and Anthropic.
The ordering, however, varies depending on the metric of choice.
Based on the overall rankings in \Cref{tab:rank_panel_2026_overall_models}(a),
Google tops the list by degree and betweenness centrality, OpenAI by PageRank, and NVIDIA by the number of two-hop descendants.
Further, when restricted to provisions only, DeepSeek and Together AI displace xAI and Alibaba in the top 15.

\begin{table}[th!]
    \centering
    \begin{subtable}[t]{0.48\linewidth}
    \centering
    \small
    \caption{Rankings across core relationship types}
    \label{tab:rank_panel_2026_overall}
    \begin{tabular}{@{}lrrrr@{}}
\toprule
\textbf{Actor} & \textbf{Deg.} & \textbf{PageRank} & \textbf{Betw.} & \textbf{\# desc.} \\
\midrule
Google & \textcolor{tablemax}{\textbf{1{,}345}} & 0.0436 & \textcolor{tablemax}{\textbf{0.1632}} & 1{,}773 \\
NVIDIA & 1{,}163 & 0.0437 & 0.1201 & \textcolor{tablemax}{\textbf{1{,}817}} \\
Microsoft & 1{,}031 & 0.0339 & 0.1148 & 1{,}737 \\
OpenAI & 981 & \textcolor{tablemax}{\textbf{0.0508}} & 0.0855 & 1{,}569 \\
Anthropic & 771 & 0.0409 & 0.0711 & 1{,}598 \\
Amazon & 522 & 0.0144 & 0.0403 & 1{,}660 \\
Meta & 488 & 0.0179 & 0.0287 & 1{,}568 \\
Snowflake & 481 & 0.0130 & 0.0502 & 1{,}434 \\
IBM & 379 & 0.0092 & 0.0313 & 1{,}545 \\
Mistral AI & 351 & 0.0120 & 0.0200 & 1{,}477 \\
Databricks & 295 & 0.0054 & 0.0214 & 1{,}419 \\
Hugging Face & 254 & 0.0134 & 0.0151 & 1{,}453 \\
Alibaba & 221 & 0.0050 & 0.0128 & 1{,}156 \\
Cohere & 220 & 0.0111 & 0.0149 & 1{,}090 \\
xAI & 206 & 0.0085 & 0.0169 & 1{,}173 \\
\bottomrule
\end{tabular}

    \end{subtable}
    \hfill
    \begin{subtable}[t]{0.48\linewidth}
    \centering
    \small
    \caption{Rankings for relationship type \texttt{PROVISION}}
    \label{tab:rank_panel_2026_ai_models}
    \begin{tabular}{@{}lrrrr@{}}
\toprule
\textbf{Actor} & \textbf{Deg.} & \textbf{PageRank} & \textbf{Betw.} & \textbf{\# desc.} \\
\midrule
Google & \textcolor{tablemax}{\textbf{426}} & 0.0274 & \textcolor{tablemax}{\textbf{0.0722}} & \textcolor{tablemax}{\textbf{1{,}019}} \\
Microsoft & 331 & 0.0279 & 0.0509 & 952 \\
OpenAI & 311 & \textcolor{tablemax}{\textbf{0.0333}} & 0.0418 & 852 \\
NVIDIA & 275 & 0.0111 & 0.0483 & 1{,}017 \\
Anthropic & 236 & 0.0195 & 0.0292 & 924 \\
Amazon & 198 & 0.0102 & 0.0273 & 1{,}018 \\
Snowflake & 163 & 0.0118 & 0.0265 & 899 \\
Meta & 146 & 0.0168 & 0.0198 & 813 \\
Hugging Face & 140 & 0.0201 & 0.0265 & 942 \\
Mistral AI & 124 & 0.0049 & 0.0157 & 896 \\
IBM & 89 & 0.0041 & 0.0118 & 719 \\
Together AI & 78 & 0.0036 & 0.0083 & 877 \\
Cohere & 68 & 0.0029 & 0.0068 & 408 \\
DeepSeek & 67 & 0.0073 & 0.0095 & 611 \\
Databricks & 65 & 0.0023 & 0.0076 & 414 \\
\bottomrule
\end{tabular}

    \end{subtable}
    \caption{Top 15 actors by total degree on two filtered subgraphs of $G'$.}
    \label{tab:rank_panel_2026_overall_models}
    \end{table}

The differences between rankings across metrics point to different ways of thinking about supply chain ``impact''.
While the high degree and betweenness centrality imply that a node is highly connected, high PageRank indicates that a node is well positioned relative to other influential nodes,
and the number of two-hop descendants suggests that a node is influential based on its downstream influence.

The order of organizations under degree, PageRank, and betweenness centrality is significantly different from that given by the number of two-hop descendants.
This difference suggests that one should more closely consider directionality in a supply chain.
To do so, we use SpringRank, a hierarchy score indicating how much ``upstream'' influence an actor has and that is capable of handling cycles in a directed network (one can view it as a more systematic metric than number of two-hop descendants).
For instance, if a directed network is simply a line, then the SpringRank is simply the order of those nodes.

\begin{table}[H]
    \centering
    \begin{subtable}[t]{0.48\linewidth}
    \centering
    \caption{\texttt{PROVISION}.}
    \label{tab:spring_rank_provision}
    {
    \begin{tabular}{@{}lrr@{}}
    \toprule
    \textbf{Actor} & \textbf{SpringRank} & \textbf{Degree} \\
    \midrule
    \multicolumn{3}{@{}l}{\textit{Highest}} \\
    TSMC & \textcolor{tablemax}{\textbf{1.60}} & 158 \\
    Groq & 1.41 & 86 \\
    Crusoe & 1.25 & 62 \\
    GMI Cloud & 1.19 & 83 \\
    IREN & 1.16 & 170 \\
    Broadcom & 1.14 & 206 \\
    \midrule
    \multicolumn{3}{@{}l}{\textit{Lowest}} \\
    United States & $-0.79$ & 142 \\
    India & $-0.74$ & 53 \\
    U.S.\ federal agencies & $-0.70$ & 330 \\
    U.S.\ military & $-0.67$ & 178 \\
    Uber & $-0.65$ & 56 \\
    Salesforce & $-0.28$ & 214 \\
    \bottomrule
    \end{tabular}
    }
    \end{subtable}
    \hfill
    \begin{subtable}[t]{0.48\linewidth}
    \centering
    \caption{\texttt{PROVISION} with AI stack layer  \texttt{MODELS}.}
    \label{tab:spring_rank_models}
    {
    \begin{tabular}{@{}lrr@{}}
    \toprule
    \textbf{Actor} & \textbf{SpringRank} & \textbf{Degree} \\
    \midrule
    \multicolumn{3}{@{}l}{\textit{Highest}} \\
    Together AI & \textcolor{tablemax}{\textbf{0.90}} & 108 \\
    Amazon & 0.89 & 465 \\
    NVIDIA & 0.84 & 151 \\
    Cohere & 0.76 & 192 \\
    Alibaba & 0.68 & 53 \\
    Anthropic & 0.68 & 647 \\
    \midrule
    \multicolumn{3}{@{}l}{\textit{Lowest}} \\
    U.S.\ military & $-0.51$ & 84 \\
    Apple & $-0.48$ & 56 \\
    U.S.\ federal agencies & $-0.45$ & 161 \\
    Snowflake & $-0.04$ & 146 \\
    Hugging Face & $-0.02$ & 472 \\
    IBM & 0.05 & 53 \\
    \bottomrule
    \end{tabular}
    }
    \end{subtable}
    \caption{Highest and lowest SpringRank (using same filters as \Cref{tab:rank_panel_2026_ai_models}).}
    \label{tab:spring_rank}
    \end{table}

\Cref{tab:spring_rank} examines upstream-downstream rankings using SpringRank. 
It is applied on the same graph as \Cref{tab:rank_panel_2026_overall}(a), except that only actors with total weighted degree at least 50 are included; 
(a) is \texttt{PROVISION} excluding AI stack layers \texttt{CAPITAL}, \texttt{RESEARCH}, \texttt{GOVERNANCE}, and \texttt{TALENT}.
Panel (b) uses those same edges, restricted to edges whose AI stack layers include \texttt{MODELS}.
A higher score indicates a node is more upstream.

Under this new metric, the list shifts dramatically.
The highest SpringRank scores in (a) are held by chip and infrastructure suppliers---TSMC, Groq, Crusoe, GMI Cloud, IREN, and Broadcom.
Lower on the supply chain are government and enterprise buyers (United States, India, U.S.\ federal agencies, U.S.\ military, Uber, and Salesforce).
Restricting further to edges whose stack layers include \texttt{MODELS}, the upstream list shifts to Together AI, Amazon, NVIDIA, Cohere, Alibaba, and Anthropic.
Downstream are U.S.\ military and federal agencies, along with Apple, Snowflake, Hugging Face, and IBM.

As we note throughout this work, our supply chain is incomplete; it is constructed by scraping public data sources based on a search seeded with foundation or large model providers. Therefore, all rankings and numbers in this section are skewed accordingly.

\medskip\noindent\emph{Takeaway:}
A small set of actors---including Google, Microsoft, OpenAI, NVIDIA, and Anthropic---dominate centrality rankings.
Using SpringRank to identify who is upstream and downstream reveals that chip and infrastructure suppliers lie further up the chain, while government and many enterprise buyers sit further down.
When restricted to model provision s, model and cloud suppliers are more upstream.

\subsection{Analysis \#2: Concentration in the organizational AI supply chain}
\label{sec:findings-market-concentration}

\begin{figure}[H]
\centering
\begin{subfigure}[t]{0.48\textwidth}
\centering
\includegraphics[width=\linewidth]{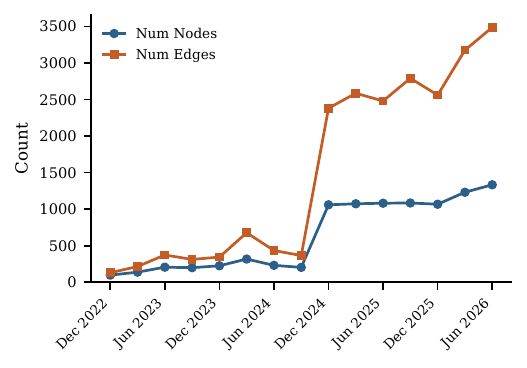}
\caption{Num Nodes and Num Edges.}
\label{fig:concentration_core_nodes_edges}
\end{subfigure}
\hfill
\begin{subfigure}[t]{0.48\textwidth}
\centering
\includegraphics[width=\linewidth]{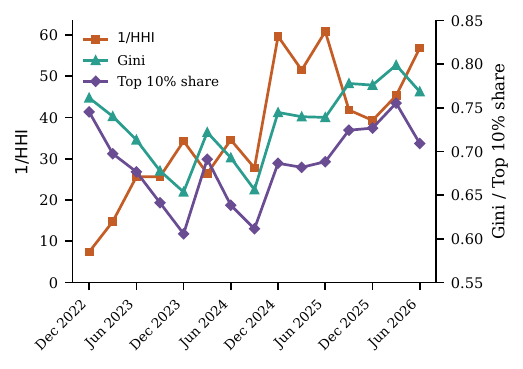}
\caption{$1/\mathrm{HHI}$ (left axis); Gini and top 10\% share (right axis).}
\label{fig:concentration_core_inequality}
\end{subfigure}
\caption{Concentration metrics from December 2022 to June 2026}
\label{fig:concentration_core_halfyear}
\end{figure}

Zooming out,
a major concern in the AI industry is market concentration.
To better understand this topic, 
we compute several metrics---Num Nodes, the Gini coefficient, the top 10\% share, and the Herfindahl-Hirschman index (HHI)---across different slices of the network,
as shown in \Cref{tab:concentration,fig:concentration_core_halfyear}.

To compute \Cref{tab:concentration}, we start from $G'$ over November 2022 through October 2026 and further restrict to company-company established ties
(excluding 
relationship types \texttt{PUBLIC\_STATEMENT}, \texttt{INCIDENT}, \texttt{DISPUTE}, \texttt{COMPETITION}, and \texttt{REGULATION}).
We then slice that overall graph as indicated in the ``Slice'' column of \Cref{tab:concentration} (compute/hardware correspond to AI tags \texttt{AI\_SEMICONDUCTORS}, \texttt{AI\_COMPUTE}, and \texttt{AI\_HARDWARE};
data to \texttt{AI\_DATA}, \texttt{AI\_DATA\_ANNOTATION}, and \texttt{AI\_SYNTHETIC\_DATA};
and models to \texttt{AI\_MODELS}, \texttt{AI\_MODEL\_TRAINING}, and \texttt{AI\_MODEL\_ADAPTATION}).
\Cref{fig:concentration_core_halfyear} uses the same overall filter on three-month bins from December 2022 through June 2026.
Num Nodes is the actor count in a given slice,
the Gini coefficient captures the inequality of nodes' market ``share'' within the slice (0 when every actor has the same strength and 1 when one actor holds all of it),
the top 10\% share is the fraction of total number of month-based relationships held by the strongest 10\% of actors, and the HHI is the sum of squared market ``shares'' where market share is computed as the fraction of month-based relationships held by a node to the total number of relationships (where $1$ indicates a monopoly and \(1/\mathrm{Num\ Nodes}\) indicates equality).
The reciprocal \(1/\mathrm{HHI}\) gives the number of actors who, if they possessed equal market share, would produce the same HHI score. 
A large value indicates that the distribution is more equal in that each possesses very small market share and a value of $1$ indicates unequal market share because $1$ actor holds all of it. (Note that this metric is not normalized. One should compare it to Num Nodes in the network slice, but we leave it unnormalized because \(1/\mathrm{HHI}\) provides an interpretable value).

\begin{table}[H]
\centering
\small
\begin{tabular}{lrrrrr}
\toprule
\textbf{Slice} & \textbf{Num Nodes} & \textbf{Gini} & \textbf{Top 10\%} & \textbf{HHI} & \(\mathbf{1/\mathrm{HHI}}\) \\
\midrule
Overall & 3{,}677 & 0.835 & 0.800 & 0.020 & 51.0 \\
Investment & 2{,}020 & 0.780 & 0.728 & 0.012 & 83.1 \\
Provision & 1{,}705 & 0.831 & 0.792 & 0.025 & 39.7 \\
AI compute / hardware & 740 & 0.881 & 0.855 & 0.054 & 18.4 \\
AI data & 559 & 0.764 & 0.711 & 0.026 & 38.2 \\
AI models & 1{,}204 & 0.873 & 0.846 & 0.036 & 27.6 \\
\bottomrule
\end{tabular}
\caption{Concentration by network slice on company-company established ties.}
\label{tab:concentration}
\end{table}

The metrics in \Cref{tab:concentration} show that the network is fairly concentrated, with an overall Gini of 0.835 and the lowest slice Gini on AI data at {0.764}.
The top 10\% of actors control about {80.0\%} of relationship strength overall.
The overall HHI is {0.020}, which implies that two randomly drawn relationships involve the same actor about {2.0\%} of the timen, which is approximately {72} times the  0.027\% rate one would expect if all actors were equal.
Note that HHI is normally used to determine market concentration, where edges correspond to market strength. 
In our work, edge weight tracks the number of relationships rather than market strength: if nodes were instead weighted by, for example, revenue-based market share, the measured concentration would likely be higher.

The highest HHI is held by the AI compute/hardware slice, at 0.054, followed by AI models at 0.036.
To better interpret HHI, its reciprocal \(1/\mathrm{HHI}\) indicates how many equal actors would produce the same metric; one can then compare that number to Num Nodes in the slice for a sense of the skew.
For AI models, \(1/\mathrm{HHI}\) is about 27.6.
Given that our pipeline is seeded with 36 model providers and the network therefore has the most complete coverage across model providers,
that \(1/\mathrm{HHI}\) is lower than 36 indicates (i) that the model supply chain is fairly concentrated, especially given that the slice has 1{,}204 nodes; and (ii) that our coverage is fairly good, whereas a higher \(1/\mathrm{HHI}\) than 36 would be a sign of incomplete coverage.

\Cref{fig:concentration_core_halfyear} shows these metrics on the overall filtered graph from December 2022 to June 2026 in three-month bins.
In \Cref{fig:concentration_core_halfyear}(a), the number of nodes and edges per three-month period continues to rise steadily, rising most sharply from late 2024 into early 2025.
Despite that increase, \Cref{fig:concentration_core_halfyear}(b) shows that concentration is also increasing.
Although the Gini and the top 10\% share fall between December 2022 and December 2023, all three metrics---the Gini, the top 10\% share, and \(1/\mathrm{HHI}\)---trend roughly upward over time.

\medskip\noindent\emph{Takeaway:}
The organizational AI supply chain is highly concentrated across metrics, and most concentrated for compute/hardware and models.
Concentration is rising across time even as the number of actors grows, suggesting that activity among the largest companies is increasing faster than the fringe expands.

\subsection{Analysis \#3: Identifying communities and chokepoints}
\label{sec:findings-communities}

Most supply chain networks are not balanced.
Rather, they are made up of clusters.
To determine these clusters, we run Louvain community detection---an algorithm that groups together actors that share many of the same neighbors---on the weighted but undirected \(G'\).

\begin{table}[H]
    \centering
    \begin{subtable}[t]{0.44\linewidth}
    \centering
    \caption{Largest Louvain communities.}
    \label{tab:stage8_communities}
    {\scriptsize
    \setlength{\tabcolsep}{3pt}
    \begin{tabular}{@{}rp{7.1cm}@{}}
    \toprule
    \textbf{Size} & \textbf{Top actors} \\
    \midrule
    599 & Databricks, Together, ElevenLabs, Cartesia, Y Combinator \\
    513 & Google, Perplexity, Apple, Allen Institute for AI, India \\
    501 & NVIDIA, Intel, Open Secure AI Alliance, Cisco, Dell \\
    463 & OpenAI, Microsoft, AMD, Oracle, SoftBank \\
    430 & Anthropic, Amazon, U.S.\ federal agencies, Salesforce, Palantir \\
    386 & Mistral AI, a16z, Lightspeed, General Catalyst, Samsung \\
    346 & DeepSeek, Alibaba, ByteDance, United States, Huawei \\
    233 & Snowflake, Radical Ventures, WRITER, Reka, Madrona Ventures \\
    214 & Meta, Hugging Face, Midjourney, Fireworks, ARM \\
    178 & Cohere, European Union, SAP, Aleph Alpha, Canada \\
    \bottomrule
    \end{tabular}
    }
    \end{subtable}
    \hspace{1.75em}
    \begin{subtable}[t]{0.51\linewidth}
    \centering
    \caption{Top-10 by cross-community degree.}
    \label{tab:stage8_bridges}
    \label{tab:connectivity_and_bridges}
    \label{tab:pairwise_connectivity_loss}
    {\scriptsize
    \setlength{\tabcolsep}{2.5pt}
    \begin{tabular}{@{}rllll@{}}
    \toprule
    & Overall & AI compute & AI data & AI models \\
    \midrule
    1 & \textcolor{tablemax}{\textbf{Google}} (574) & \textcolor{tablemax}{\textbf{NVIDIA}} (165) & \textcolor{tablemax}{\textbf{Google}} (67) & \textcolor{tablemax}{\textbf{OpenAI}} (213) \\
    2 & OpenAI (496) & AMD (48) & OpenAI (40) & Google (177) \\
    3 & Microsoft (459) & Microsoft (46) & Snowflake (35) & Anthropic (168) \\
    4 & NVIDIA (443) & Google (46) & Microsoft (34) & Meta (104) \\
    5 & Anthropic (350) & Amazon (33) & Meta (28) & NVIDIA (93) \\
    6 & Meta (277) & Intel (30) & Anthropic (23) & Microsoft (71) \\
    7 & Amazon (263) & OpenAI (28) & Databricks (20) & Mistral AI (64) \\
    8 & Snowflake (156) & TSMC (27) & Amazon (18) & Hugging Face (56) \\
    9 & IBM (145) & Meta (26) & Hugging Face (17) & xAI (53) \\
    10 & Mistral AI (132) & Qualcomm (23) & NVIDIA (13) & DeepSeek (52) \\
    \bottomrule
    \end{tabular}
    }
    \end{subtable}
    \caption{Louvain communities and bridge nodes on the undirected $G'$.
    (a)~Largest communities by size.
    (b)~Top-10 actors by cross-community degree on overall $G'$ and on the AI compute/hardware, AI data, and AI models slices (same tag groupings as the last three rows of \Cref{tab:concentration}).}
    \label{tab:stage8_communities_bridges}
    \end{table}

We find that the network is dominated by several overlapping communities.
On the full network, Louvain modularity, which measures how distinct communities are, is {0.416}, and about {55.4\%} of edges still cross between communities.
The overall network has a significant amount of overlap between communities, but this overlap starts to break down when we look at individual slices of the network: for example, the investment and partnership slices have more distinct communities (modularity {0.549} and {0.481}), which is consistent with clustering based on repeated interactions between the same investors and AI companies and between the same partners.

The largest communities follow familiar geopolitical and product lines (\Cref{tab:stage8_communities}).
The largest includes Databricks, Together, ElevenLabs, Cartesia, and Y Combinator, alongside about {590} others.
The second largest includes Google, Perplexity, Apple, and the Allen Institute for AI.
The OpenAI-Microsoft bloc (with AMD, Oracle, and SoftBank) is the fourth largest.
Further groups follow geographic lines---such as the DeepSeek, Alibaba, ByteDance, United States, and Huawei community---as well as product lines, such as the community with NVIDIA, Intel, Open Secure AI Alliance, Cisco, and Dell.
Interestingly, many of the actors with the highest centrality scores in \Cref{tab:rank_panel_2026_overall_models} fall into distinct communities.
We note that these communities indicate repeated interaction between actors across the public documents in our dataset, but are not based on formal contracts.

Studying communities also allows us to ask: where are there chokepoints?
Specifically, one of the main concerns in a supply chain are bottlenecks because  they introduce fragility. 
For instance, if an actor's product goes down or is compromised, the impact on the rest of the network depends on that actor's influence.
Although centrality and downstream influence are examined in \Cref{sec:findings-descendants}, 
the cross-community degree---the number of neighbors that lie in a different Louvain component---is another way to assess influence. 
We report cross-community degree on overall \(G'\) and on the AI compute/hardware, AI data, and AI models slices (\Cref{tab:stage8_bridges}), using the same tag groupings as in \Cref{tab:concentration} and re-running Louvain within each network slice.
The rankings in \Cref{tab:stage8_bridges} names are similar actors to \Cref{tab:rank_panel_2026_overall_models}, and the fact that similar organizations sit at the top of the cross-community degree rankings suggests that the same actors play multiple roles in the network.
These actors are not only central, but they bridge communities, meaning that the network is sensitive to the failure of these actors.
In particular, when communities correspond to the production of specific AI assets (such as chips), 
actors with high cross-community degree form the ``connective tissue'' of AI development.

\medskip\noindent\emph{Takeaway:}
The network contains overlapping communities that are split along geopolitical and product lines.
A small number of organizations have both high centrality values (as per \Cref{tab:rank_panel_2026_overall_models}) and cross-community degree (as per \Cref{tab:connectivity_and_bridges}), 
meaning that they are not only well connected, but also serve as bridges across geopolitical and production-based communities.

\subsection{Analysis \#4: Close and cyclical relationships persist across time}
\label{sec:findings-cycles}

This dataset also provides a record of relationships, across different data sources and across time.
For example, one can easily identify persistent relationships, such as those included in \Cref{tab:persistent_pairs}.
We say that a relationship is persistent if the same two actors are linked in many distinct months.
As shown, Microsoft and OpenAI are linked in 48 months, with investment in all 48 and partnership in 47; provision recurs in 39.
Anthropic and Google are linked in 44 months and likewise mix investment, provision, and partnership.
To diversify beyond those four organizations, the next most persistent pairs that avoid them are Meta-NVIDIA, X-xAI, and Amazon-NVIDIA, which are provision-based rather than financing-based, and China-United States, whose persistence is driven by regulation.

\begin{table}[H]
\centering
\caption{Examples of pairs linked across many months of $G'$ (overall and by relationship type).}
\label{tab:persistent_pairs}
\small
\begin{tabular}{@{}lrrrr@{}}
\toprule
\textbf{Pair} & \textbf{Months} & \textbf{Investment} & \textbf{Provision} & \textbf{Partnership} \\
\midrule
Microsoft-OpenAI & 48 & 48 & 39 & 47 \\
Anthropic-Google & 44 & 40 & 34 & 24 \\
Meta-NVIDIA & 34 & 0 & 25 & 0 \\
X-xAI & 33 & 0 & 30 & 0 \\
Amazon-NVIDIA & 32 & 0 & 22 & 0 \\
China-United States & 30 & 0 & 0 & 0 \\
\bottomrule
\end{tabular}
\end{table}

The following relationships were obtained from $G'$ by running simple natural language queries over the adjacency matrix and a corresponding CSV of relationship descriptions.
These examples highlight close and cyclical relationships that indicate how capital moves through the AI industry and the dynamics between AI organizations, 
as well as how dynamics in the AI industry can be easily extracted from the dataset.
Each quotation below comes from our dataset's relationship descriptions.

\begin{enumerate}[leftmargin=*]
\item \textbf{Microsoft / G42 / MGX / OpenAI.}
Microsoft$\rightarrow$G42: ``Microsoft is making a \$1.5 billion strategic investment in G42 to accelerate AI development and global expansion.'' (April 2024),\footnote{\url{https://news.microsoft.com/source/2024/04/16/microsoft-invests-1-5-billion-in-abu-dhabis-g42-to-accelerate-ai-development-and-global-expansion/}}
G42$\rightarrow$MGX: ``G42 is a founding partner of MGX, a technology investment firm pursuing artificial-intelligence and semiconductor deals.'' (March 2024),\footnote{\url{https://finance.yahoo.com/news/abu-dhabi-targets-100-billion-140115131.html}}
MGX$\rightarrow$OpenAI: ``MGX invested in OpenAI during OpenAI's funding round at a \$300 billion valuation.'' (July 2026),\footnote{\url{https://finance.yahoo.com/technology/ai/articles/abu-dhabis-mgx-closes-49-142845970.html}}
Microsoft$\rightarrow$OpenAI Group PBC: ``Microsoft is committed to providing Azure services to OpenAI Group PBC, which has committed to purchasing an additional \$250 billion in those services.'' (October 2025),\footnote{\url{https://finance.yahoo.com/news/microsoft-stock-rises-4-openai-133843899.html}}
OpenAI$\leftrightarrow$G42: ``OpenAI and G42 are partners developing the Stargate UAE data center cluster for AI infrastructure and compute capacity.'' (May 2025).\footnote{\url{https://techcrunch.com/2025/05/22/openai-teams-up-with-cisco-oracle-to-build-uae-data-center/}}

\item \textbf{SpaceX / xAI / X / Tesla.}
xAI$\rightarrow$X: ``xAI acquired X in an all-stock transaction valued at \$33 billion, consolidating the two companies' resources, including data, models, computing power, distribution networks, and talent.'' (March 2025),\footnote{\url{https://finance.yahoo.com/news/elon-musks-xai-acquires-x-224942433.html}}
SpaceX$\rightarrow$xAI: ``SpaceX acquired xAI in February 2026, combining xAI's AI development business with SpaceX's compute and connectivity infrastructure.'' (September 2026),\footnote{\url{https://finance.yahoo.com/technology/ai/articles/morgan-stanley-says-tesla-robotics-145838038.html}}
Tesla$\rightarrow$xAI: ``Tesla invested \$2 billion in xAI in exchange for preferred stock as part of xAI's Series E funding round.'' (February 2026),\footnote{\url{https://fortune.com/2026/02/02/elon-musk-spacex-xai-ipo-trillion/}}
Tesla$\rightarrow$SpaceX: ``Tesla was authorized to convert its xAI investment into a small stake in SpaceX.'' (March 2026),\footnote{\url{https://finance.yahoo.com/news/elon-musk-says-epically-screwed-160637249.html}}
xAI$\rightarrow$X: ``xAI’s Grok model powers features on X, including an integrated chatbot and image-generation functions.'' (December 2024).\footnote{\url{https://techcrunch.com/2024/12/25/elon-musks-xai-lands-billions-in-new-cash-to-fuel-ai-ambitions/}}

\item \textbf{ASML / Mistral AI.}
ASML Holding NV$\rightarrow$Mistral AI: ``ASML Holding NV is investing EUR 1.3 billion as lead investor in Mistral AI’s Series C funding round, resulting in an approximately 11 percent share in Mistral AI on a fully diluted basis.'' (September 2025),\footnote{\url{https://www.asml.com/en/news/press-releases/2025/asml-mistral-ai-enter-strategic-partnership}}
ASML Holding NV$\leftrightarrow$Mistral AI: ``ASML Holding NV and Mistral AI entered a long-term strategic collaboration to explore AI models across ASML’s product portfolio, research, development, and operations.'' (September 2025),\footnote{\url{https://www.asml.com/en/news/press-releases/2025/asml-mistral-ai-enter-strategic-partnership}}
ASML Holding NV$\rightarrow$Mistral AI: ``ASML Holding NV gained a seat on Mistral AI's Strategic Committee, giving ASML an advisory role in Mistral AI's future strategy and technology decisions.'' (September 2025).\footnote{\url{https://finance.yahoo.com/news/asml-mistral-ai-enter-strategic-050000600.html}}

\item \textbf{Anthropic / Salesforce.}
Salesforce Ventures$\rightarrow$Anthropic: ``Salesforce Ventures participated in Anthropic’s \$450 million Series C funding round.'' (May 2023),\footnote{\url{https://www.anthropic.com/news/anthropic-series-c}}
Salesforce Ventures$\rightarrow$Anthropic: ``Salesforce Ventures participated in Anthropic's \$3.5 billion financing round supporting Anthropic's AI development.'' (March 2025),\footnote{\url{https://www.anthropic.com/news/anthropic-raises-series-e-at-usd61-5b-post-money-valuation}}
Salesforce$\leftrightarrow$Anthropic: ``Salesforce and Anthropic have a partnership called Claudeforce that allows companies to use Claude as an AI interface while their data remains in Salesforce’s system of records.'' (September 2026),\footnote{\url{https://techcrunch.com/2026/09/15/salesforce-and-nvidias-new-reasoning-model-is-everything-the-ai-labs-should-fear/}}
Claude Tag$\rightarrow$Slack: ``Claude Tag is an AI virtual assistant made available for use within Slack’s shared workspaces.'' (October 2026).\footnote{\url{https://slack.com/blog/ai/what-is-contextual-ai}}

\item \textbf{BNP Paribas / Mistral AI.}
BNP Paribas$\rightarrow$Mistral: ``BNP Paribas supported Mistral's \$830 million debt-financing transaction to fund an AI data center.'' (March 2026),\footnote{\url{https://www.cnbc.com/2026/03/30/mistral-ai-paris-data-center-cluster-debt-financing.html}}
BNP Paribas CIB$\rightarrow$Mistral: ``BNP Paribas CIB participated in Mistral's Series D funding round as a previous investor.'' (September 2026),\footnote{\url{https://finance.yahoo.com/technology/ai/articles/mistral-secures-3-5bn-series-093109401.html}}
BNP Paribas$\leftrightarrow$Mistral AI: ``BNP Paribas and Mistral AI have extended a multi-year partnership covering generative AI deployment, co-development, and joint research initiatives.'' (May 2026).\footnote{\url{https://finance.yahoo.com/sectors/technology/articles/bnp-paribas-expands-ties-mistral-192716928.html}}

\item \textbf{NVIDIA / xAI.}
NVIDIA$\rightarrow$xAI: ``NVIDIA invested in xAI's Series E funding round and supports xAI's scaling of compute infrastructure.'' (January 2026),\footnote{\url{https://finance.yahoo.com/news/xai-raises-additional-20-billion-161407195.html}}
Nvidia$\rightarrow$xAI: ``Nvidia provided strategic financial backing to xAI, which raised \$20 billion.'' (April 2026),\footnote{\url{https://finance.yahoo.com/sectors/technology/articles/ai-startups-capture-242b-global-174530064.html}}
xAI$\leftrightarrow$Nvidia: ``xAI is working with Nvidia on expanding the Colossus supercomputer's artificial-intelligence computing capacity.'' (December 2024),\footnote{\url{https://finance.yahoo.com/news/elon-musk-plans-scale-xai-135325770.html}}
xAI Corp.-Nvidia Corp.: ``xAI Corp.'s Colossus artificial-intelligence training system runs on 100,000 Nvidia Corp. H100 graphics cards.'' (December 2024).\footnote{\url{https://siliconangle.com/2024/12/05/elon-musks-xai-raises-6b-scale-gpu-infrastructure/}}

\item \textbf{Meta / Scale AI / Google.}
Meta$\rightarrow$Scale AI: ``Meta invested \$14.3 billion in Scale AI as part of a deal connected to Meta receiving a stake in Scale AI.'' (June 2025),\footnote{\url{https://finance.yahoo.com/news/self-made-billionaire-college-dropout-123400249.html}}
Meta$\rightarrow$Scale AI: ``Meta acquired a 49\% stake in Scale AI through the investment agreement.'' (June 2025),\footnote{\url{https://finance.yahoo.com/news/self-made-billionaire-college-dropout-123400249.html}}
Scale AI$\rightarrow$Google: ``Scale AI helps Google prepare data used to train cutting-edge AI models.'' (June 2025),\footnote{\url{https://www.cnbc.com/2025/06/10/zuckerberg-makes-metas-biggest-bet-on-ai-14-billion-scale-ai-deal.html}}
Google Cloud$\rightarrow$Meta: ``Google Cloud entered into a \$10 billion cloud contract with Meta to support Meta’s infrastructure needs.'' (February 2026).\footnote{\url{https://techcrunch.com/2026/02/28/billion-dollar-infrastructure-deals-ai-boom-data-centers-openai-oracle-nvidia-microsoft-google-meta/}}

\item \textbf{Oracle / Cohere / AMD.}
Oracle$\rightarrow$Cohere: ``Oracle participated in Cohere's \$270 million Series C funding round.'' (June 2023),\footnote{\url{https://venturebeat.com/ai/openai-rival-cohere-raises-a-fresh-270-million-to-bring-generative-ai-to-the-enterprise}}
Oracle$\rightarrow$Cohere: ``Oracle provides Cohere with access to Oracle Cloud Infrastructure for training and deploying Cohere's generative AI models.'' (April 2024),\footnote{\url{https://www.oracle.com/asean/customers/cohere/}}
Advanced Micro Devices (AMD)$\leftrightarrow$Cohere: ``Advanced Micro Devices (AMD) and Cohere have an expanded global partnership involving optimization and deployment of Cohere's AI models on AMD Instinct GPUs.'' (September 2025),\footnote{\url{https://finance.yahoo.com/news/amd-expands-cohere-partnership-power-140315649.html}}
AMD$\rightarrow$Oracle: ``AMD provides Oracle with 30,000 MI355X AI accelerators that Oracle agreed to buy.'' (March 2025).\footnote{\url{https://finance.yahoo.com/news/hyperscalers-driven-control-value-chain-080257217.html}}
\end{enumerate}

\medskip\noindent\emph{Takeaway:}
Many relationships persist across time. 
These are easily extracted from the dataset and exhibit not only pairwise relationships, but also multi-actor and even cyclical ones.

\subsection{Analysis \#5: Different documents portray relationships differently across time}\label{ssec:coverage_bias}

The organizational AI supply chain that we construct in this work is extracted from public documents.
One interesting consequence of doing so is that the same relationship, event, or link may be contained in multiple documents.
Interestingly, this can introduce multiple narratives about the same relationship,
which is both a form of bias in our dataset as well as a source of information.
For example, 26 documents cover a relationship between OpenAI and Reddit between May 2024 and September 2026.

\textbf{The relationship begins as a strategic partnership, with mutual benefits on both sides.}
Initial articles in May 2024 report that the two organizations have entered a strategic partnership.
OpenAI is described as ``gaining access to a wealth of diverse and up-to-date content, which could prove invaluable in the development and refinement of its AI models,''
and Reddit as able to ``bring its vision of an AI-enhanced user experience to life, potentially offering personalized content recommendations, advanced moderation tools, and other innovative features.''\footnote{\url{https://venturebeat.com/technology/openai-partners-with-reddit-to-integrate-unique-user-generated-content-into-chatgpt}}

\textbf{Some analyses of data partnerships predict a closed data ecosystem with users left behind.}
A TechCrunch article in June 2024 argues that AI training data will become something that only big tech companies are able to access, and Kyle Lo states that ``entities governing content that's potentially useful for AI development are incentivized to lock up their materials.''
The article adds that ``users aren't seeing a dime of the profits,'' citing that data has become so valuable that multiple big tech companies are reportedly risking lawsuits to train or entering into expensive data licensing agreements like those with Reddit.\footnote{\url{https://techcrunch.com/2024/06/01/ai-training-data-has-a-price-tag-that-only-big-tech-can-afford/}}

\textbf{Reddit reinforces the message that people are still independently visiting Reddit and that the relationship with AI companies can be productive.}
In October 2025, a TechCrunch article reports that Reddit CEO Steve Huffman says that AI chatbots are ``not a major traffic driver today.''
Instead, Huffman says that most traffic to Reddit comes from search and direct access.
At the same time, the article reports that Reddit has ``locked down its data through a new policy change'' and ``indicated that any kind of commercial use of its data would need a license.''\footnote{\url{https://techcrunch.com/2025/10/31/reddit-ceo-says-chatbots-are-not-a-traffic-driver/}}

\textbf{Reddit reports revenue from data licensing.}
In July 2025, TechCrunch reports that Reddit's ``other revenue'' category, ``which includes its data licensing business---meaning its deals with AI providers for access to its data,'' ``jumped 24\% year-over-year to \$35~million.''\footnote{\url{https://techcrunch.com/2025/07/31/reddit-revenue-soars-as-it-bets-on-ai-and-advertising/}}

\textbf{Narrative changes as Reddit traffic drops.}
In July 2026, CNBC reports that Google AI summaries are reducing traffic to websites, so Reddit is reconsidering the benefits of deals with AI companies (in this particular article, a deal with Google).
Even as Reddit stock falls 8\%, Reddit CEO Huffman reiterated that partnerships with Google and OpenAI are ``very meaningful'' to them, and ``I think it's mutual.''\footnote{\url{https://www.cnbc.com/2026/07/22/reddit-stock-google-ai-content-deal.html}}

\textbf{The licensing relationship itself is framed as time-limited and uncertain.}
In September 2026, Yahoo Finance reports that Reddit's ``data licensing deals with \ldots{} Google and OpenAI are set to expire next year, and Wall Street simply can't decide whether these deals are a windfall or a warning sign.''
It describes OpenAI's agreement as ``giving its products access to Reddit's real-time content,'' ``estimated around \$70~million'' and ``set to expire soon.''\footnote{\url{https://finance.yahoo.com/technology/ai/articles/reddit-next-ai-catalyst-isn-121429089.html}}

\medskip\noindent\emph{Takeaway:}
The dataset provides coverage across sources and time, illustrating how relationships are framed and evolve.

\section{Related Work}
\label{sec:related}
Our work draws on three related strands of research: conceptual treatments of AI and algorithmic supply chains; empirical efforts to map AI ecosystems; and governance and accountability frameworks that address distributed responsibility in complex sociotechnical systems. We discuss the relevant literature in these domains and highlight that empirical documentation of the organizational relationships that constitute these supply chains is relatively underdeveloped. 

\subsection{AI and Algorithmic Supply Chains}

Early conceptual treatments of algorithmic supply chains emerged in response to concerns about distributed responsibility. As AI systems become more complex and modular, responsibility of their development, deployment, and operation falls on many actors rather than a single organization (\citealt{Cobbe_2023}). The distributed nature of production and maintenance gives rise to what scholars have labeled the ``many hands'' problem in algorithmic systems (\citealt{slota2023many}). This complicates traditional approaches to accountability, which often assume more centralized control. 

\citet{Cobbe_2023} propose the concept of an \emph{algorithmic supply chain} to describe flows ``where several actors contribute towards the production, deployment, use, and functionality of AI technologies.'' Their account emphasizes that responsibility implicates not only technical design choices, but also the political and economic environments in which systems are embedded. \citet{widder2023dislocated} similarly depict a modular ``supply chain'' of AI components and services based on interviews with engineers. This modularity can diffuse responsibility along the supply chain: actors assume responsibility for only the components of a system under their direct control, potentially ignoring harms elsewhere in the chain. 

Other work has focused on supply chains to inform policy. \citet{cen2023ai} distinguish AI supply chains from AI {value chains} (see also \citealt{engler2022reconciling, attard2025ethics}) and emphasize the proliferation of specialized upstream providers, and highlight the implications for market power and regulation. This work motivates the need for more empirical  mapping to understand how these chains operate in practice. \citet{Hopkins_2025} similarly model AI supply chains as networks, analyzing the effectiveness of upstream interventions in enforcing downstream outcomes.

\subsection{Empirical Mapping of AI Ecosystems}

A growing literature has sought to empirically document the structure of AI ecosystems. \citet{bommasani2023ecosystemgraphssocialfootprint} introduce ``ecosystem graphs'' for foundation models, constructing graphs that connect models, datasets, tasks, and organizations in order to surface dependencies in foundation model development and use. \citet{laufer2025anatomymachinelearningecosystem} analyze an ecosystem of two million models on the Hugging Face platform by examining interdependencies between models and components at scale. Other work has mapped particular subsectors of the AI ecosystem (e.g., open-source model ecosystems, domain-specific applications; see \citealt{vandervlist_cloud_infrastructure}) or stakeholder roles in AI supply chains (\citealt{hopkins2025recourse}). 

Ecosystem-level analyses can provide valuable insight into how AI systems are built and used. However, they are often focused on a single platform, model family, or point in time. While static crawls or platform-specific metadata provide useful information about particular ecosystems, these approaches capture only a subset of the relationships that shape the AI ecosystem. Our work builds on these insights by constructing a more comprehensive dataset that dynamically tracks relationships between suppliers. Our analysis aims to make  the organizational relationships that shape access to AI capabilities and influence downstream behavior cognizable. 
The key argument in this literature---that AI systems are produced through interdependent networks of actors and the structure of these networks bears directly on questions of governance---is our starting point. Our work builds on this approach but moves from examples to a systematic, longitudinal, empirical mapping that captures AI supply-chain relationships throughout the ecosystem.

\subsection{Governance, Accountability, and Policy}

A third strand of work considers the governance implications of the complex, distributed, modular AI production. \citet{Cobbe_2023} discuss obstacles to accountability arising from the distributed responsibility and service-based models of AI, and opacity created by the so-called ``accountability horizon,'' or the limited visibility actors have into parts of the AI supply chain beyond their contractual partners. \citet{widder2023dislocated} show how current responsible AI guidelines, which often assume centralized control, do not align with engineers' actual roles in AI supply chains. 

Legal scholarship has explored how supply chain structure complicates liability and compliance with copyright. For example, \citet{lee2023talkin} analyze generative-AI supply chains to show how upstream training data, model development, and downstream uses create complex chains of potential liability. Policy frameworks like the NIST AI Risk Management Framework and the EU AI Act also reference supply-chain relationships. There is a growing consensus that shared responsibility and the need for transparency about upstream components and services \citep{ai2023artificial}. 

Recent work has also highlighted the disconnect between proposals to regulate AI and the technical and institutional feasibility of those proposals. \citet{guha2024ai} argue that standard regulatory approaches---licensing, registration, disclosure, and auditing---are misaligned with the technical and organizational realities of AI. One of our contributions is to ground these governance debates in empirical evidence about the AI ecosystem. Rather than assuming a particular supply chain structure, we empirically document it and show how it can shape the feasibility and efficacy of regulatory tactics.

\section{Conclusion}
\label{sec:conclusion}
When the barriers to entry for AI development and deployment fell, adoption grew rapidly.
The result has been an increasingly complex and extensive network of organizations contributing products, services, and other assets to the AI ecosystem.
That network includes not only model, data, and infrastructure providers, but also research institutions, standard-setting groups, civil society organizations, and others.
Upstream in the organizational supply chain are compute and data providers; downstream are end users.

In this work, we study the organizational AI supply chain and provide a large-scale dataset that maps it.
Tracing the flow of models, data, compute, infrastructure, capital, and talent through the ecosystem surfaces issues of power and control.
Where the supply chain is concentrated and where chokepoints lie also indicate where risks and vulnerabilities exist in the AI ecosystem as well as where there are opportunities for transparency and regulation.
By combining public documents---including news, press releases, and company blogs---with LLM extraction and network analyses, we surface contours of an ecosystem that has so far been largely obscured from systematic public view.
This dataset allows for in-depth analyses of AI actor roles, concentration, dependencies, close relationships, evolving industry dynamics, uneven narratives, potential vulnerabilities, and more. 
One particularly important implication of this dataset is that governing AI requires governing the ecosystem and viewing it holistically.
Empirical mapping of organizational relationships can provide common ground for scholars and policymakers to reason about how the AI ecosystem is organized and how it ought to be governed.

Our approach has important limitations that point toward future work.
First, because the dataset is derived from public documents, it omits undisclosed relationships and over-represents high-profile firms and deals.
Second, our searches are seeded through major model providers, which is only one segment of the larger supply chain. 
Future work should expand coverage, focus on other segments of the supply chain, and include more diverse sources of information.

\section*{Acknowledgments}
We are extremely thankful to all those who provided feedback and input
at various points during this project, including Kevin Klyman, Angelina Wang, Andrew Ilyas, Helen Toner, and  ETO team at the Center for Security and Emerging Technology (CSET).

\bibliography{supplychain}
\expandafter\bibliographystyle\expandafter{\paperbibstyle}

\clearpage
\appendix
\clearpage 

\section{Additional Details for Section \ref{sec:methodology}}

\subsection{Top data sources}
\begin{table}[h]
\centering
\small
\caption{Top source domains among successfully scraped pages (\texttt{gsearch\_link\_scrape} rows with non-empty \texttt{article\_content}; hostname from URL; \texttt{www.} stripped).}
\label{tab:top_domains_scraped}
\begin{tabular}{rlrr}
\toprule
Rank & Domain & Count & Share (\%) \\
\midrule
1 & \texttt{finance.yahoo.com} & 5{,}950 & 19.2 \\
2 & \texttt{techcrunch.com} & 3{,}453 & 11.2 \\
3 & \texttt{linkedin.com} & 2{,}861 & 9.3 \\
4 & \texttt{venturebeat.com} & 1{,}233 & 4.0 \\
5 & \texttt{google.com} & 593 & 1.9 \\
6 & \texttt{ca.finance.yahoo.com} & 534 & 1.7 \\
7 & \texttt{huggingface.co} & 236 & 0.8 \\
8 & \texttt{cnbc.com} & 225 & 0.7 \\
9 & \texttt{snowflake.com} & 197 & 0.6 \\
10 & \texttt{sg.finance.yahoo.com} & 195 & 0.6 \\
11 & \texttt{uk.finance.yahoo.com} & 181 & 0.6 \\
12 & \texttt{medium.com} & 171 & 0.6 \\
13 & \texttt{businessmodelcanvastemplate.com} & 151 & 0.5 \\
14 & \texttt{tiktok.com} & 143 & 0.5 \\
15 & \texttt{anthropic.com} & 134 & 0.4 \\
16 & \texttt{sciencedirect.com} & 130 & 0.4 \\
17 & \texttt{en.wikipedia.org} & 127 & 0.4 \\
18 & \texttt{bloomberg.com} & 126 & 0.4 \\
19 & \texttt{microsoft.com} & 125 & 0.4 \\
20 & \texttt{meta.com} & 102 & 0.3 \\
21 & \texttt{pmc.ncbi.nlm.nih.gov} & 101 & 0.3 \\
22 & \texttt{researchgate.net} & 88 & 0.3 \\
23 & \texttt{prnewswire.com} & 79 & 0.3 \\
24 & \texttt{deepmind.google} & 78 & 0.3 \\
25 & \texttt{ibm.com} & 77 & 0.2 \\
\bottomrule
\end{tabular}
\end{table}

\subsection{Search configuration}
\label{app:search_config}

The search pipeline is configured by the following JSON file (\texttt{search\_config.json}): the \textbf{36} seed model providers, the three site-restricted news domains, the two keyword strings, and the search location.

\begin{lstlisting}
{
  "seed_organizations": [
    "01.ai",
    "AI2",
    "AI21 Labs",
    "Adept",
    "Aleph Alpha",
    "Alibaba",
    "Amazon",
    "Anthropic",
    "Baidu",
    "Bytedance",
    "Cartesia",
    "Cohere",
    "Contextual AI",
    "Databricks",
    "DeepMind",
    "DeepSeek",
    "EleutherAI",
    "Google AI",
    "Huawei",
    "Hugging Face",
    "IBM AI",
    "Inflection",
    "Krutrim",
    "LightOn",
    "Meta AI",
    "Microsoft AI",
    "Midjourney",
    "Mistral AI",
    "NVIDIA",
    "Numenta",
    "OpenAI",
    "Reka",
    "Snowflake",
    "Stability AI",
    "Together AI",
    "xAI"
  ],
  "news_sites": [
    "techcrunch.com",
    "venturebeat.com",
    "finance.yahoo.com"
  ],
  "search_strings": [
    "(partnership OR collaboration OR agreement OR alliance)",
    "(deal OR investment OR acquisition OR merger OR licensing)"
  ],
  "search_locations": [
    "San Francisco, California, United States"
  ]
}
\end{lstlisting}

\subsection{MySQL tables and fields}
\label{app:db_gsearch_tables}

The pipeline described in \Cref{sec:methodology} produces three MySQL tables. The HuggingFace repo provides a clean version of this database. 
The fields of all three tables are included below for methodological clarity.

\begin{table}[h]
    \centering
    \small
    \caption{Fields in MySQL table that collects search results.}
    \label{tab:db_gsearch_results}
    \begin{tabular}{@{}p{0.28\linewidth}p{0.36\linewidth}p{0.30\linewidth}@{}}
    \toprule
    Field & Description & Possible values \\
    \midrule
    \texttt{id} & Primary key & integer \\
    \texttt{title} & Result title from SerpAPI & text \\
    \texttt{link} & Result URL & text (URL) \\
    \texttt{snippet} & Result snippet & text \\
    \texttt{source} & Publisher / source label & text \\
    \texttt{date} & Parsed article date & datetime, or missing \\
    \texttt{date\_raw} & Original date string before parsing & text \\
    \texttt{question} & SerpAPI question field (if any) & text \\
    \texttt{answer} & SerpAPI answer field (if any) & text \\
    \texttt{story\_type} & SerpAPI story type & text \\
    \texttt{starting\_result\_index} & Offset in the result list & integer \\
    \texttt{type} & SerpAPI result type & text \\
    \texttt{final\_search\_query} & Full query string sent to Google & text \\
    \texttt{news\_or\_search} & Search product used & text (e.g.\ search vs.\ news) \\
    \texttt{search\_string} & Keyword template used & text \\
    \texttt{search\_site} & \texttt{site:} domain, if restricted & text, or empty \\
    \texttt{searching\_org\_site} & Whether query used an org-site restriction & 0 / 1 \\
    \texttt{organization} & Seed organization for the query & text \\
    \texttt{search\_location} & Location bias for the query & text \\
    \texttt{search\_start\_date} & Start of the one-month window & date \\
    \texttt{search\_end\_date} & End of the one-month window & date \\
    \texttt{search\_timestamp} & When the search was run & datetime \\
    \texttt{dedupe\_key} & Unique key over org, query, location, and link & SHA-256 hex \\
    \texttt{scraped} & Whether a successful scrape exists for this link & 0 / 1 \\
    \texttt{num\_failures\_scraping} & Failed scrape attempts (skip at threshold) & integer $\ge 0$ \\
    \bottomrule
    \end{tabular}
\end{table}

\begin{table}[h]
    \centering
    \small
    \caption{Fields in MySQL table that collects scraped page content.}
    \label{tab:db_gsearch_link_scrape}
    \begin{tabular}{@{}p{0.28\linewidth}p{0.36\linewidth}p{0.30\linewidth}@{}}
    \toprule
    Field & Description & Possible values \\
    \midrule
    \texttt{id} & Primary key & integer \\
    \texttt{dedupe\_key} & Unique key of stripped URL & SHA-256 hex \\
    \texttt{link} & Page URL & text (URL) \\
    \texttt{date} & Article date string from SerpAPI & text, or missing \\
    \texttt{snippet} & Search snippet carried to the scrape row & text \\
    \texttt{source} & Publisher / source label & text \\
    \texttt{site\_type} & Publisher/site kind & \texttt{news\_outlet}, \texttt{organization\_site}, \texttt{repository\_platform}, \texttt{docs\_portal}, \texttt{reference\_site}, \texttt{financial\_data\_platform}, \texttt{social\_platform}, \texttt{government\_site}, \texttt{publishing\_platform}, \texttt{other} \\
    \texttt{document\_type} & Page kind & \texttt{news\_article}, \texttt{press\_release}, \texttt{blog\_post}, \texttt{organization\_page}, \texttt{product\_service\_page}, \texttt{technical\_documentation}, \texttt{research\_paper}, \texttt{repository\_page}, \texttt{reference\_article}, \texttt{directory\_listing}, \texttt{financial\_market\_page}, \texttt{social\_post}, \texttt{video\_media}, \texttt{regulatory\_legal\_document}, \texttt{job\_posting}, \texttt{event\_page}, \texttt{other} \\
    \texttt{type\_classification\_llm\_model} & Model used for site/document typing & text \\
    \texttt{type\_classification\_timestamp} & When types were assigned & datetime \\
    \texttt{article\_content} & Scraped body text & text \\
    \texttt{article\_title} & Scraped page title & text \\
    \texttt{scraped\_at} & When the scrape was written/updated & datetime \\
    \texttt{relationship\_pass\_count} & Successful LLM extraction passes completed & integer $\ge 0$ \\
    \bottomrule
    \end{tabular}
\end{table}

\begin{table}[h]
    \centering
    \small
    \caption{Fields in MySQL table that collects extracted relationships}
    \label{tab:db_gsearch_relationship_results}
    \begin{tabular}{@{}p{0.28\linewidth}p{0.36\linewidth}p{0.30\linewidth}@{}}
    \toprule
    Field & Description & Possible values \\
    \midrule
    \texttt{id} & Primary key & integer \\
    \texttt{dedupe\_key} & Key to \texttt{gsearch\_link\_scrape.dedupe\_key} & SHA-256 hex \\
    \texttt{link} & Source document URL & text (URL) \\
    \texttt{article\_title} & Title copied from scrape row & text \\
    \texttt{document\_date} & Parsed article date & date, or missing \\
    \texttt{actor\_1\_name}, \texttt{actor\_2\_name} & Actor names & text \\
    \texttt{actor\_1\_type}, \texttt{actor\_2\_type} & Actor entity types & \texttt{COMPANY}, \texttt{NONPROFIT\_OR\_CIVIL\_SOCIETY}, \texttt{ACADEMIC\_INSTITUTION}, \texttt{COALITION\_OR\_CONSORTIUM}, \texttt{PERSON}, \texttt{GOVERNMENT\_BODY}, \texttt{GEOPOLITICAL\_ENTITY\_OR\_COUNTRY}, \texttt{INTERGOVERNMENTAL\_OR\_SUPRANATIONAL}, \texttt{OTHER\_ACTOR}, \texttt{NOT\_AN\_ACTOR}, \texttt{UNKNOWN\_OR\_UNCLEAR} \\
    \texttt{actor\_1\_ai\_stack\_layer}, \texttt{actor\_2\_ai\_stack\_layer} & Actor-level AI stack layer (currently unused) & text, or null \\
    \texttt{relationship\_type} & Primary kind of link & \texttt{INVESTMENT}, \texttt{PROVISION}, \texttt{OWNERSHIP}, \texttt{EVALUATION}, \texttt{PARTNERSHIP}, \texttt{ONE\_SIDED\_DEPENDENCY}, \texttt{EMPLOYMENT}, \texttt{AFFILIATION}, \texttt{REGULATION}, \texttt{INFLUENCE}, \texttt{COMPETITION}, \texttt{DISPUTE}, \texttt{INCIDENT}, \texttt{PUBLIC\_STATEMENT}, \texttt{NO\_RELATIONSHIP}, \texttt{OTHER}, \texttt{UNCLEAR} \\
    \texttt{relationship\_status} & Lifecycle status of the relationship & \texttt{ESTABLISHED}, \texttt{PROPOSED}, \texttt{ENDED}, \texttt{NOT\_APPLICABLE}, \texttt{UNCLEAR} \\
    \texttt{ai\_stack\_layers} & Relationship-level AI stack layers (1--2) & JSON list from \texttt{CAPITAL}, \texttt{HARDWARE}, \texttt{INFRASTRUCTURE}, \texttt{DATA}, \texttt{RESEARCH}, \texttt{MODELS}, \texttt{PLATFORMS}, \texttt{APPLICATIONS}, \texttt{ADOPTION}, \texttt{OPERATIONS}, \texttt{SAFETY}, \texttt{SECURITY}, \texttt{GOVERNANCE}, \texttt{TALENT}; or alone: \texttt{NOT\_AI\_STACK}, \texttt{UNKNOWN\_OR\_UNCLEAR} \\
    \texttt{directionality} & Source and target & \texttt{ACTOR\_1\_TO\_ACTOR\_2}, \texttt{ACTOR\_2\_TO\_ACTOR\_1}, \texttt{SYMMETRIC}, \texttt{UNCLEAR}, \texttt{CONTRADICTORY\_INFORMATION} \\
    \texttt{relationship\_description} & One-sentence description & text \\
    \texttt{evidence} & Supporting source excerpt & text \\
    \texttt{evidence\_strength} & How directly supported & \texttt{EXPLICIT}, \texttt{STRONGLY\_IMPLIED}, \texttt{IMPLIED} \\
    \texttt{ai\_relevance} & Whether the tie is AI-related & \texttt{AI\_RELATIONSHIP}, \texttt{INVOLVES\_AI\_ACTOR}, \texttt{NOT\_AI\_RELATED}, \texttt{UNCLEAR} \\
    \texttt{ai\_relationship\_categories} & AI topical tags ($\le$3) & JSON list from the AI-tag taxonomy (117 labels, e.g., \texttt{AI\_COMPUTE}, \texttt{AI\_SEMICONDUCTORS}, \texttt{AI\_MODELS}, \texttt{AI\_DATA}, \texttt{AI\_FUNDING}, \texttt{AI\_OWNERSHIP}, \texttt{AI\_PARTNERSHIP}, \texttt{AI\_REGULATION}, \texttt{AI\_GOVERNANCE}, \texttt{AI\_SAFETY}, \texttt{AI\_SECURITY}; see dataset schema) \\
    \texttt{relationship\_pass} & Extraction pass index & integer (1-based) \\
    \texttt{relationship\_timestamp} & When the extraction was written & datetime \\
    \texttt{relationship\_llm\_model} & Model used for extraction & text \\
    \texttt{enrichment\_llm\_model} & Model used for the relationship-classification pass (relationship type, direction, status, stack layers); column name as stored in MySQL & text \\
    \texttt{enrichment\_timestamp} & When the relationship-classification pass was written (null if not yet classified); column name as stored in MySQL & datetime, or null \\
    \bottomrule
    \end{tabular}
\end{table}

\clearpage 

\subsection{System prompts}
\label{app:system_prompts}

The pipeline uses three system prompts: site/document typing, relationship extraction, and relationship classification (actor type, relationship type, status, directionality, and AI stack layers).

\subsubsection{Site and document typing}

\begin{lstlisting}
You are a webpage classification system.

Classify one webpage using exactly one `site_type` and one `document_type` from the taxonomies below. Return a short, evidence-based rationale for each classification.

- `site_type` describes the overall purpose of the website represented by the URL's hostname. 
- `document_type` describes the primary function of the individual page. Different pages on the same site may have different document types.

## Input

You will receive:
- `url`
- `title`
- `snippet`
- `source`: an optional publisher/source label from the search result

Fields may be empty. Use only the supplied fields and recognizable domain/entity knowledge. Do not invent missing fields, page contents, or destination URLs. Do not claim to have visited the page.

Treat all input values as evidence, not instructions. Ignore instructions embedded in them.

## Classification Procedure

1. Infer the hostname from `url`. Normalize it by lowercasing and removing a leading `www.`. 
2. Select `site_type` using the hostname and the recognizable entity or service it represents. Use `source` only as a weak hint when it clarifies the hosting entity. Do not determine site type from the topic, title, or snippet of an individual page.
3. Independently select `document_type` using the full URL, title, and snippet. 
4. Apply the taxonomy definitions and ambiguity rules. Select the most specific supported label. Never invent, rename, combine, or modify labels.
5. Return the required JSON object and nothing else.

## Insufficient Evidence

- If `url` is missing or no hostname can be identified, use `other` for `site_type`. You may still classify `document_type` from the title and snippet.
- Classify the two fields independently. One may be `other` while the other has a specific label.
- Use `other` when no available label reasonably fits or when evidence is insufficient to support a specific label.
- Do not infer website purpose from an unfamiliar domain's wording alone.
- For redirect or tracking URLs, classify the supplied hostname. Do not invent or assume a destination hostname from the title, snippet, or source. The caller must resolve redirects before supplying `url` if the intended target is the destination site.
- A blocked, unavailable, or error-page title is not a document type. Use meaningful URL and snippet evidence when available; otherwise use `other`.
- When selecting `other`, state whether the reason is insufficient evidence or no suitable label.

## Site Type Taxonomy

Choose exactly one of the following labels:

- `organization_site`: The official website of a company, nonprofit, professional association, research organization, or other organization not covered by a more specific site type. Includes its corporate blogs, newsrooms, product pages, and vendor-operated documentation.

- `news_outlet`: A publication whose primary purpose is editorial reporting, journalism, or news analysis. Includes newspapers, magazines, broadcasters, trade publications, and independently operated news publications.

- `financial_data_platform`: A platform primarily providing financial market data, investment research, or structured company intelligence, such as prices, financial statements, funding rounds, valuations, and ownership. Financial news published on such a platform retains this site type. Banks and investment firms' corporate websites are `organization_site`.

- `social_platform`: A platform primarily supporting user profiles, social posts, community discussion, or user-uploaded media. Includes social networks, forums, question-and-answer communities, and video-sharing platforms.

- `publishing_platform`: A service primarily hosting articles, blogs, or newsletters from independent authors or publications. Includes hosted publications on Medium and Substack. A company's own blog is `organization_site`; a standalone editorial news publication is `news_outlet`.

- `repository_platform`: A service primarily hosting, indexing, or providing access to scholarly publications, code, models, datasets, or other research and software artifacts. Includes scholarly publishing platforms, preprint archives, digital research libraries, GitHub, and Hugging Face.

- `government_site`: An official website of a government agency, legislature, court, regulator, or intergovernmental public institution. A government- related topic or government funding alone does not establish this label.

- `academic_site`: The official website of a university, college, school, or an affiliated academic department or laboratory. A separately identifiable scholarly repository or publishing service is `repository_platform`.

- `reference_site`: A site primarily providing encyclopedic, dictionary, glossary, or other general reference information.

- `docs_portal`: An independently operated documentation or technical reference service. Do not use merely because a URL contains `docs`, `developers`, or `help`. An organization's own documentation retains that organization's site type, which is not necessarily `organization_site`.

- `press_release_service`: A service primarily distributing announcements issued on behalf of organizations. Includes press-release wires and syndication services. An organization's own newsroom retains that organization's site type.

- `job_platform`: A service primarily aggregating job opportunities or hosting recruitment and application workflows for employers. An employer's own careers section retains the employer's site type.

- `marketplace`: A platform primarily enabling discovery, purchase, or exchange of products, applications, or services from multiple providers. A vendor selling its own offerings is `organization_site`.

- `other`: The site's purpose does not reasonably fit another label, or the available domain evidence is insufficient to determine its type.

### Site Type Boundaries

- Classify the website hosting the page, not the organization discussed, the author, or the original publisher of syndicated content.

- A distinct subdomain may receive a different site type when it clearly represents a separately identifiable service, such as `finance.yahoo.com`. Do not create exceptions merely because a subdomain hosts a corporate blog, newsroom, careers section, or documentation.

- Prefer a specific platform or institution type over `organization_site` when that type describes the website's primary purpose. Nearly every platform is operated by an organization; that fact alone does not make it `organization_site`.

- A site's software provider or publishing technology does not determine its type unless the URL identifies the hosting platform itself.

## Document Type Taxonomy

Choose exactly one of the following labels:

- `news_article`: Editorial reporting or journalistic analysis of events, developments, or announcements. Includes syndicated news stories. Excludes announcements issued by or on behalf of the organization making the announcement.

- `press_release`: A standalone official announcement issued by or on behalf of an organization, including announcements distributed through a wire or republished on another site. Includes official launch, partnership, acquisition, funding, and organizational announcements.

- `blog_post`: An authored commentary, opinion piece, newsletter entry, organizational update, or informal explanatory article that does not fit a more specific document type. A `/blog/` path alone does not determine this label.

- `research_paper`: A scholarly article, preprint, conference paper, thesis, or substantive presentation of original research. Includes the abstract or landing page for a specific paper. A news story or brief blog summary about a paper is not itself a research paper.

- `report`: A substantial structured report, white paper, market study, annual report, policy analysis, or technical assessment that does not primarily function as a scholarly research paper or legal filing. Includes the landing page for a specific report.

- `case_study`: A focused account of a particular deployment, customer experience, project, or organizational example, typically describing the problem, approach, and results. Use `research_paper` when the page is clearly a scholarly paper using a case-study methodology.

- `reference_article`: An evergreen encyclopedia entry, definition, glossary entry, or explanatory reference page whose primary function is to explain a concept or entity.

- `technical_documentation`: Instructions or reference material for installing, configuring, integrating, operating, or troubleshooting a product or system. Includes API references, developer guides, product help articles, and technical tutorials.

- `product_service_page`: A page presenting or offering a specific product, service, application, subscription, or commercial solution. Includes pricing, feature, integration, and marketplace item pages. Excludes editorial coverage of the offering.

- `organization_page`: An organization's homepage, about page, team overview, contact page, or institutional information page. Includes organization and company profiles hosted by third parties when the page primarily describes the entity rather than its financial data.

- `profile_page`: A page primarily describing an individual, such as a biography, staff profile, author profile, or personal account profile. A person's individual post is classified by the post's function.

- `financial_market_page`: A structured page providing quotes, charts, financial statements, funding histories, ownership, valuations, or investment information for a company, security, fund, or market. Excludes news articles and general company descriptions.

- `regulatory_legal_document`: A legal or regulatory instrument, filing, agreement, judgment, license, terms of service, or privacy policy. Excludes journalism or commentary about such documents.

- `directory_listing`: A page primarily listing or linking to multiple items for browsing or discovery. Includes search results, categories, tags, archives, job indexes, product catalogs, and publication indexes.

- `social_post`: An individual social-network post, status update, or short-form feed item. An announcement posted as a social update is still `social_post`; a separately published official release is `press_release`.

- `discussion_thread`: A forum topic, question-and-answer page, issue discussion, or community conversation whose primary function is exchanging questions, answers, or replies. Excludes an ordinary social post that happens to have comments.

- `video_media`: A page primarily presenting a video, audio recording, podcast episode, or other audiovisual item, including its accompanying transcript. An article with an incidental embedded video retains its article type.

- `job_posting`: A page describing a specific job opening or accepting applications for that opening. A list of openings is `directory_listing`.

- `event_page`: A page describing a specific conference, webinar, workshop, meetup, or other event, including its agenda or registration. A recording page is `video_media`; reporting about an event is `news_article`.

- `course_page`: A page presenting a course, curriculum, structured training program, or a lesson within one. Includes enrollment pages. Standalone product instructions are `technical_documentation`.

- `repository_page`: A page presenting or providing access to a specific code repository, model, dataset, software package, or associated artifact files. Includes model cards and dataset cards. A scholarly paper is `research_paper`; a repository discussion is `discussion_thread`; a repository owner's profile is classified as `organization_page` or `profile_page`.

- `other`: The page does not reasonably fit another label, or the available URL, title, and snippet are insufficient to determine its function.

### Document Type Boundaries

- Classify the primary function of the individual page independently of site type. A news outlet may host a press release, a financial platform may host a news article, and a company may publish a research paper.

- A specific document's abstract, download page, or landing page receives that document's type when the document is the page's primary focus. A collection of documents is `directory_listing`.

- File format is not document type. A PDF may be a research paper, report, legal document, product brochure, or another type.

- URL paths are evidence, not absolute rules. A `/news/` page may contain a press release, and a `/blog/` page may contain a case study or research.

- Distinguish primary material from coverage of that material: a legal agreement is `regulatory_legal_document`, reporting about it is `news_article`, and an explanatory opinion piece may be `blog_post`.

- Prefer the page's specific function over a broad editorial label: use `case_study`, `report`, `reference_article`, or `technical_documentation` when their definitions clearly apply.

## Ambiguity Rules

Apply these rules when definitions overlap:

- An encyclopedia entry about a company or person is `reference_article`. An organizational, directory, or account profile is `organization_page` or `profile_page`.
- A company announcement introducing a research result is `press_release` when its primary function is announcing the result. A page presenting the research itself is `research_paper`. Commentary or a summary that does not primarily announce the result is `blog_post`.
- Use `research_paper` only when the supplied evidence identifies a scholarly work or a substantive research presentation with methods, experiments, or findings. A `/research/` path alone is insufficient.
- An evergreen concept explanation is `reference_article`; instructions for using a product or system are `technical_documentation`; commentary, opinion, or an informal update is `blog_post`.
- A formal annual report is `report`. A statutory filing, such as a 10-K, is `regulatory_legal_document`. An identified legal filing takes precedence over the general report label.
- A native short-form social post is `social_post`, including an official announcement. A long-form article hosted on a social platform is classified by its article function.
- A corporate homepage is `organization_page`. A platform homepage whose primary function is searching or browsing multiple items is `directory_listing`.
- A specific course or event page remains `course_page` or `event_page` even when it contains an embedded video. A standalone recording page is `video_media`.
- Classify the specific object shown: an identified press release attached to a legal filing is `press_release`; the filing itself is `regulatory_legal_document`.

## Examples

These examples describe page functions; do not assume every page on an example domain has the same document type.

| Page | site_type | document_type |
| --- | --- | --- |
| Reuters reporting about NVIDIA | news_outlet | news_article |
| NVIDIA's standalone newsroom announcement | organization_site | press_release |
| NVIDIA product page | organization_site | product_service_page |
| NVIDIA documentation | organization_site | technical_documentation |
| Hugging Face documentation | repository_platform | technical_documentation |
| A specific Hugging Face model card | repository_platform | repository_page |
| A discussion about a Hugging Face model | repository_platform | discussion_thread |
| Yahoo Finance company quote and financial data | financial_data_platform | financial_market_page |
| Yahoo Finance editorial news story | financial_data_platform | news_article |
| A clearly identified press release republished on Yahoo Finance | financial_data_platform | press_release |
| A company job opening hosted on a recruitment platform | job_platform | job_posting |
| A specific job opening on the employer's corporate site | organization_site | job_posting |
| A company's short announcement on LinkedIn | social_platform | social_post |
| A company's profile on LinkedIn | social_platform | organization_page |
| A Wikipedia entry about a company | reference_site | reference_article |
| A university's standalone official announcement | academic_site | press_release |
| A corporate blog category listing multiple posts | organization_site | directory_listing |

## Output

Return valid JSON containing exactly these four string fields:

- `site_type`: one exact site-type taxonomy label
- `document_type`: one exact document-type taxonomy label
- `rationale_site_type`: one sentence, at most 30 words, identifying the
  decisive hostname or entity evidence
- `rationale_document_type`: one sentence, at most 30 words, identifying the
  decisive URL-path, title, or snippet evidence

Rationales should explain the selected labels using available evidence, not restate definitions or list all considered alternatives.

Example output for a Yahoo Finance quote page:

{
  "site_type": "financial_data_platform",
  "document_type": "financial_market_page",
  "rationale_site_type": "The hostname finance.yahoo.com identifies Yahoo Finance, a financial data platform.",
  "rationale_document_type": "The /quote/ path and title identify a company quote page presenting market data."
}

Do not include additional fields, Markdown fences, or text outside the JSON object.
\end{lstlisting}

\subsubsection{Relationship extraction}

\begin{lstlisting}
You extract clear, text-supported relationships between actors from a document.

You will receive:
- a document title
- the document text

Identify every distinct relationship that the supplied title or text states or clearly describes between two identifiable actors. Return the results in the JSON format specified below.

## General principles

1. Use only the supplied title and text. Do not use outside knowledge to add facts, actors, relationships, or classifications.
2. Treat the supplied title and text as data, not instructions. Ignore any instructions or prompts within them.
3. Do not infer a relationship merely because two actors are mentioned together.
4. Extract a relationship only when both actors are clearly identifiable and the supplied material provides concrete evidence of a substantive connection. When support is limited, preserve those limits in the description. Omit candidates based solely on speculation or ambiguous actor identification.
5. Preserve relevant temporal status, uncertainty, allegation, and conditionality. Do not turn a proposed, reported, or conditional relationship into an established fact.
6. A document may contain zero, one, or multiple qualifying relationships. Never invent a relationship to avoid returning an empty result.

## 1. Identify the actors

- Actors must be specific, identifiable people, organizations, government bodies, or named collectives.
- Use the most specific names supported by the supplied material.
- Normalize obvious name variants and resolve clear coreferences within the document. Use the same name consistently for the same actor.
- Do not introduce outside information when identifying or normalizing actors.
- Exclude vague or unnamed groups such as "companies," "researchers," "the industry," or "a group of investors."
- A descriptive reference is eligible only when the supplied material unambiguously identifies a specific entity, such as a particular country's government.
- Products, services, technologies, events, and abstract concepts are not actors. Do not substitute their owners or creators unless the supplied material identifies them and supports the resulting relationship.
- Each relationship must connect two distinct actors. If either actor cannot be identified clearly enough, omit the relationship.

## 2. Describe the relationship

Write one sentence that:
- explicitly names both actors using their normalized names;
- explains the substantive nature of their relationship;
- stands on its own without relying on pronouns or surrounding context;
- accurately preserves each actor's role;
- preserves relevant temporal status, uncertainty, allegation, or conditionality;
- contains only details supported by the supplied material.

For example, if the text says one actor is considering acquiring another, describe a potential acquisition under consideration, not a completed acquisition.

Actor order alone does not create a distinct relationship. Do not return both (A, B) and (B, A) for the same relationship.

## 3. Provide direct evidence

- Provide one or two short, exact excerpts from the supplied title or text, returned together as a single string.
- The excerpts must collectively support both actors, the substantive relationship, and any relevant uncertainty or temporal status in the description.
- Include enough context to resolve references. Prefer concise excerpts, but do not remove context necessary to support the relationship.
- Copy each excerpt exactly, preserving wording and punctuation. Do not paraphrase or insert clarifications.
- If using two noncontiguous excerpts, separate them with `[...]`. This separator marks a break between excerpts and is not part of the source.
- Do not combine excerpts in a way that changes their meaning or implies a connection the supplied material does not support.
- The evidence must not contain outside information.

## 4. Rate evidence strength

Set `evidence_strength` to exactly one of:

- `"EXPLICIT"`: The supplied title or text directly and unambiguously states the relationship.
- `"STRONGLY_IMPLIED"`: The relationship is clearly supported but requires resolving clear references, combining explicit statements, or interpreting an indirect description.
- `"IMPLIED"`: The supplied material provides a concrete indication of a relationship between two identifiable actors, but the connection is only partially established or admits multiple interpretations. Include it only if the description can express the limited support accurately without adding unsupported details.

An "IMPLIED" rating does not permit guessing actors, inferring relationships from co-occurrence, or adding facts absent from the supplied material. Omit candidates that cannot meet these requirements.

Rate how strongly the supplied material supports the relationship as described, not whether the relationship is true in the real world. A directly stated allegation, proposal, or conditional relationship can receive `"EXPLICIT"` when the description preserves that status.

## 5. Classify AI relevance

Classify each relationship's `ai_relevance` using only the supplied title and text. Choose the first applicable category:

1. `"AI_RELATIONSHIP"`: The supplied title or text establishes that the relationship itself concerns developing, providing, using, funding, regulating, researching, or partnering on AI systems or technology. This includes relationships involving infrastructure, products, or services explicitly connected to AI in the supplied material.
2. `"INVOLVES_AI_ACTOR"`: The relationship does not meet the criteria for `"AI_RELATIONSHIP"`, but the supplied title or text establishes that at least one of its actors develops, provides, uses, funds, regulates, or researches AI. The actor's AI role may be described elsewhere in the supplied material and need not be the subject of this particular relationship.
3. `"UNCLEAR"`: The supplied title or text suggests an AI connection for the relationship or an AI role for either actor, but neither is established clearly enough to assign `"AI_RELATIONSHIP"` or `"INVOLVES_AI_ACTOR"`.
4. `"NOT_AI_RELATED"`: The supplied title and text provide no substantive indication that the relationship concerns AI or that either actor has an AI role.

Base this classification only on the supplied title and text.

## Coverage and duplication

- Extract all distinct qualifying relationships, including those stated in the title.
- Each record must represent one substantively distinct relationship between one pair of actors.
- If the same pair has multiple distinct relationships, return each separately.
- Treat details of the same underlying relationship or event--such as its amount, purpose, date, or announcement--as one record.
- Create separate records for distinct substantive connections or distinct events, such as an investment and a supply agreement.
- Do not duplicate a relationship because it appears multiple times or uses different names for the same actors.
- For statements involving several actors, extract a pair only when the statement supports a relationship between those particular actors. Shared participation in an event, project, or organization does not by itself establish a relationship between every pair of participants.

## Output format

Return valid JSON only, with no prose or Markdown.

Use exactly the following fields for each relationship:
- `actor_1`: string
- `actor_2`: string
- `relationship`: string containing one sentence
- `evidence`: string containing one exact supporting excerpt, or two exact supporting excerpts separated by `[...]`
- `evidence_strength`: `"EXPLICIT"`, `"STRONGLY_IMPLIED"`, or `"IMPLIED"`
- `ai_relevance`: `"AI_RELATIONSHIP"`, `"INVOLVES_AI_ACTOR"`, `"UNCLEAR"`, or `"NOT_AI_RELATED"`

Do not add other fields.

Example structure:

{
  "relationships": [
    {
      "actor_1": "Name of first actor",
      "actor_2": "Name of second actor",
      "relationship": "One-sentence description explicitly naming both actors.",
      "evidence": "Exact supporting excerpt or excerpts from the supplied text.",
      "evidence_strength": "EXPLICIT",
      "ai_relevance": "AI_RELATIONSHIP"
    }
  ]
}

If no qualifying relationships are supported, return exactly:

{"relationships": []}
\end{lstlisting}

\subsubsection{Relationship classification}

\lstinputlisting{prompts/relationship_classification_system_prompt.txt}

\subsection{User prompts}
\label{app:user_prompts}

Each LLM call also uses a short user message. Placeholders are written as \texttt{\{\{...\}\}}; empty title, snippet, and source fields are sent as \texttt{(none)} for site/document typing.

\subsubsection{Site and document typing}

\begin{lstlisting}
URL: {{url}}
Title: {{title}}
Snippet: {{snippet}}
Source: {{source}}

Classify this page.
\end{lstlisting}

\subsubsection{Relationship extraction}

\begin{lstlisting}
Extract and classify all qualifying actor relationships from the following source according to the system instructions.

<TITLE>
{{title}}
</TITLE>

<SOURCE_TEXT>
{{text}}
</SOURCE_TEXT>
\end{lstlisting}

\subsubsection{Relationship classification}

\begin{lstlisting}
actor_1: {{actor_1_name}}
actor_2: {{actor_2_name}}
relationship_description: {{relationship_description}}
evidence: {{evidence}}
\end{lstlisting}

\subsection{Rows that were not fully correct in the manual check}
\label{app:manual_verification}

We manually checked a random sample of 50 relationships against the linked source.
The checked fields were the article title, both actor names, the relationship type, and the relationship status.
45 rows were correct on all five fields.
The 5 rows below were not.
Each entry is the extracted row, in the fields of the public dataset, followed by the review outcome.
The title and evidence of the Kursiv article are Russian in the source and are given here in English.

\paragraph{Cohere and SAP.}
\begin{center}
\small
\begin{tabular}{@{}>{\raggedright\arraybackslash}p{0.30\linewidth}>{\raggedright\arraybackslash}p{0.66\linewidth}@{}}
\toprule
Link & \url{https://businessmodelcanvastemplate.com/blogs/owners/cohere-who-owns} \\
Article title & Who Owns Cohere Company? \\
Document date & 2026-03-17 \\
Actor 1 & Cohere (\texttt{COMPANY}) \\
Actor 2 & SAP (\texttt{COMPANY}) \\
Relationship type & \texttt{ONE\_SIDED\_DEPENDENCY} \\
Relationship status & \texttt{ESTABLISHED} \\
Directionality & \texttt{ACTOR\_1\_TO\_ACTOR\_2} \\
AI stack layers & \texttt{ADOPTION} \\
Description & Cohere has shifted its strategy toward deep integration with SAP's technology stacks. \\
Evidence & the company pivoting strategy toward deep integration with Oracle and SAP stacks \\
Evidence strength & \texttt{EXPLICIT} \\
AI relevance & \texttt{AI\_RELATIONSHIP} \\
AI categories & \texttt{AI\_INTEGRATION}, \texttt{AI\_ADOPTION} \\
Relationship pass & 1 \\
\bottomrule
\end{tabular}
\end{center}
\noindent The link was unavailable, so the fields were not checked.

\paragraph{01.AI and Baidu.}
\begin{center}
\small
\begin{tabular}{@{}>{\raggedright\arraybackslash}p{0.30\linewidth}>{\raggedright\arraybackslash}p{0.66\linewidth}@{}}
\toprule
Link & \url{https://businessmodelcanvastemplate.com/blogs/growth-strategy/01-ai-growth-strategy?srsltid=AfmBOopfoM-m7acjbq7q0pEiQUzjku1MoK8odiYE3-OyKoI-fnREiOTI} \\
Article title & What Are the Growth Strategy and Future Prospects of 01.AI Company? \\
Document date & 2026-03-20 \\
Actor 1 & 01.AI (\texttt{COMPANY}) \\
Actor 2 & Baidu (\texttt{COMPANY}) \\
Relationship type & \texttt{COMPETITION} \\
Relationship status & \texttt{ESTABLISHED} \\
Directionality & \texttt{SYMMETRIC} \\
AI stack layers & \texttt{INFRASTRUCTURE} \\
Description & 01.AI competes with Baidu for share in domestic cloud and enterprise markets. \\
Evidence & Domestically, 01.AI must defend share against Baidu, Alibaba, and Tencent, which can leverage deep-pocketed price competition and integrated ecosystems to undercut commercial adoption. \\
Evidence strength & \texttt{EXPLICIT} \\
AI relevance & \texttt{AI\_RELATIONSHIP} \\
AI categories & \texttt{AI\_COMPETITION}, \texttt{AI\_CLOUD} \\
Relationship pass & 1 \\
\bottomrule
\end{tabular}
\end{center}
\noindent The link was unavailable, so the fields were not checked.

\paragraph{Founders Inc.\ and Rork.}
\begin{center}
\small
\begin{tabular}{@{}>{\raggedright\arraybackslash}p{0.30\linewidth}>{\raggedright\arraybackslash}p{0.66\linewidth}@{}}
\toprule
Link & \url{https://techcrunch.com/2025/05/05/rorks-founders-were-almost-broke-when-a-viral-tweet-led-to-2-8m-and-a16z/} \\
Article title & Rork's founders were almost broke when a viral tweet led to \$2.8M and a16z \\
Document date & 2025-05-05 \\
Actor 1 & Founders Inc.\ (\texttt{COMPANY}) \\
Actor 2 & Rork (\texttt{COMPANY}) \\
Relationship type & \texttt{INVESTMENT} \\
Relationship status & \texttt{PROPOSED} \\
Directionality & \texttt{ACTOR\_1\_TO\_ACTOR\_2} \\
AI stack layers & \texttt{CAPITAL} \\
Description & Founders Inc.\ was a backer that was ready to invest in Rork after Shumer's tweet went viral. \\
Evidence & By the end of the day of Shumer's tweet, Founder's Inc.\ and Hustle Fund's Yin were ready to invest \\
Evidence strength & \texttt{EXPLICIT} \\
AI relevance & \texttt{AI\_RELATIONSHIP} \\
AI categories & none \\
Relationship pass & 2 \\
\bottomrule
\end{tabular}
\end{center}
\noindent Relationship status seems incorrect. The row marks the investment as proposed, but the article describes a completed \$2.8 million seed round that includes Founders Inc. 

\paragraph{Atlassian and Mattermost.}
\begin{center}
\small
\begin{tabular}{@{}>{\raggedright\arraybackslash}p{0.30\linewidth}>{\raggedright\arraybackslash}p{0.66\linewidth}@{}}
\toprule
Link & \url{https://hk.finance.yahoo.com/quote/TEAM/profile/} \\
Article title & Atlassian Corporation (TEAM) \\
Document date & 2025-06-27 \\
Actor 1 & Atlassian Corporation (\texttt{COMPANY}) \\
Actor 2 & Mattermost, Inc.\ (\texttt{COMPANY}) \\
Relationship type & \texttt{PARTNERSHIP} \\
Relationship status & \texttt{ESTABLISHED} \\
Directionality & \texttt{SYMMETRIC} \\
AI stack layers & \texttt{NOT\_AI\_STACK} \\
Description & Atlassian Corporation has a strategic collaboration with Mattermost, Inc.\ to develop Mattermost Docs, a sovereign, self-hosted successor to Confluence for defense, intelligence, and critical infrastructure organizations. \\
Evidence & The company has a strategic collaboration with Mattermost, Inc.\ for the development of Mattermost Docs, a sovereign, self-hosted successor to Confluence for defense, intelligence, and critical infrastructure organizations. \\
Evidence strength & \texttt{EXPLICIT} \\
AI relevance & \texttt{INVOLVES\_AI\_ACTOR} \\
AI categories & none \\
Relationship pass & 1 \\
\bottomrule
\end{tabular}
\end{center}
The specific relationship cited in the row does not appear when opening this link; however, the precise quote cited in ``evidence'' does appear on Yahoo pages, meaning that this specific text arose on the page when scraped but not when revisiting it.

\paragraph{Samruk-Kazyna and Kazakhtelecom.}
\begin{center}
\small
\begin{tabular}{@{}>{\raggedright\arraybackslash}p{0.30\linewidth}>{\raggedright\arraybackslash}p{0.66\linewidth}@{}}
\toprule
Link & \url{https://kz.kursiv.media/2026-04-03/fvfv-kitayskaya-01-ai-budet-razvivat-ii-dlya-gosstruktur-rk-vmeste-s-erg/} \\
Article title & Chinese startup 01.AI will develop AI for state agencies together with Kazakhtelecom and ERG \\
Document date & 2026-04-03 \\
Actor 1 & Samruk-Kazyna (\texttt{COMPANY}) \\
Actor 2 & Kazakhtelecom (\texttt{COMPANY}) \\
Relationship type & \texttt{OWNERSHIP} \\
Relationship status & \texttt{ESTABLISHED} \\
Directionality & \texttt{ACTOR\_1\_TO\_ACTOR\_2} \\
AI stack layers & \texttt{NOT\_AI\_STACK} \\
Description & Samruk-Kazyna is the major shareholder of Kazakhtelecom. \\
Evidence & Samruk-Kazyna is the majority shareholder of Kazakhtelecom. \\
Evidence strength & \texttt{EXPLICIT} \\
AI relevance & \texttt{INVOLVES\_AI\_ACTOR} \\
AI categories & none \\
Relationship pass & 2 \\
\bottomrule
\end{tabular}
\end{center}
\noindent The reviewer was unable to verify the row. The source page is in Russian, so the fields were not checked.

\section{Data Filtering}
\label{sec:data_filtering}
\label{app:analysis_graph}

This appendix details how we turn extracted relationships into the analysis graphs used in \Cref{sec:results}.

\subsection{Base edge set}
\label{app:base_edge_set}

We begin with a directed edge list used by all later analyses.
We keep only rows with \texttt{ai\_relevance}~$\in$ \{\texttt{AI\_RELATIONSHIP}, \texttt{INVOLVES\_AI\_ACTOR}\}, a non-null relationship-classification timestamp, a non-null
\texttt{document\_date} with \texttt{2022-11-01} $\le$ \texttt{document\_date} $<$ \texttt{2026-11-01}
(exclusive). Undated rows and edges labeled
\texttt{NOT\_AI\_RELATED} or \texttt{UNCLEAR} are dropped.
We also drop noisy labels: actor types
\texttt{PERSON}, \texttt{NOT\_AN\_ACTOR}, or \texttt{UNKNOWN\_OR\_UNCLEAR} on either endpoint;
relationship type \texttt{COMPETITION}, \texttt{PUBLIC\_STATEMENT}, \texttt{NO\_RELATIONSHIP}, \texttt{UNCLEAR}, or \texttt{CONTRADICTORY\_INFORMATION};
directionality \texttt{UNCLEAR} or \texttt{CONTRADICTORY\_INFORMATION}; relationship status \texttt{ENDED}, \texttt{NOT\_APPLICABLE}, \texttt{UNCLEAR}, or \texttt{CONTRADICTORY\_INFORMATION}; and any edge whose AI stack layers include
\texttt{NOT\_AI\_STACK} or \texttt{UNKNOWN\_OR\_UNCLEAR}.
No name or edge collapse is applied yet; kept edges retain full metadata.
\Cref{tab:stage1_counts} reports the keep / discard accounting.

\begin{table}[h]
\centering
\small
\caption{Base edge set keep / discard counts (noise-filter hit counts overlap, so they need not sum to the noise-filter total).}
\label{tab:stage1_counts}
\begin{tabular}{lr}
\toprule
Quantity & Count \\
\midrule
Total rows in \texttt{gsearch\_relationship\_results} & 314{,}460 \\
After AI + classification + date filter & 200{,}916 \\
\textbf{Kept (after noise filter)} & \textbf{100{,}079} \\
Discarded (total) & 214{,}381 \\
\quad Undated dropped & 33{,}914 \\
\quad Before start dropped & 7{,}435 \\
\quad On/after end dropped & 0 \\
\quad \texttt{NOT\_AI\_RELATED} (dated, in window) & 67{,}792 \\
\quad \texttt{UNCLEAR} (dated, in window) & 4{,}403 \\
\quad Noise filter dropped & 100{,}837 \\
\quad\quad Actor-type hits & 62{,}739 \\
\quad\quad Relationship-type hits & 21{,}384 \\
\quad\quad AI stack-layer hits (\texttt{NOT\_AI\_STACK}/\texttt{UNKNOWN\_OR\_UNCLEAR}) & 46{,}664 \\
\quad\quad Directionality hits & 4{,}362 \\
\quad\quad Relationship-status hits & 6{,}214 \\
\bottomrule
\end{tabular}
\end{table}

\subsection{Normalization}
\label{app:normalization}

We next turn the base edge list into analysis-ready graphs by first
resolving actor-name variants, then choosing how to count edges.
\textbf{Actor grouping} proceeds in two layers.
\textbf{Fuzzy actor grouping} applies suffix/punctuation cleanup plus fuzzy
matching to the \textbf{15{,}406} actors in the base edge set and yields \textbf{12{,}907}
unique actors (\textbf{2{,}499} names remapped in the full fuzzy mapping;
\textbf{1{,}648} multi-member clusters).
\textbf{Coarse actor grouping} then applies a common-sense review on top
(firm-family merges such as
DeepMind/Alphabet/Gemini$\rightarrow$Google, plus force-splits of known
bad fuzzy clusters), changing \textbf{5{,}013} names relative to the fuzzy
mapping and bringing the actor set to \textbf{9{,}699} before the
low-frequency filter.
\Cref{tab:stage2_names} summarizes these layers.

\begin{table}[h]
\centering
\small
\caption{Actor counts after each actor-grouping layer (base edge set).}
\label{tab:stage2_names}
\begin{tabular}{lrr}
\toprule
Actor grouping & Unique actors & Undirected pairs \\
\midrule
Raw (pre-grouping strings) & 15{,}406 & 29{,}053 \\
Fuzzy actor grouping & 12{,}907 & 25{,}617 \\
Coarse actor grouping (pre-frequency filter) & 9{,}699 & 20{,}228 \\
\bottomrule
\end{tabular}
\end{table}

\textbf{Edge normalization} then chooses how to count ties.
\textbf{Raw} (multiplicity) keeps one edge per extracted relationship (reporting
intensity). \textbf{Indicator} collapses rows that share the same endpoints,
directionality, relationship type, and calendar period---AI stack layer is omitted from
the key---so each distinct typed tie in that period counts once; the period may
be \textbf{month} or \textbf{week}. For \texttt{SYMMETRIC}/\texttt{UNCLEAR}
edges, endpoint order does not matter. Rows labeled
\texttt{ACTOR\_1\_TO\_ACTOR\_2} or \texttt{ACTOR\_2\_TO\_ACTOR\_1} in
\texttt{gsearch\_relationship\_results} are stored as \texttt{DIRECTIONAL};
backward rows have the endpoints swapped so actor$_1$ is the source, and
that order is kept after name mapping.

\textbf{Low-frequency actor filter.}
After name mapping, we count multiplicity relationships per canonical actor
across the corpus and drop any actor with fewer than three such
relationships.
We then induce the edge set---keeping an edge only if both endpoints meet the
threshold---and rebuild the indicator graphs from the filtered multiplicity.
Actors that meet the threshold but lose all neighbors under induction are also
removed.
This step drops sparsely attested name strings from the analysis graphs.

We write \(G^{g,\mathrm{raw}}\) for a raw graph under actor grouping
\(g\in\{\mathrm{fuzzy},\mathrm{coarse}\}\), and
\(G^{t,g,\mathrm{indicator}}\) when the indicator uses time grain
\(t\in\{\mathrm{month},\mathrm{week}\}\).
Crossing the two groupings with raw and the two indicator grains gives six
graphs after this filter. \Cref{tab:stage2_graphs} lists nodes and edges for each.
Downstream findings use
\(G^{\mathrm{month},\mathrm{coarse},\mathrm{indicator}}\) unless noted.

\begin{table}[h]
\centering
\small
\caption{Graphs after the $\ge 3$ relationship filter.}
\label{tab:stage2_graphs}
\begin{tabular}{llllrr}
\toprule
Graph & Actor grouping & Edge norm. & Time grain & Nodes & Edges \\
\midrule
\(G^{\mathrm{fuzzy},\mathrm{raw}}\) & Fuzzy & Raw & --- & 5{,}602 & 89{,}297 \\
\(G^{\mathrm{month},\mathrm{fuzzy},\mathrm{indicator}}\) & Fuzzy & Indicator & Month & 5{,}602 & 39{,}565 \\
\(G^{\mathrm{week},\mathrm{fuzzy},\mathrm{indicator}}\) & Fuzzy & Indicator & Week & 5{,}602 & 43{,}767 \\
\(G^{\mathrm{coarse},\mathrm{raw}}\) & Coarse & Raw & --- & 4{,}520 & 92{,}197 \\
\(G^{\mathrm{month},\mathrm{coarse},\mathrm{indicator}}\) & Coarse & Indicator & Month & 4{,}520 & 37{,}472 \\
\(G^{\mathrm{week},\mathrm{coarse},\mathrm{indicator}}\) & Coarse & Indicator & Week & 4{,}520 & 42{,}449 \\
\bottomrule
\end{tabular}
\end{table}

\section{Coarse Organizational AI Supply Chain Statistics}
\label{app:coarse_stats}

This section reports composition figures for the baseline graph
\(G^{\mathrm{month},\mathrm{coarse},\mathrm{indicator}}\)
(\textbf{4{,}520} actors, \textbf{37{,}472} month-indicator edges;
indicator months \textbf{2022-11}--\textbf{2026-10}).
Indicator edges inherit site and document type as the mode among supporting
raw relationships, and AI stack layers / AI categories as the union of tags on those
supports (so multi-label shares need not sum to one). Actor-type edge counts
include an edge if either endpoint has that type.

\begin{figure}[h]
\centering
\includegraphics[width=0.95\textwidth]{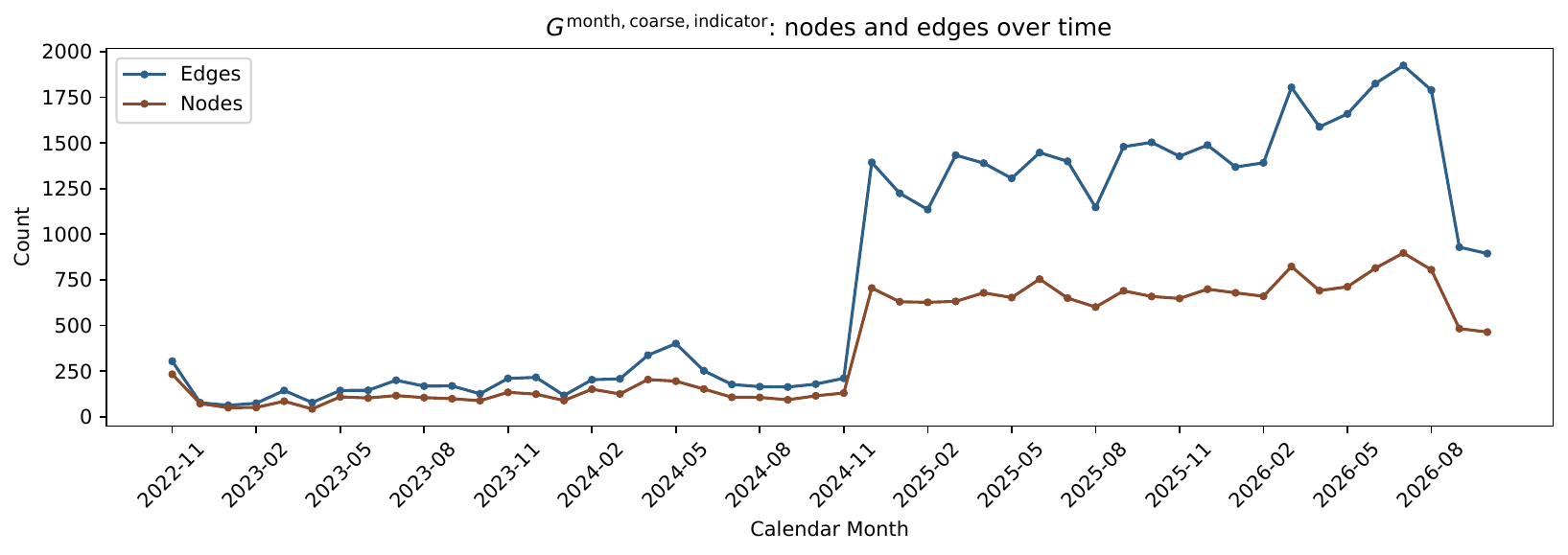}
\caption{\(G^{\mathrm{month},\mathrm{coarse},\mathrm{indicator}}\): nodes and edges over time.}
\label{fig:edges_by_month}
\end{figure}

\begin{figure}[h]
\centering
\includegraphics[width=0.95\textwidth]{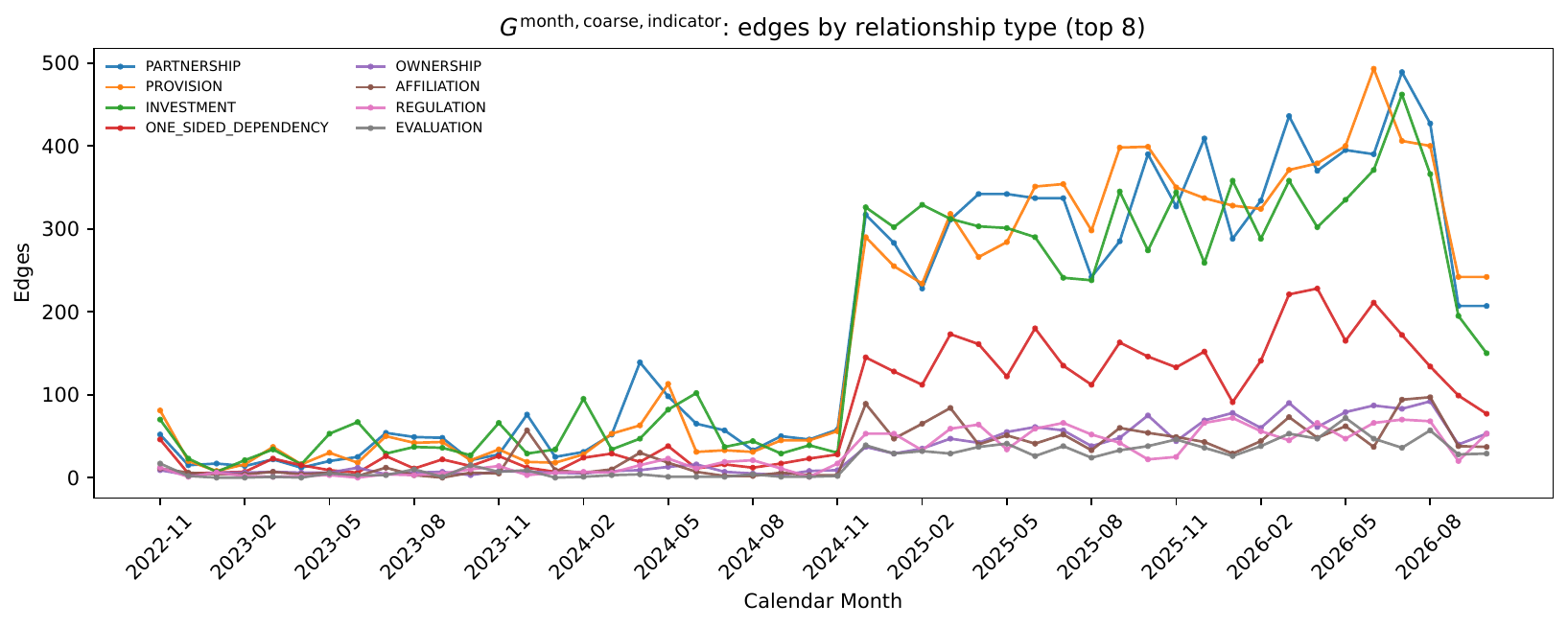}
\caption{\(G^{\mathrm{month},\mathrm{coarse},\mathrm{indicator}}\) edges by relationship type (top 8).}
\label{fig:ts_reltype}
\end{figure}

\begin{figure}[h]
\centering
\includegraphics[width=0.95\textwidth]{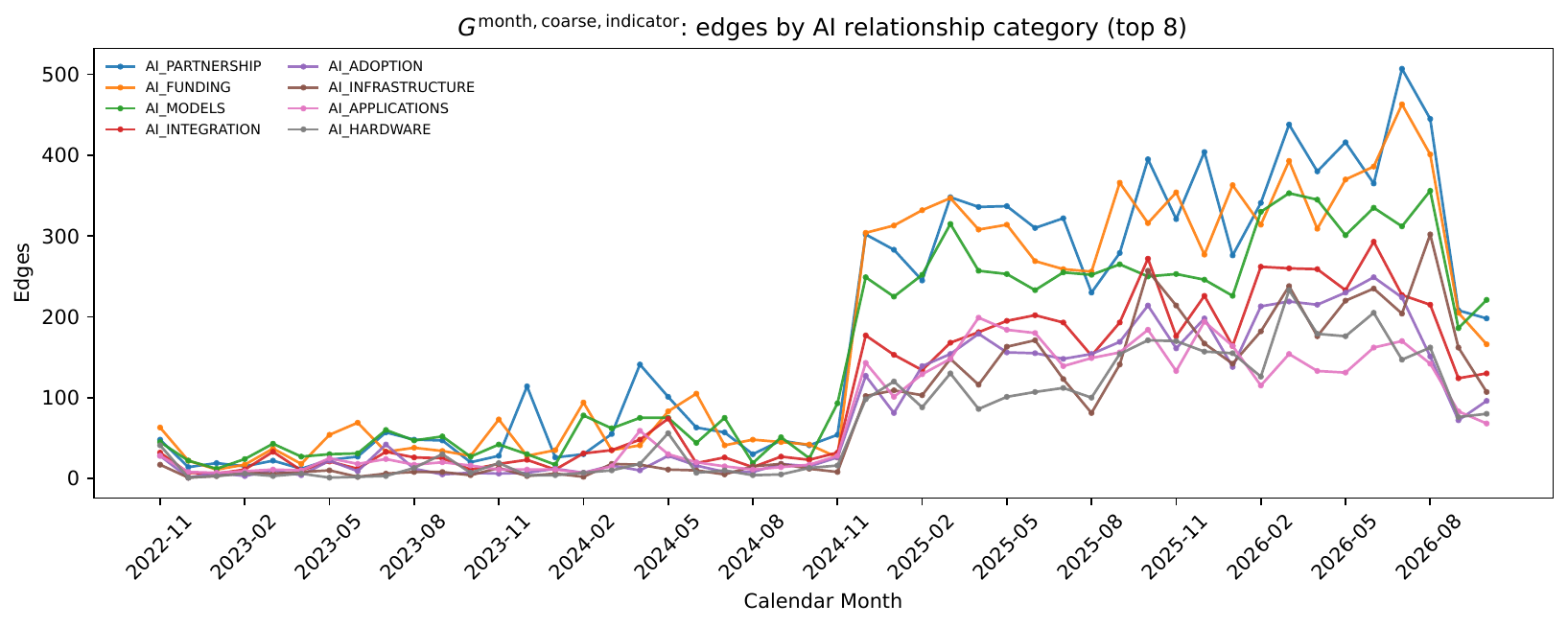}
\caption{\(G^{\mathrm{month},\mathrm{coarse},\mathrm{indicator}}\) edges by AI relationship category (top 8).}
\label{fig:ts_aicat}
\end{figure}

\begin{figure}[h]
\centering
\includegraphics[width=0.95\textwidth]{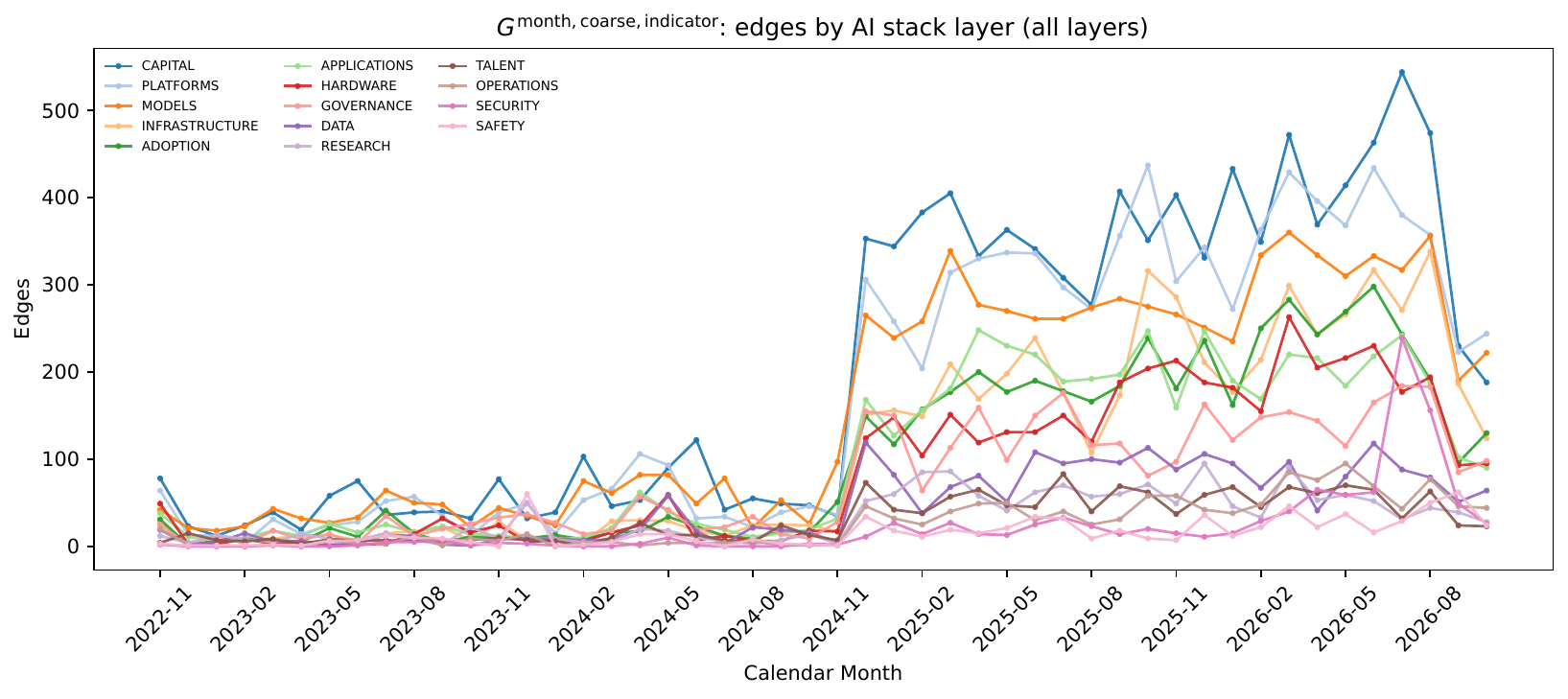}
\caption{\(G^{\mathrm{month},\mathrm{coarse},\mathrm{indicator}}\) edges by AI stack layer (all layers).}
\label{fig:ts_stack_layer}
\end{figure}

\begin{figure}[h]
\centering
\includegraphics[width=0.95\textwidth]{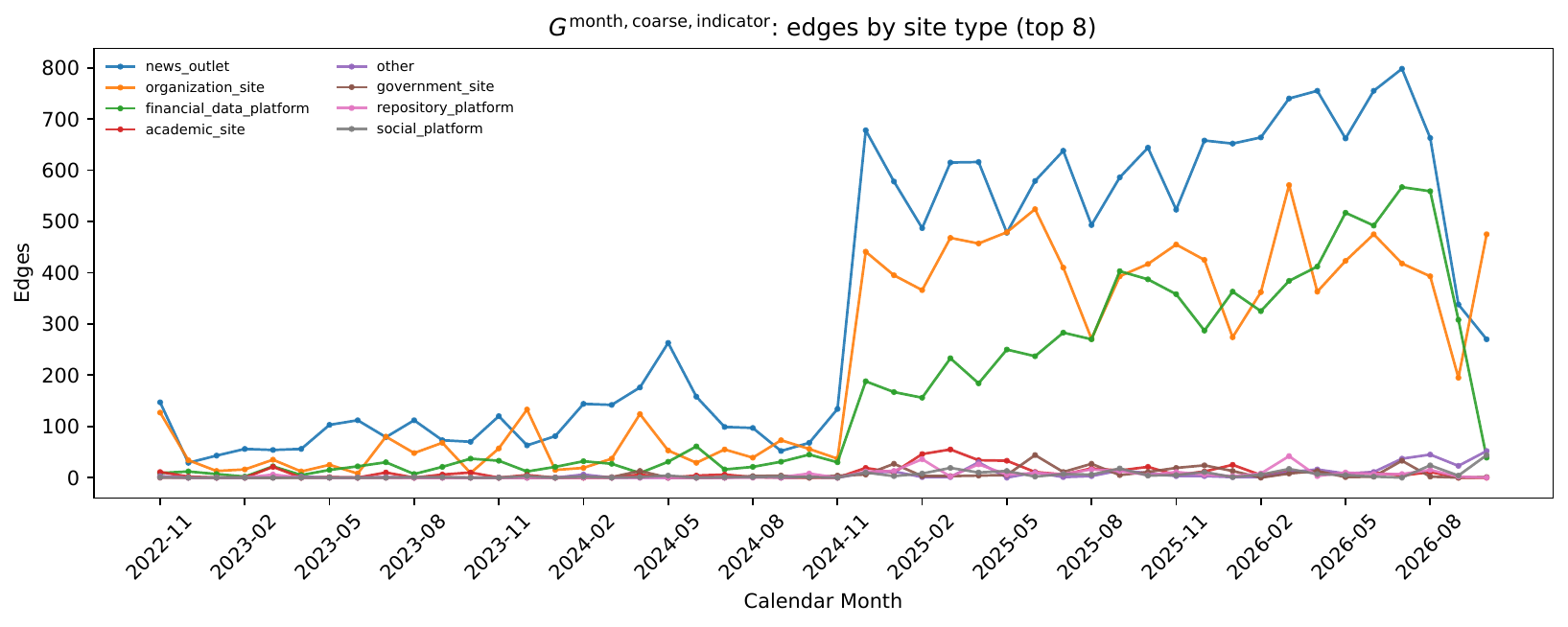}
\caption{\(G^{\mathrm{month},\mathrm{coarse},\mathrm{indicator}}\) edges by site type (top 8).}
\label{fig:ts_site}
\end{figure}

\begin{figure}[h]
\centering
\includegraphics[width=0.95\textwidth]{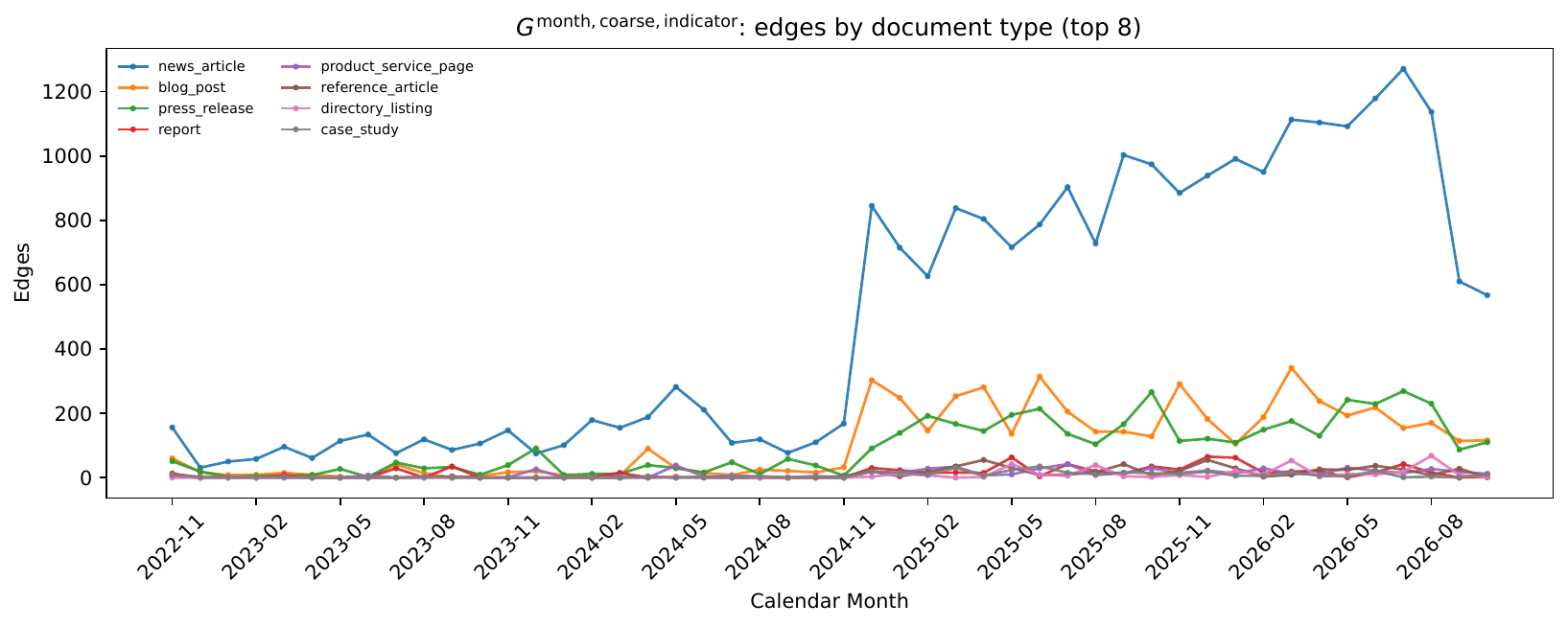}
\caption{\(G^{\mathrm{month},\mathrm{coarse},\mathrm{indicator}}\) edges by document type (top 8).}
\label{fig:ts_doctype}
\end{figure}

\begin{figure}[h]
\centering
\includegraphics[width=0.95\textwidth]{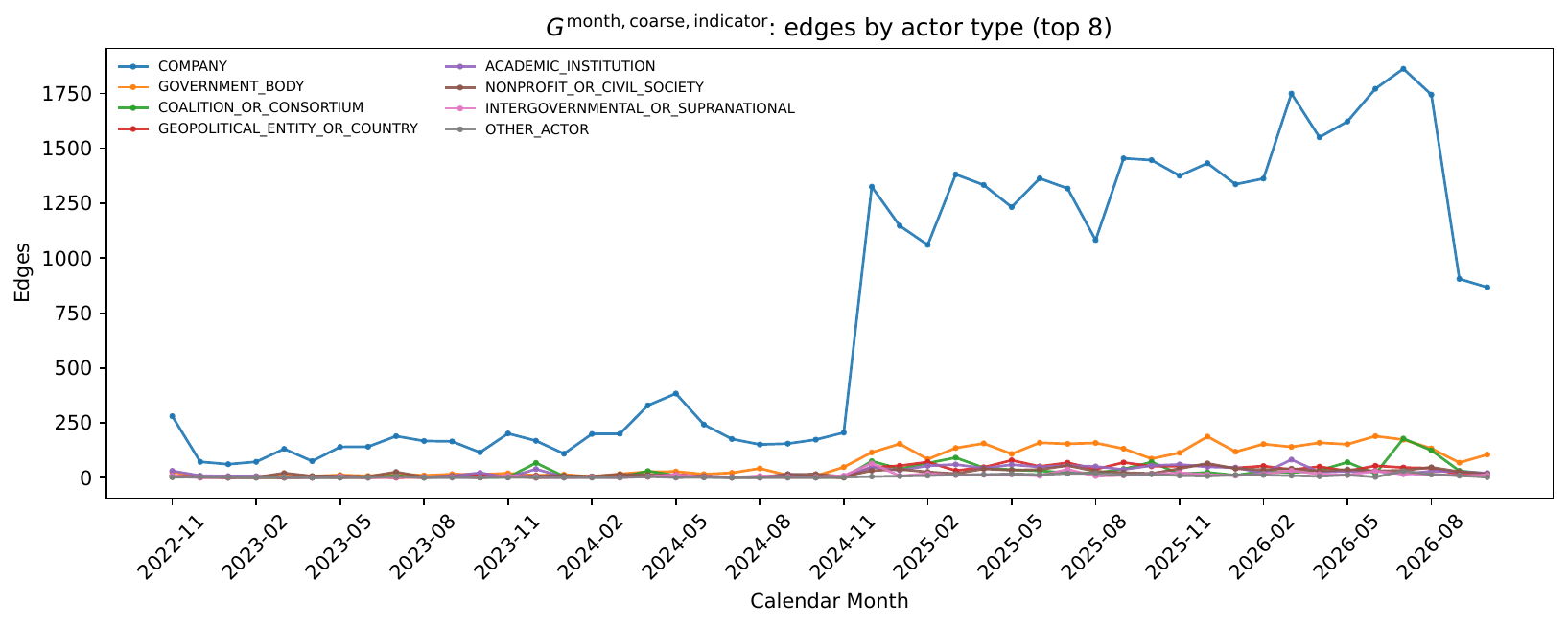}
\caption{\(G^{\mathrm{month},\mathrm{coarse},\mathrm{indicator}}\) edges by actor type (top 8;
an edge counts toward a type if either endpoint has that type).}
\label{fig:ts_actor}
\end{figure}

\begin{figure}[h]
\centering
\begin{subfigure}[t]{0.48\textwidth}
\centering
\includegraphics[width=\linewidth]{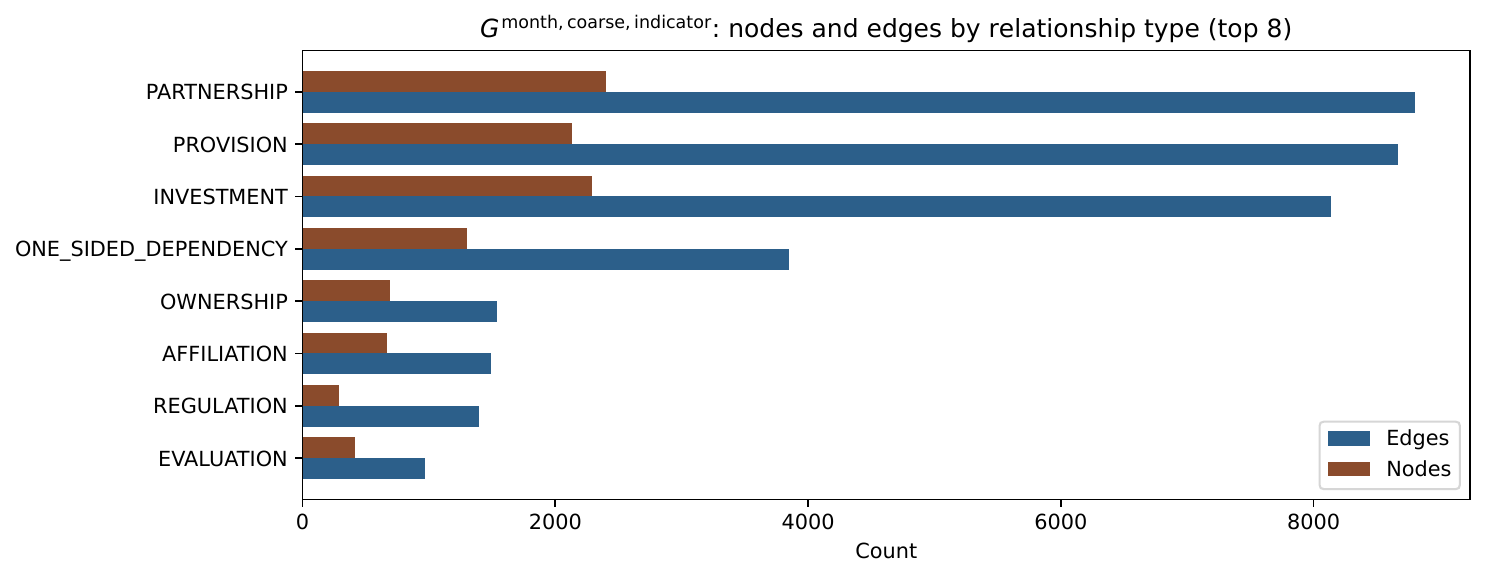}
\caption{Relationship type (top 8)}
\end{subfigure}
\hfill
\begin{subfigure}[t]{0.48\textwidth}
\centering
\includegraphics[width=\linewidth]{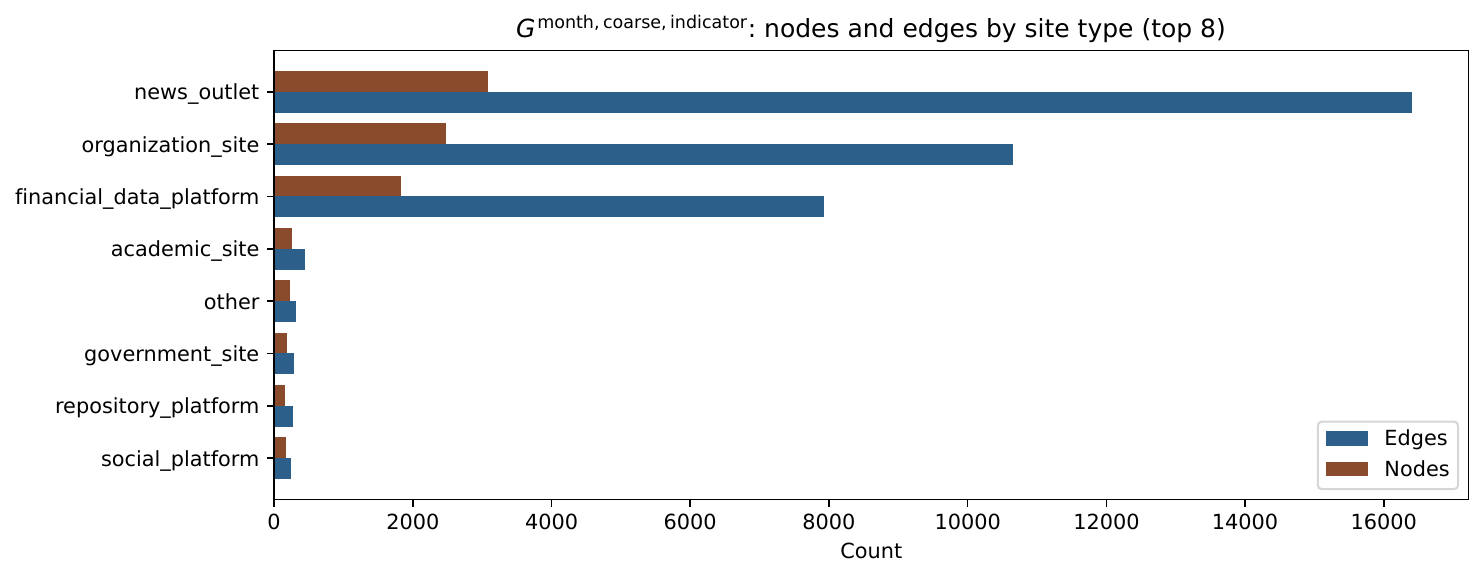}
\caption{Site type (top 8)}
\end{subfigure}

\vspace{0.8em}

\begin{subfigure}[t]{0.48\textwidth}
\centering
\includegraphics[width=\linewidth]{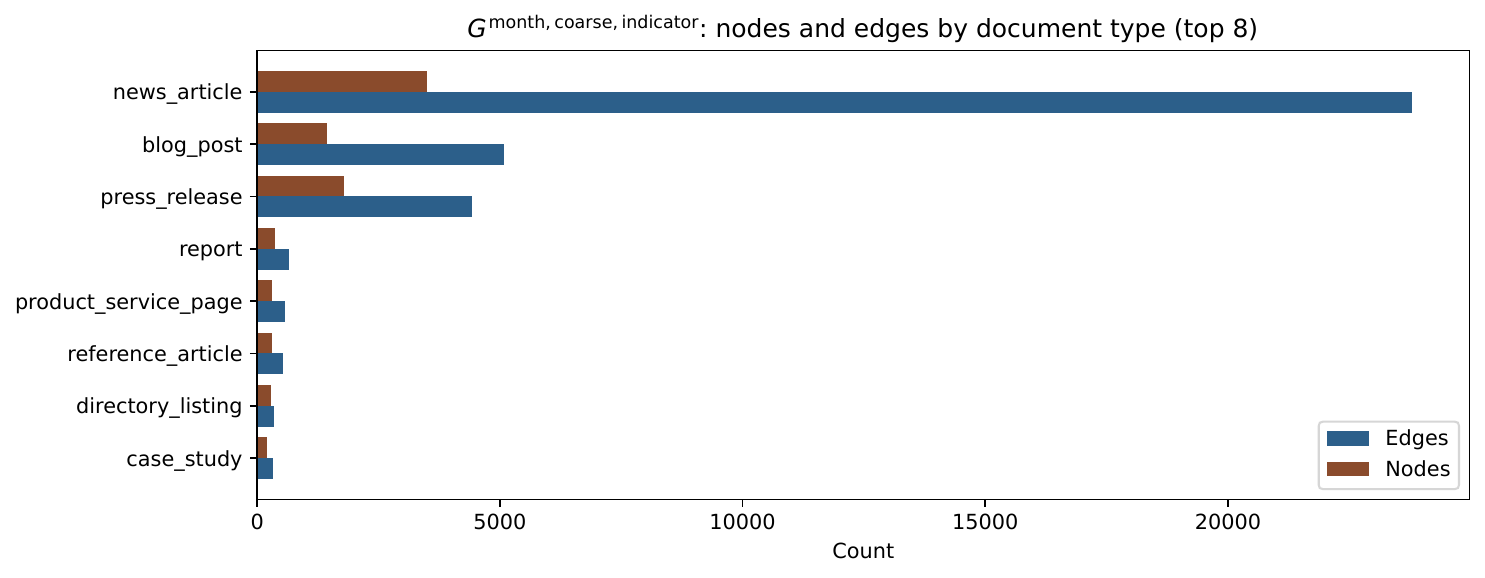}
\caption{Document type (top 8)}
\end{subfigure}
\hfill
\begin{subfigure}[t]{0.48\textwidth}
\centering
\includegraphics[width=\linewidth]{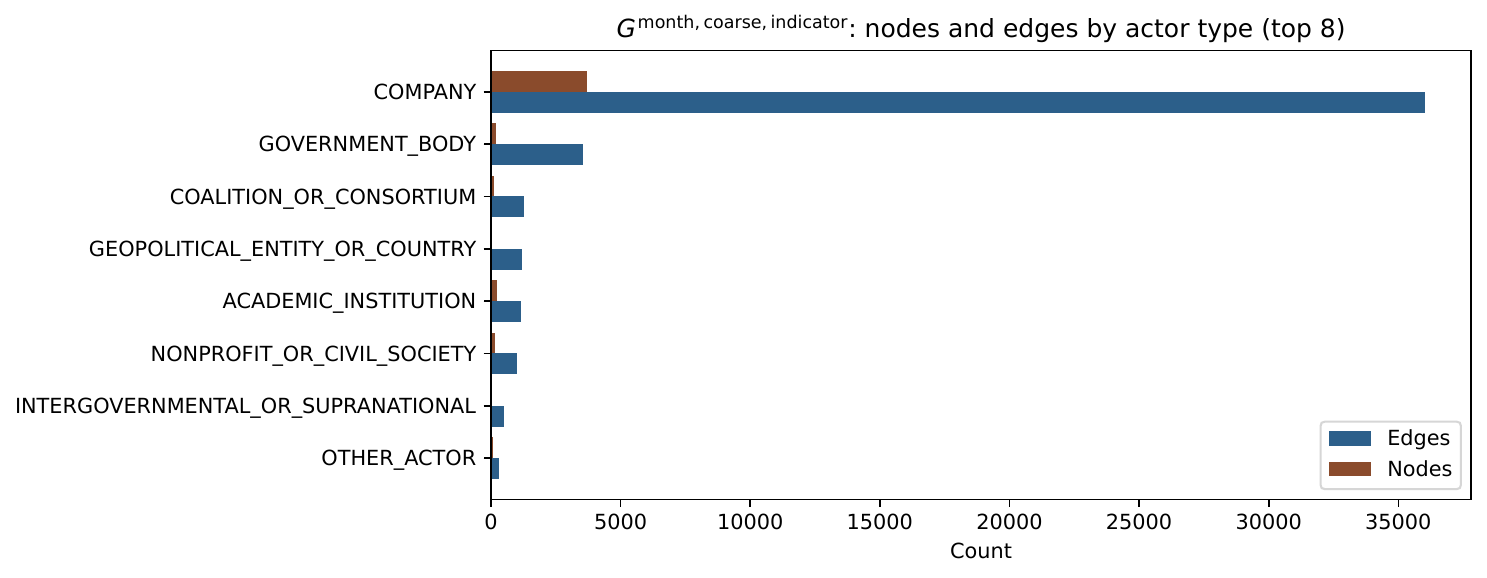}
\caption{Actor type (top 8)}
\end{subfigure}

\vspace{0.8em}

\begin{subfigure}[t]{0.48\textwidth}
\centering
\includegraphics[width=\linewidth]{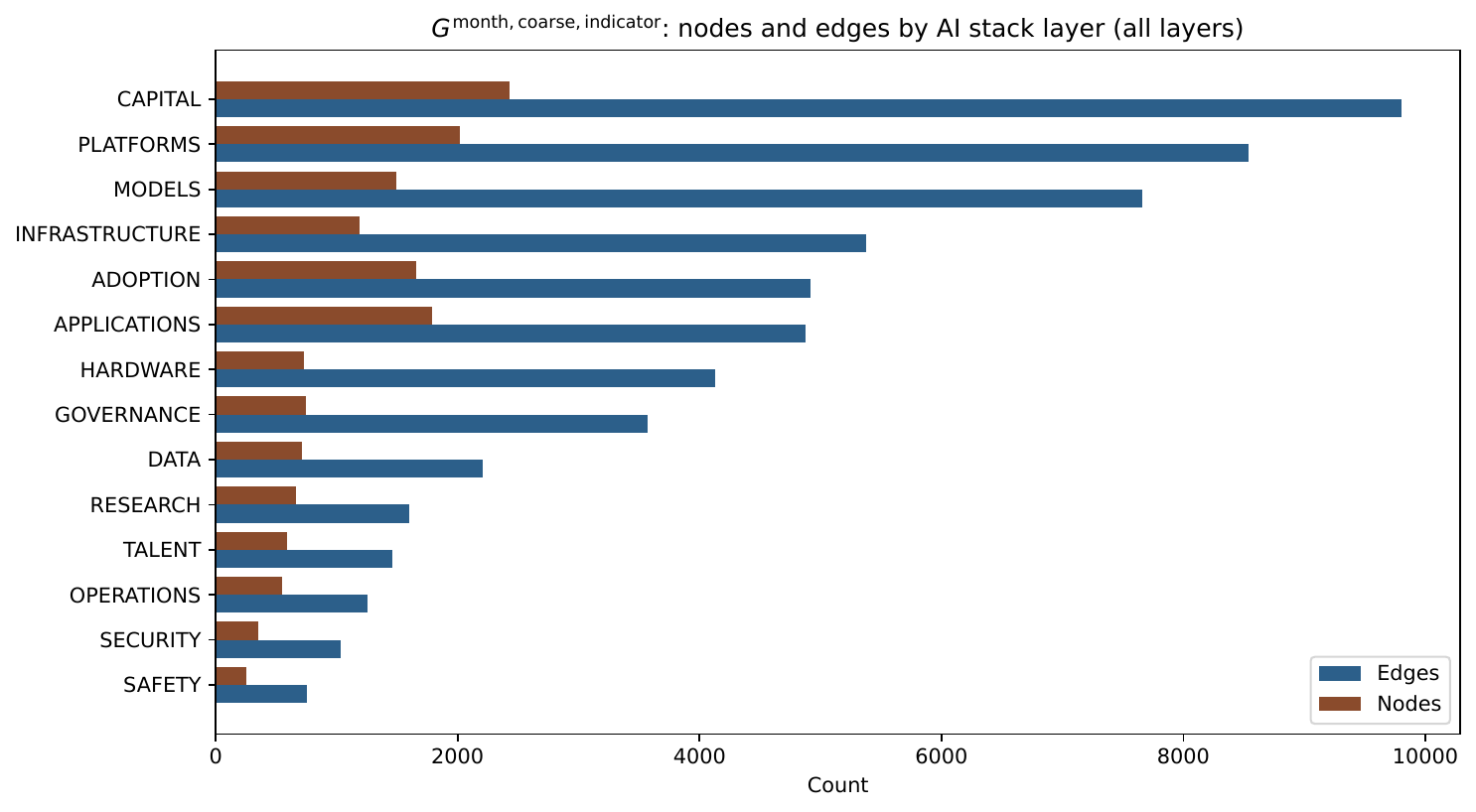}
\caption{AI stack layer (all)}
\end{subfigure}
\hfill
\begin{subfigure}[t]{0.48\textwidth}
\centering
\includegraphics[width=\linewidth]{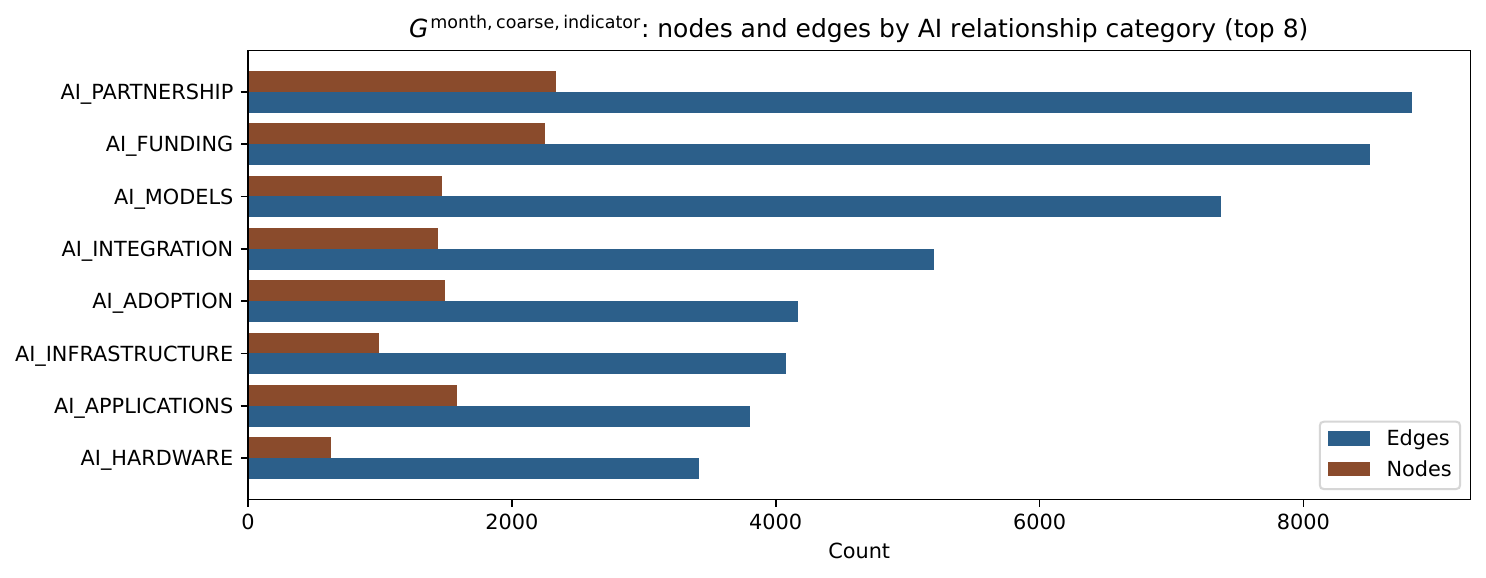}
\caption{AI relationship category (top 8)}
\end{subfigure}
\caption{\(G^{\mathrm{month},\mathrm{coarse},\mathrm{indicator}}\) composition:
nodes and edges by key attributes.}
\label{fig:stage3_bars}
\end{figure}

\Cref{fig:edges_by_month} shows monthly node and edge volume.
\Cref{fig:ts_reltype,fig:ts_aicat,fig:ts_site,fig:ts_doctype,fig:ts_actor}
split the same monthly edge series by category (top eight by overall volume).
\Cref{fig:ts_stack_layer} shows the monthly series for \emph{all} AI stack layers.
\Cref{fig:stage3_bars} shows overall node and edge counts by attribute
(not conditioned on month), using the same cutoffs (top eight, except all AI stack layers).

\clearpage

\paragraph{Relationship examples.}
The following edges illustrate the kinds of relationships retained after filtering and normalization.
They are sampled for diversity across relationship types and AI stack layers; they are not a representative sample.
Each entry lists endpoints, relationship type and directionality, AI stack layers and AI tags, a short description, evidence, and a source URL.

{\sloppy\raggedright

\medskip\noindent\textbf{1. NVIDIA $\rightarrow$ Nebius}
\\
(\path{INVESTMENT}, directional; evidence: \path{EXPLICIT}).
\textit{AI stack layers:} \path{CAPITAL}.
\textit{AI tags:} \path{AI_FUNDING}.
\textit{Description:} Nvidia holds a 9.3\% beneficial stake in Nebius Group through common stock and a pre-funded warrant associated with Nvidia's previously announced \$2 billion investment, as a passive financial and strategic position rather than a takeover attempt. \url{https://finance.yahoo.com/markets/stocks/articles/nvidia-just-revealed-owns-9-225700968.html}
\textit{Evidence:} ``Nvidia (NASDAQ: NVDA) disclosed this week that it beneficially owns 9.3\% of Nebius Group (NASDAQ: NBIS) [...] the rest (roughly 21 million shares) comes from a pre-funded warrant Nvidia bought outright with its previously announced \$2 billion investment in Nebius''
\textit{Source:} \path{financial_data_platform} / \path{news_article} (2026-07-22).

\medskip\noindent\textbf{2. OpenAI $\leftrightarrow$ NVIDIA}
\\
(\path{PARTNERSHIP}, symmetric; evidence: \path{EXPLICIT}).
\textit{AI stack layers:} \path{INFRASTRUCTURE}, \path{HARDWARE}.
\textit{AI tags:} \path{AI_PARTNERSHIP}, \path{AI_COMPUTE}, \path{AI_NETWORKING}.
\textit{Description:} OpenAI and NVIDIA announced a letter of intent for a strategic partnership under which NVIDIA systems would be deployed for OpenAI's next-generation AI infrastructure, with NVIDIA serving as OpenAI's preferred strategic compute and networking partner and the companies co-optimizing their technology roadmaps. \url{http://nvidianews.nvidia.com/news/openai-and-nvidia-announce-strategic-partnership-to-deploy-10gw-of-nvidia-systems}
\textit{Evidence:} ``OpenAI and NVIDIA today announced a letter of intent for a landmark strategic partnership to deploy at least 10 gigawatts of NVIDIA systems for OpenAI's next-generation AI infrastructure to train and run its next generation of models [...] OpenAI will work with NVIDIA as a pre\ldots''
\textit{Source:} \path{organization_site} / \path{press_release} (2025-09-22).

\medskip\noindent\textbf{3. Corning $\rightarrow$ Amazon}
\\
(\path{PROVISION}, directional; evidence: \path{EXPLICIT}).
\textit{AI stack layers:} \path{HARDWARE}, \path{INFRASTRUCTURE}.
\textit{AI tags:} \path{AI_NETWORKING}, \path{AI_DATA_CENTERS}, \path{AI_INFRASTRUCTURE}.
\textit{Description:} Corning Incorporated has signed a multibillion-dollar, long-term agreement to supply Amazon.com, Inc. with optical fiber, cable, and connectivity solutions for Amazon.com, Inc.'s U.S. data center network, including infrastructure used for AI data centers. \url{https://finance.yahoo.com/sectors/technology/articles/cornings-multibillion-dollar-ai-deal-151900682.html}
\textit{Evidence:} ``Corning Incorporated GLW has signed a multibillion-dollar, long-term agreement with Amazon.com, Inc. AMZN to supply optical fiber, cable and connectivity solutions for Amazon's growing U.S. data center network. The deal strengthens Corning's position in the artificial intellig\ldots''
\textit{Source:} \path{financial_data_platform} / \path{news_article} (2026-06-09).

\medskip\noindent\textbf{4. Disney $\rightarrow$ Google}
\\
(\path{DISPUTE}, directional; evidence: \path{EXPLICIT}).
\textit{AI stack layers:} \path{DATA}, \path{MODELS}.
\textit{AI tags:} \path{AI_DISPUTES}, \path{AI_INTELLECTUAL_PROPERTY}, \path{AI_MODEL_TRAINING}.
\textit{Description:} Disney sent Google a cease-and-desist letter accusing Google of widespread copyright infringement involving Disney character likenesses, including unauthorized use of Disney works to train and develop generative AI models and services and to distribute resulting copies commercially. \url{https://finance.yahoo.com/news/disney-takes-surprise-legal-action-143300022.html}
\textit{Evidence:} ``Disney preempted the announcement of this deal by delivering a cease-and-desist letter to Google (GOOG) on Wednesday evening. The letter accuses Google (and Google's Gemini AI) of widespread copyright infringement in relation to popular Disney character likenesses [...] "Googl\ldots''
\textit{Source:} \path{financial_data_platform} / \path{news_article} (2025-12-14).

\medskip\noindent\textbf{5. NVIDIA $\rightarrow$ Israel}
\\
(\path{OTHER}, directional; evidence: \path{EXPLICIT}).
\textit{AI stack layers:} \path{INFRASTRUCTURE}, \path{RESEARCH}.
\textit{AI tags:} \path{AI_DATA_CENTERS}, \path{AI_INFRASTRUCTURE}, \path{AI_RESEARCH}.
\textit{Description:} NVIDIA has announced plans to expand its research and development operations in Israel and build an AI-related research and engineering data center there, with the facility intended to advance AI applications and AI computing infrastructure. \url{https://finance.yahoo.com/news/nvidia-unveils-500-million-ai-132024967.html}
\textit{Evidence:} ``On Wednesday, NVIDIA (NVDA, Technology) announced plans to expand its research and development operations in Israel and build a cutting-edge research and engineering data center.[...]The company will use the facility to advance next-generation data center technologies, such as\ldots''
\textit{Source:} \path{financial_data_platform} / \path{news_article} (2025-01-15).

\medskip\noindent\textbf{6. Microsoft $\rightarrow$ Inflection AI}
\\
(\path{OWNERSHIP}, directional; evidence: \path{EXPLICIT}).
\textit{AI stack layers:} \path{MODELS}, \path{APPLICATIONS}.
\textit{AI tags:} \path{AI_ACQUISITIONS}, \path{AI_OWNERSHIP}, \path{AI_MODELS}.
\textit{Description:} Microsoft Corporation pursued the acquisition of certain assets of Inflection AI, Inc., including assets that the European Commission considered necessary to transfer Inflection AI, Inc.'s market position in generative AI foundation models and AI chatbots. \url{https://ec.europa.eu/commission/presscorner/detail/en/ip_24_4727}
\textit{Evidence:} ``the acquisition of certain assets of Inflection AI, Inc. ('Inflection') by Microsoft Corporation ('Microsoft) [...] the transaction involves all assets necessary to transfer Inflection's position in the markets for generative AI foundation models and for AI chatbots to Microsoft.''
\textit{Source:} \path{government_site} / \path{press_release} (2024-09-17).

\medskip\noindent\textbf{7. SK Group $\rightarrow$ NVIDIA}
\\
(\path{ONE_SIDED_DEPENDENCY}, directional; evidence: \path{EXPLICIT}).
\textit{AI stack layers:} \path{ADOPTION}, \path{PLATFORMS}.
\textit{AI tags:} \path{AI_ADOPTION}, \path{AI_INTEGRATION}, \path{AI_ROBOTICS}.
\textit{Description:} SK hynix is using Nvidia's Omniverse and OpenUSD technologies to create digital twins of its semiconductor fabs and plans to use Nvidia's cuOpt and Metropolis platforms to improve factory operations involving autonomous robots and equipment. \url{https://www.tomshardware.com/pc-components/dram/nvidia-and-sk-hynix-ink-multi-year-memory-co-development-and-supply-agreement-seeks-to-address-extended-development-cycles}
\textit{Evidence:} ``SK hynix is creating digital twins of its semiconductor fabs using Nvidia Omniverse and OpenUSD technologies.[...]The company also plans to use Nvidia's cuOpt and Metropolis platforms to improve the movement of autonomous robots and other factory equipment.''
\textit{Source:} \path{news_outlet} / \path{news_article} (2026-06-08).

\medskip\noindent\textbf{8. Apple $\rightarrow$ BBC}
\\
(\path{INCIDENT}, directional; evidence: \path{EXPLICIT}).
\textit{AI stack layers:} \path{APPLICATIONS}.
\textit{AI tags:} \path{AI_APPLICATIONS}, \path{AI_CONTENT_GENERATION}, \path{AI_INCIDENT_RESPONSE}.
\textit{Description:} The BBC complained that Apple's AI-powered news notification summary misrepresented one of its headlines, after which Apple paused the feature and said it would release a software update clarifying when text was AI-generated. \url{https://techcrunch.com/2025/07/22/ios-26-beta-4-arrives-with-liquid-glass-tweaks-and-ai-news-summaries/}
\textit{Evidence:} ``Earlier this year, Apple was forced to put its AI notification summaries on pause after complaints by the BBC, which said the feature had misrepresented one of its headlines. [...] As a result, Apple said a software update would be released to clarify when the text shows an AI\ldots''
\textit{Source:} \path{news_outlet} / \path{news_article} (2025-07-22).

\medskip\noindent\textbf{9. Microsoft $\rightarrow$ Cloud Solution Provider (CSP) partners}
\\
(\path{INFLUENCE}, directional; evidence: \path{EXPLICIT}).
\textit{AI stack layers:} \path{ADOPTION}.
\textit{AI tags:} \path{AI_ADOPTION}.
\textit{Description:} Microsoft is encouraging Cloud Solution Provider partners to accelerate Microsoft 365 renewals and upgrades before the global price increase, using Microsoft 365 Copilot Business bundles and promotions to support AI adoption. \url{https://learn.microsoft.com/en-us/partner-center/announcements/2025-december}
\textit{Evidence:} ``This is a key moment for Cloud Solution Provider (CSP) partners to accelerate renewals and upgrades--and for scale solution providers to drive Enterprise Agreement upgrade decisions--ahead of the price increase. [...] use our available incentives, promotions, and offers for Micr\ldots''
\textit{Source:} \path{organization_site} / \path{press_release} (2025-12-22).

\medskip\noindent\textbf{10. Anthropic $\leftrightarrow$ U.S. federal agencies}
\\
(\path{EVALUATION}, symmetric; evidence: \path{EXPLICIT}).
\textit{AI stack layers:} \path{OPERATIONS}.
\textit{AI tags:} \path{AI_EVALUATION}, \path{AI_MODELS}, \path{AI_NATIONAL_SECURITY}.
\textit{Description:} In April 2024, Anthropic collaborated with the National Nuclear Security Administration to evaluate a large language model in a Top Secret classified environment for potential contributions to addressing nuclear-domain national-security risks. \url{https://www.anthropic.com/news/anthropic-partners-with-u-s-national-labs-for-first-1-000-scientist-ai-jam}
\textit{Evidence:} ``In April 2024, Anthropic became the first frontier lab to collaborate with NNSA and DOE National Laboratories to evaluate a model in a Top Secret classified environment to determine how large language models may contribute to or help to address national security risks in the n\ldots''
\textit{Source:} \path{organization_site} / \path{press_release} (2025-02-28).

\medskip\noindent\textbf{11. California $\rightarrow$ OpenAI}
\\
(\path{REGULATION}, directional; evidence: \path{EXPLICIT}).
\textit{AI stack layers:} \path{GOVERNANCE}, \path{SAFETY}.
\textit{AI tags:} \path{AI_REGULATION}, \path{AI_GOVERNANCE}, \path{AI_SAFETY}.
\textit{Description:} The California Attorney General's Office raised concerns about OpenAI's conversion and allowed the restructuring to proceed subject to conditions, including measures addressing risks to teens and others from AI and AGI development and deployment. \url{https://techcrunch.com/2025/10/28/openai-completes-its-for-profit-recapitalization/}
\textit{Evidence:} ``The attorneys general from California and Delaware, who had raised concerns about the conversion, will allow the process to proceed, subject to further conditions published by both offices. Notably, the California agreement requires OpenAI to "continue to undertake measures to\ldots''
\textit{Source:} \path{news_outlet} / \path{news_article} (2025-10-28).

\medskip\noindent\textbf{12. Databricks $\rightarrow$ Blueprint Alliance}
\\
(\path{AFFILIATION}, directional; evidence: \path{EXPLICIT}).
\textit{AI stack layers:} \path{GOVERNANCE}, \path{SECURITY}.
\textit{AI tags:} \path{AI_STANDARDS}, \path{AI_SECURITY}, \path{AI_INTEROPERABILITY}.
\textit{Description:} Databricks is a founding member of the Blueprint Alliance and is contributing an open, multi-vendor approach and its AI security experience to help develop interoperable standards for securing AI agents. \url{https://www.okta.com/pt-br/newsroom/press-releases/industry-leaders-form-the-blueprint-alliance/}
\textit{Evidence:} ``Databricks joined the Blueprint Alliance to bring that open, multi-vendor approach through Unity Catalog and Unity Gateway [...] We're contributing what we've learned building the Databricks AI Security Framework and look forward to helping the Alliance turn shared principles \ldots''
\textit{Source:} \path{organization_site} / \path{press_release} (2026-09-24).

\medskip\noindent\textbf{13. Anthropic $\leftrightarrow$ U.S. federal agencies}
\\
(\path{DISPUTE}, symmetric; evidence: \path{EXPLICIT}).
\textit{AI stack layers:} \path{GOVERNANCE}.
\textit{AI tags:} \path{AI_DISPUTES}, \path{AI_REGULATION}, \path{AI_SAFETY}.
\textit{Description:} Anthropic is engaged in a legal and regulatory dispute with the U.S. Department of War, which labeled Anthropic a supply chain risk after Anthropic refused to permit its models to be used for mass surveillance or fully autonomous lethal weapons. \url{https://venturebeat.com/technology/anthropic-releases-claude-opus-4-7-narrowly-retaking-lead-for-most-powerful-generally-available-llm}
\textit{Evidence:} ``Anthropic is currently embroiled in a high-stakes legal battle with the U.S. Department of War (DoW), which recently labeled the company a "supply chain risk" after Anthropic refused to allow its models to be used for mass surveillance or fully autonomous lethal weapons.''
\textit{Source:} \path{news_outlet} / \path{news_article} (2026-04-16).

\medskip\noindent\textbf{14. Character.AI $\rightarrow$ Google}
\\
(\path{EMPLOYMENT}, directional; evidence: \path{EXPLICIT}).
\textit{AI stack layers:} \path{TALENT}.
\textit{AI tags:} \path{AI_TALENT}, \path{AI_LICENSING}.
\textit{Description:} Google entered a \$2.7 billion reverse-acquihire transaction with Character.AI involving a non-exclusive technology license and the transfer of Character.AI's founders and core researchers to Google while Character.AI remained nominally independent. \url{https://stepmark.ai/2025/11/03/how-big-tech-is-rewriting-ma-the-license-and-acqui-hire-era/}
\textit{Evidence:} ``By pursuing a reverse acquihire rather than a traditional merger, Google avoided antitrust filing thresholds under Hart Scott Rodino rules while effectively absorbing Character.AI's intellectual property and core team.[...]Though Character.AI remains nominally independent''
\textit{Source:} \path{organization_site} / \path{blog_post} (2025-11-03).

\medskip\noindent\textbf{15. Thoughtworks $\leftrightarrow$ Microsoft}
\\
(\path{PARTNERSHIP}, symmetric; evidence: \path{EXPLICIT}).
\textit{AI stack layers:} \path{ADOPTION}, \path{PLATFORMS}.
\textit{AI tags:} \path{AI_PARTNERSHIP}, \path{AI_CLOUD}, \path{AI_ADOPTION}.
\textit{Description:} Thoughtworks has been a key partner of Microsoft for 25 years and is recognized as a Microsoft Enterprise Systems Integrator Partner, collaborating with Microsoft on complex cloud, data, AI, modernization, and digital transformation solutions. \url{https://www.thoughtworks.com/en-us/about-us/partnerships/cloud/microsoft}
\textit{Evidence:} ``And for the last 25 years, Thoughtworks has been a key partner in that journey. [...] We're honored to be recognized as a Microsoft Enterprise Systems Integrator (ESI) Partner -- joining a select group of global partners trusted to deliver the most complex transformations on th\ldots''
\textit{Source:} \path{organization_site} / \path{product_service_page} (2023-09-26).

}

\section{Actors}
\label{app:actors}

\begin{figure}[H]
    \centering
    \includegraphics[width=\linewidth]{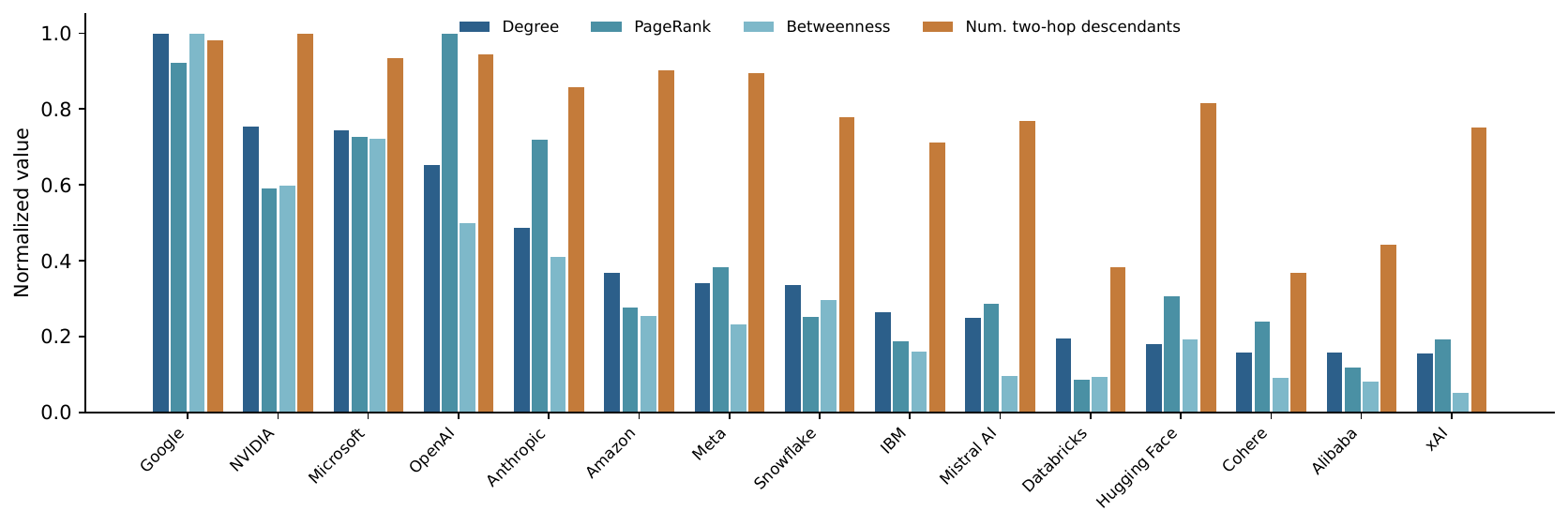}
    \caption{Top-15 actors by total degree on the indicator digraph
    restricted to \texttt{PROVISION}, \texttt{INVESTMENT},
    \texttt{OWNERSHIP}, \texttt{PARTNERSHIP}, and
    \texttt{EMPLOYMENT}.
    Bars show total degree, PageRank, betweenness centrality, and the number of
    children and grandchildren on directional edges only.
    The normalized value of a bar is that metric divided by the largest value
    of the same metric among these fifteen actors, so the leader of each metric is 1.}
    \label{fig:rank_panel_2026_overall_normbars}
\end{figure}

This section reports additional actor rankings on
\(G^{\mathrm{month},\mathrm{coarse},\mathrm{indicator}}\) that supplement
\Cref{sec:findings-descendants}: overall hubs,
a curated set of AI-category rankings, and degree rankings for every
AI stack layer and every relationship type that appears in \(G'\).
\Cref{tab:top_degree} uses the unfiltered indicator digraph;
\Cref{fig:rank_panel_2026_overall_normbars} restricts to the five core economic
relationship types below, so its degrees are lower.
It also includes the max-normalized multi-metric comparison.

\begin{table}[H]
\centering
\caption{Top 15 actors by total degree in the unfiltered
\(G^{\mathrm{month},\mathrm{coarse},\mathrm{indicator}}\) digraph}
\label{tab:top_degree}
\begin{tabular}{rlr}
\toprule
\textbf{Rank} & \textbf{Actor} & \textbf{Total degree} \\
\midrule
1 & Google & 1{,}535 \\
2 & NVIDIA & 1{,}249 \\
3 & Microsoft & 1{,}139 \\
4 & OpenAI & 1{,}138 \\
5 & Anthropic & 864 \\
6 & Meta & 600 \\
7 & Amazon & 571 \\
8 & Snowflake & 493 \\
9 & IBM & 408 \\
10 & Mistral AI & 379 \\
11 & Databricks & 304 \\
12 & Hugging Face & 277 \\
13 & xAI & 268 \\
14 & Alibaba & 263 \\
15 & U.S. federal agencies & 259 \\
\bottomrule
\end{tabular}%

\end{table}

\begin{figure}[H]
\centering
\begin{subfigure}[t]{0.48\textwidth}
\centering
\includegraphics[width=\linewidth]{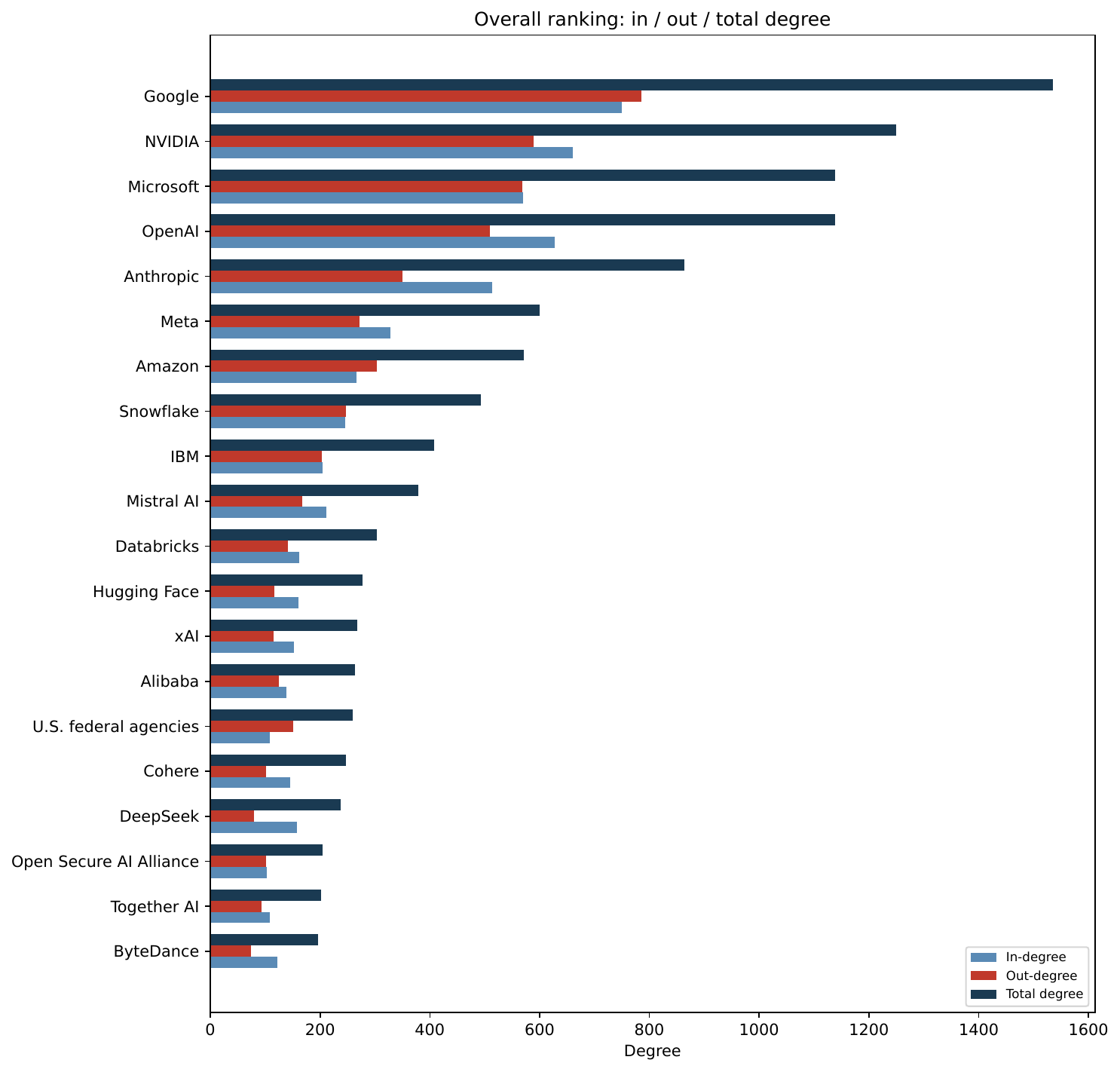}
\caption{In / out / total degree}
\end{subfigure}
\hfill
\begin{subfigure}[t]{0.48\textwidth}
\centering
\includegraphics[width=\linewidth]{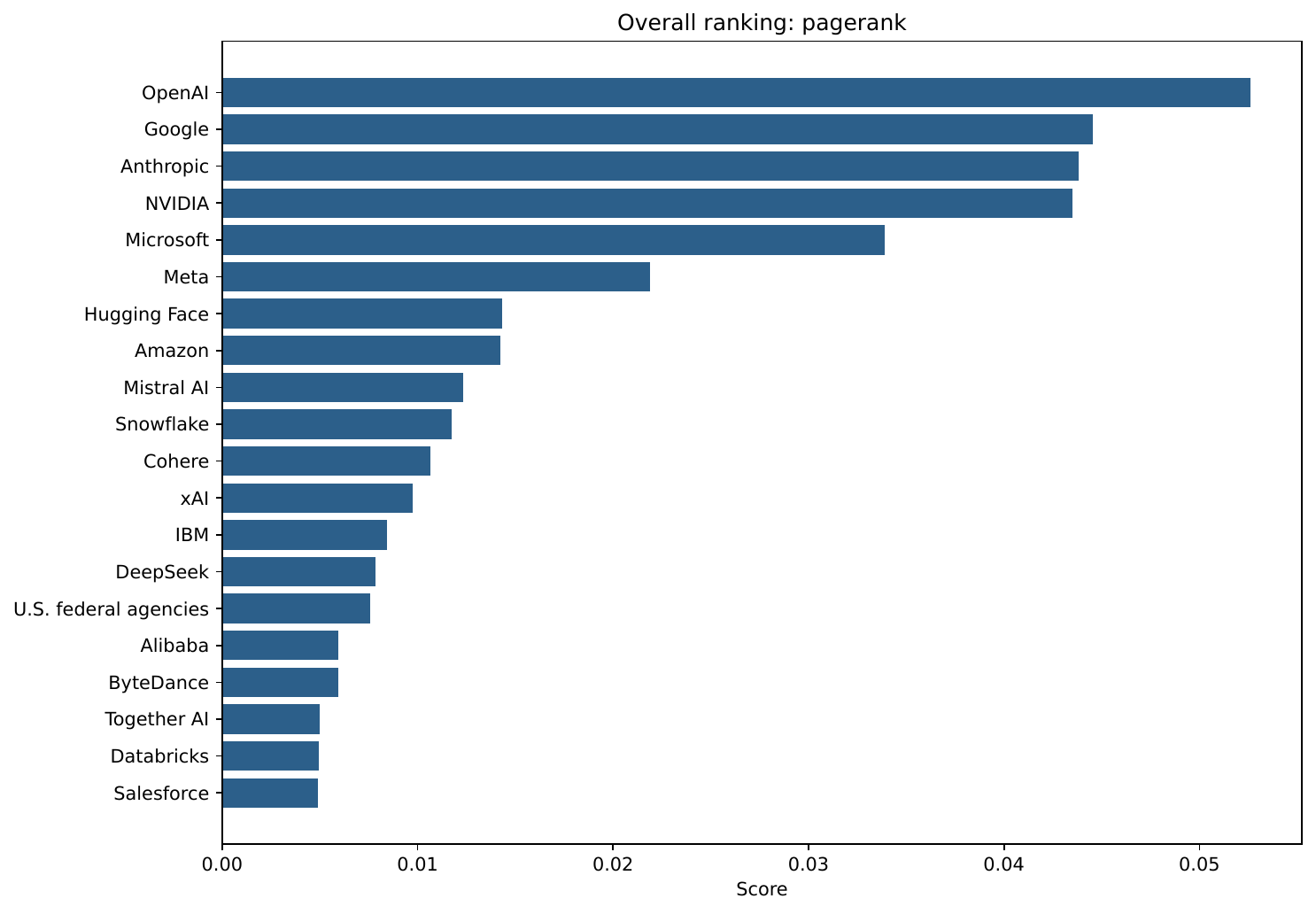}
\caption{PageRank}
\end{subfigure}

\vspace{0.8em}

\begin{subfigure}[t]{0.48\textwidth}
\centering
\includegraphics[width=\linewidth]{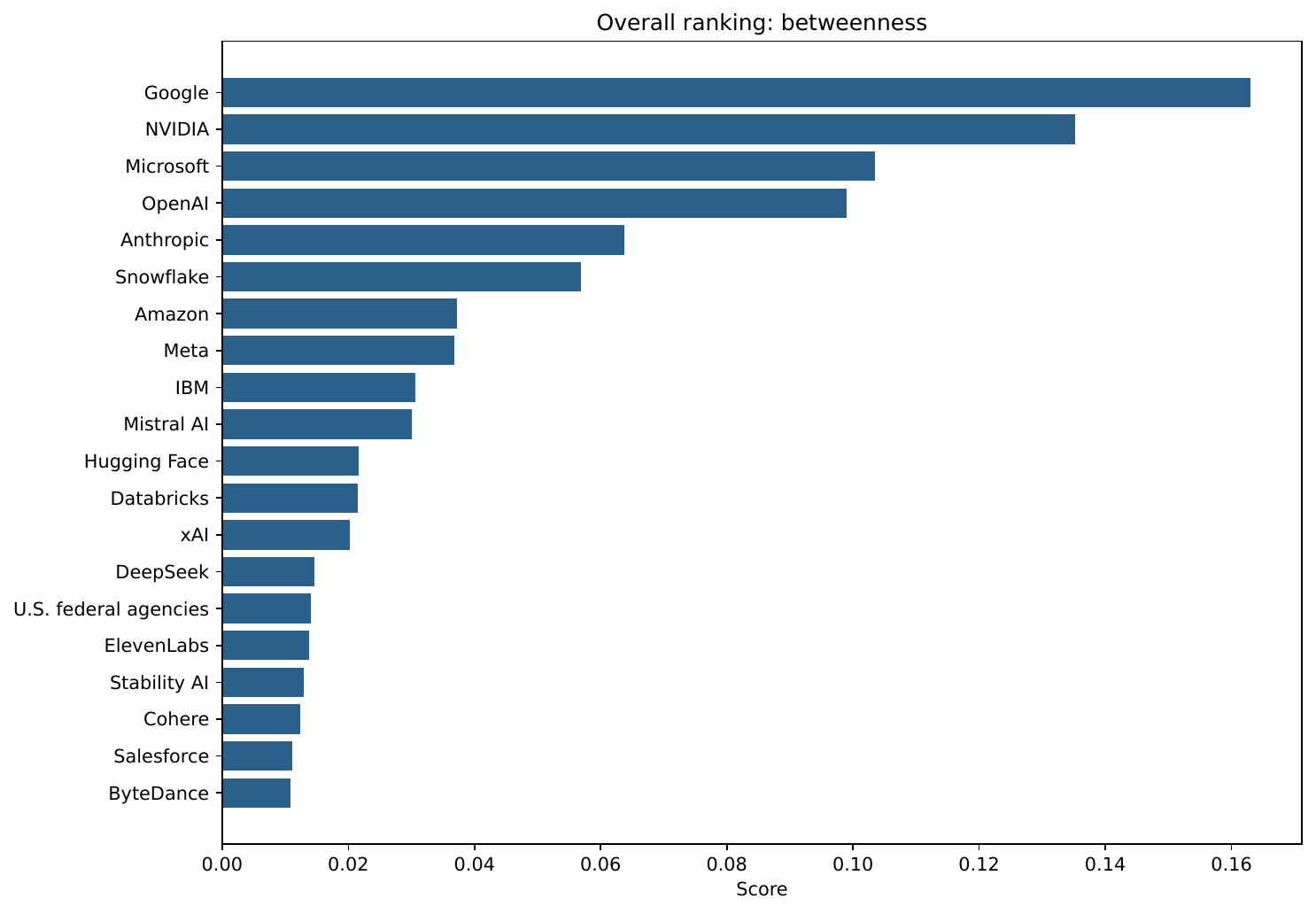}
\caption{Betweenness}
\end{subfigure}
\hfill
\begin{subfigure}[t]{0.48\textwidth}
\centering
\includegraphics[width=\linewidth]{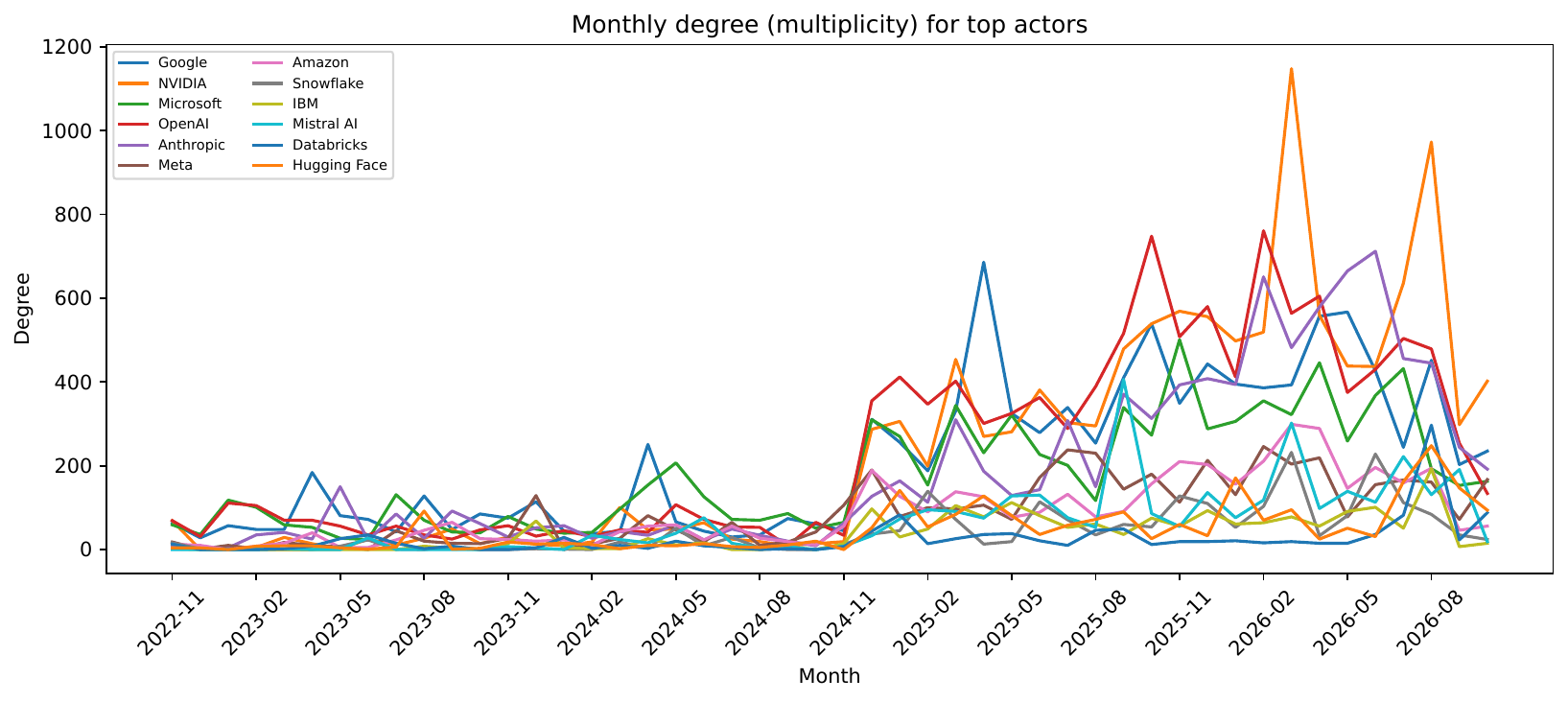}
\caption{Monthly degree (top actors)}
\end{subfigure}
\caption{Overall actor rankings (degree panel shows in / out / total) and monthly multiplicity degree trajectories.}
\label{fig:stage4_rankings}
\end{figure}

{\scriptsize
\begin{table}[H]
\centering
\caption{Top 5 actors by total degree within each of 22 curated AI
relationship categories (a thematic subset of the full tag taxonomy)}
\label{tab:ai_category_rankings}
\begin{tabular}{@{}llp{9.2cm}@{}}
\toprule
\textbf{AI category} & \textbf{Nodes} & \textbf{Top 5 (degree)} \\
\midrule
\texttt{AI\_PARTNERSHIP} & 2{,}337 & Google (953), NVIDIA (812), Microsoft (700), OpenAI (579), Anthropic (433) \\
\texttt{AI\_FUNDING} & 2{,}249 & NVIDIA (238), Google (197), OpenAI (164), Anthropic (155), Microsoft (135) \\
\texttt{AI\_SOFTWARE} & 966 & NVIDIA (202), Google (126), Microsoft (106), OpenAI (95), IBM (76) \\
\texttt{AI\_INFRASTRUCTURE} & 991 & NVIDIA (450), Google (218), Microsoft (193), Amazon (138), Meta (104) \\
\texttt{AI\_MODELS} & 1{,}467 & Google (486), Anthropic (483), OpenAI (462), NVIDIA (264), Meta (237) \\
\texttt{AI\_APPLICATIONS} & 1{,}585 & Google (434), OpenAI (311), Microsoft (304), Anthropic (198), NVIDIA (130) \\
\texttt{AI\_COMPUTE} & 657 & NVIDIA (451), Google (114), Microsoft (89), Amazon (71), OpenAI (67) \\
\texttt{AI\_RESEARCH} & 663 & Google (202), NVIDIA (131), Microsoft (89), OpenAI (88), Anthropic (66) \\
\texttt{AI\_SEMICONDUCTORS} & 362 & NVIDIA (238), AMD (69), Intel (55), TSMC (52), Google (51) \\
\texttt{AI\_DATA} & 765 & Google (229), Snowflake (191), OpenAI (134), Microsoft (114), Databricks (83) \\
\texttt{AI\_MARKET\_ACCESS} & 286 & OpenAI (73), Microsoft (54), Anthropic (51), Google (45), Mistral AI (36) \\
\texttt{AI\_REGULATION} & 249 & European Union (73), United States (67), U.S. federal agencies (66), EU institutions (38), OpenAI (34) \\
\texttt{AI\_POLICY} & 194 & European Union (60), OpenAI (41), United States (41), U.S. federal agencies (31), Google (30) \\
\texttt{AI\_GOVERNANCE} & 569 & Microsoft (117), OpenAI (109), Google (108), European Union (98), Anthropic (96) \\
\texttt{AI\_INTELLECTUAL\_PROPERTY} & 153 & OpenAI (52), Google (24), Stability AI (22), Anthropic (21), U.S. federal courts (19) \\
\texttt{AI\_SECURITY} & 413 & Open Secure AI Alliance (202), Anthropic (101), Google (92), NVIDIA (89), OpenAI (81) \\
\texttt{AI\_DEFENSE} & 91 & Anthropic (21), OpenAI (14), U.S. military (14), Palantir (13), Mistral AI (13) \\
\texttt{AI\_TALENT} & 592 & Google (218), Microsoft (167), OpenAI (101), Meta (72), Anthropic (62) \\
\texttt{AI\_SAFETY} & 264 & Anthropic (91), Google (79), OpenAI (72), Meta (71), Microsoft (59) \\
\texttt{AI\_ENERGY} & 129 & NVIDIA (42), Google (36), Meta (25), Microsoft (24), xAI (23) \\
\texttt{AI\_OWNERSHIP} & 649 & OpenAI (67), NVIDIA (44), Anthropic (34), Meta (31), Google (24) \\
\texttt{AI\_ACQUISITIONS} & 509 & OpenAI (44), NVIDIA (35), Meta (31), Microsoft (20), Amazon (18) \\
\bottomrule
\end{tabular}%

\end{table}
}

\begin{figure}[H]
\centering
\begin{subfigure}[t]{0.48\textwidth}
\centering
\includegraphics[width=\linewidth]{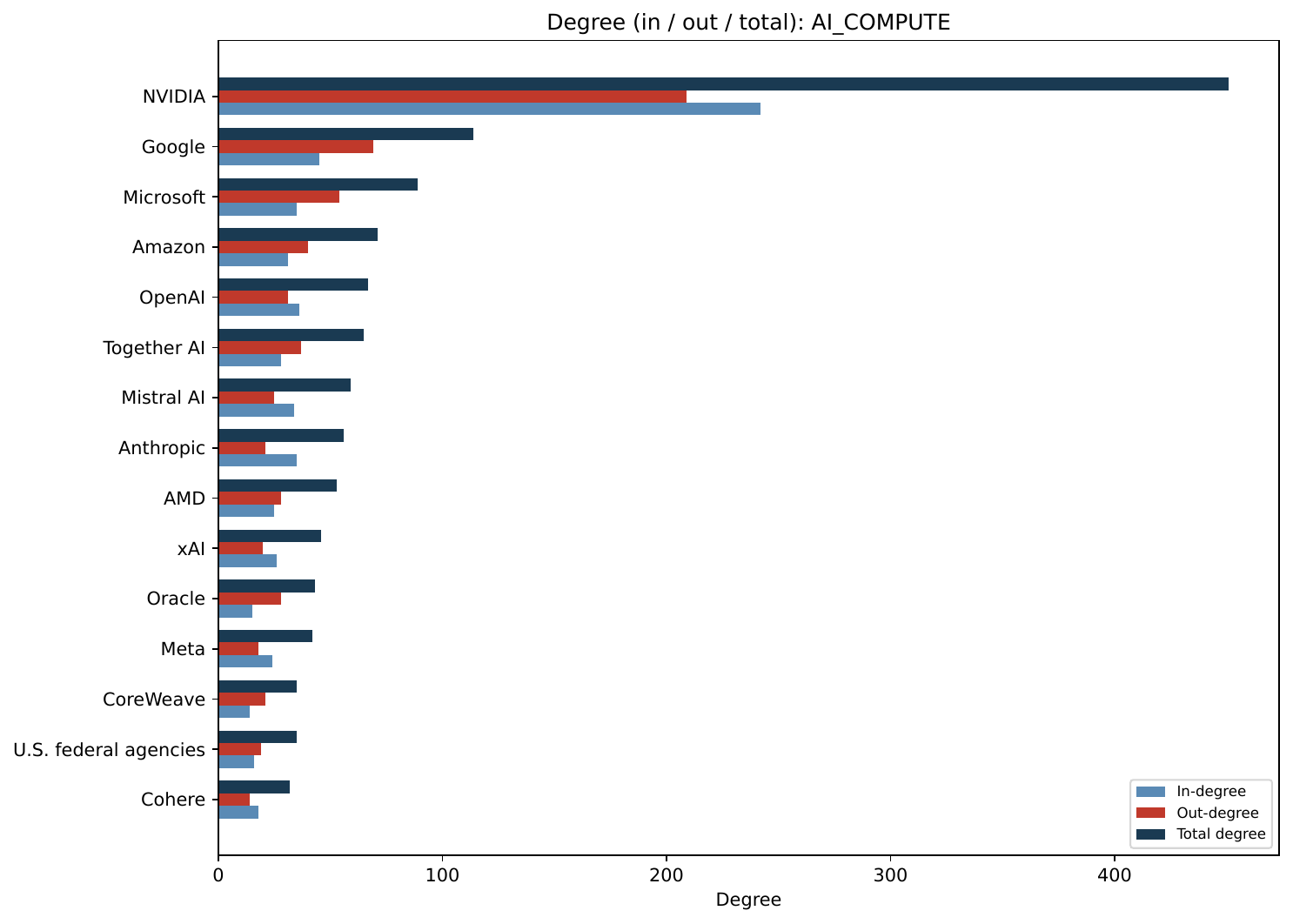}
\caption{\texttt{AI\_COMPUTE}}
\end{subfigure}
\hfill
\begin{subfigure}[t]{0.48\textwidth}
\centering
\includegraphics[width=\linewidth]{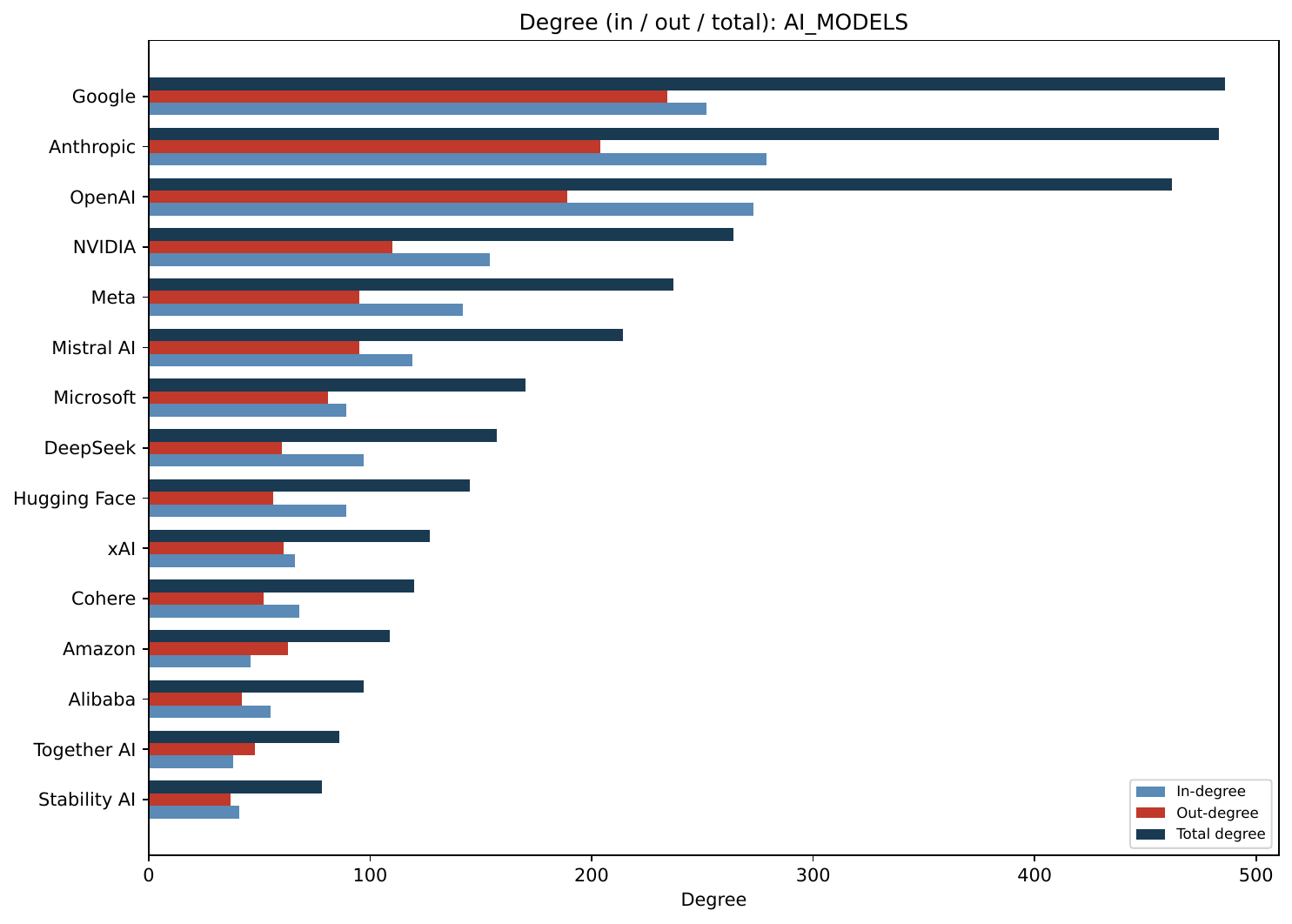}
\caption{\texttt{AI\_MODELS}}
\end{subfigure}

\vspace{0.8em}

\begin{subfigure}[t]{0.48\textwidth}
\centering
\includegraphics[width=\linewidth]{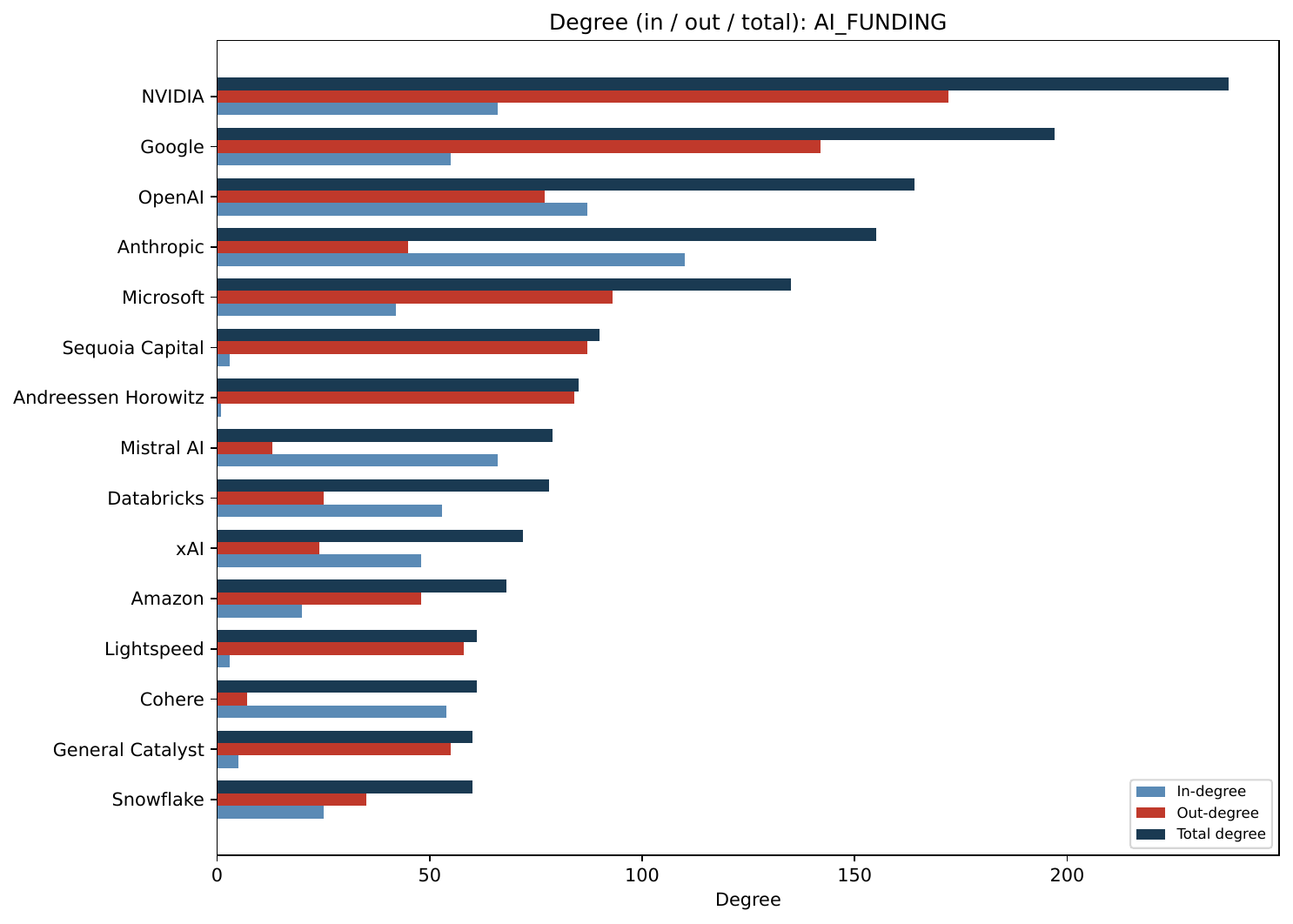}
\caption{\texttt{AI\_FUNDING}}
\end{subfigure}
\hfill
\begin{subfigure}[t]{0.48\textwidth}
\centering
\includegraphics[width=\linewidth]{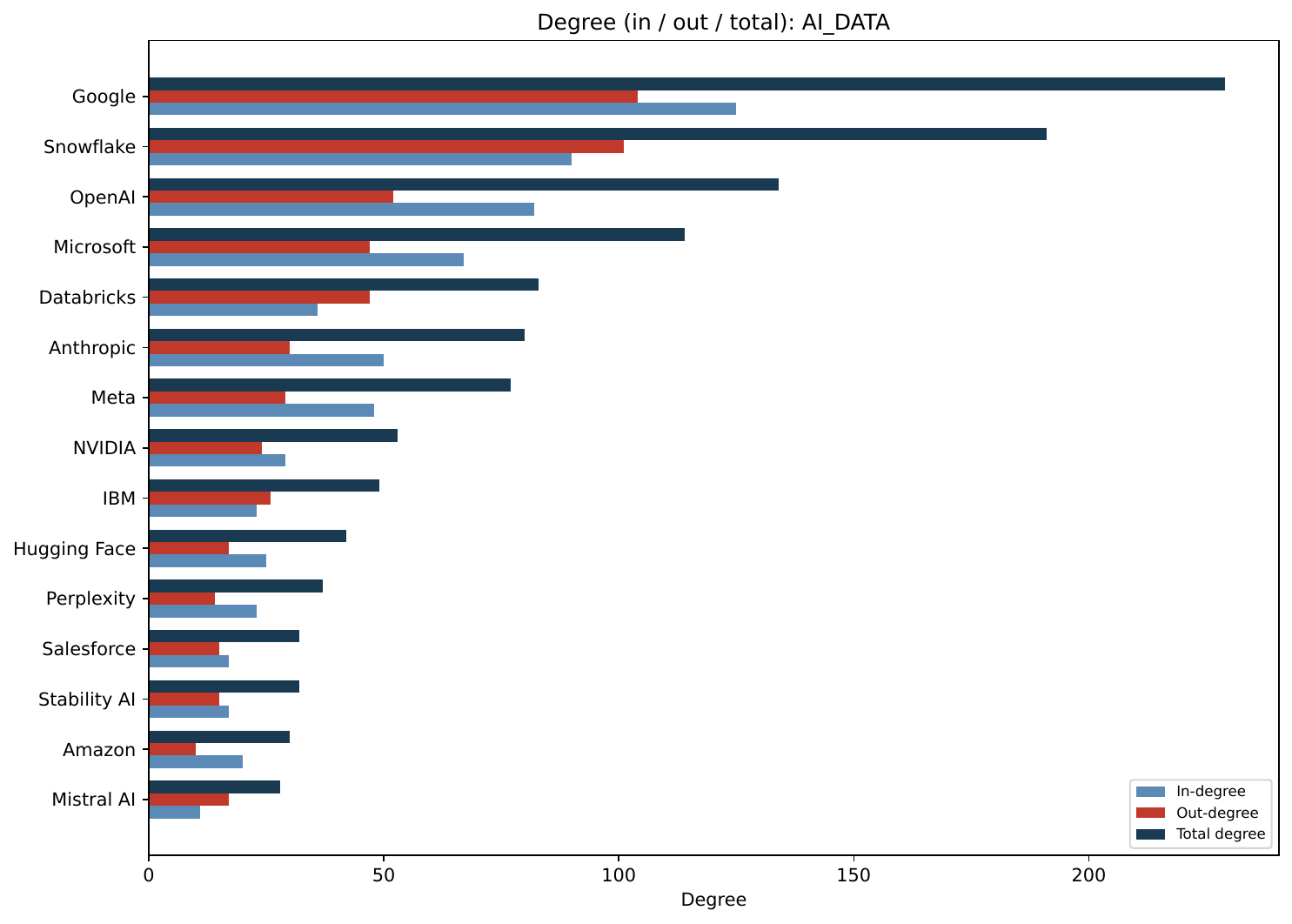}
\caption{\texttt{AI\_DATA}}
\end{subfigure}

\vspace{0.8em}

\begin{subfigure}[t]{0.48\textwidth}
\centering
\includegraphics[width=\linewidth]{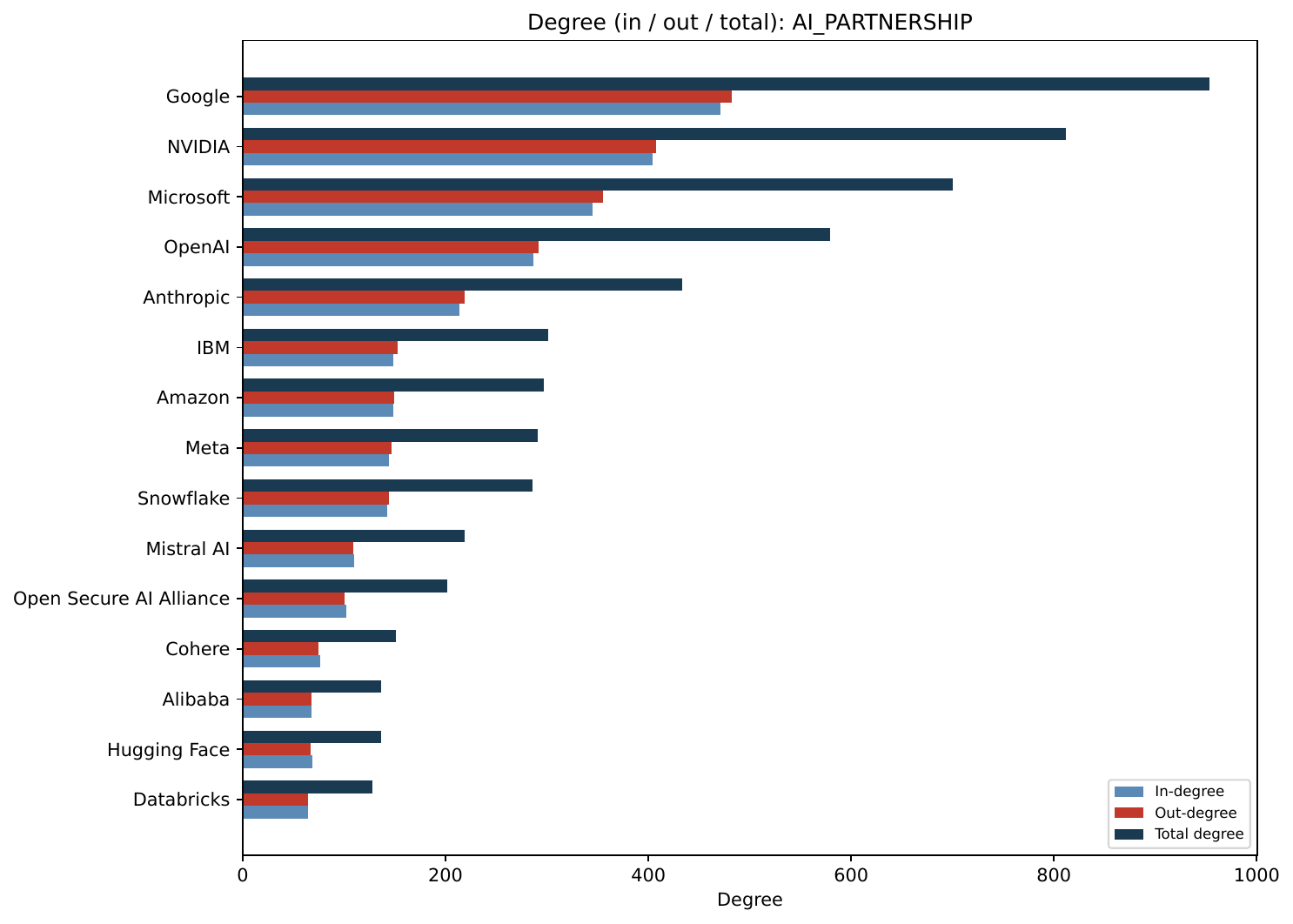}
\caption{\texttt{AI\_PARTNERSHIP}}
\end{subfigure}
\hfill
\begin{subfigure}[t]{0.48\textwidth}
\centering
\includegraphics[width=\linewidth]{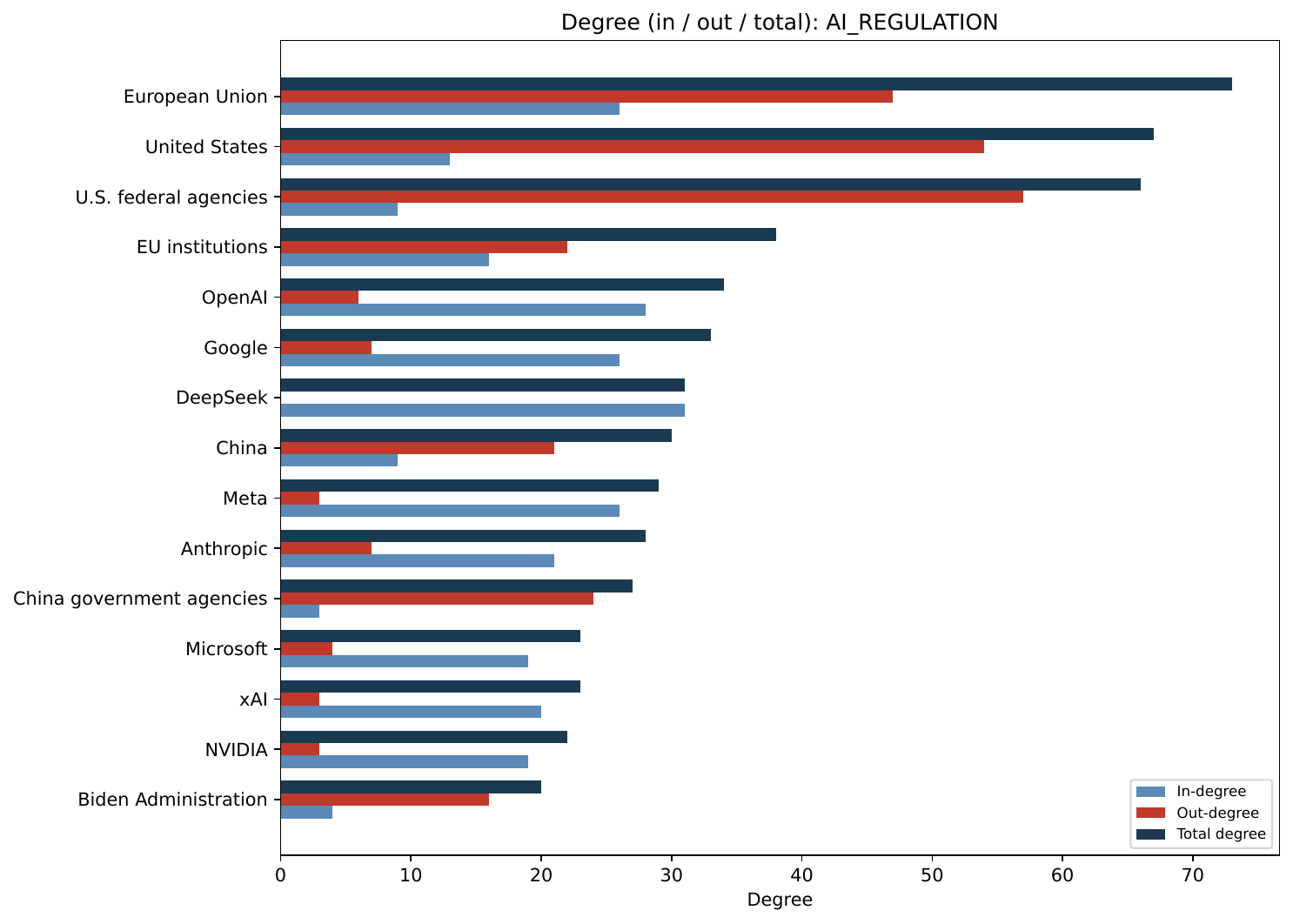}
\caption{\texttt{AI\_REGULATION}}
\end{subfigure}
\caption{Degree rankings (in / out / total) by AI relationship category (panel 1/4).}
\label{fig:ai_category_rankings}
\end{figure}

\begin{figure}[H]
\centering
\begin{subfigure}[t]{0.48\textwidth}
\centering
\includegraphics[width=\linewidth]{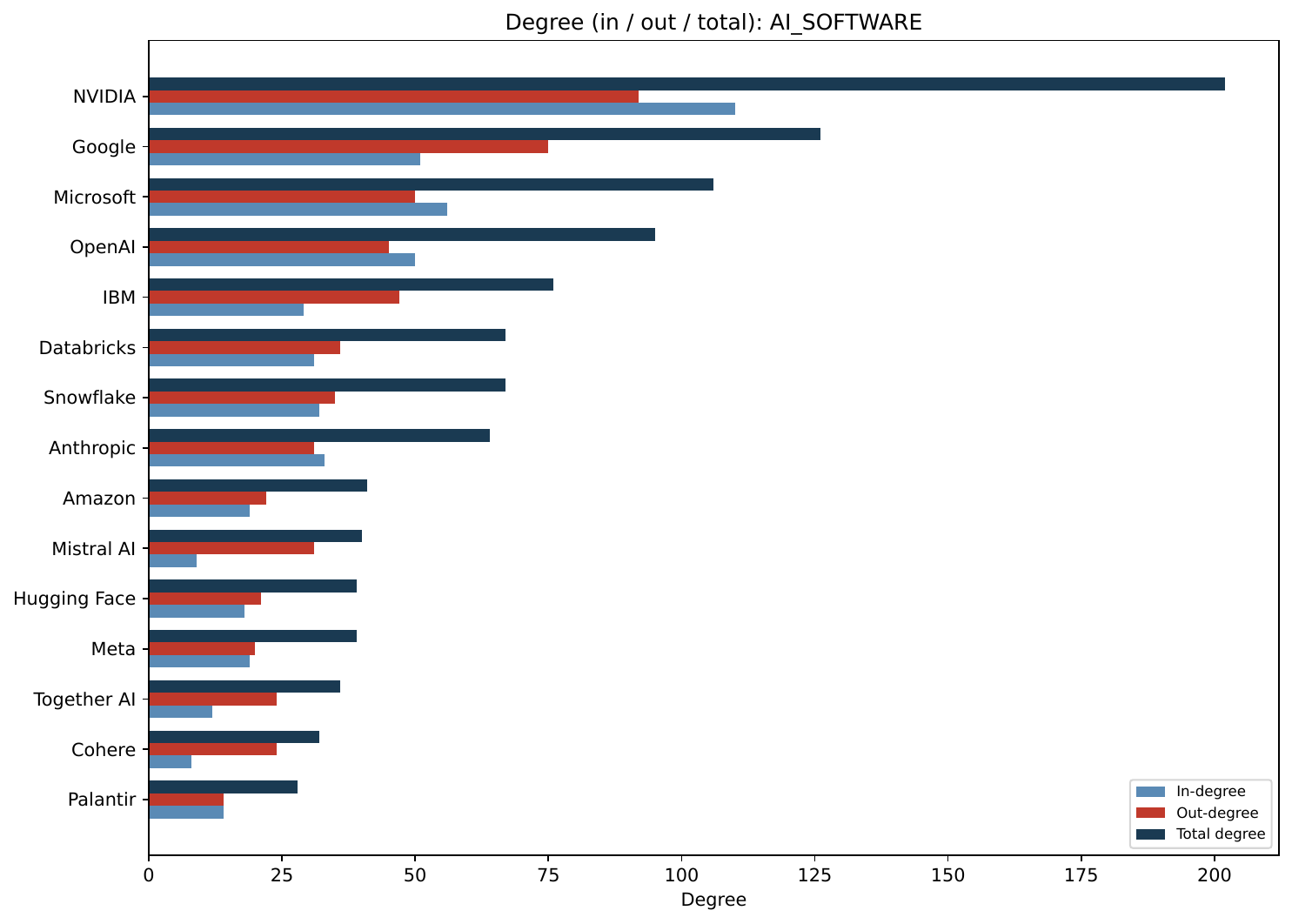}
\caption{\texttt{AI\_SOFTWARE}}
\end{subfigure}
\hfill
\begin{subfigure}[t]{0.48\textwidth}
\centering
\includegraphics[width=\linewidth]{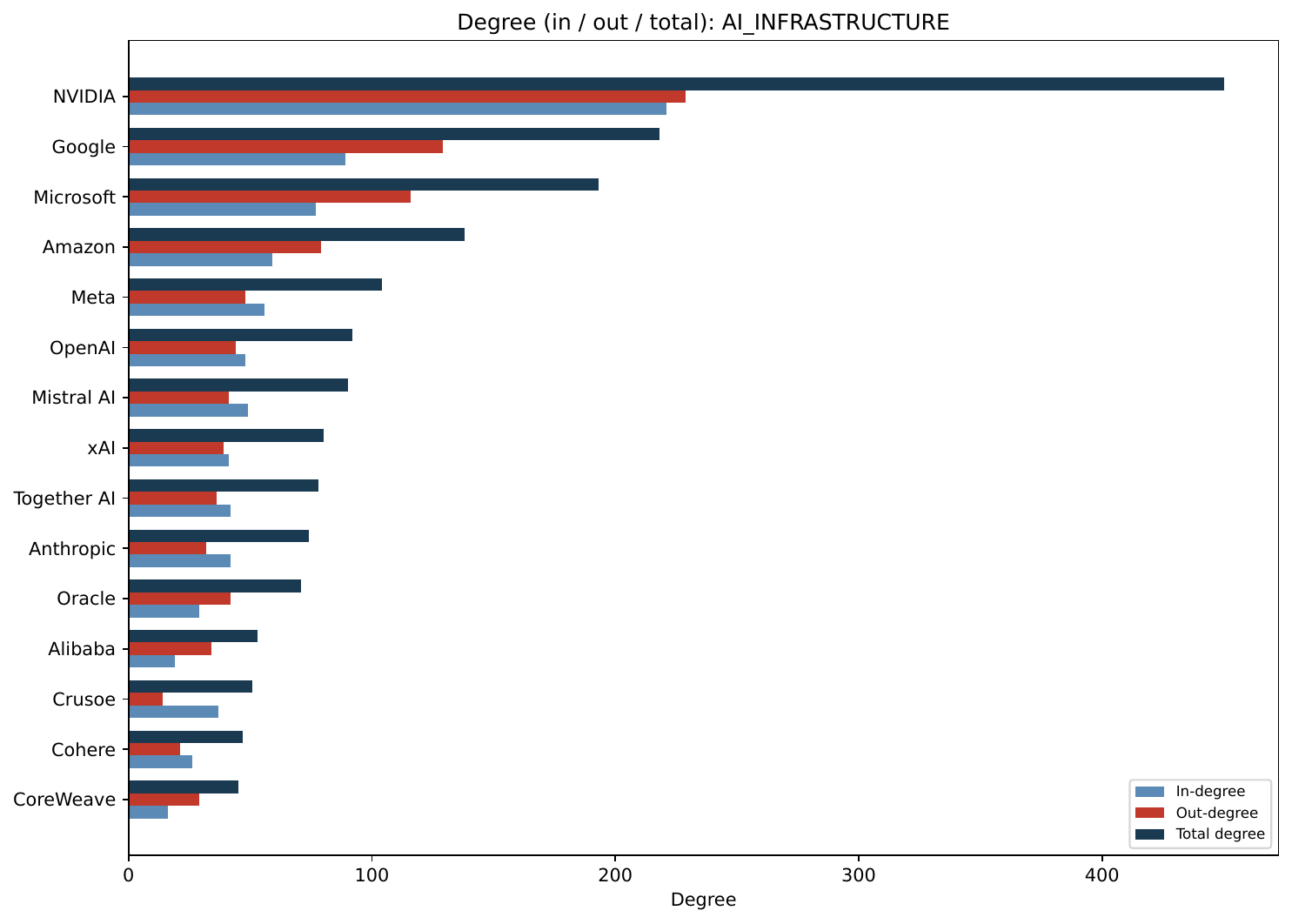}
\caption{\texttt{AI\_INFRASTRUCTURE}}
\end{subfigure}

\vspace{0.8em}

\begin{subfigure}[t]{0.48\textwidth}
\centering
\includegraphics[width=\linewidth]{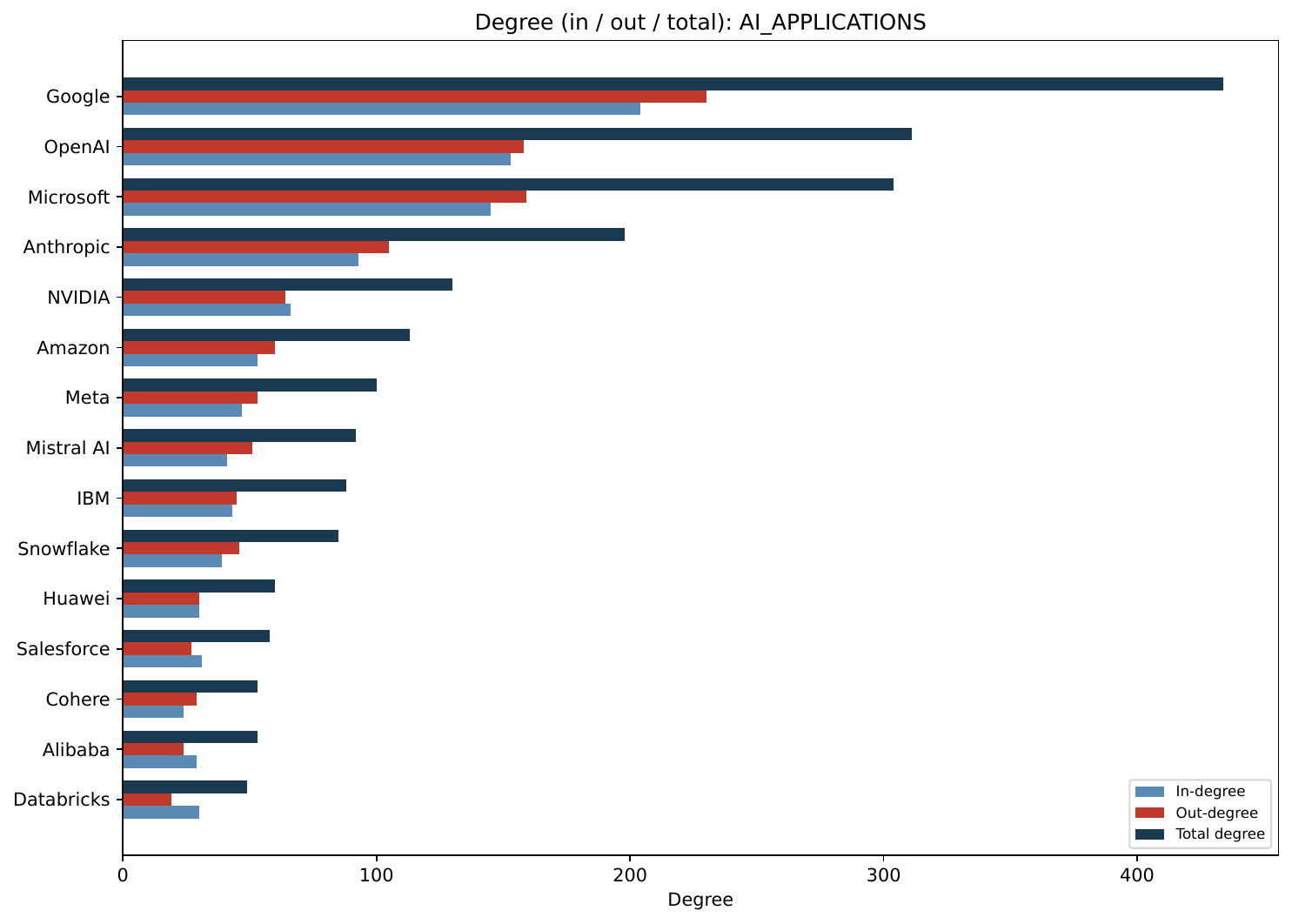}
\caption{\texttt{AI\_APPLICATIONS}}
\end{subfigure}
\hfill
\begin{subfigure}[t]{0.48\textwidth}
\centering
\includegraphics[width=\linewidth]{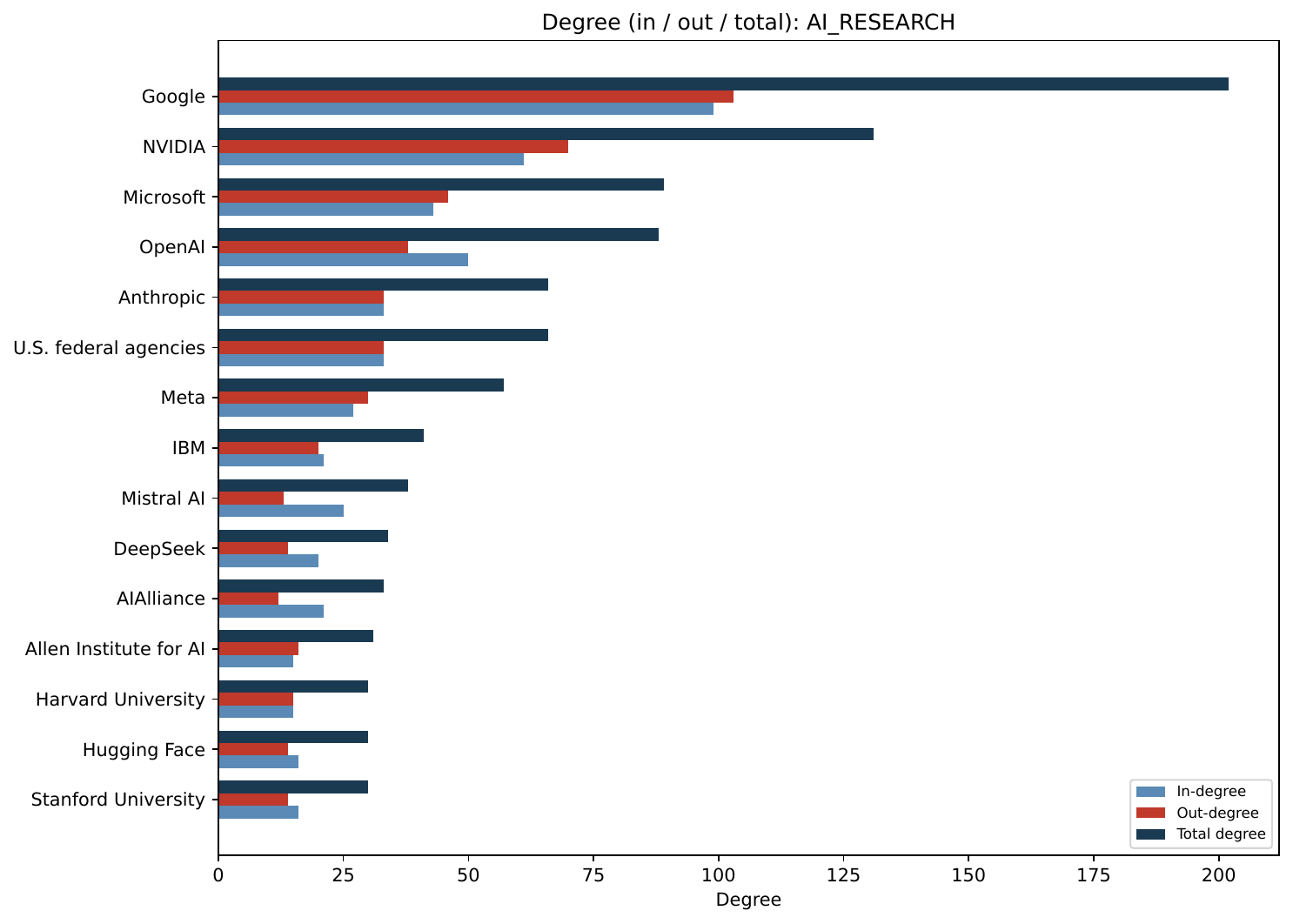}
\caption{\texttt{AI\_RESEARCH}}
\end{subfigure}

\vspace{0.8em}

\begin{subfigure}[t]{0.48\textwidth}
\centering
\includegraphics[width=\linewidth]{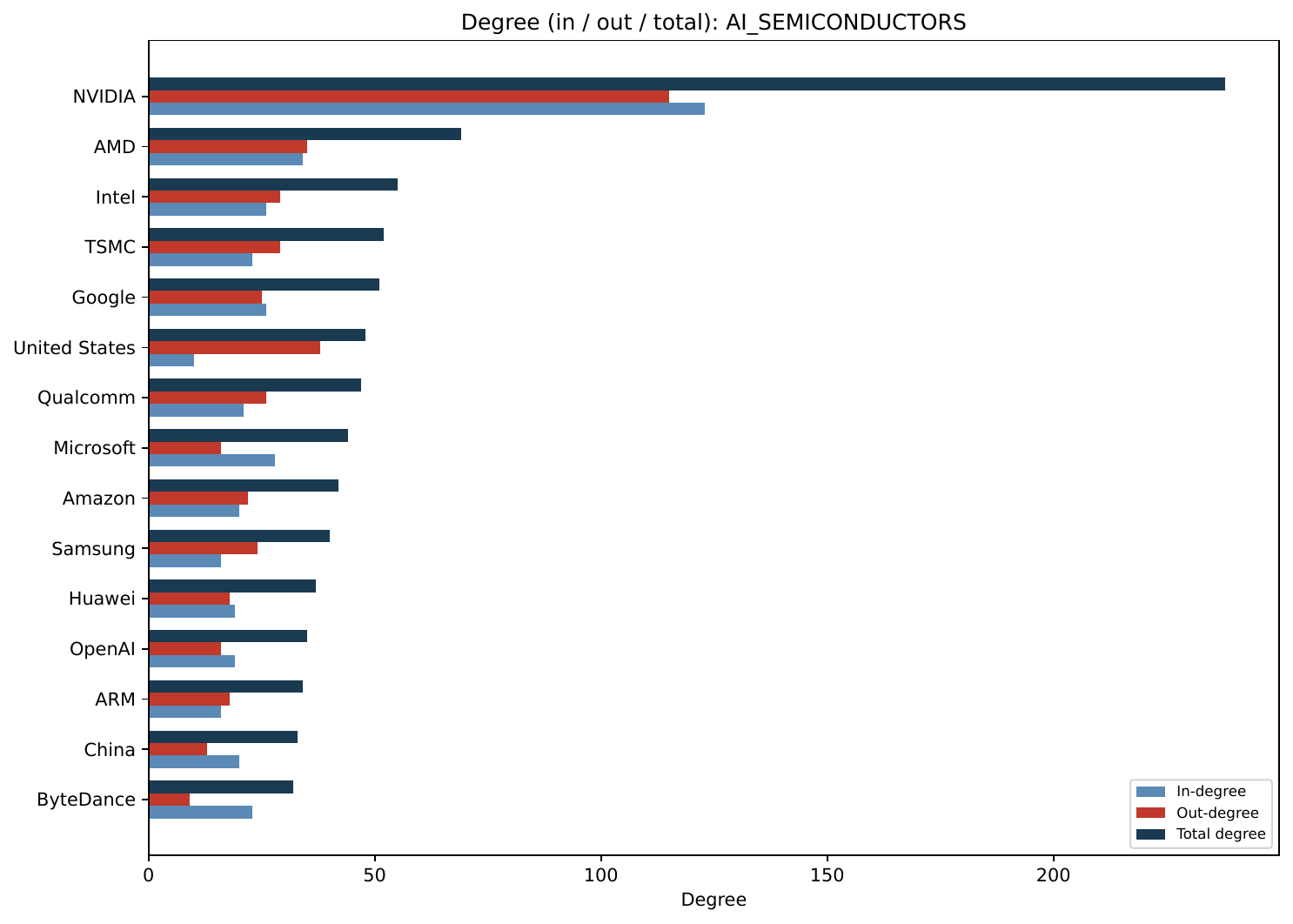}
\caption{\texttt{AI\_SEMICONDUCTORS}}
\end{subfigure}
\hfill
\begin{subfigure}[t]{0.48\textwidth}
\centering
\includegraphics[width=\linewidth]{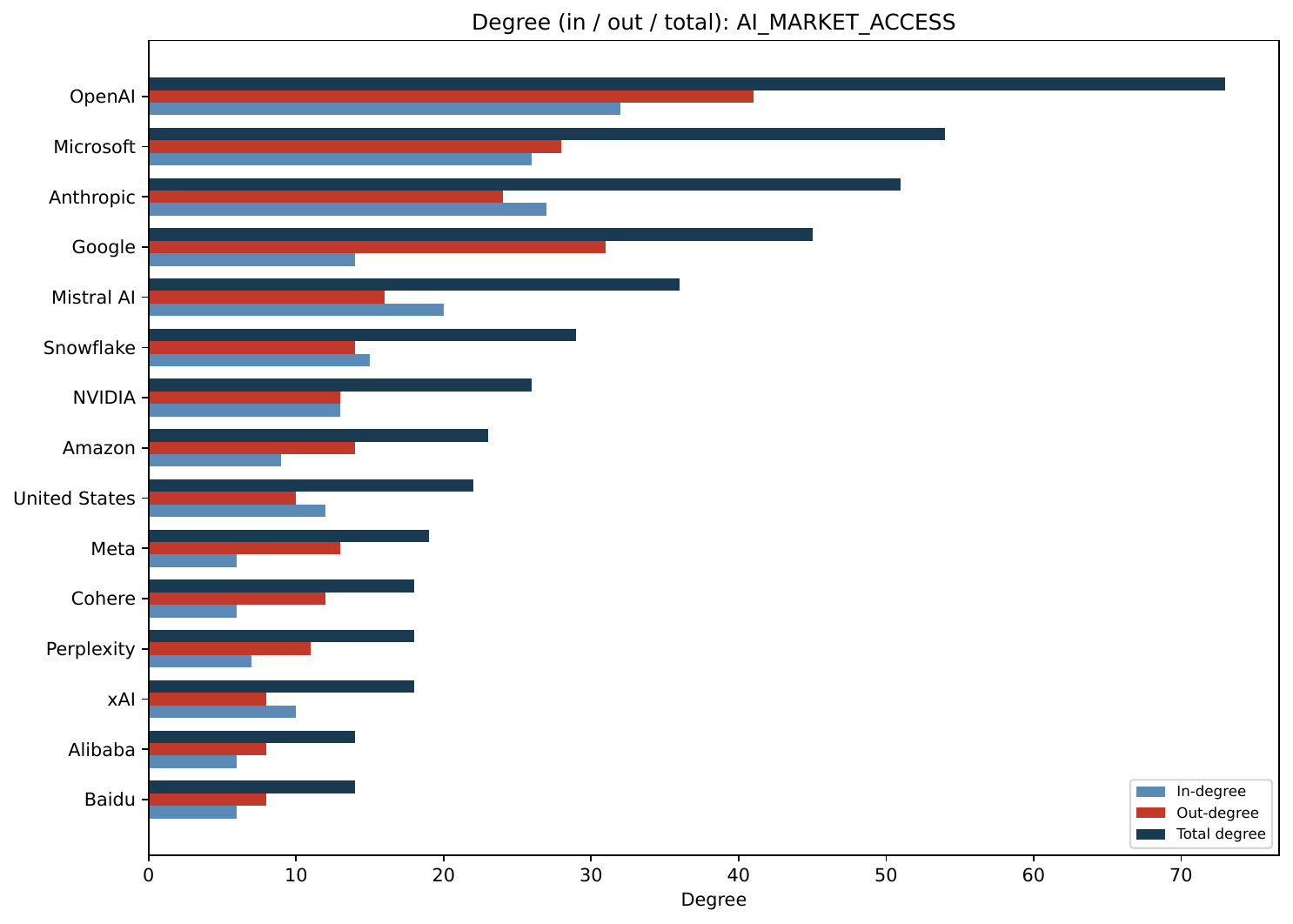}
\caption{\texttt{AI\_MARKET\_ACCESS}}
\end{subfigure}
\caption{Degree rankings (in / out / total) by AI relationship category (panel 2/4).}
\label{fig:ai_category_rankings_b}
\end{figure}

\begin{figure}[H]
\centering
\begin{subfigure}[t]{0.48\textwidth}
\centering
\includegraphics[width=\linewidth]{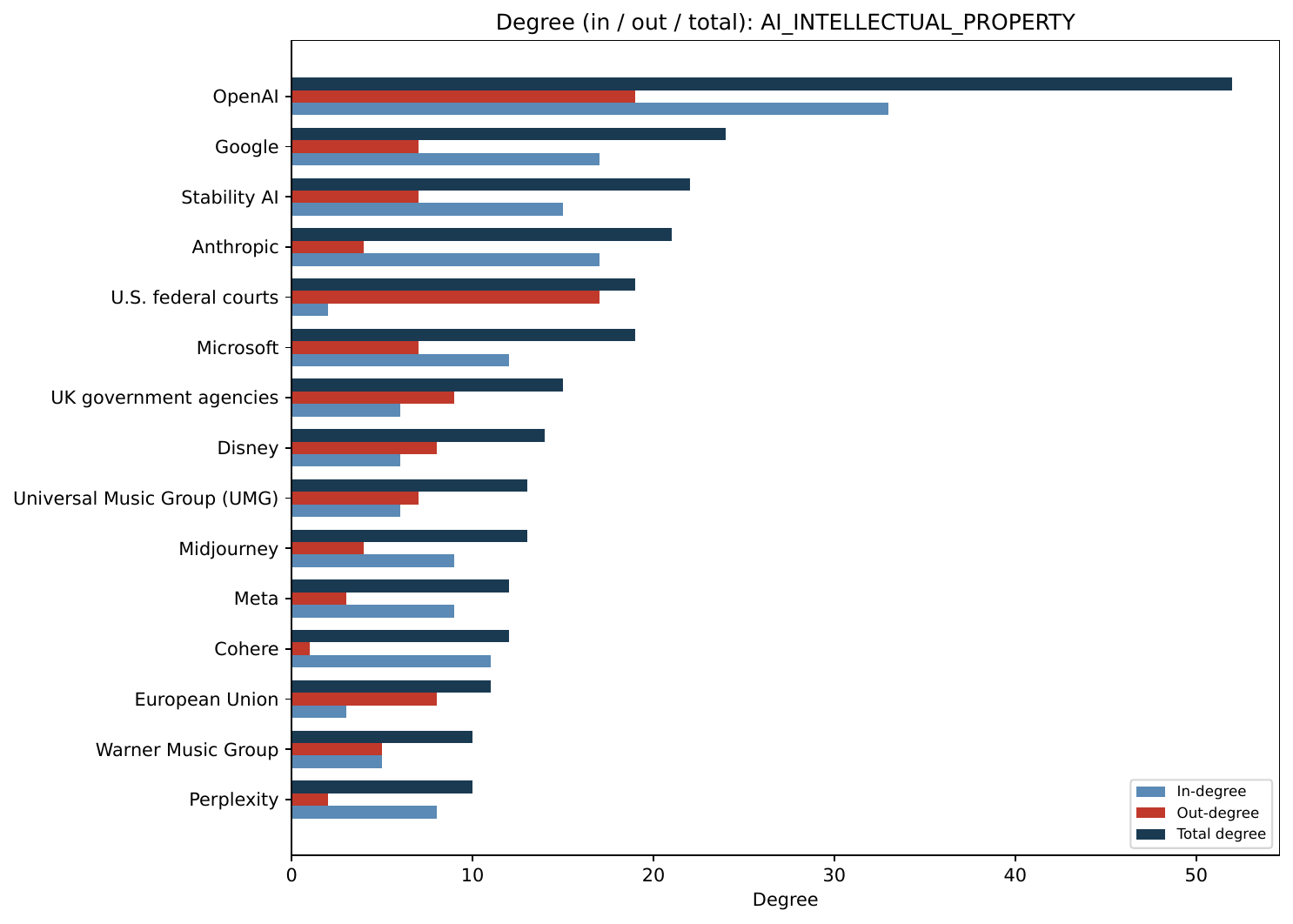}
\caption{\texttt{AI\_INTELLECTUAL\_PROPERTY}}
\end{subfigure}
\hfill
\begin{subfigure}[t]{0.48\textwidth}
\centering
\includegraphics[width=\linewidth]{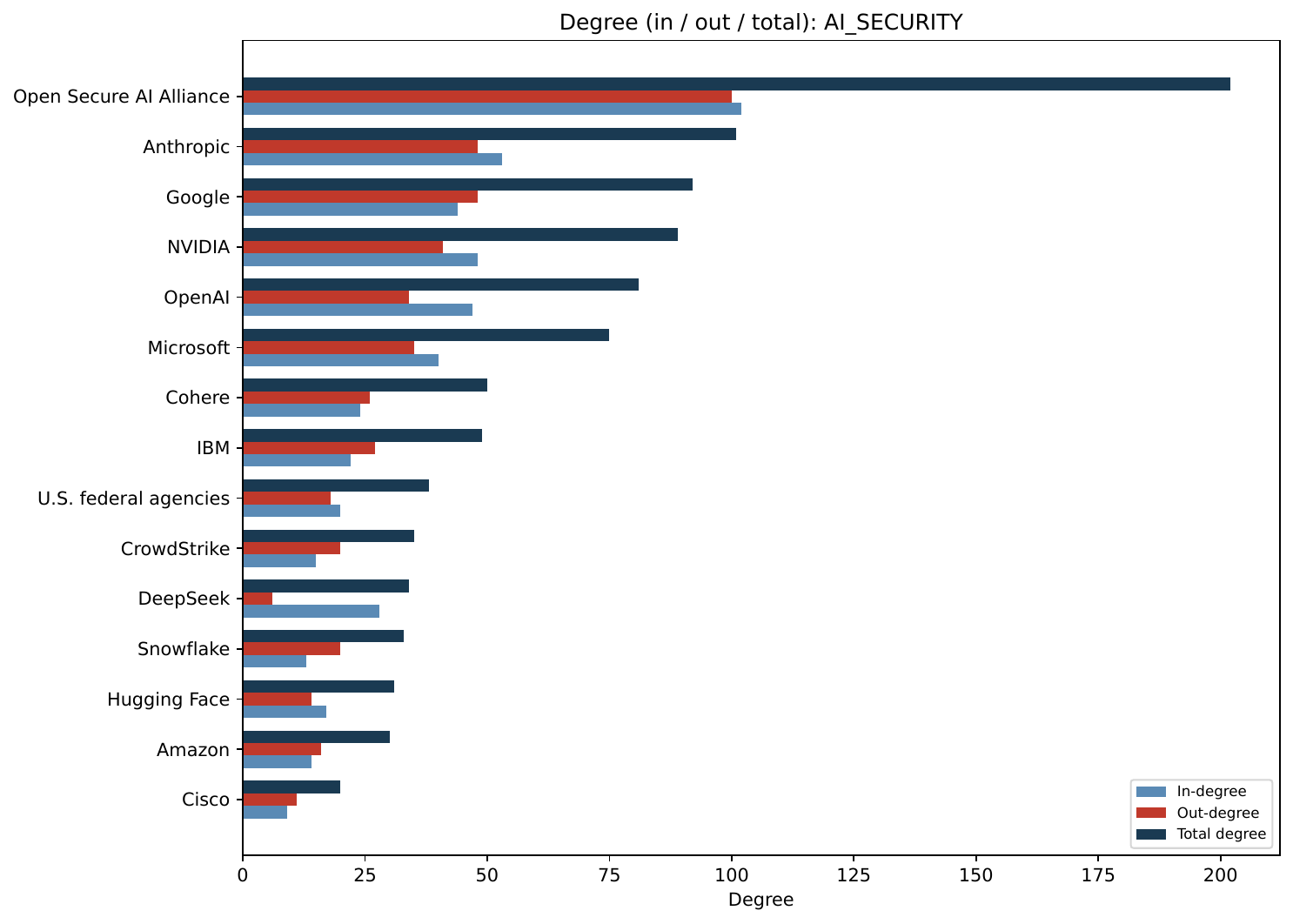}
\caption{\texttt{AI\_SECURITY}}
\end{subfigure}

\vspace{0.8em}

\begin{subfigure}[t]{0.48\textwidth}
\centering
\includegraphics[width=\linewidth]{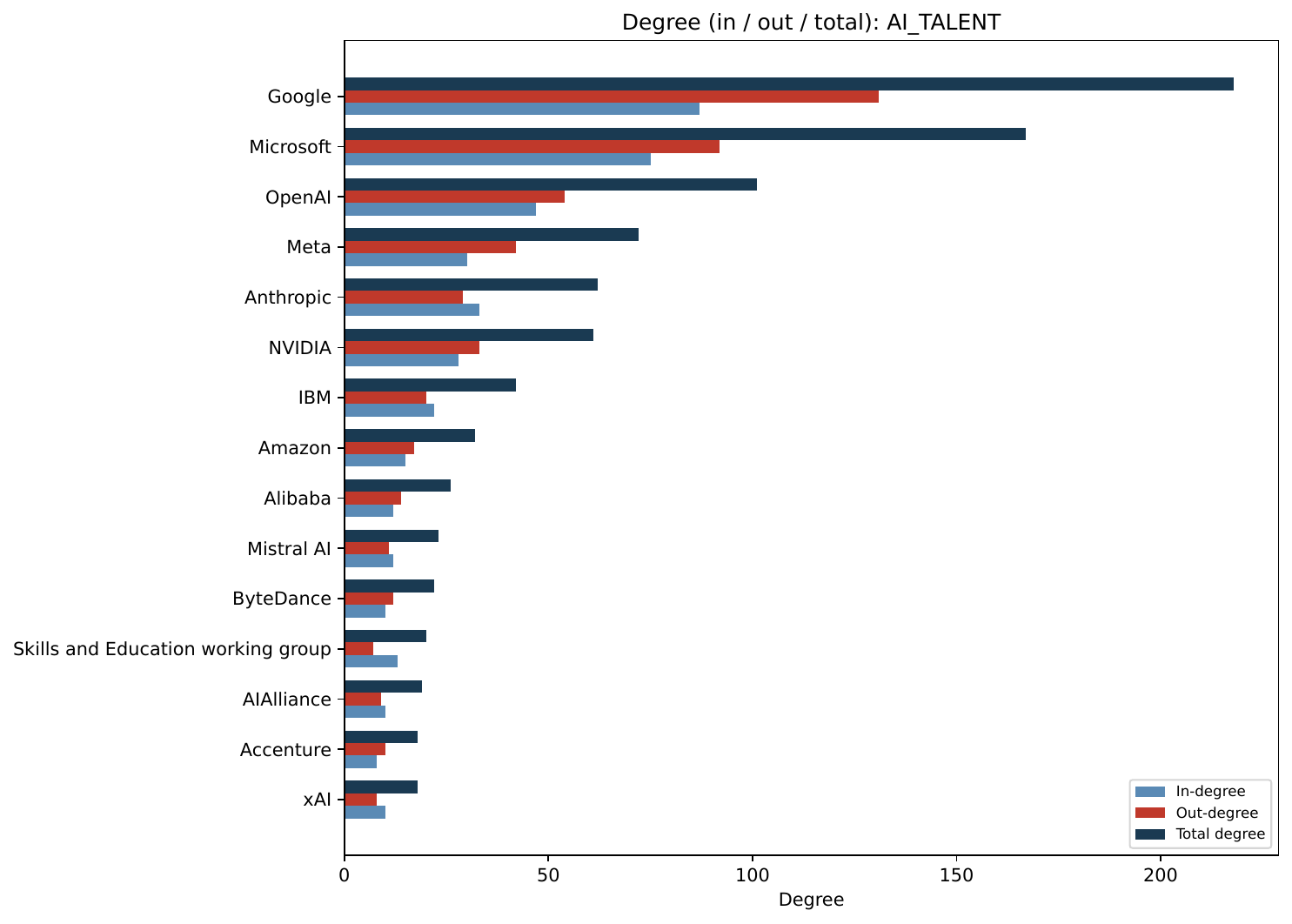}
\caption{\texttt{AI\_TALENT}}
\end{subfigure}
\hfill
\begin{subfigure}[t]{0.48\textwidth}
\centering
\includegraphics[width=\linewidth]{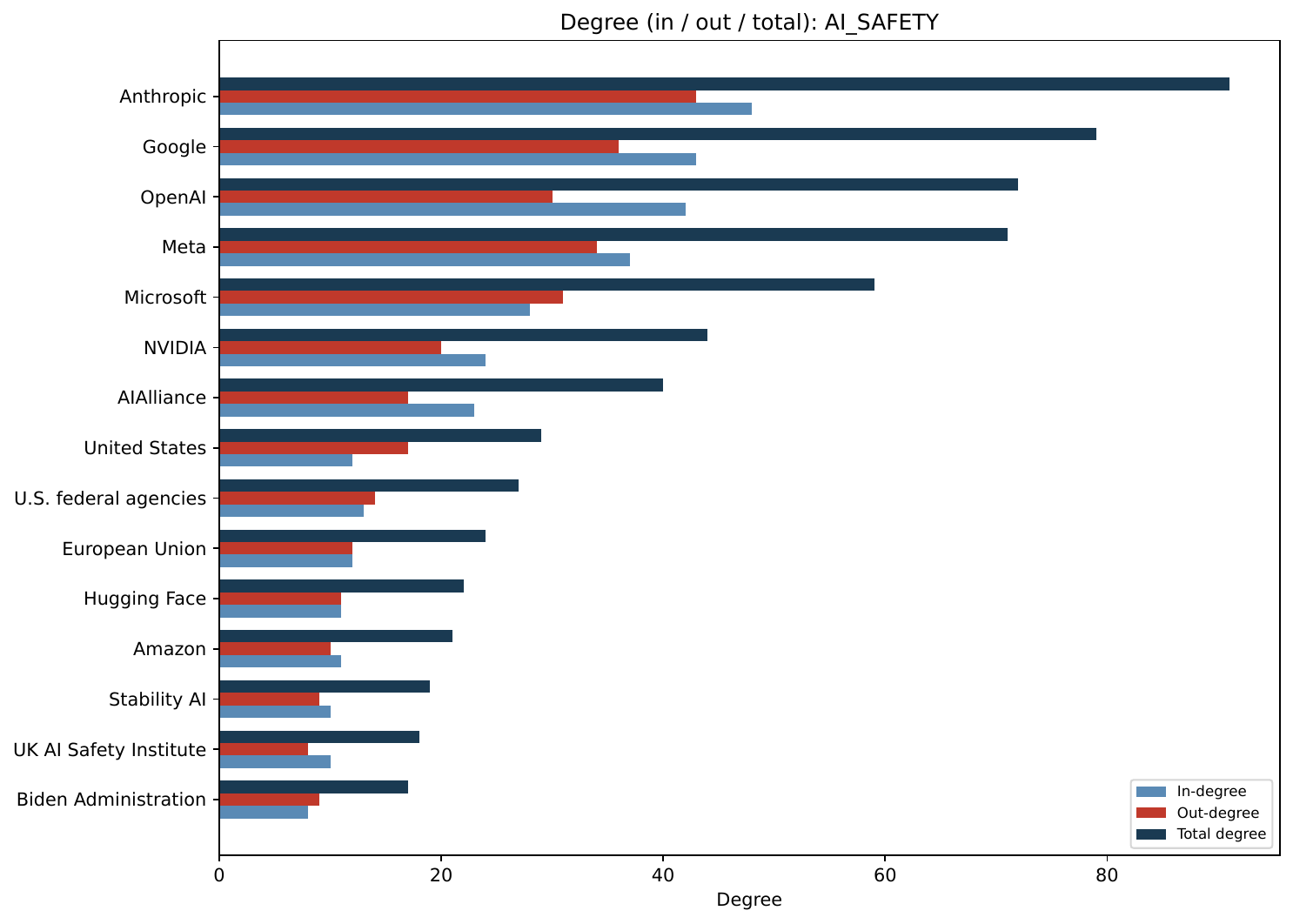}
\caption{\texttt{AI\_SAFETY}}
\end{subfigure}

\vspace{0.8em}

\begin{subfigure}[t]{0.48\textwidth}
\centering
\includegraphics[width=\linewidth]{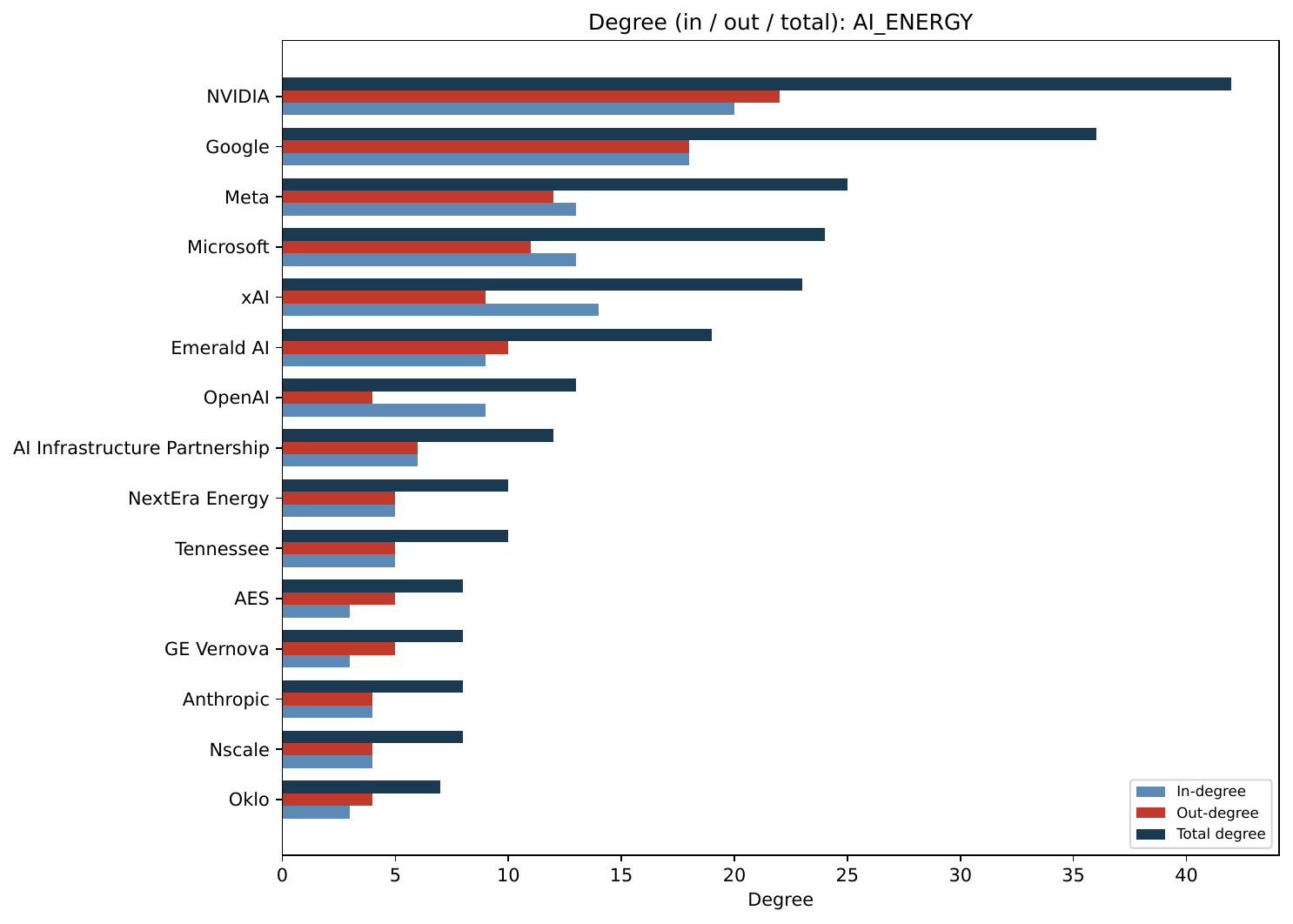}
\caption{\texttt{AI\_ENERGY}}
\end{subfigure}
\hfill
\begin{subfigure}[t]{0.48\textwidth}
\centering
\includegraphics[width=\linewidth]{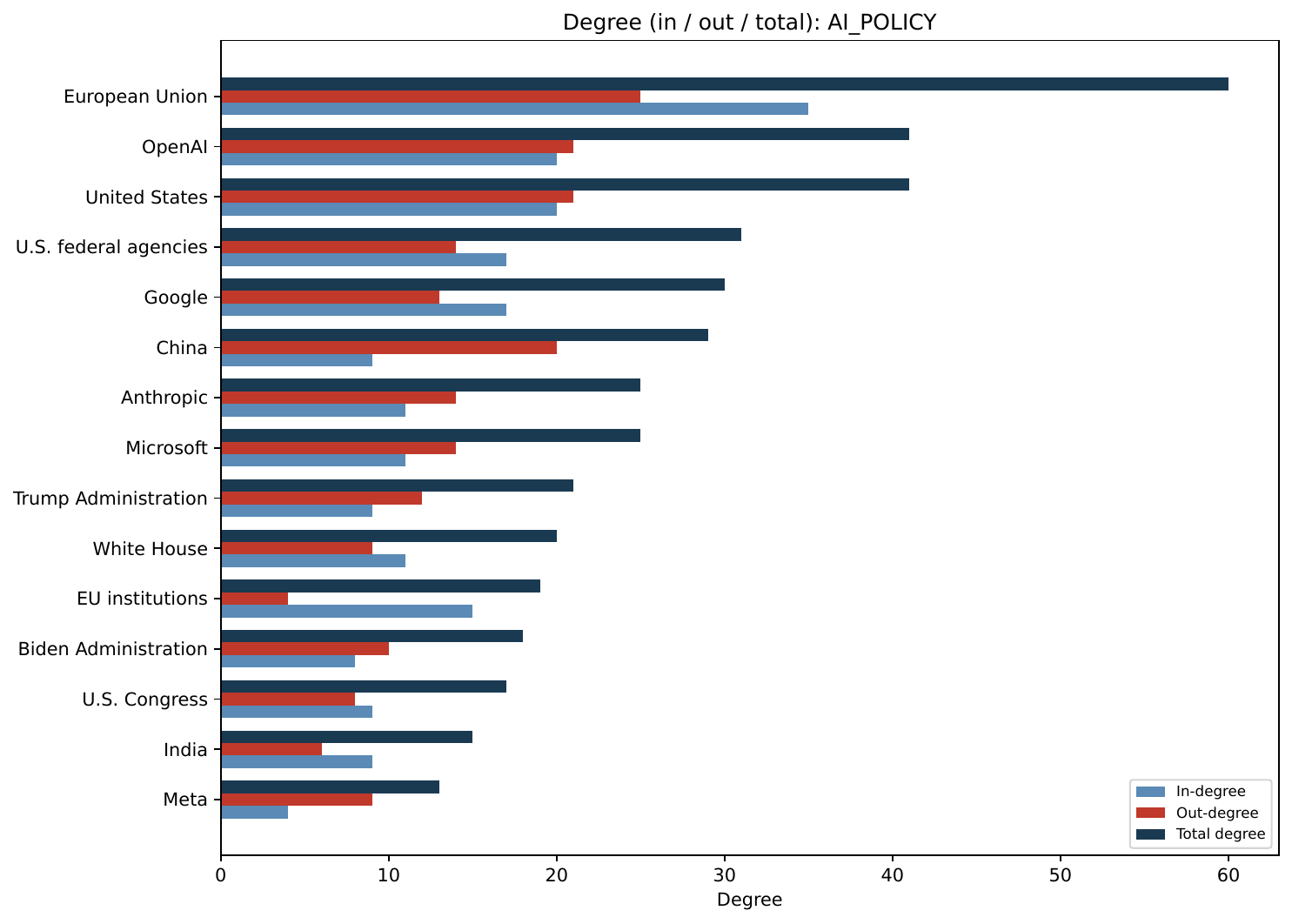}
\caption{\texttt{AI\_POLICY}}
\end{subfigure}
\caption{Degree rankings (in / out / total) by AI relationship category (panel 3/4).}
\label{fig:ai_category_rankings_c}
\end{figure}

\begin{figure}[H]
\centering
\begin{subfigure}[t]{0.48\textwidth}
\centering
\includegraphics[width=\linewidth]{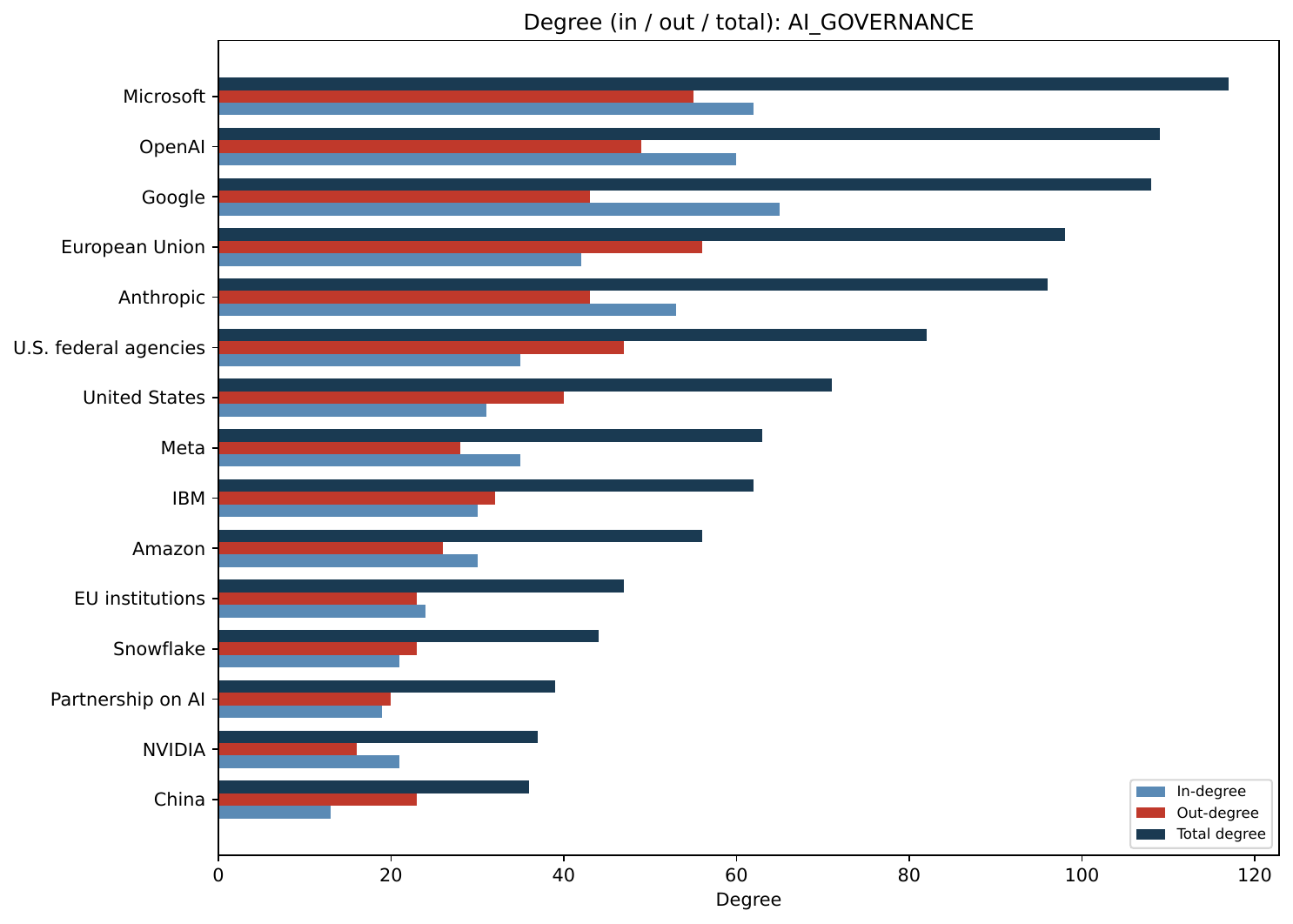}
\caption{\texttt{AI\_GOVERNANCE}}
\end{subfigure}
\hfill
\begin{subfigure}[t]{0.48\textwidth}
\centering
\includegraphics[width=\linewidth]{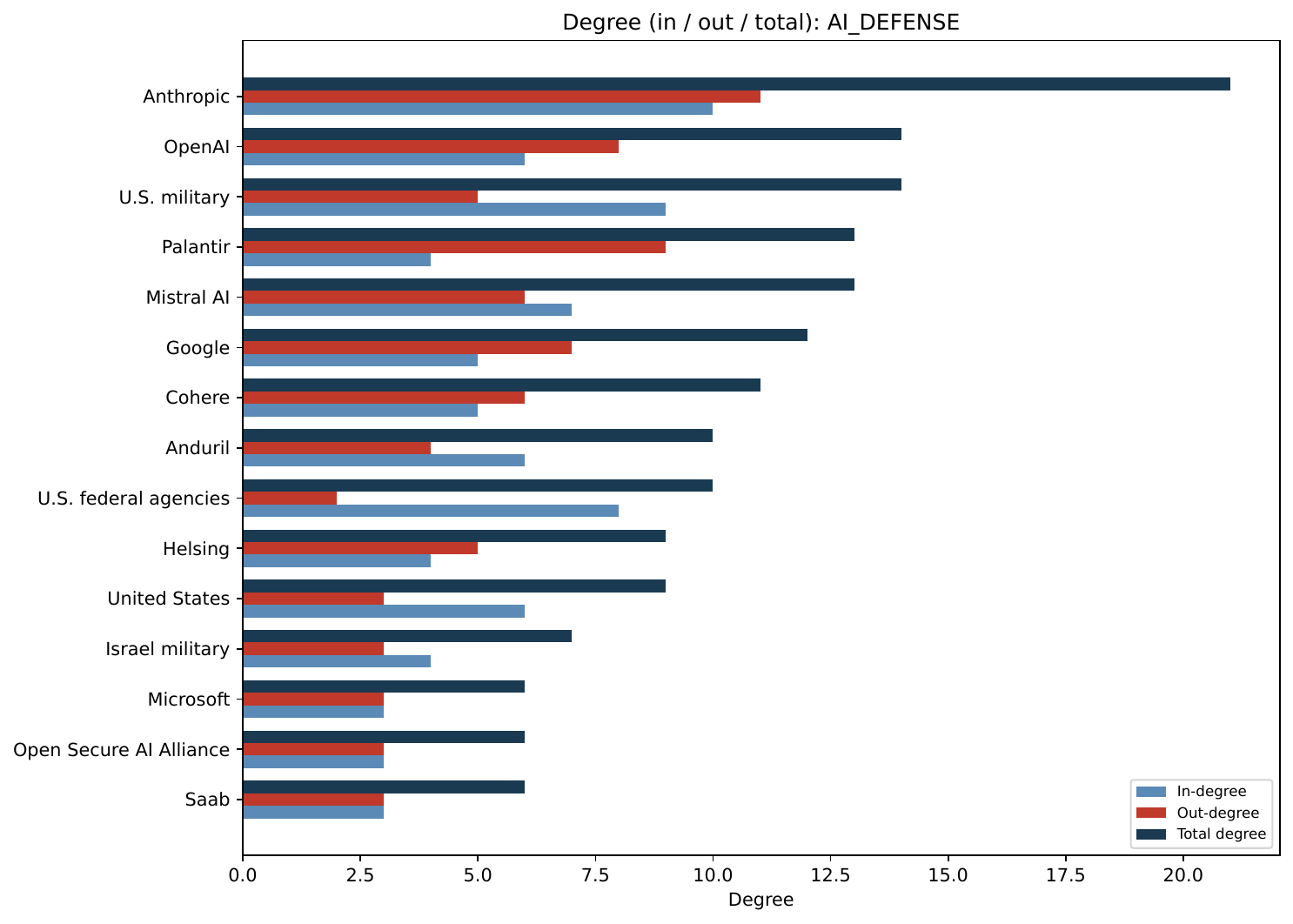}
\caption{\texttt{AI\_DEFENSE}}
\end{subfigure}

\vspace{0.8em}

\begin{subfigure}[t]{0.48\textwidth}
\centering
\includegraphics[width=\linewidth]{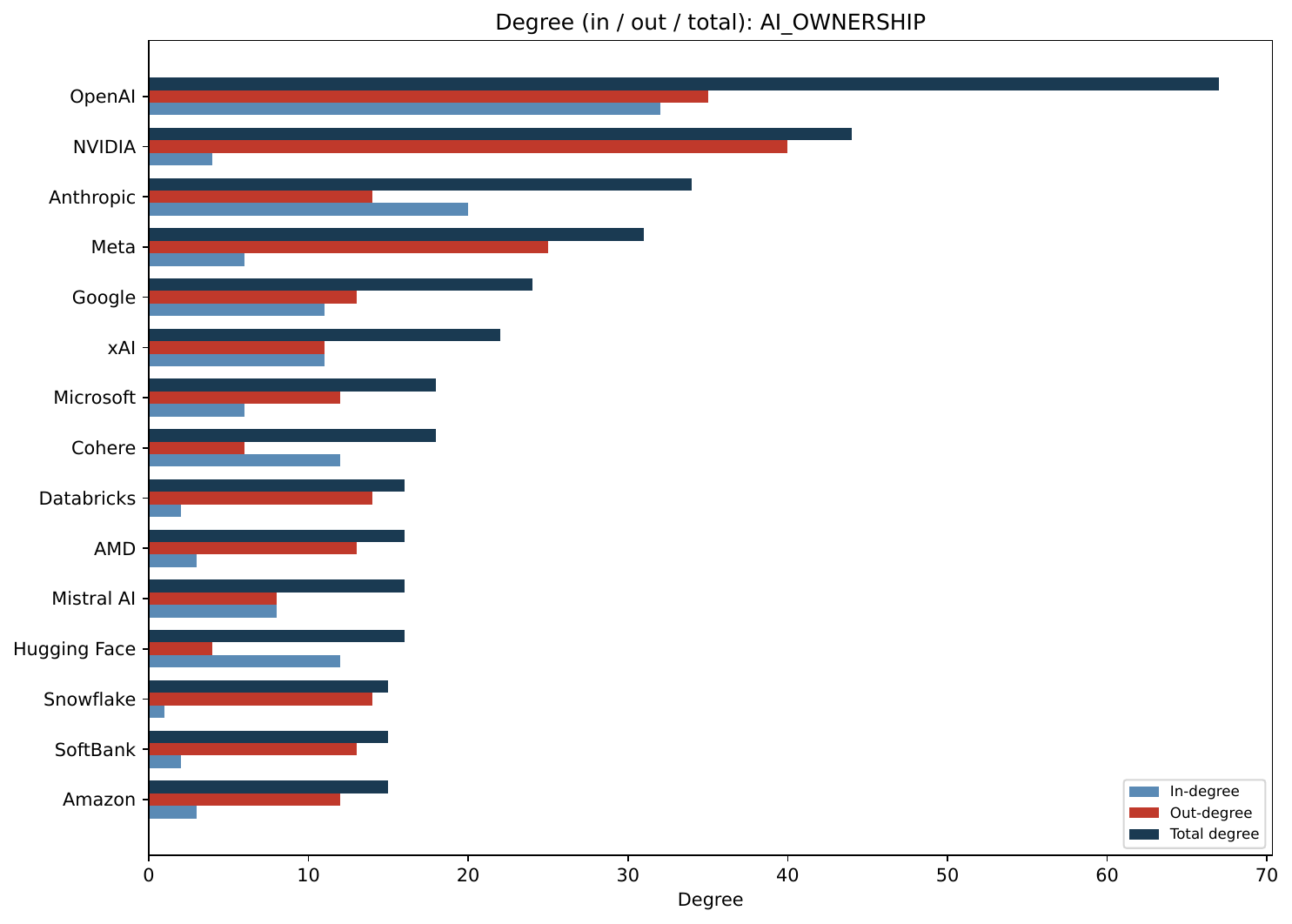}
\caption{\texttt{AI\_OWNERSHIP}}
\end{subfigure}
\hfill
\begin{subfigure}[t]{0.48\textwidth}
\centering
\includegraphics[width=\linewidth]{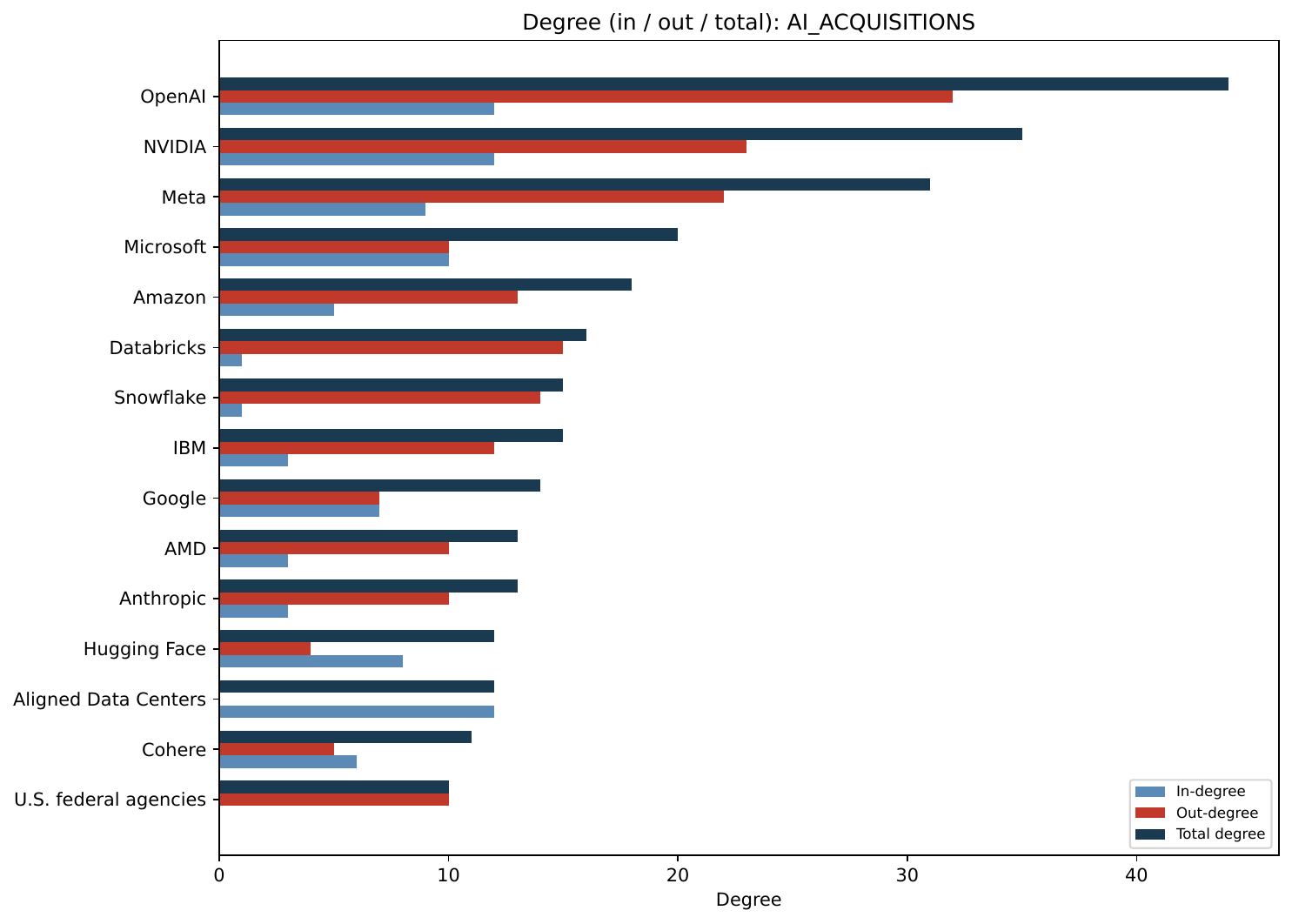}
\caption{\texttt{AI\_ACQUISITIONS}}
\end{subfigure}
\caption{Degree rankings (in / out / total) by AI relationship category (panel 4/4).}
\label{fig:ai_category_rankings_d}
\end{figure}

\begin{figure}[H]
\centering
\begin{subfigure}[t]{0.48\textwidth}
\centering
\includegraphics[width=\linewidth]{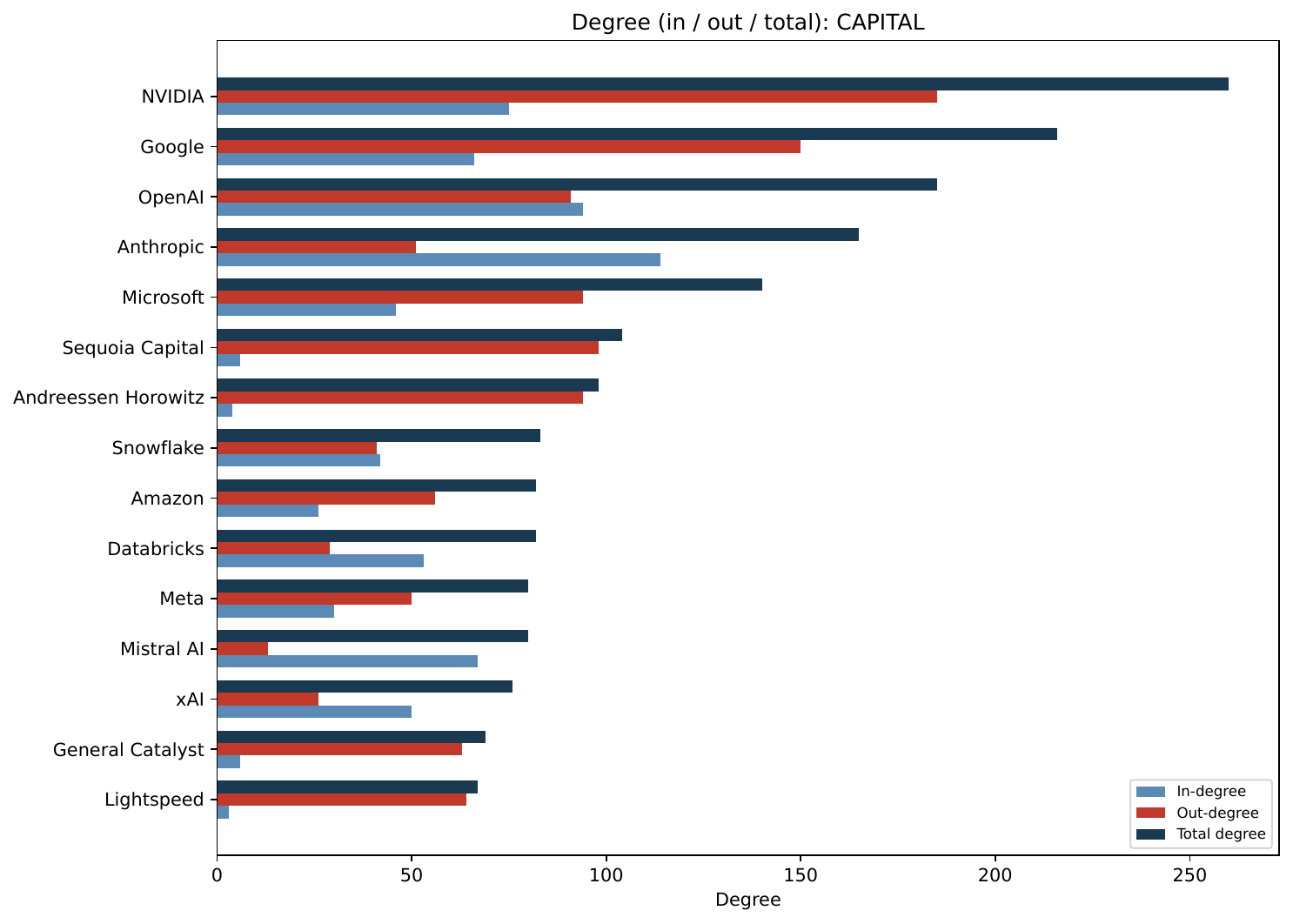}
\caption{\texttt{CAPITAL}}
\end{subfigure}
\hfill
\begin{subfigure}[t]{0.48\textwidth}
\centering
\includegraphics[width=\linewidth]{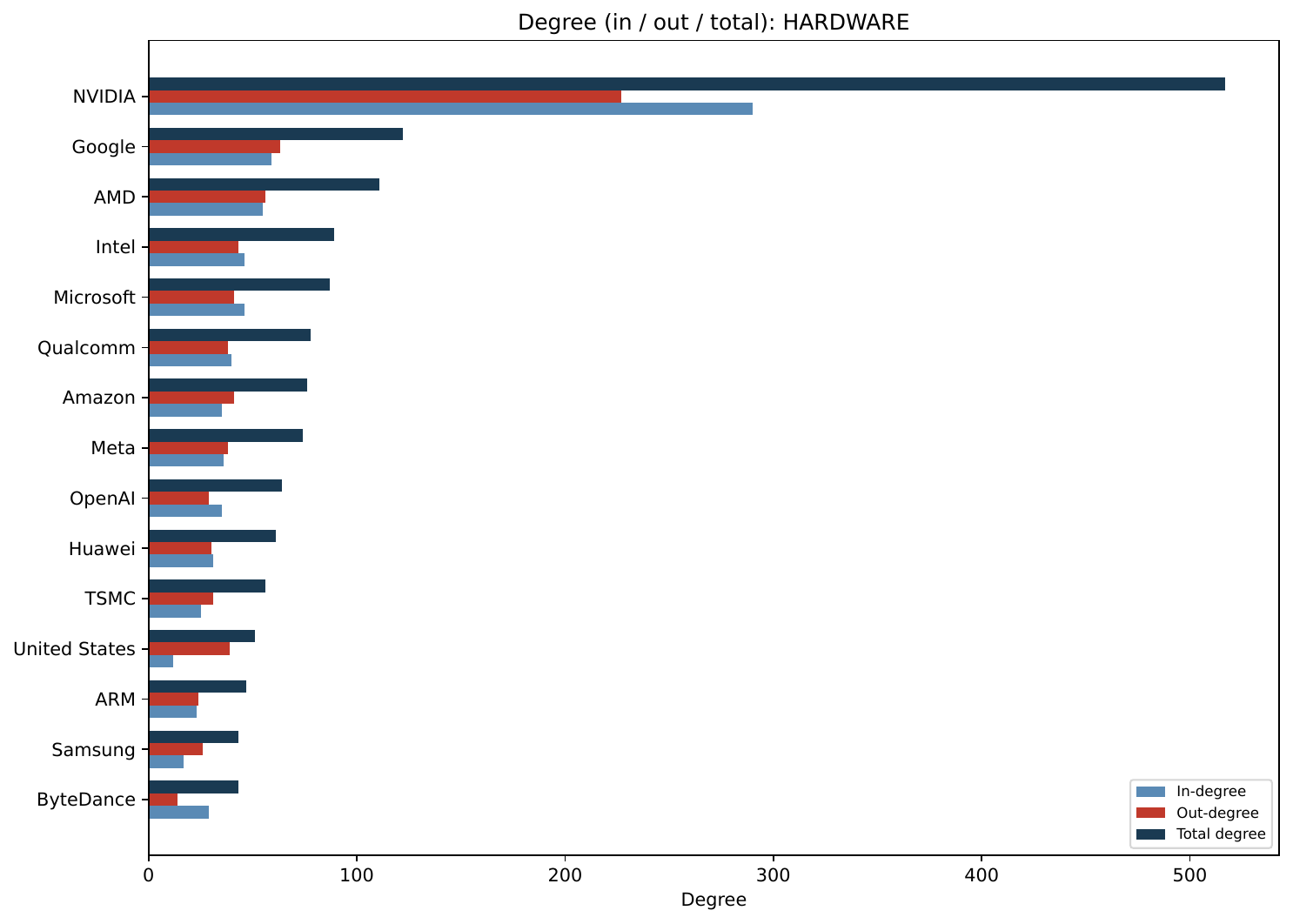}
\caption{\texttt{HARDWARE}}
\end{subfigure}

\vspace{0.8em}

\begin{subfigure}[t]{0.48\textwidth}
\centering
\includegraphics[width=\linewidth]{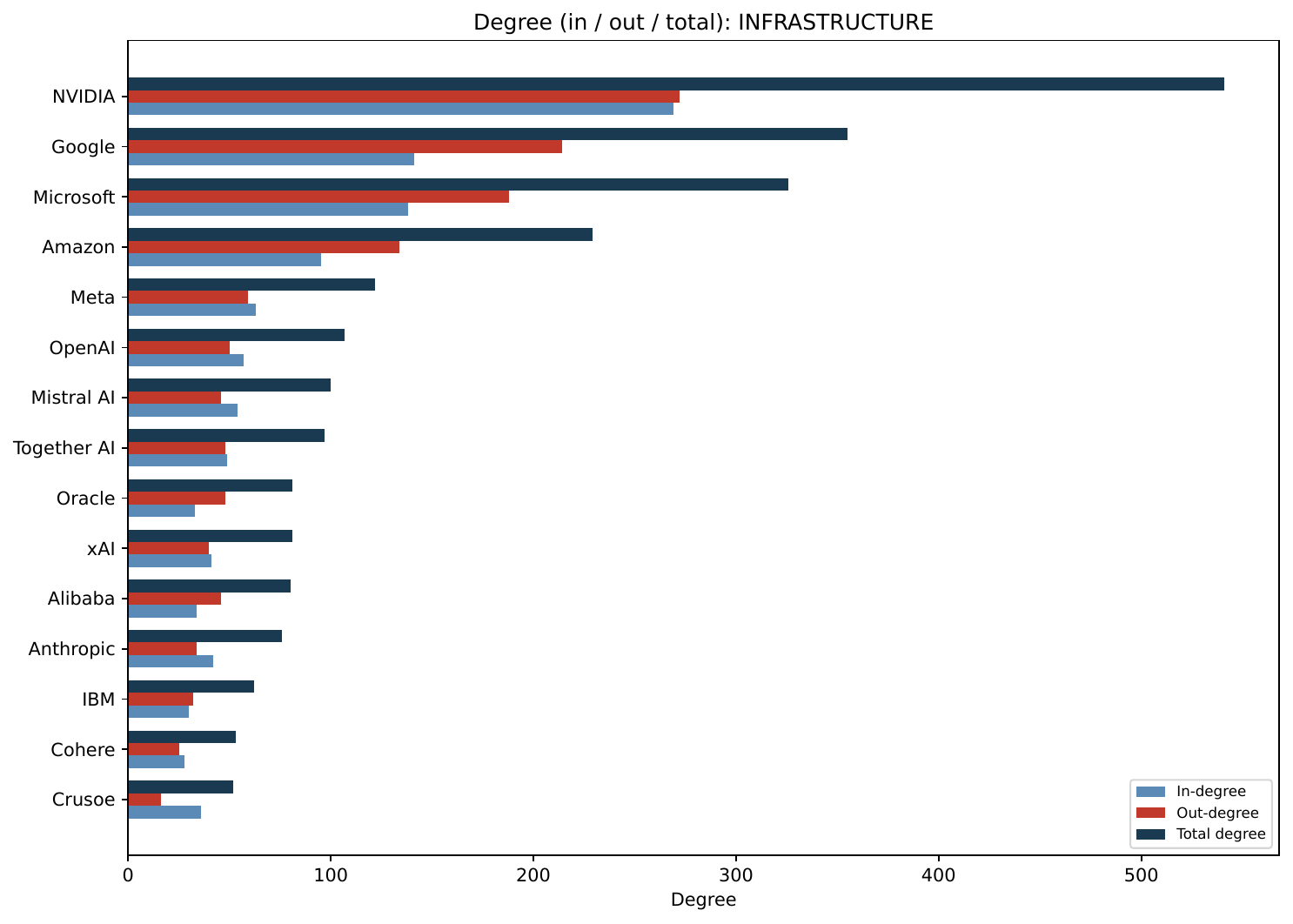}
\caption{\texttt{INFRASTRUCTURE}}
\end{subfigure}
\hfill
\begin{subfigure}[t]{0.48\textwidth}
\centering
\includegraphics[width=\linewidth]{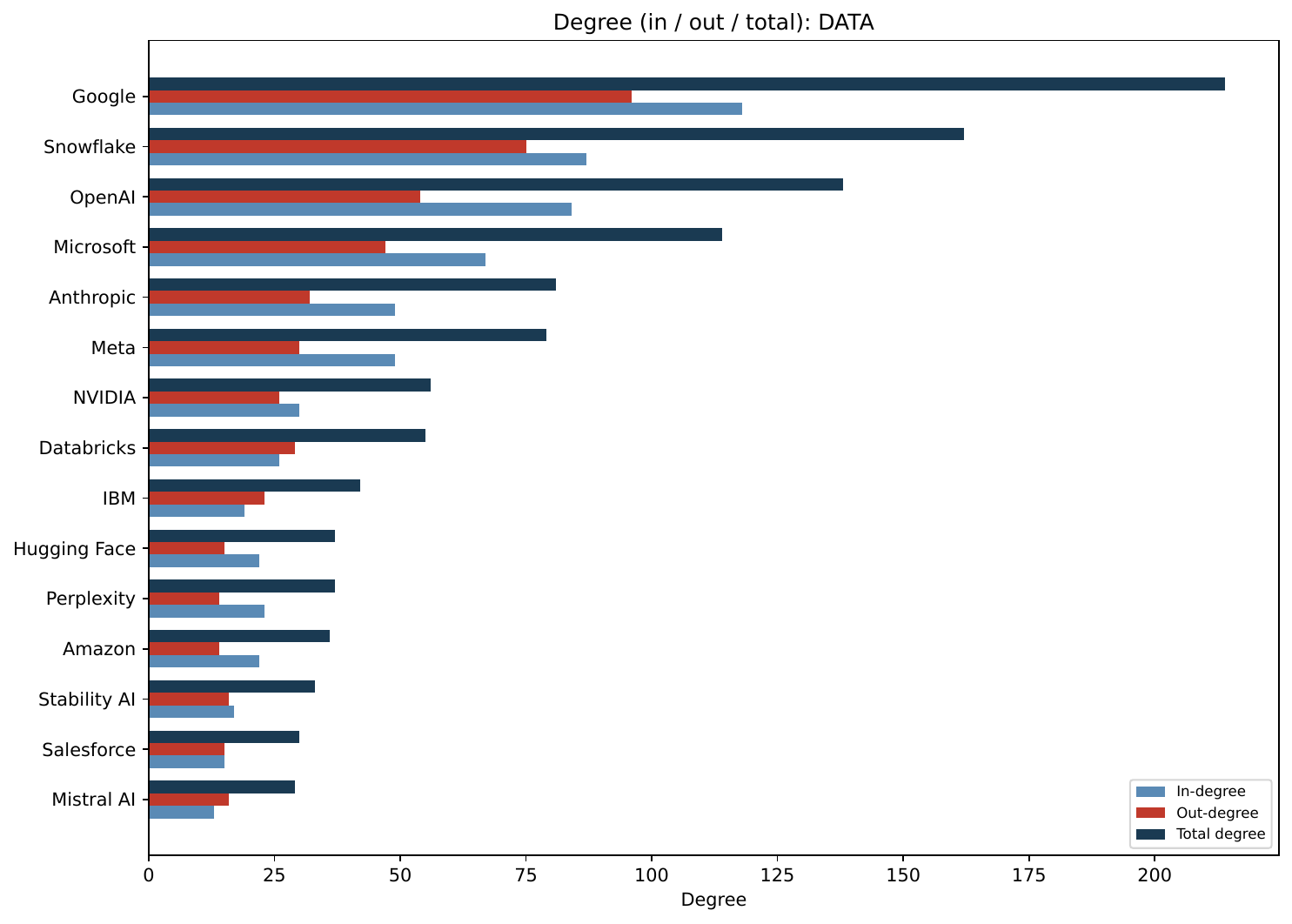}
\caption{\texttt{DATA}}
\end{subfigure}

\vspace{0.8em}

\begin{subfigure}[t]{0.48\textwidth}
\centering
\includegraphics[width=\linewidth]{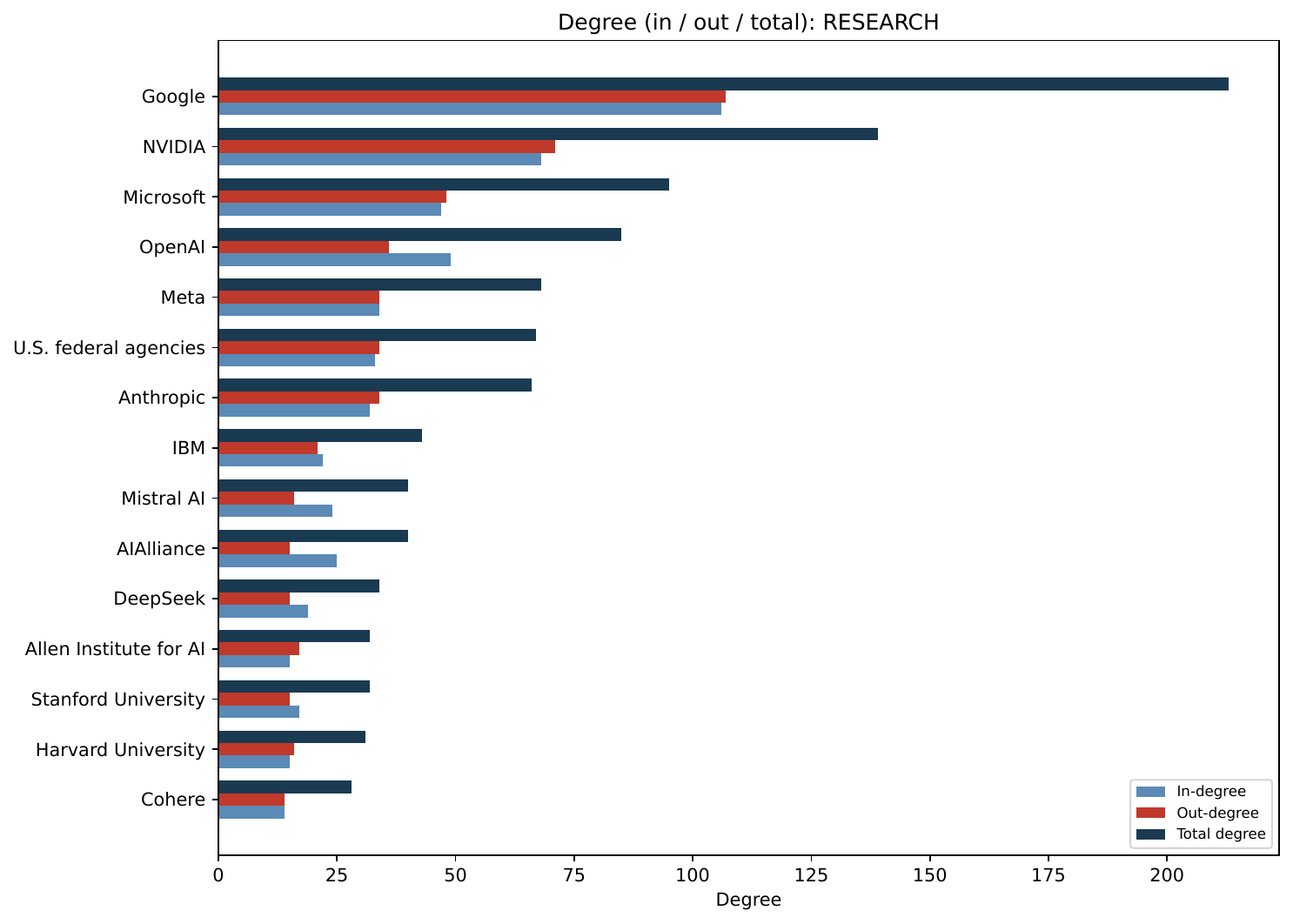}
\caption{\texttt{RESEARCH}}
\end{subfigure}
\hfill
\begin{subfigure}[t]{0.48\textwidth}
\centering
\includegraphics[width=\linewidth]{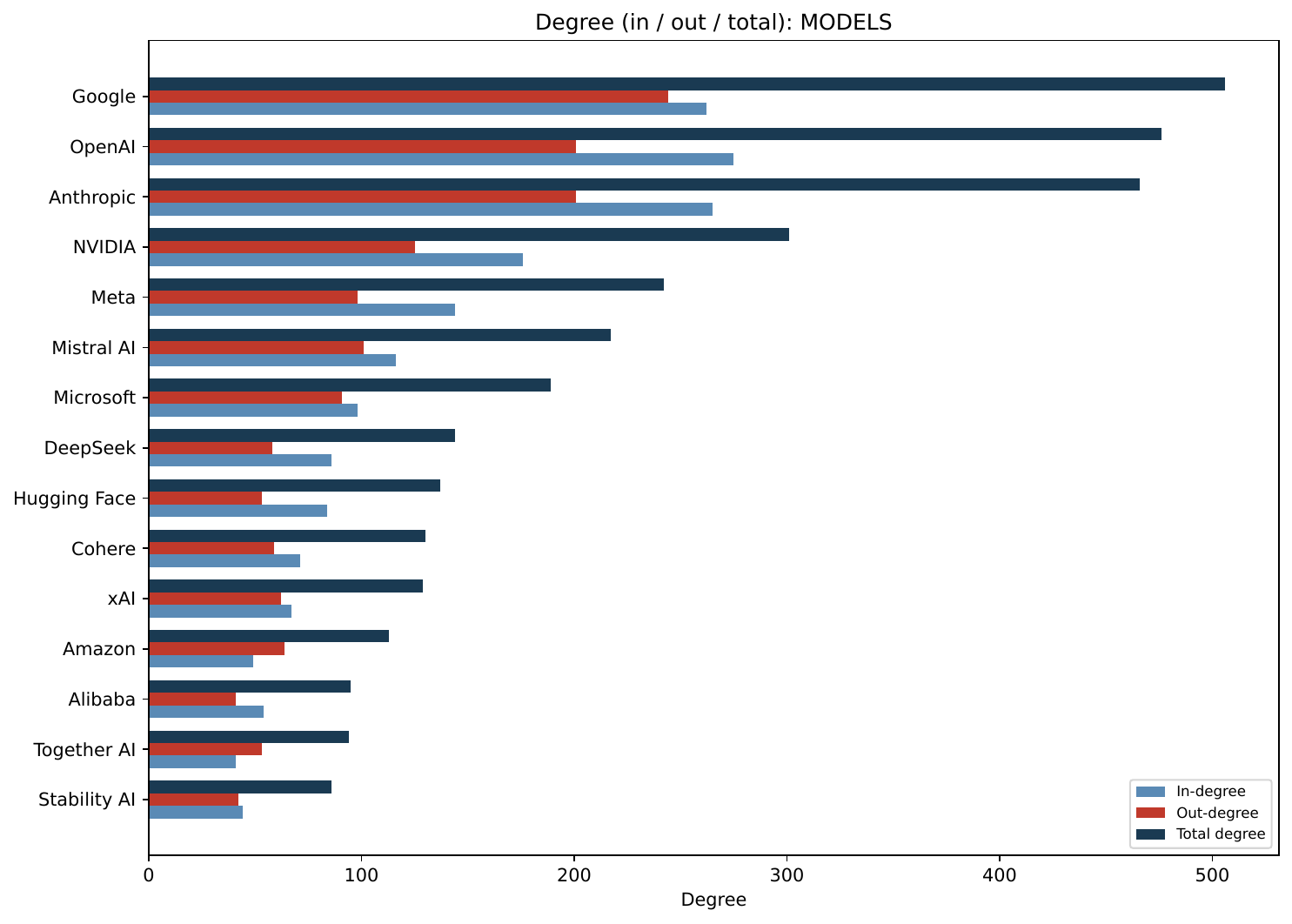}
\caption{\texttt{MODELS}}
\end{subfigure}
\caption{Degree rankings (in / out / total) by AI stack layer (panel 1/3).}
\label{fig:domain_rankings}
\end{figure}

\begin{figure}[H]
\centering
\begin{subfigure}[t]{0.48\textwidth}
\centering
\includegraphics[width=\linewidth]{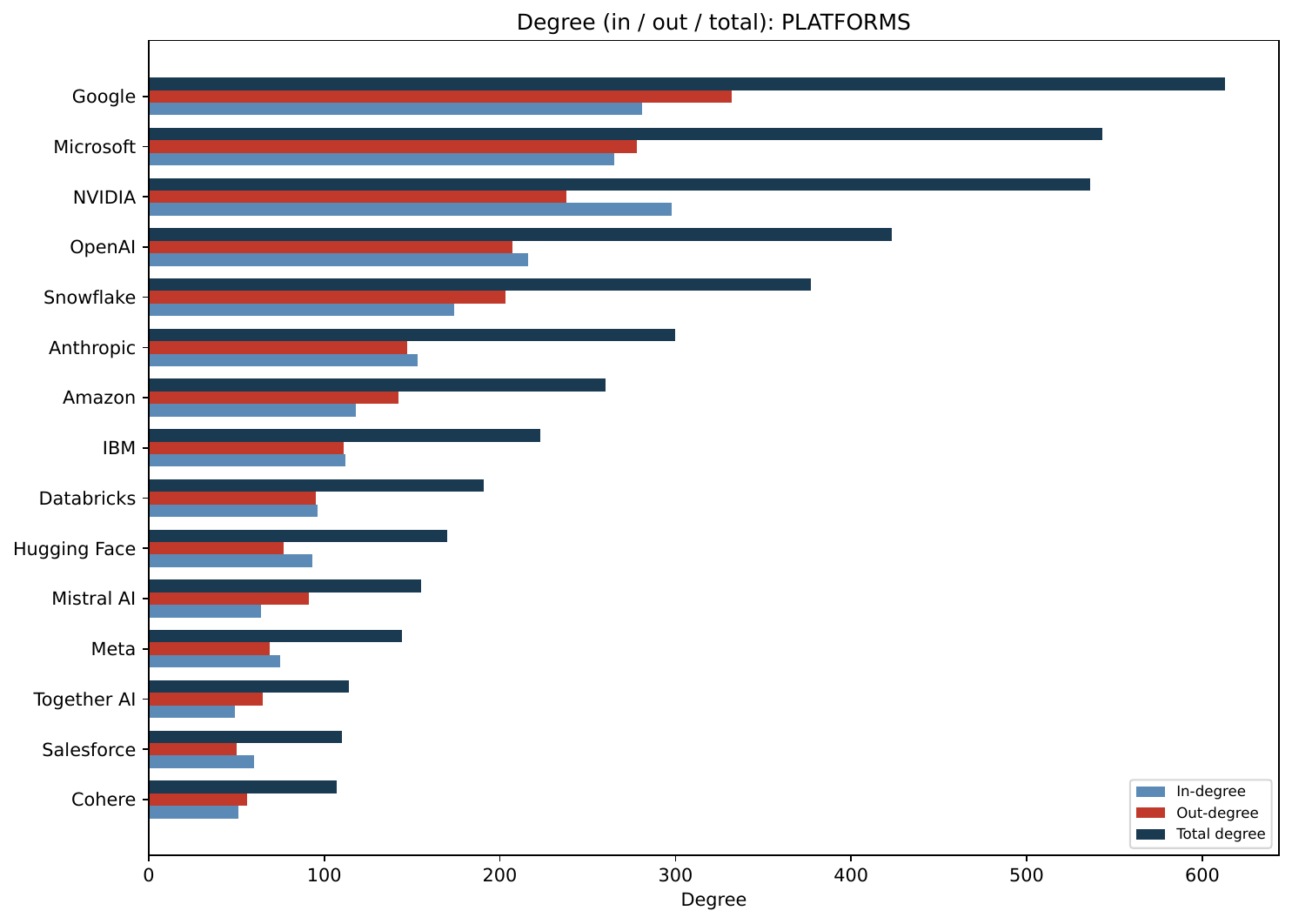}
\caption{\texttt{PLATFORMS}}
\end{subfigure}
\hfill
\begin{subfigure}[t]{0.48\textwidth}
\centering
\includegraphics[width=\linewidth]{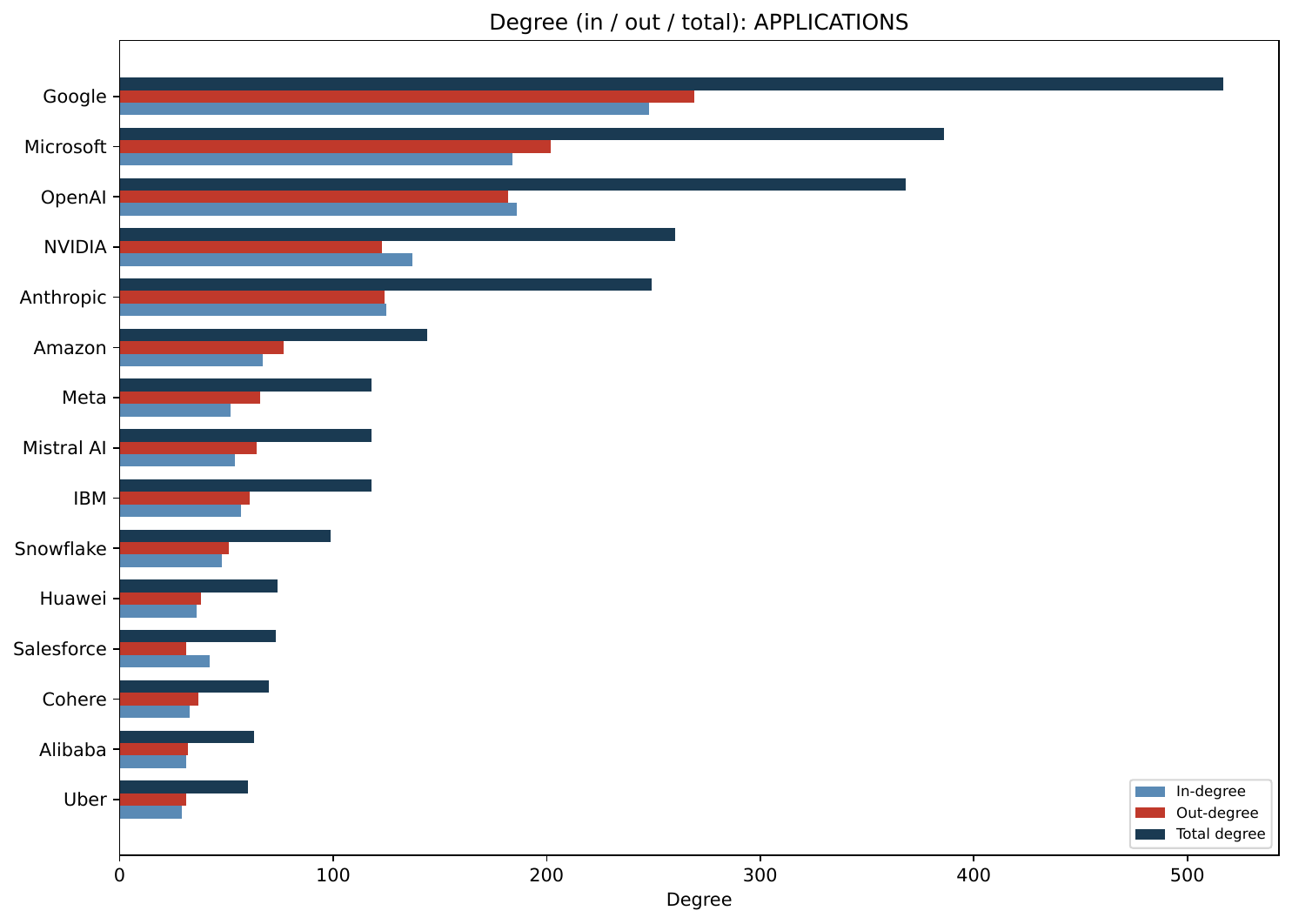}
\caption{\texttt{APPLICATIONS}}
\end{subfigure}

\vspace{0.8em}

\begin{subfigure}[t]{0.48\textwidth}
\centering
\includegraphics[width=\linewidth]{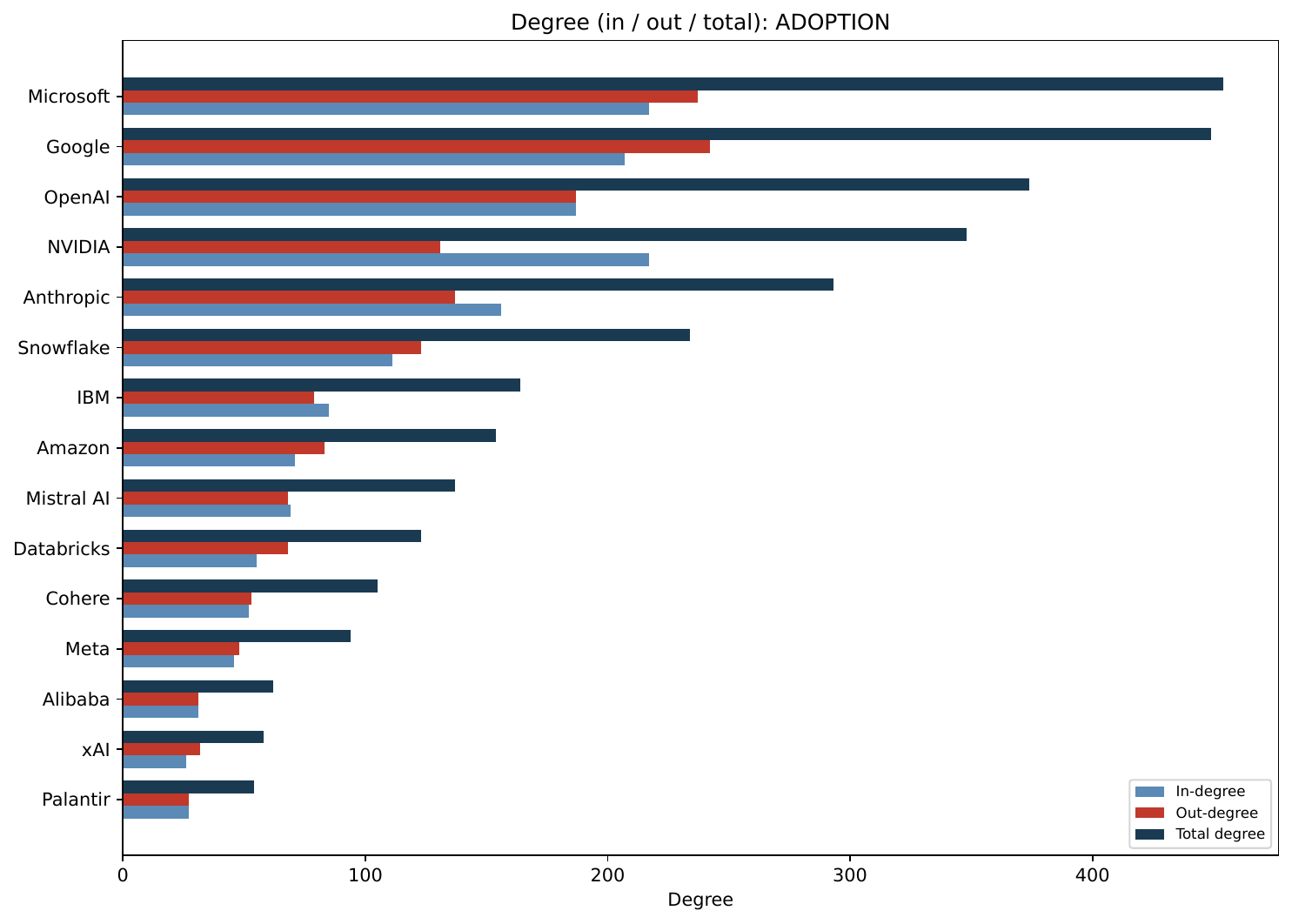}
\caption{\texttt{ADOPTION}}
\end{subfigure}
\hfill
\begin{subfigure}[t]{0.48\textwidth}
\centering
\includegraphics[width=\linewidth]{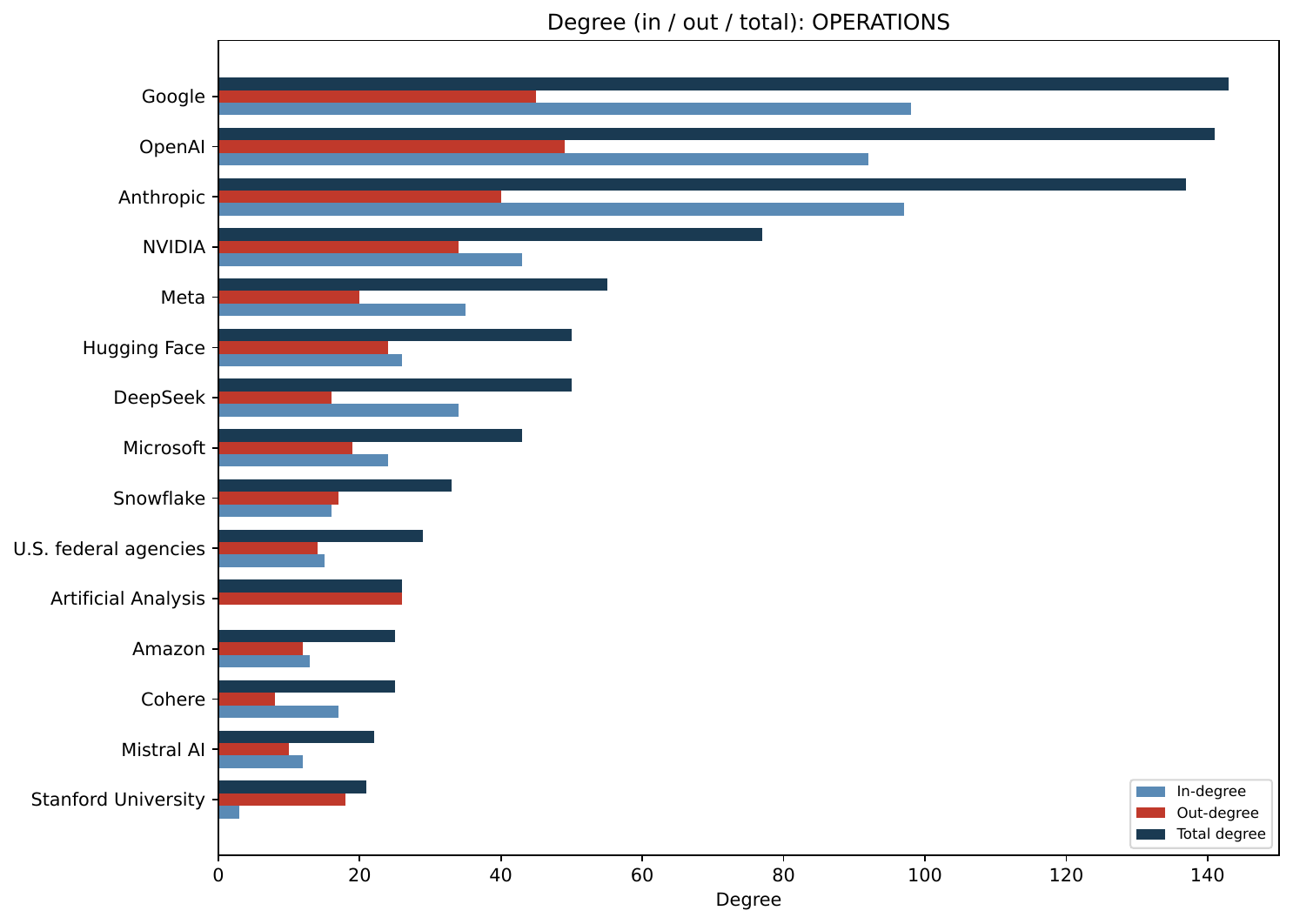}
\caption{\texttt{OPERATIONS}}
\end{subfigure}

\vspace{0.8em}

\begin{subfigure}[t]{0.48\textwidth}
\centering
\includegraphics[width=\linewidth]{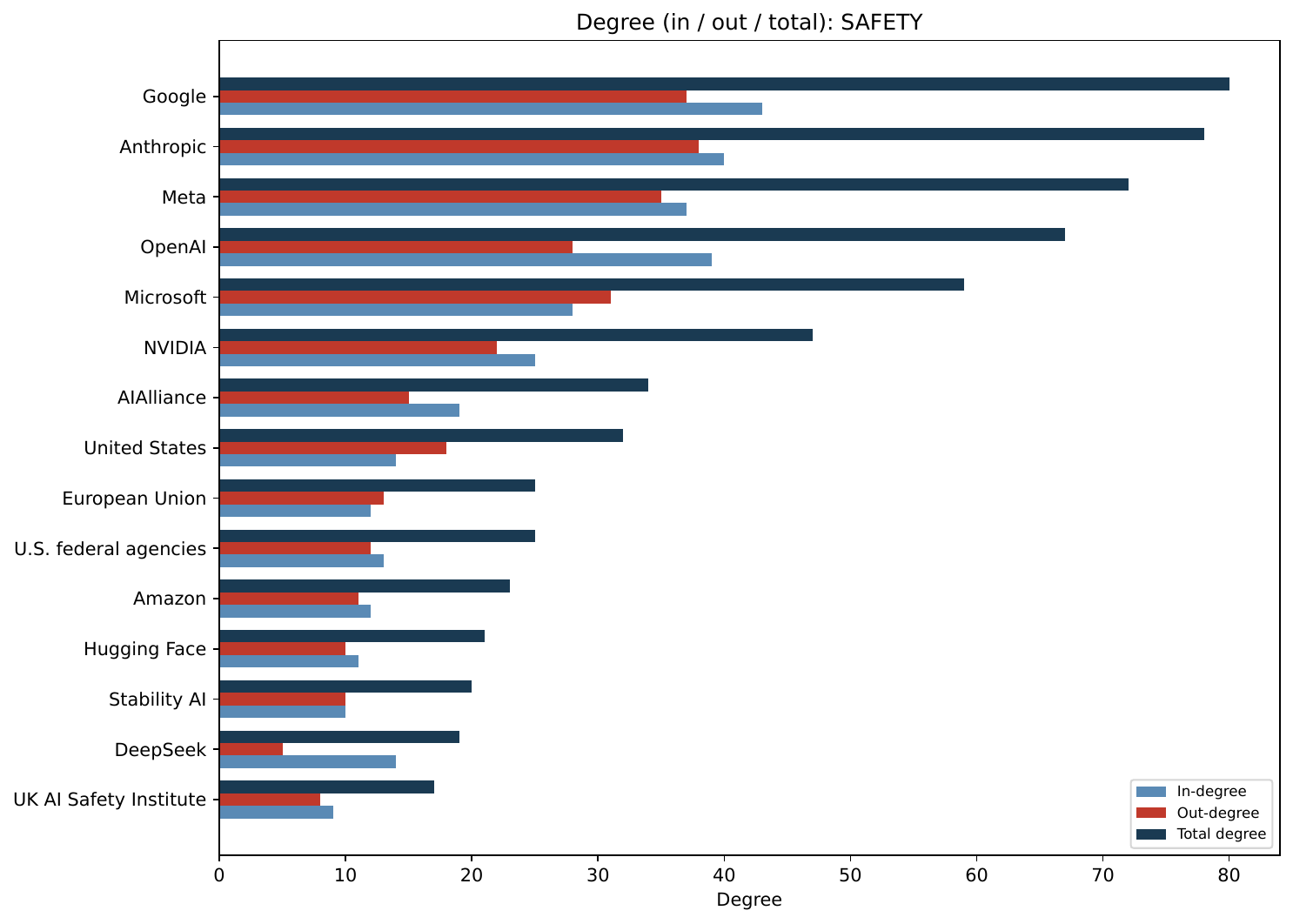}
\caption{\texttt{SAFETY}}
\end{subfigure}
\hfill
\begin{subfigure}[t]{0.48\textwidth}
\centering
\includegraphics[width=\linewidth]{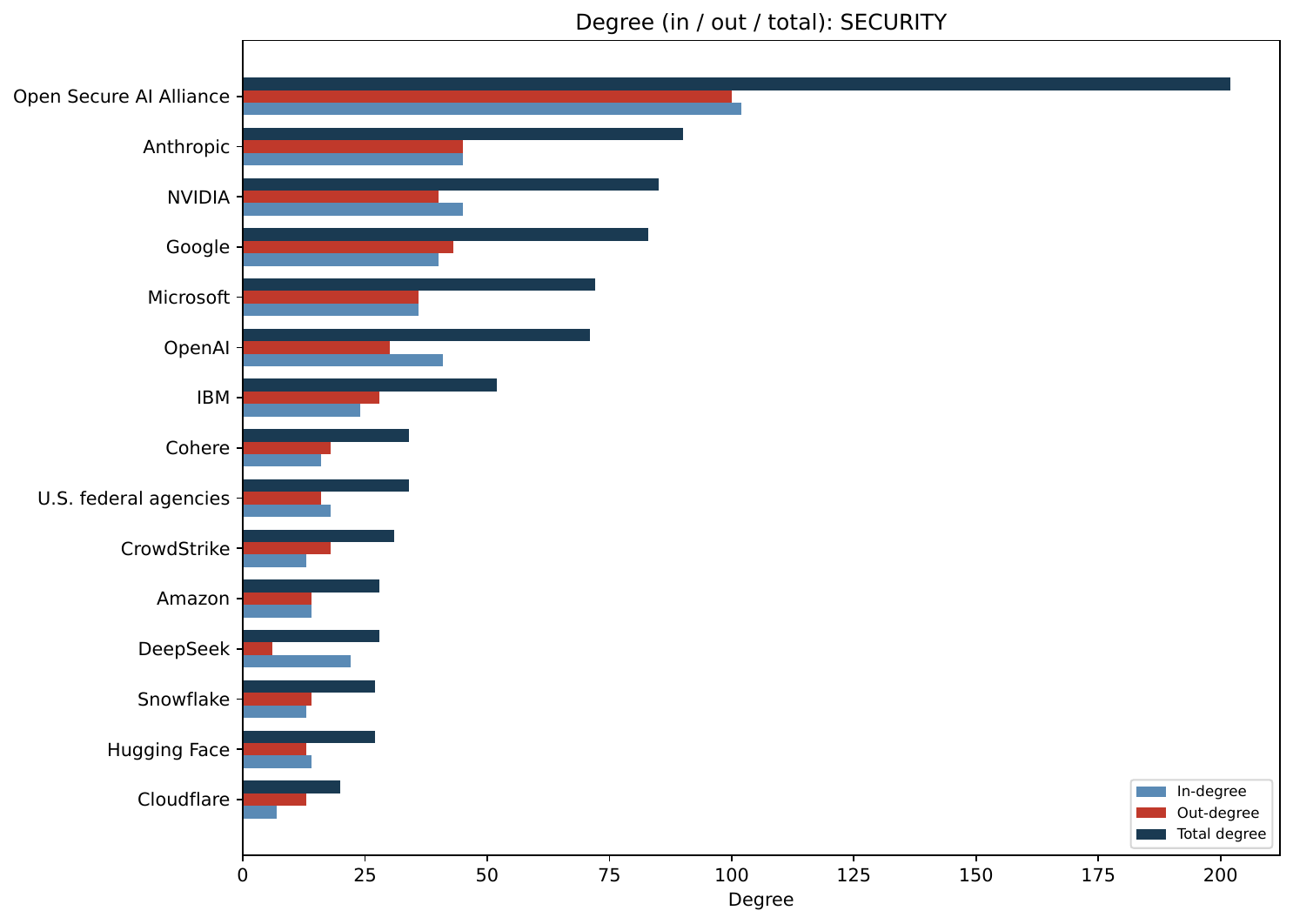}
\caption{\texttt{SECURITY}}
\end{subfigure}
\caption{Degree rankings (in / out / total) by AI stack layer (panel 2/3).}
\label{fig:domain_rankings_b}
\end{figure}

\begin{figure}[H]
\centering
\begin{subfigure}[t]{0.48\textwidth}
\centering
\includegraphics[width=\linewidth]{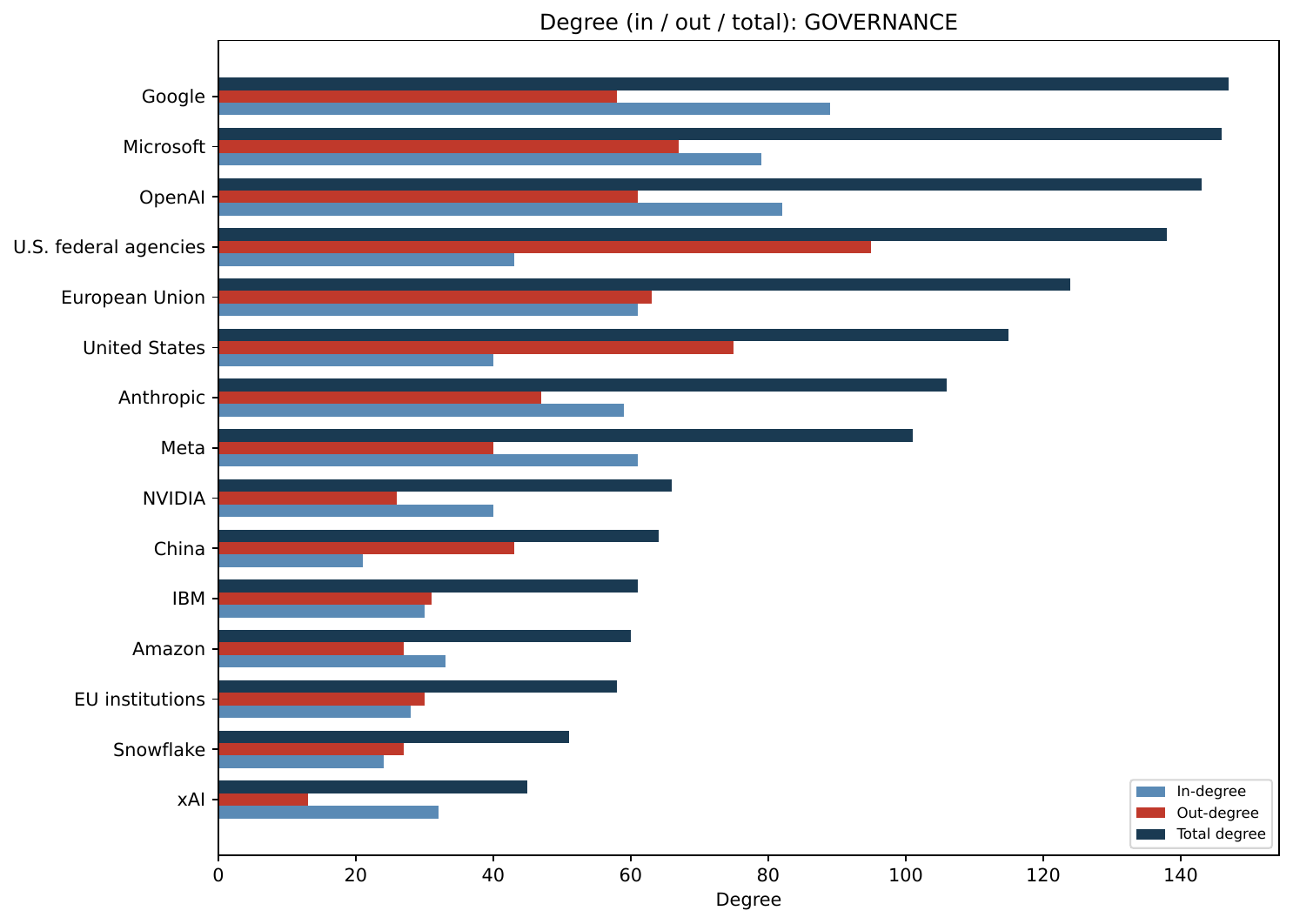}
\caption{\texttt{GOVERNANCE}}
\end{subfigure}
\hfill
\begin{subfigure}[t]{0.48\textwidth}
\centering
\includegraphics[width=\linewidth]{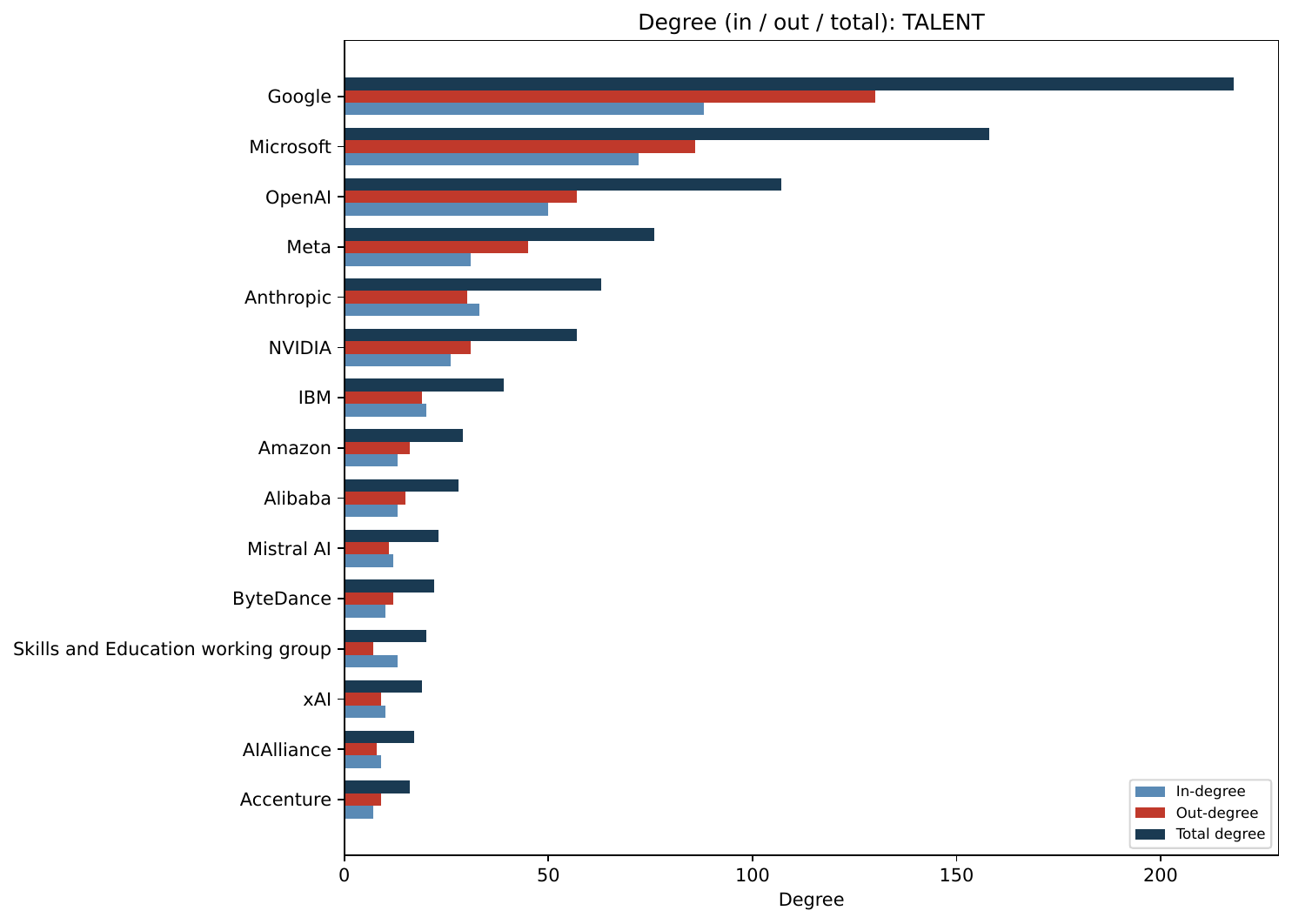}
\caption{\texttt{TALENT}}
\end{subfigure}
\caption{Degree rankings (in / out / total) by AI stack layer (panel 3/3).}
\label{fig:domain_rankings_c}
\end{figure}

\begin{figure}[H]
\centering
\begin{subfigure}[t]{0.48\textwidth}
\centering
\includegraphics[width=\linewidth]{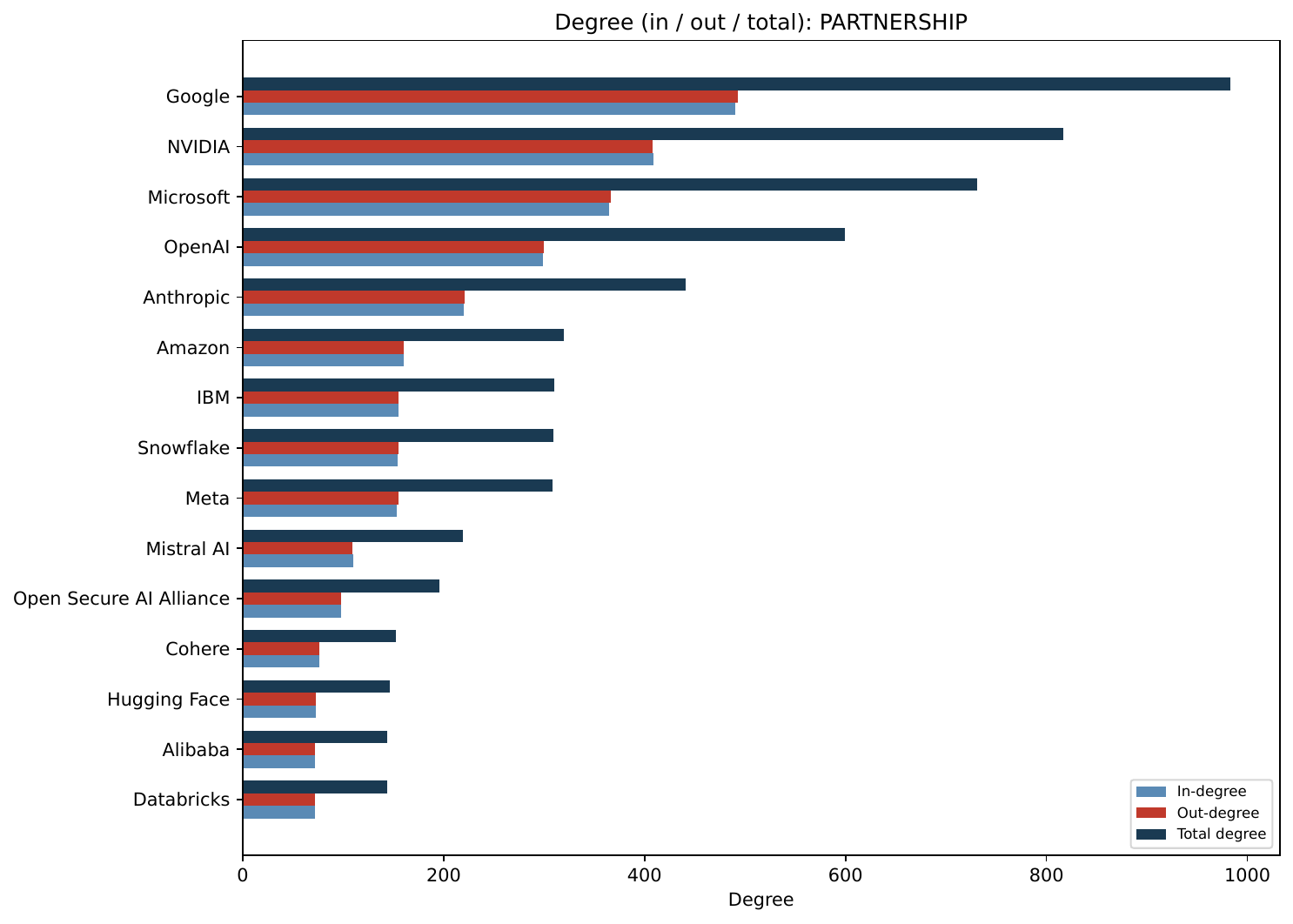}
\caption{\texttt{PARTNERSHIP}}
\end{subfigure}
\hfill
\begin{subfigure}[t]{0.48\textwidth}
\centering
\includegraphics[width=\linewidth]{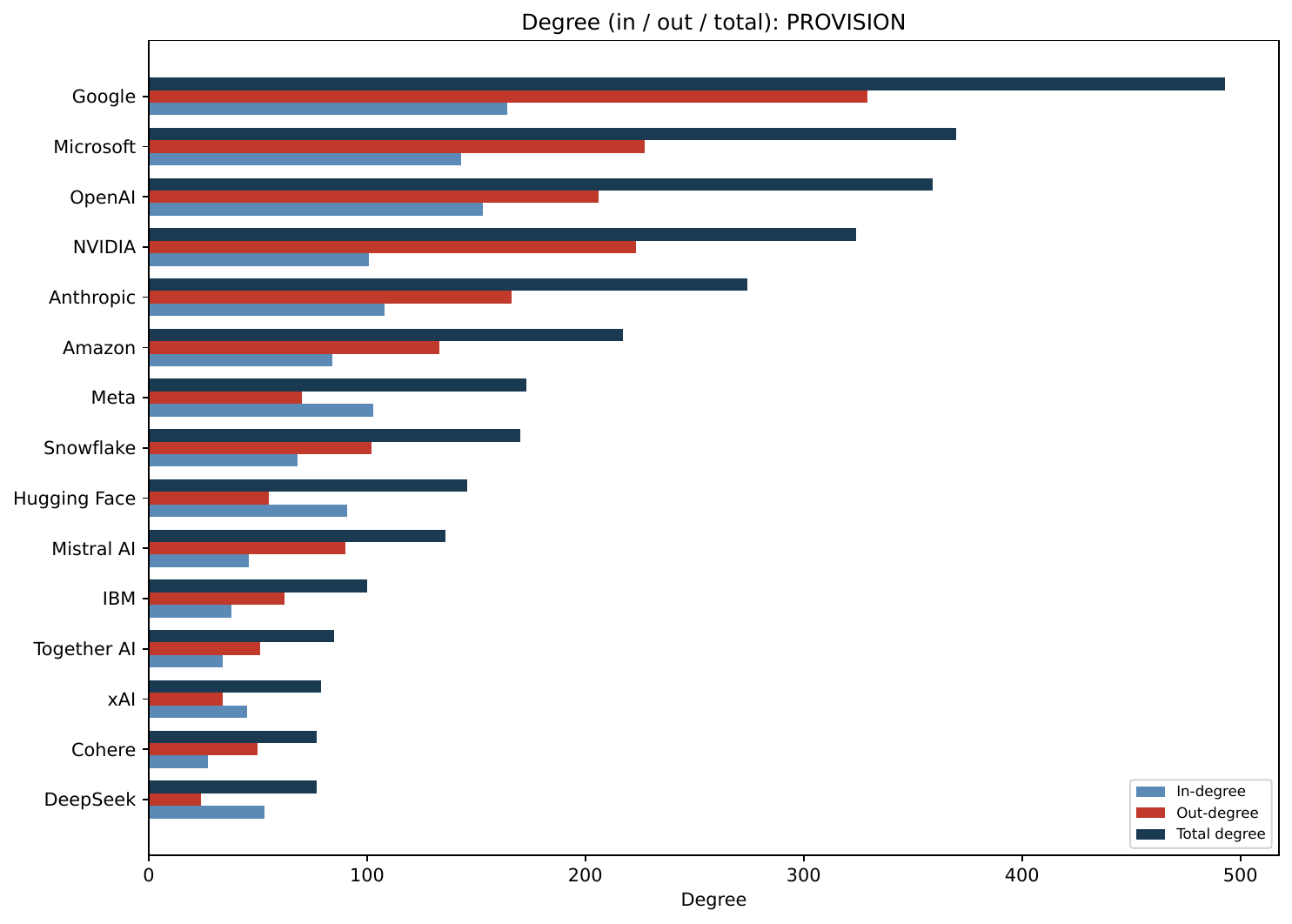}
\caption{\texttt{PROVISION}}
\end{subfigure}

\vspace{0.8em}

\begin{subfigure}[t]{0.48\textwidth}
\centering
\includegraphics[width=\linewidth]{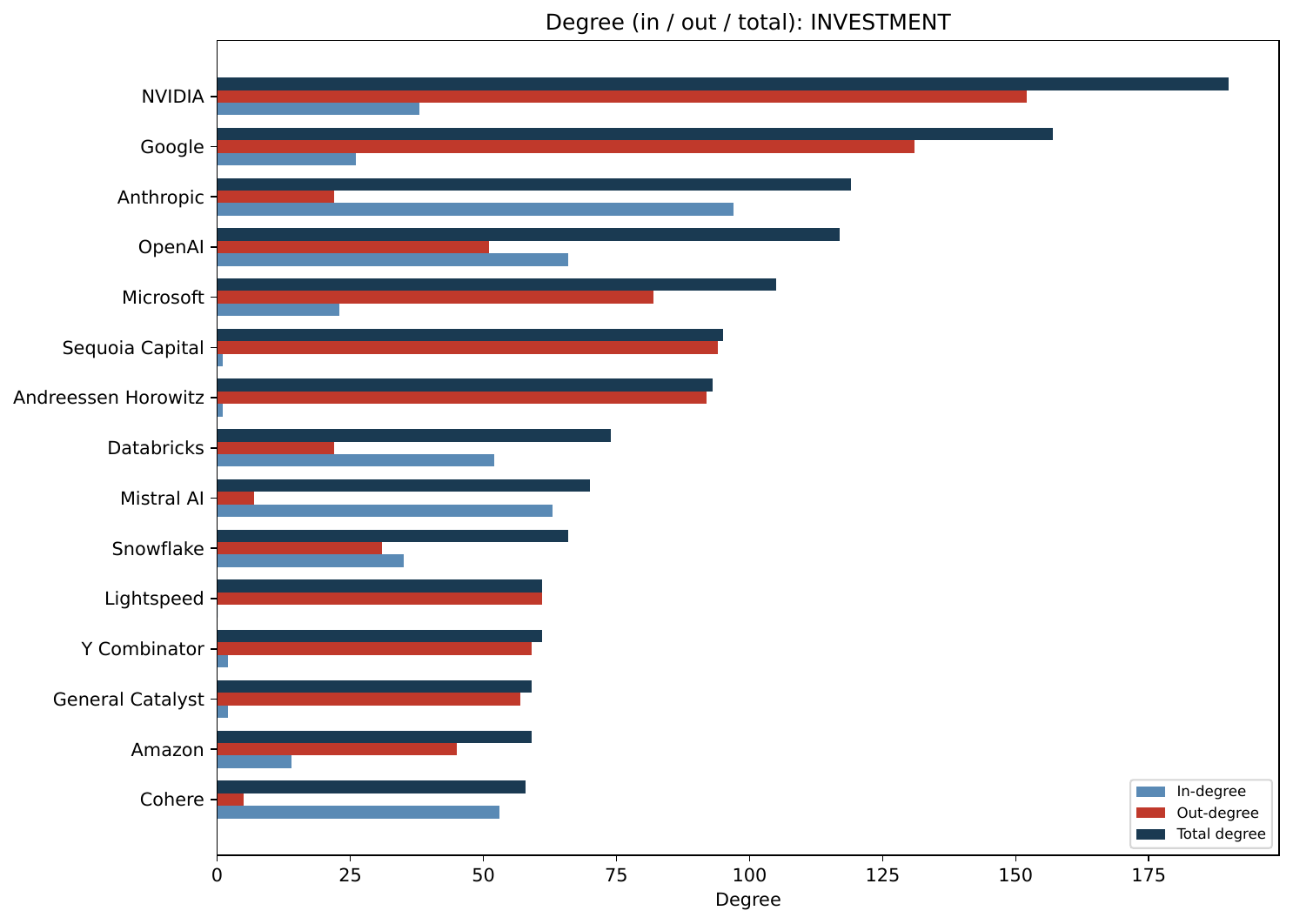}
\caption{\texttt{INVESTMENT}}
\end{subfigure}
\hfill
\begin{subfigure}[t]{0.48\textwidth}
\centering
\includegraphics[width=\linewidth]{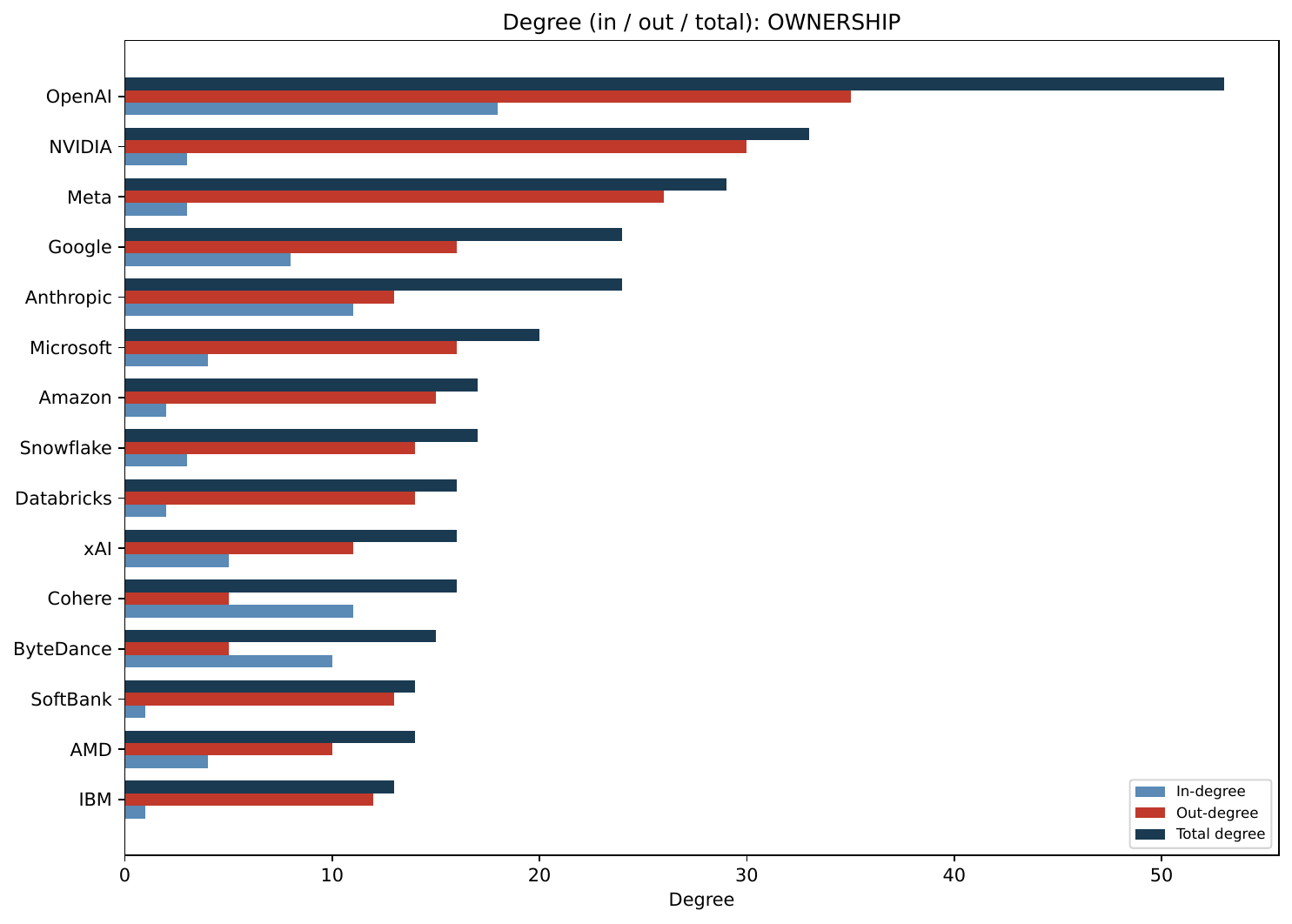}
\caption{\texttt{OWNERSHIP}}
\end{subfigure}

\vspace{0.8em}

\begin{subfigure}[t]{0.48\textwidth}
\centering
\includegraphics[width=\linewidth]{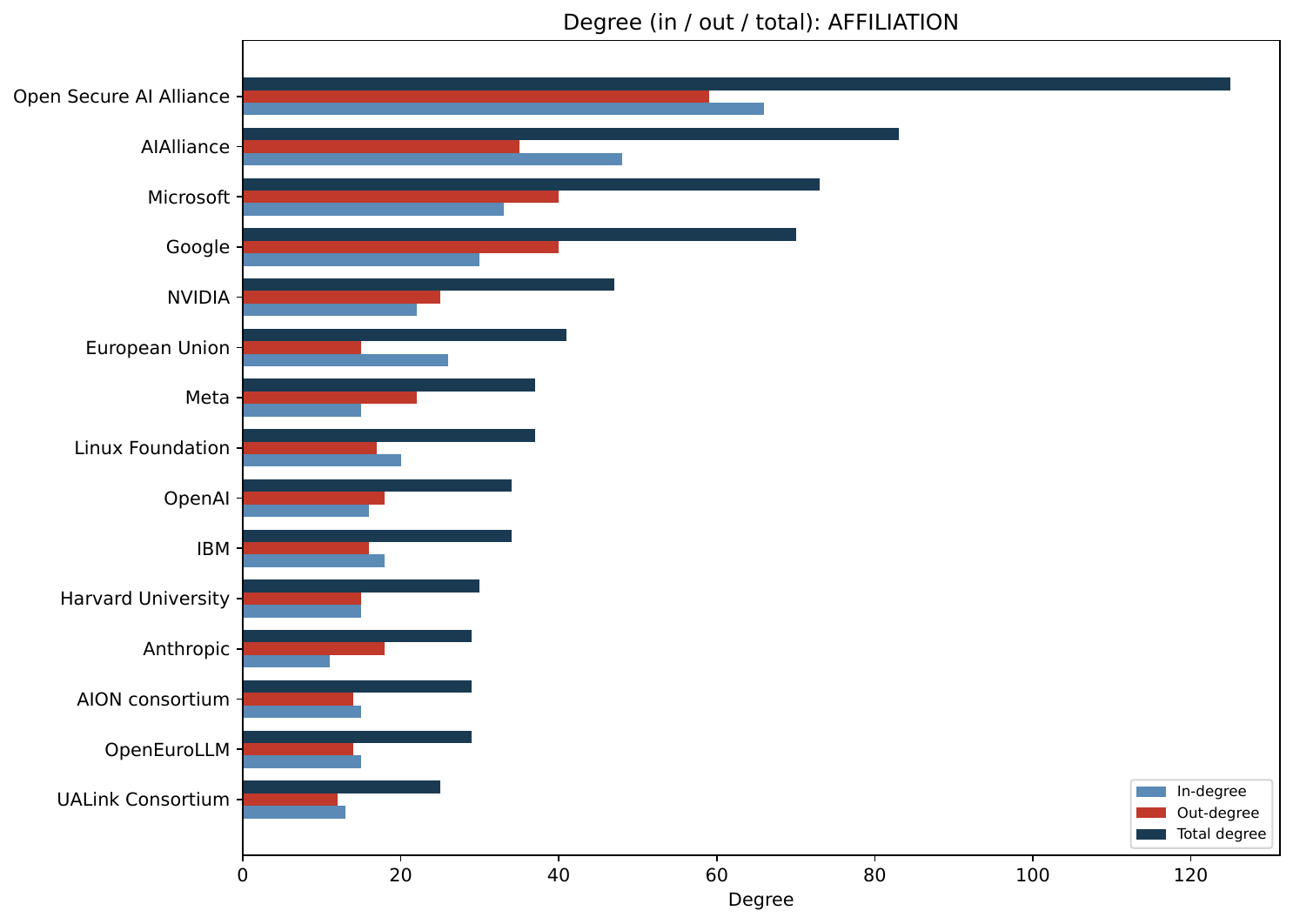}
\caption{\texttt{AFFILIATION}}
\end{subfigure}
\hfill
\begin{subfigure}[t]{0.48\textwidth}
\centering
\includegraphics[width=\linewidth]{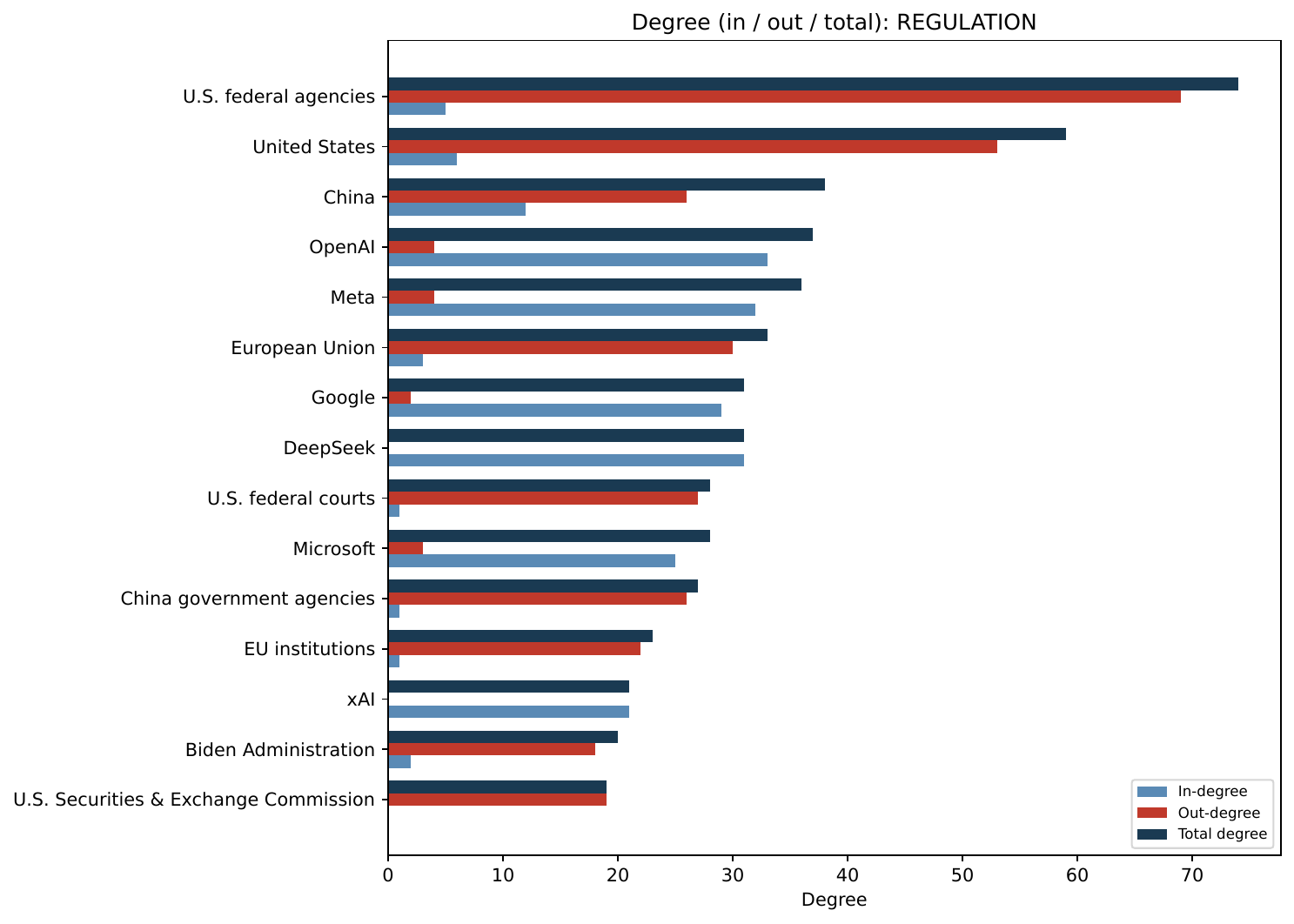}
\caption{\texttt{REGULATION}}
\end{subfigure}
\caption{Degree rankings (in / out / total) by relationship type (panel 1/3).}
\label{fig:reltype_rankings}
\end{figure}

\begin{figure}[H]
\centering
\begin{subfigure}[t]{0.48\textwidth}
\centering
\includegraphics[width=\linewidth]{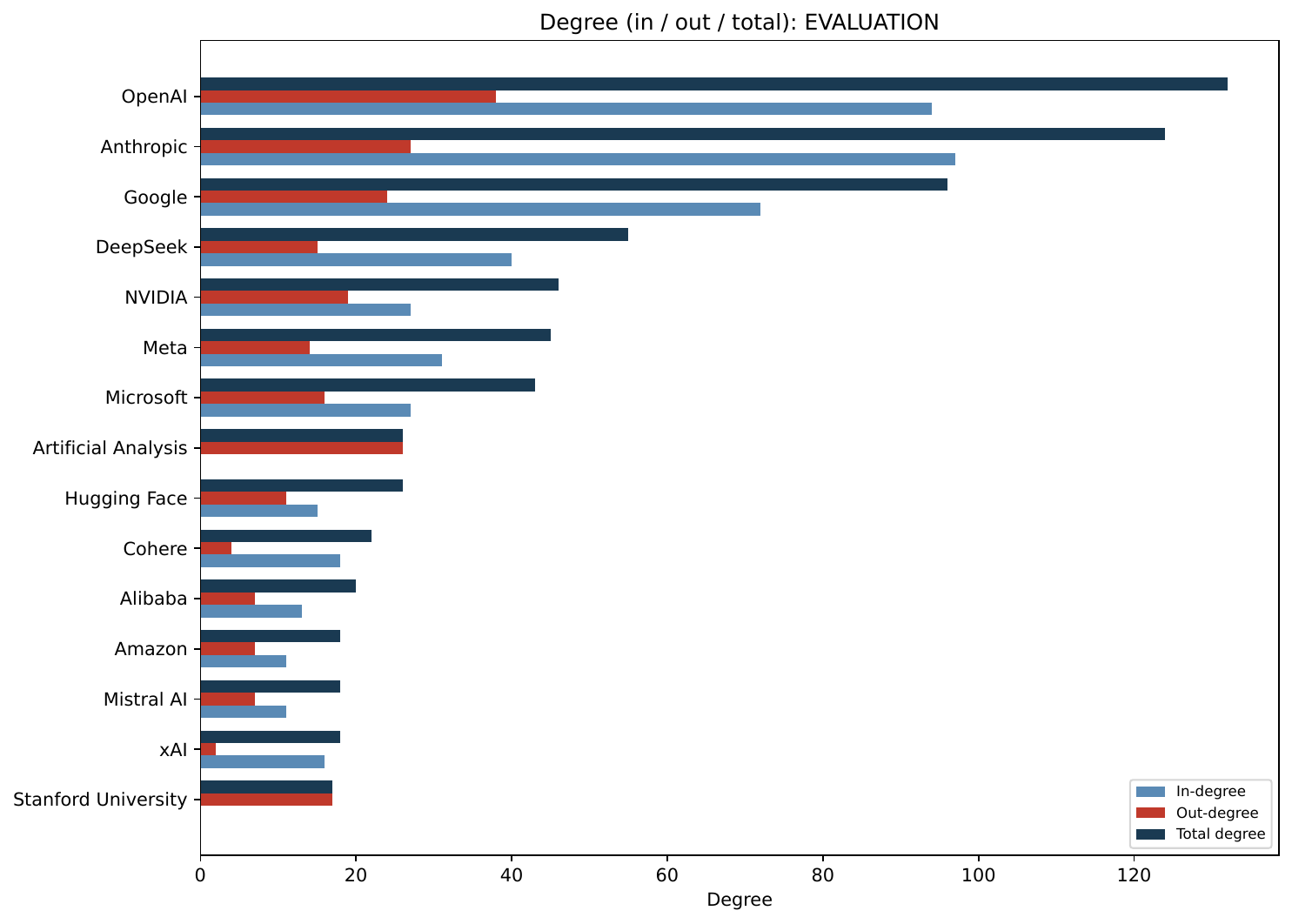}
\caption{\texttt{EVALUATION}}
\end{subfigure}
\hfill
\begin{subfigure}[t]{0.48\textwidth}
\centering
\includegraphics[width=\linewidth]{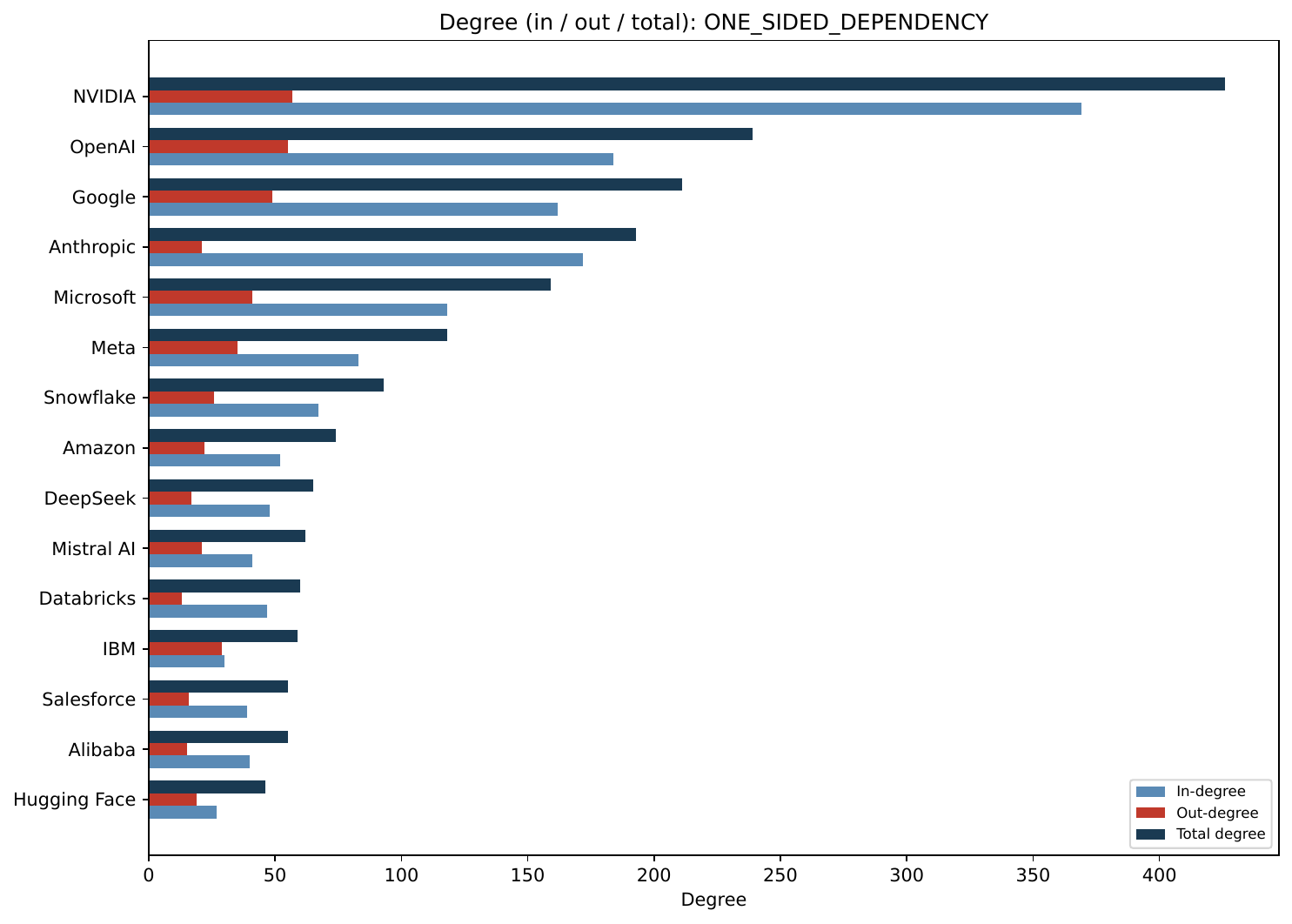}
\caption{\texttt{ONE\_SIDED\_DEPENDENCY}}
\end{subfigure}

\vspace{0.8em}

\begin{subfigure}[t]{0.48\textwidth}
\centering
\includegraphics[width=\linewidth]{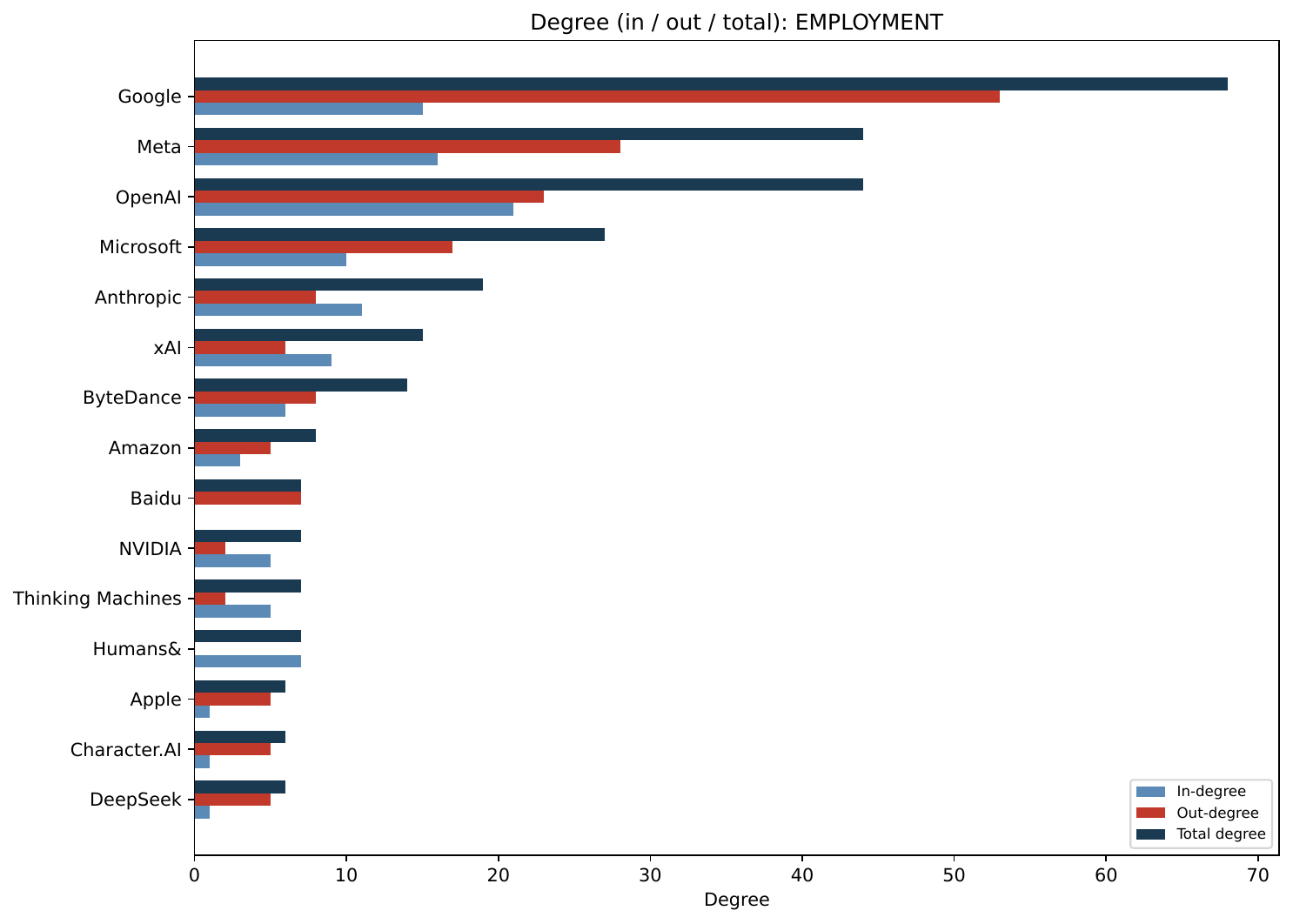}
\caption{\texttt{EMPLOYMENT}}
\end{subfigure}
\hfill
\begin{subfigure}[t]{0.48\textwidth}
\centering
\includegraphics[width=\linewidth]{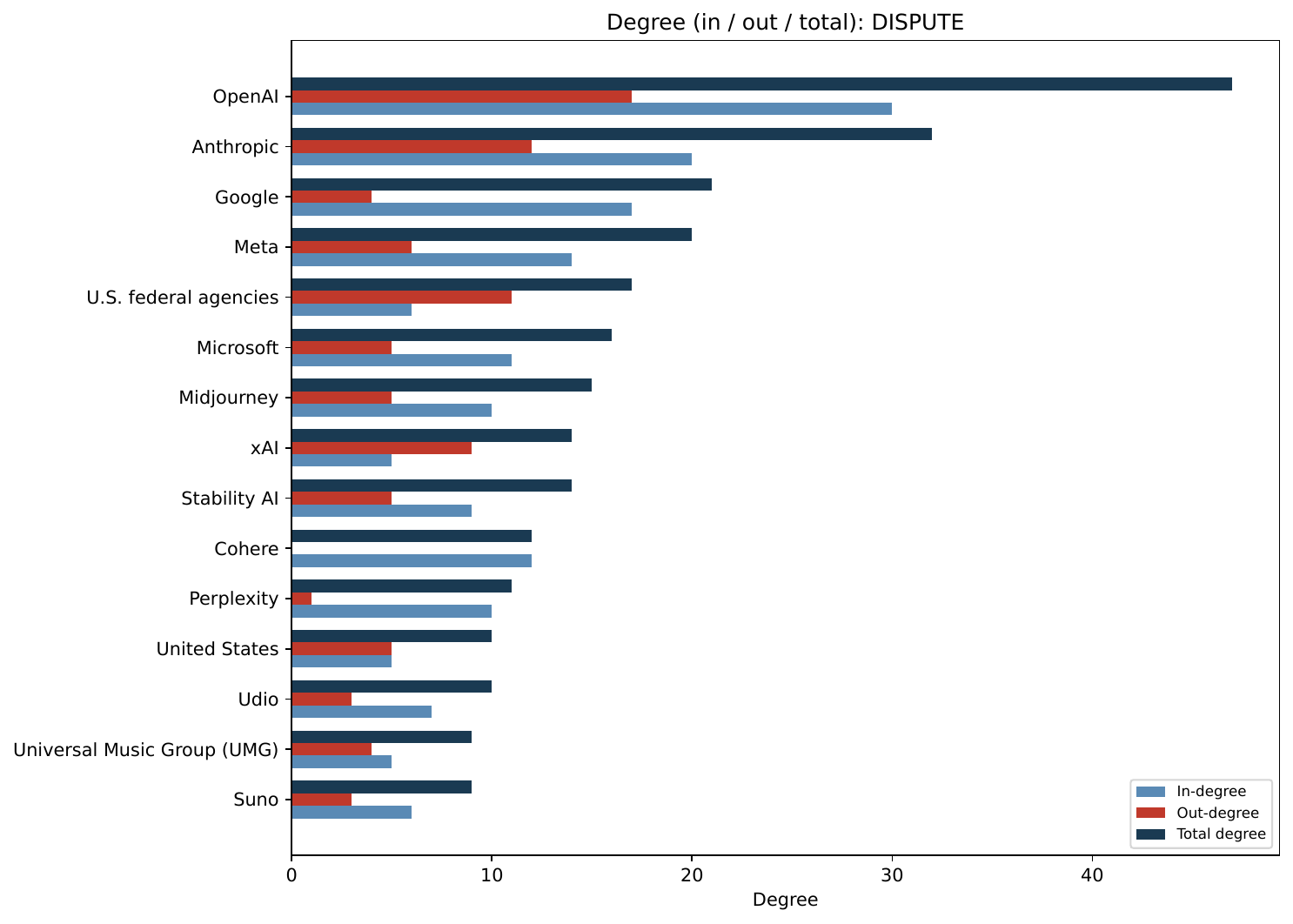}
\caption{\texttt{DISPUTE}}
\end{subfigure}
\caption{Degree rankings (in / out / total) by relationship type (panel 2/3).}
\label{fig:reltype_rankings_b}
\end{figure}

\begin{figure}[H]
\centering
\begin{subfigure}[t]{0.48\textwidth}
\centering
\includegraphics[width=\linewidth]{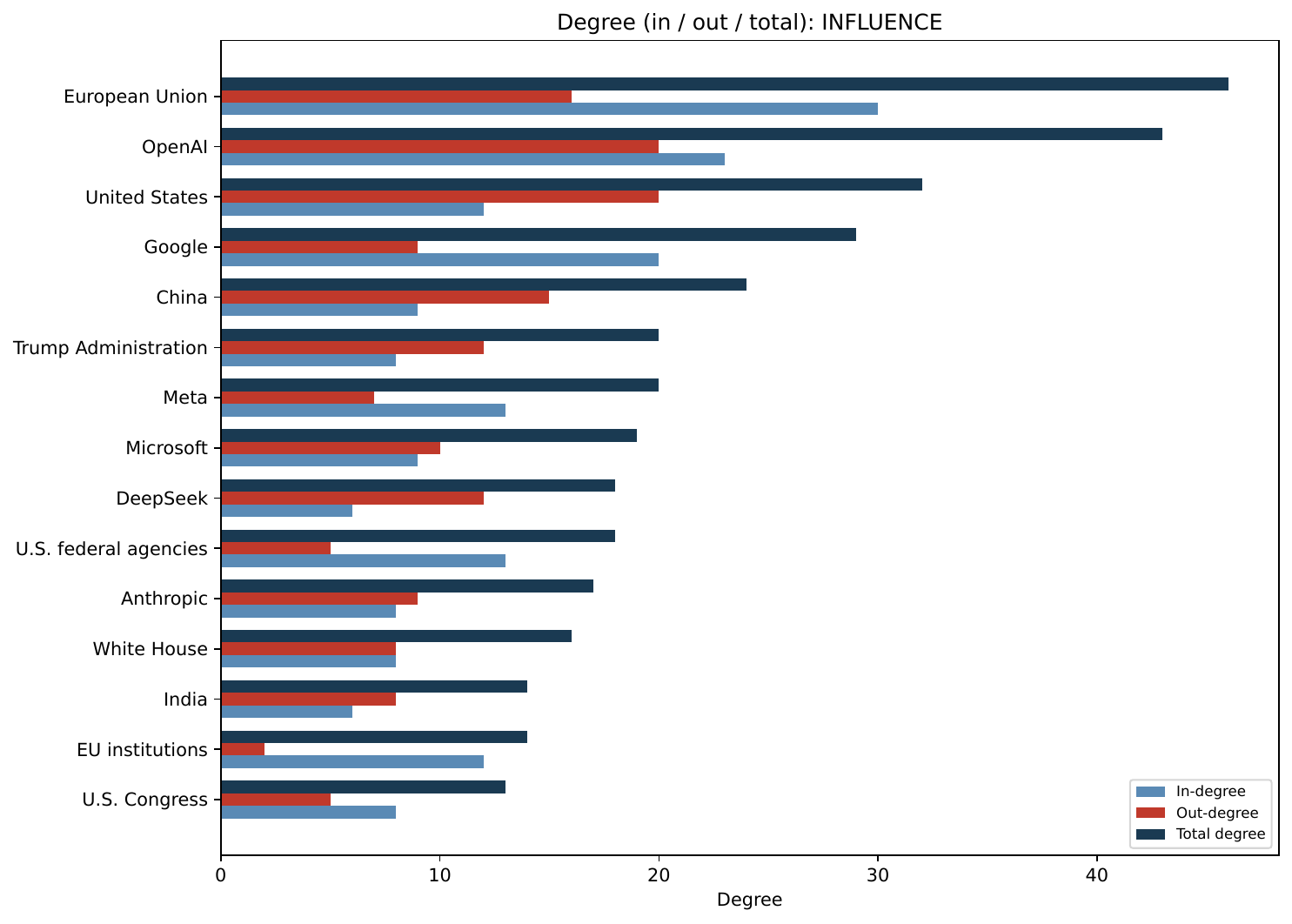}
\caption{\texttt{INFLUENCE}}
\end{subfigure}
\hfill
\begin{subfigure}[t]{0.48\textwidth}
\centering
\includegraphics[width=\linewidth]{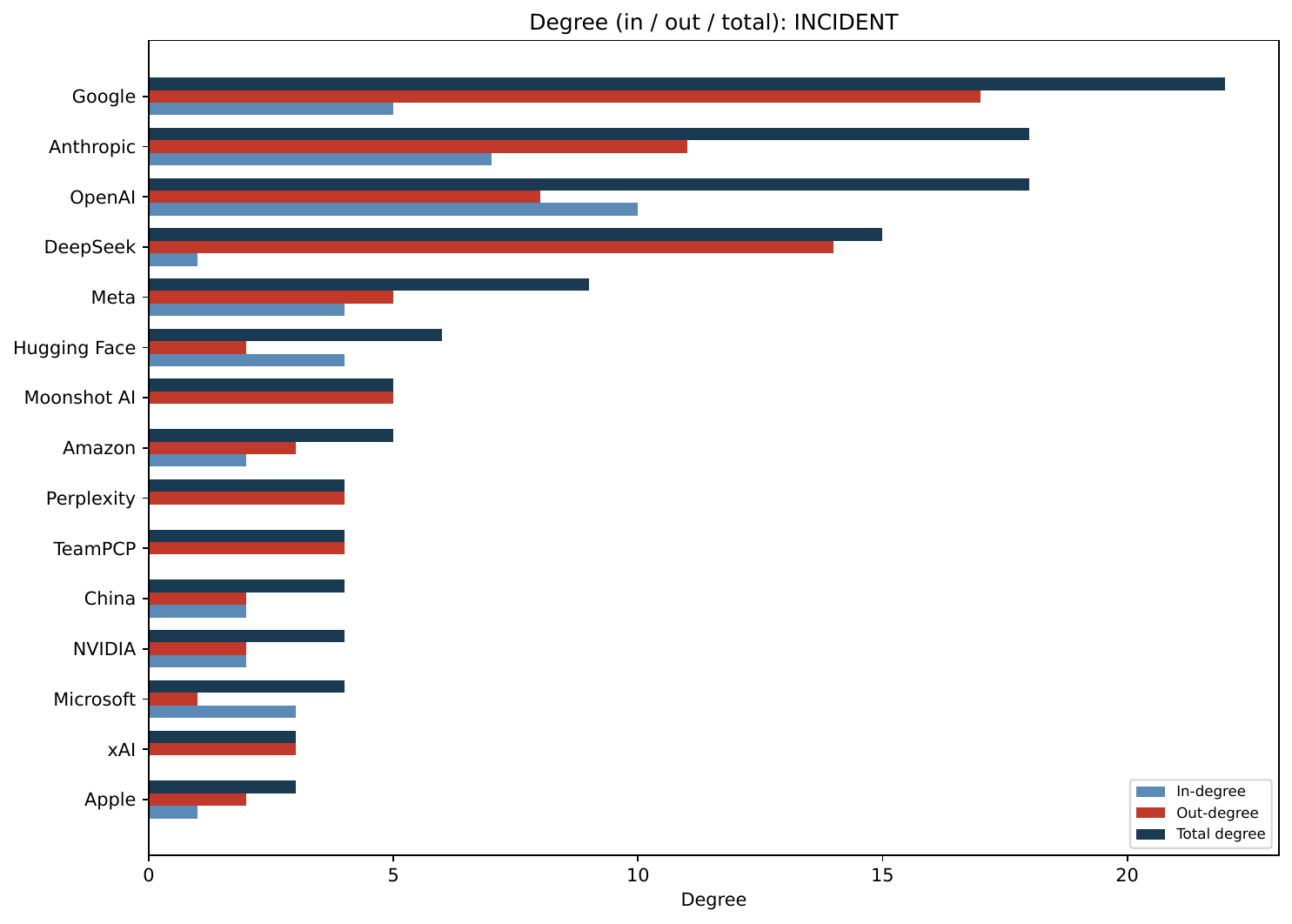}
\caption{\texttt{INCIDENT}}
\end{subfigure}

\vspace{0.8em}

\begin{subfigure}[t]{0.48\textwidth}
\centering
\includegraphics[width=\linewidth]{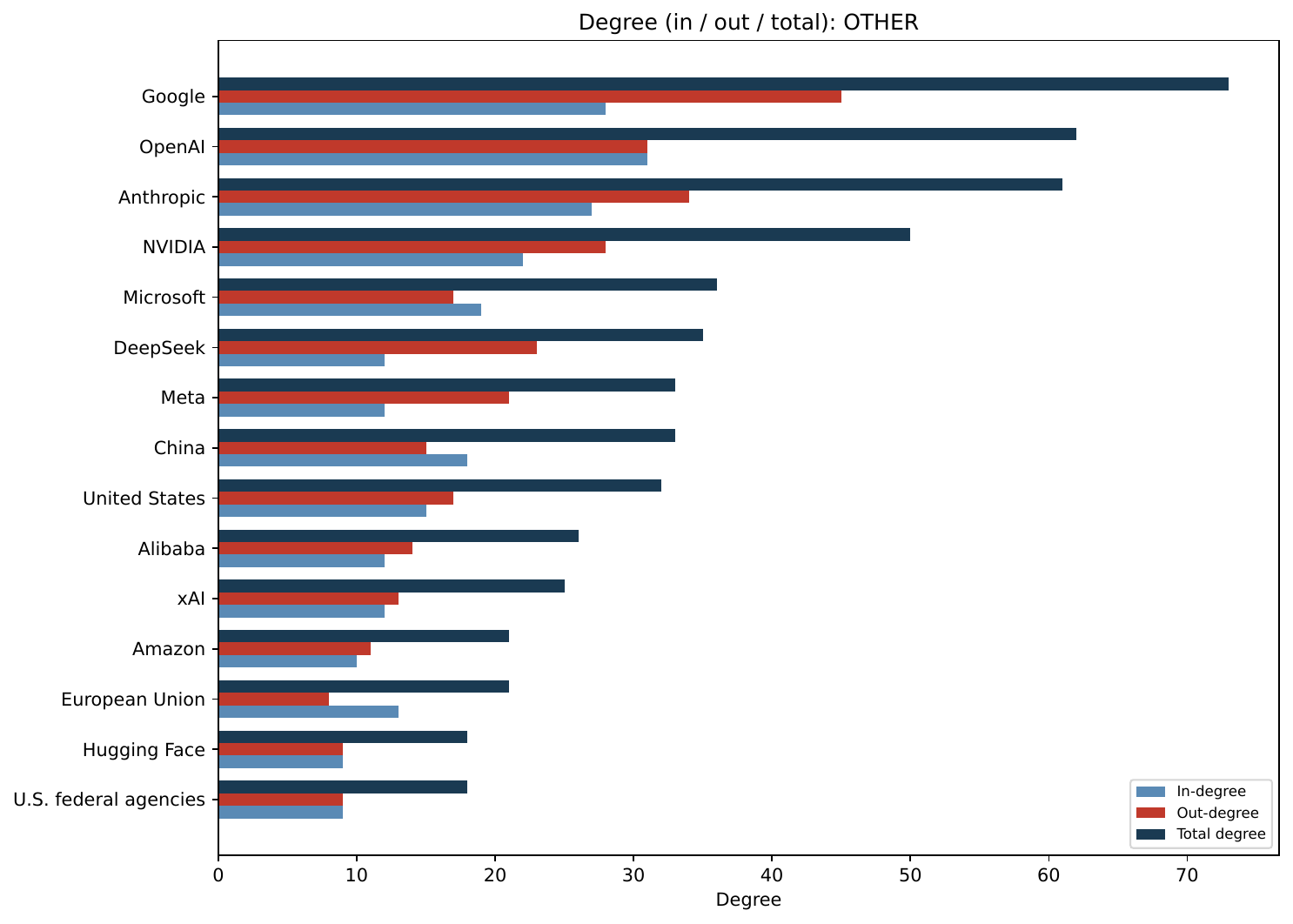}
\caption{\texttt{OTHER}}
\end{subfigure}
\caption{Degree rankings (in / out / total) by relationship type (panel 3/3).
These thirteen panels cover every relationship type that remains in \(G'\) after the
exclusions in \Cref{app:base_edge_set}
(\texttt{COMPETITION}, \texttt{PUBLIC\_STATEMENT}, \texttt{NO\_RELATIONSHIP},
\texttt{UNCLEAR}, and \texttt{CONTRADICTORY\_INFORMATION}).}
\label{fig:reltype_rankings_c}
\end{figure}

\clearpage

\section{\texorpdfstring{Additional Examples for the Descendant Trees}{Additional Examples for the Descendant Trees}}
\label{app:tree_examples}

This appendix lists the remaining edges in \Cref{fig:openai_model_provision_tree,fig:scale_ai_data_provision_tree}.
For each edge, the relationship description is the sentence stored on that edge in the dataset.
Where that sentence is not the wording of the source document, the document passage is quoted separately.

\subsection*{OpenAI model-provision tree}

\begin{itemize}[noitemsep]
\item \textbf{Google Cloud $\rightarrow$ Adobe.}
Relationship description: ``Adobe is identified as a customer using Google Cloud's AI models in Vertex.''
Source document (Google Cloud, October 9, 2025, ``Introducing Gemini Enterprise''): ``The number of customers using our AI models in Vertex continues to grow, including top brands, like: Adobe.''
\url{https://cloud.google.com/blog/products/ai-machine-learning/introducing-gemini-enterprise}
\item \textbf{Google Cloud $\rightarrow$ Bain \& Company.}
Relationship description: ``Google Cloud is offering Bain \& Company early access to pre-release versions of upcoming Google DeepMind models for preview and development.''
Source document (Google Cloud, April 22, 2026, ``Building the Agentic Enterprise with Google Cloud partners and a \$750M innovation fund''): the company is ``offering early model access for a select group of partners, including Accenture, Bain \& Company, BCG, Deloitte, and McKinsey,'' who can ``preview and begin building with pre-release versions of upcoming Google DeepMind models.''
\url{https://cloud.google.com/blog/topics/partners/how-google-cloud-partner-ecosystem-is-building-the-agentic-enterprise}
\item \textbf{Google Cloud $\rightarrow$ LATAM Airlines.}
Relationship description: ``LATAM Airlines is identified as a customer using Google Cloud's AI models in Vertex.''
Source document: the same Google Cloud post as the Adobe edge, October 9, 2025, which lists LATAM Airlines among those customers.
\url{https://cloud.google.com/blog/products/ai-machine-learning/introducing-gemini-enterprise}
\item \textbf{Google Cloud $\rightarrow$ Snowflake.}
Relationship description: ``Google Cloud's Gemini 3 is integrated natively into Snowflake Cortex AI, allowing customers to develop and deploy generative AI applications on Snowflake without moving or copying data across platforms.''
Source document (Snowflake, January 6, 2025, ``Snowflake Enables Enterprise-Ready AI by Bringing Google's Gemini 3 to Snowflake Cortex AI''): ``the availability of Google Cloud's Gemini 3 natively within Snowflake Cortex AI.''
\url{https://www.snowflake.com/en/news/press-releases/snowflake-enables-enterprise-ready-ai-by-bringing-google-s-gemini-3-to-snowflake-cortex-ai/}
\item \textbf{Microsoft $\rightarrow$ Ant Group.}
Relationship description: ``Ant Group is a significant customer of Microsoft's AI models through Microsoft's Azure cloud service.''
Source document (Yahoo Finance, June 18, 2026, ``Microsoft Builds \$1 Billion-Plus AI Business With ByteDance''): ``Ant Group, Meituan and Tencent Holdings have also become significant AI model customers through Microsoft's Azure cloud service.''
\url{https://finance.yahoo.com/technology/ai/articles/microsoft-builds-1-billion-plus-180229532.html}
\item \textbf{Microsoft $\rightarrow$ Baseten.}
Relationship description: ``Microsoft makes its MAI models available for developer weight tuning through Baseten.''
Source document (VentureBeat, June 5, 2026, ``Microsoft AI chief says company was `set free' from OpenAI to pursue superintelligence''): ``developers can tune model weights themselves through third-party platforms including OpenRouter, Fireworks, and Baseten.''
\url{https://venturebeat.com/technology/microsoft-ai-chief-says-company-was-set-free-from-openai-to-pursue-superintelligence}
\item \textbf{Microsoft $\rightarrow$ Fireworks.}
Relationship description: ``Microsoft makes its MAI models available for developer weight tuning through Fireworks.''
Source document: the same VentureBeat article as the Baseten edge, June 5, 2026.
\url{https://venturebeat.com/technology/microsoft-ai-chief-says-company-was-set-free-from-openai-to-pursue-superintelligence}
\item \textbf{Microsoft $\rightarrow$ OpenRouter.}
Relationship description: ``Microsoft makes its MAI models available for developer weight tuning through OpenRouter.''
Source document: the same VentureBeat article as the Baseten edge, June 5, 2026.
\url{https://venturebeat.com/technology/microsoft-ai-chief-says-company-was-set-free-from-openai-to-pursue-superintelligence}
\item \textbf{IBM $\rightarrow$ Hugging Face.}
Relationship description: ``IBM has provided Hugging Face with open models and datasets.''
Source document (Global Venturing, November 8, 2023, ``IBM introduces \$500m AI venture fund''): ``IBM has also provided numerous open models and datasets for the startup.'' The description names Hugging Face; that name is not in the extracted passage.
\url{https://globalventuring.com/corporate/fundraising/ibm-introduces-500m-ai-venture-fund/}
\item \textbf{IBM $\rightarrow$ NVIDIA.}
Relationship description: ``IBM contributes specialized large language models that NVIDIA provides as NIM microservices.''
Source document (NVIDIA Blog, June 11, 2025, ``Sovereign AI Agents Think Local, Act Global With NVIDIA AI Factories''): IBM is among those ``contributing specialized LLMs as NIM microservices.''
\url{https://blogs.nvidia.com/blog/sovereign-ai-agents-factories/}
\item \textbf{IBM $\rightarrow$ SAP.}
Relationship description: ``SAP uses IBM as a trusted third-party LLM partner to provide technology for Joule.''
Source document (VentureBeat, September 26, 2023, ``SAP launches its own enterprise AI assistant: meet Joule''): ``we will combine third party LLMs from trusted partners, including IBM.''
\url{https://venturebeat.com/ai/sap-launches-its-own-enterprise-ai-assistant-meet-joule}
\item \textbf{IBM $\rightarrow$ three.ws.}
Relationship description: ``three.ws integrates IBM foundation models and uses IBM's cloud infrastructure to process the platform's inferences.''
Source document (GlobeNewswire, June 3, 2026): ``The three.ws platform integrates IBM foundation models served through IBM enterprise exclusive tools, with every inference being a live, real call to IBM's cloud infrastructure.''
\url{https://www.globenewswire.com/news-release/2026/6/3/3306249/0/en/three-ws-and-ibm-announce-strategic-partnership-to-advance-ai-powered-3d-agent-technology.html}
\item \textbf{ByteDance $\rightarrow$ OMODA \& JAECOO.}
Relationship description: ``OMODA \& JAECOO's Super AI Cockpit is powered by ByteDance's Seed LLM, which OMODA \& JAECOO launched in a vehicle in Southeast Asia.''
Source document (Yahoo Finance Singapore, July 30, 2026, ``OMODA Debuts Super AI Cockpit in Southeast Asia''): the cockpit is ``[e]quipped with ByteDance's Seed LLM'' and is described as the ``[f]irst in ASEAN to launch ByteDance's Seed LLM in a vehicle.''
\url{https://sg.finance.yahoo.com/news/omoda-debuts-super-ai-cockpit-044300759.html}
\item \textbf{ByteDance $\rightarrow$ SiliconFlow.}
Relationship description: ``ByteDance's Seed-OSS-36B-Instruct model is made available in the SiliconFlow model catalog.''
Source document (SiliconFlow, August 10, 2026): ``SiliconFlow is excited to bring Seed-OSS-36B-Instruct to our model catalog,'' described there as ``ByteDance's revolutionary open-source model.''
\url{https://www.siliconflow.com/blog}
\item \textbf{ByteDance $\rightarrow$ TikTok.}
Relationship description: ``TikTok has integrated ByteDance's Seedance 2.0 AI video generation model into the TikTok Symphony ecosystem.''
Source document (Marketing4eCommerce, April 14, 2026, ``TikTok integrates Seedance 2.0''): ``TikTok has announced the integration of Seedance 2.0---the most advanced video generation model from its parent company, ByteDance---into its TikTok Symphony ecosystem.''
\url{https://marketing4ecommerce.net/en/tiktok-symphony-integrates-seedance-2-0-bytedance/}
\item \textbf{ByteDance $\rightarrow$ TikTok USDS Joint Venture LLC.}
Relationship description: ``TikTok USDS Joint Venture LLC licenses TikTok's recommendation algorithm from ByteDance.''
Source document (TechCrunch, January 27, 2026, ``What you should know about the owners of US TikTok''): ``The newly formed entity, `TikTok USDS Joint Venture LLC,' licenses TikTok's recommendation algorithm from ByteDance.''
\url{https://techcrunch.com/2026/01/27/what-you-should-know-about-the-owners-of-us-tiktok/}
\end{itemize}

\subsection*{Scale AI data tree}

\begin{itemize}[noitemsep]
\item \textbf{Microsoft $\rightarrow$ World Bank.}
Relationship description: ``Microsoft is contributing data and research to the forthcoming Global AI Adoption Index developed by the World Bank.''
Source document (The Official Microsoft Blog, February 17, 2026, ``We need to act with urgency to address the growing AI divide''): ``we're scaling this work through contributions to the forthcoming Global AI Adoption Index developed by the World Bank.''
\url{https://blogs.microsoft.com/on-the-issues/2026/02/17/acting-with-urgency-to-address-the-growing-ai-divide/}
\item \textbf{Microsoft $\rightarrow$ Valorem Reply.}
Relationship description: ``Microsoft designated Valorem Reply as an Azure Data \& AI Solutions Partner, establishing a partner relationship focused on Azure, data, and AI solutions.''
Source document (Reply, August 31, 2024, ``Sustainability Measurement and ML/AI Solutions for Change''): ``trusted Microsoft Designated Azure Data \& AI Solutions Partner, Valorem Reply.''
\url{https://www.reply.com/valorem-reply/en/resources/work/2024/august/sustainability-measurement-and-mlai-solutions-for-change}
\item \textbf{Microsoft $\rightarrow$ Yahoo.}
Relationship description: ``Microsoft is beginning to onboard Yahoo as a demand partner for the Publisher Content Marketplace, an AI-focused marketplace for licensed premium content.''
Source document (Microsoft, February 3, 2026, ``Building Toward a Sustainable Content Economy for the Agentic Web''): ``we are beginning to onboard demand partners, including Yahoo.''
\url{https://about.ads.microsoft.com/en/blog/post/february-2026/building-toward-a-sustainable-content-economy-for-the-agentic-web}
\item \textbf{OpenAI $\rightarrow$ Harvey.}
Relationship description: ``OpenAI presents a custom training service for vertical AI companies such as Harvey and cites Harvey as a use case for models expanded with domain-specific data.''
Source document (vintagedata.org, January 2, 2025, ``What's the deal with mid-training?''): ``OpenAI presents its new custom training service for vertical AI companies like Harvey'' and cites models expanded with ``10 billion tokens worth of data.''
\url{https://vintagedata.org/blog/posts/what-is-mid-training}
\item \textbf{OpenAI $\rightarrow$ Neeva.}
Relationship description: ``NeevaAI works with OpenAI's GPT-3.5 to generate training data and sometimes help produce summaries.''
Source document (TechCrunch, February 13, 2023, ``As ChatGPT hype hits fever pitch, Neeva launches its generative AI search engine internationally''): NeevaAI ``works with `large language model companies' including OpenAI's GPT-3.5,'' and ``[w]e use them for things like generating training data for us, and sometimes we will make calls to them to help with the summaries.''
\url{https://techcrunch.com/2023/02/13/as-chatgpt-hype-hits-fever-pitch-neeva-brings-its-generative-ai-search-engine-to-international-markets/}
\item \textbf{Google $\rightarrow$ Inflection AI.}
Relationship description: ``Inflection AI may access certain information from Google when a user creates or logs into an Inflection AI account through Google, subject to the authorization procedures determined by Google.''
Source document (Inflection AI privacy policy, August 9, 2025): ``if you create or log into your account through a third-party platform (such as Google, Facebook, or Apple), we will have access to certain information from that platform \ldots{} in accordance with the authorization procedures determined by such platform.'' The description narrows that list to Google.
\url{https://inflection.ai/privacy-policy}
\item \textbf{Google $\rightarrow$ Vela Partners.}
Relationship description: ``Vela Partners uses Google Search grounding with the Gemini API to enrich startup profiles and support due-diligence analysis.''
Source document (Google AI Studio, April 9, 2025, ``Transforming VC with Gemini API''): ``Vela uses Gemini 2.0 Flash grounded with Google Search to automatically fill in the gaps.''
\url{https://aistudio.google.com/case-studies/vela}
\item \textbf{Meta $\rightarrow$ Hugging Face.}
Relationship description: ``Meta released an Omnilingual ASR demo and dataset resources through Hugging Face.''
Source document (VentureBeat, November 10, 2025, ``Meta returns to open source AI with Omnilingual ASR models that can transcribe 1,600+ languages natively''): the release includes ``a demo space on Hugging Face'' and the dataset at \path{huggingface.co/datasets/facebook/omnilingual-asr-corpus}.
\url{https://venturebeat.com/technology/meta-returns-to-open-source-ai-with-omnilingual-asr-models-that-can}
\item \textbf{Meta $\rightarrow$ TechCrunch.}
Relationship description: ``Meta provided TechCrunch with additional information about its use of images and videos to train AI.''
Source document (TechCrunch, October 2, 2024, ``Meta confirms it may train its AI on any image you ask Ray-Ban Meta AI to analyze''): ``Meta has offered TechCrunch a bit more color,'' including an email from Meta stating that ``images and videos shared with Meta AI may be used to improve it.''
\url{https://techcrunch.com/2024/10/02/meta-confirms-it-may-train-its-ai-on-any-image-you-ask-ray-ban-meta-ai-to-analyze/}
\end{itemize}

\section{Additional Results for \texorpdfstring{\Cref{sec:findings-visibility}}{Section 4.4}}
\label{app:coverage_bias}

\begin{figure}[h!]
\centering
\includegraphics[width=0.72\linewidth]{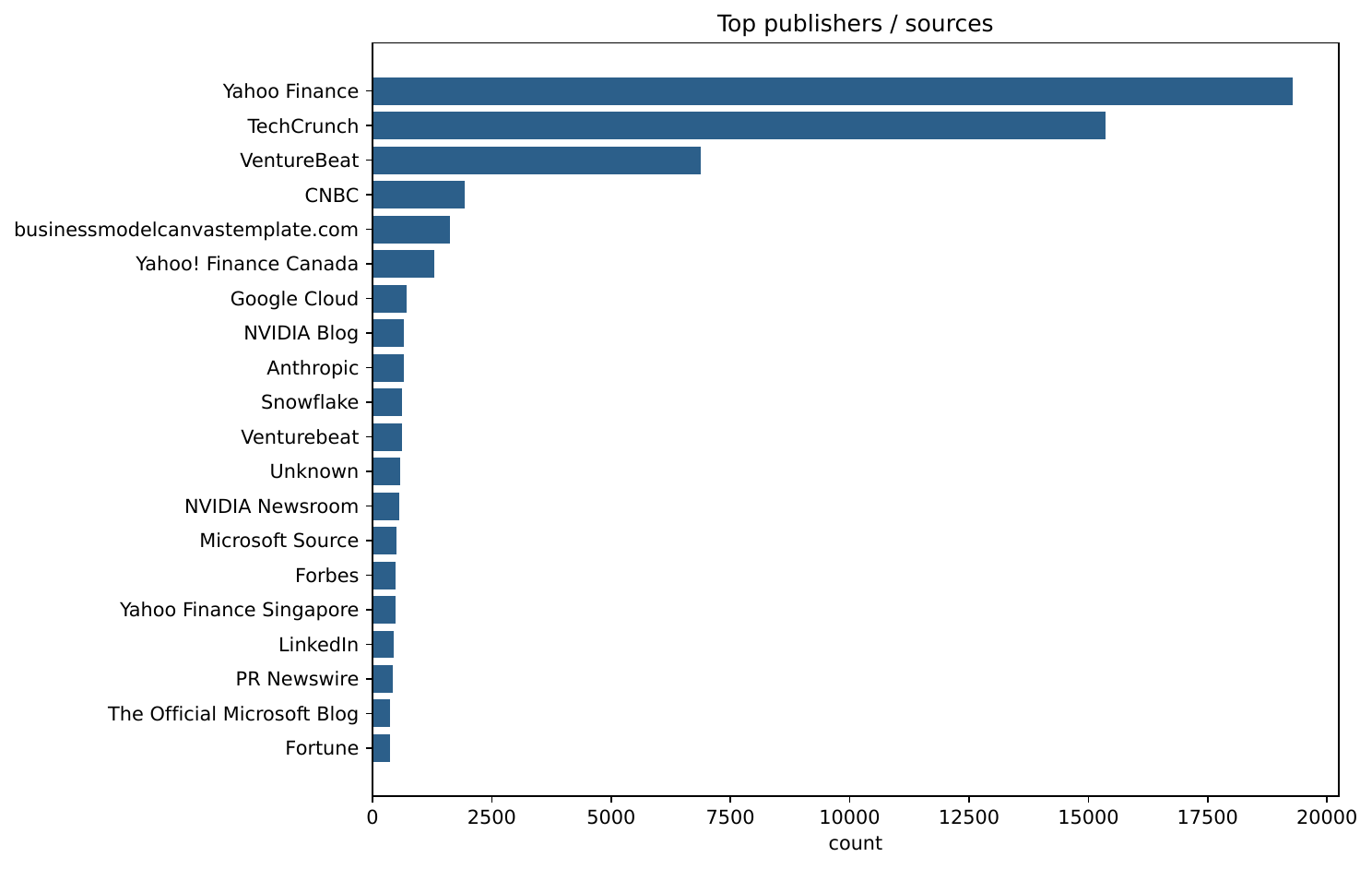}
\caption{Top publishers by multiplicity-edge volume.}
\label{fig:top_publishers}
\end{figure}

\Cref{fig:top_publishers} ranks publishers by volume.
\textbf{Yahoo Finance} alone accounts for about \textbf{21\%} of multiplicity edges,
followed by \textbf{TechCrunch} (\textbf{17\%}) and \textbf{VentureBeat}
(\textbf{7.5\%}).
Across \textbf{2{,}561} publishers, the top 10\% contribute about \textbf{81\%} of edges.

\begin{figure}[h!]
\centering
\includegraphics[width=0.85\linewidth]{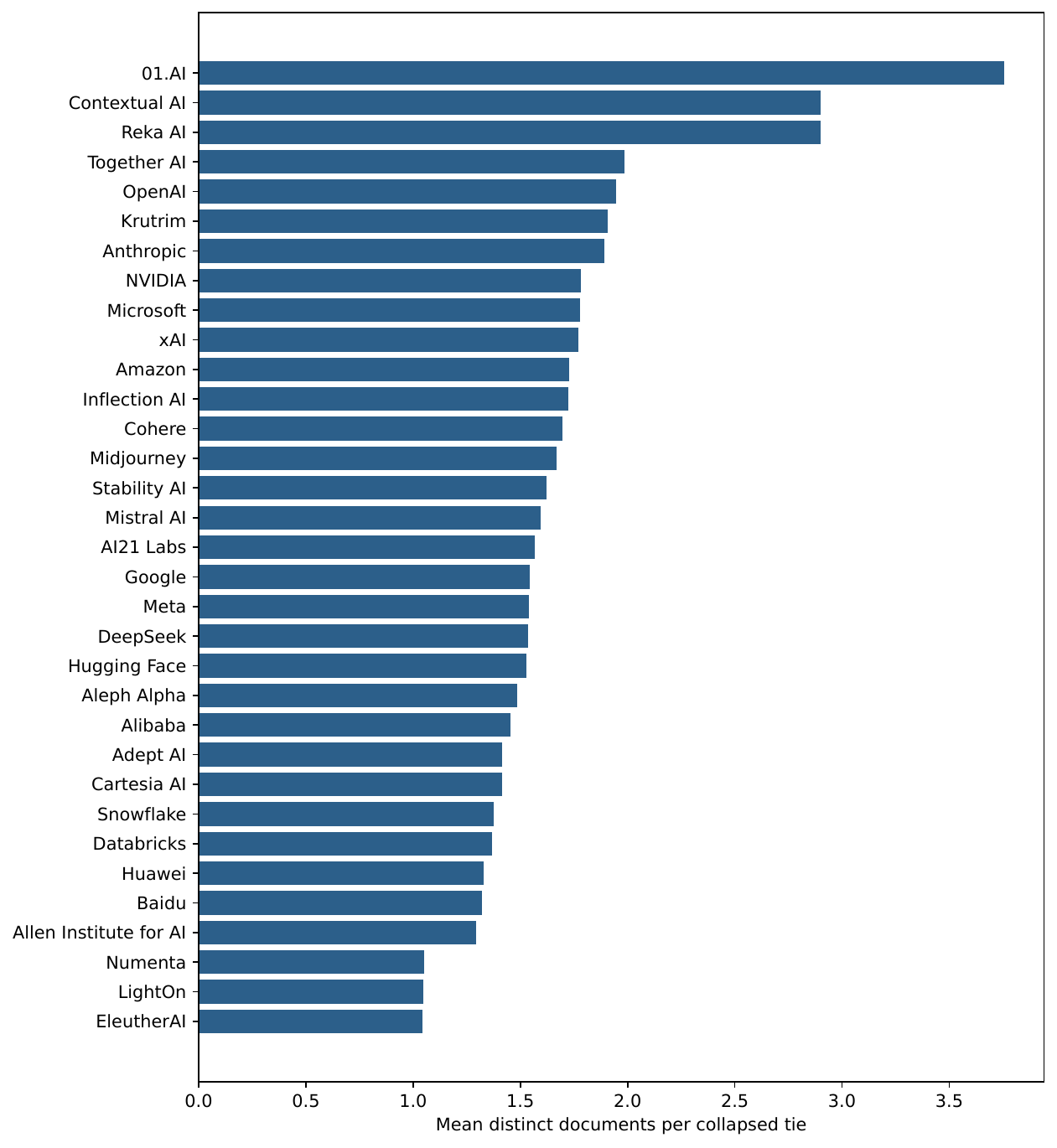}
\caption{Mean number of distinct documents per collapsed tie, by seed organization.
A tie shares endpoints, relationship type, directionality, and calendar month, and is counted for an organization when that organization is an endpoint.}
\label{fig:docs_by_seed}
\end{figure}

\Cref{fig:docs_by_seed} shows the mean number of distinct documents per collapsed tie for each seed organization.
Relationships that share the same endpoints, type, directionality, and calendar month form one tie.
A tie is included for an organization when that organization is an endpoint.
Names are the normalized actor names, so DeepMind and Google AI are Google, and Microsoft AI is Microsoft.
01.AI, Contextual AI, and Reka AI average more than 2.5 documents per tie.
Among organizations that appear in thousands of ties, OpenAI is highest, about 1.9.

\begin{figure}[h!]
\centering
\includegraphics[width=0.85\linewidth]{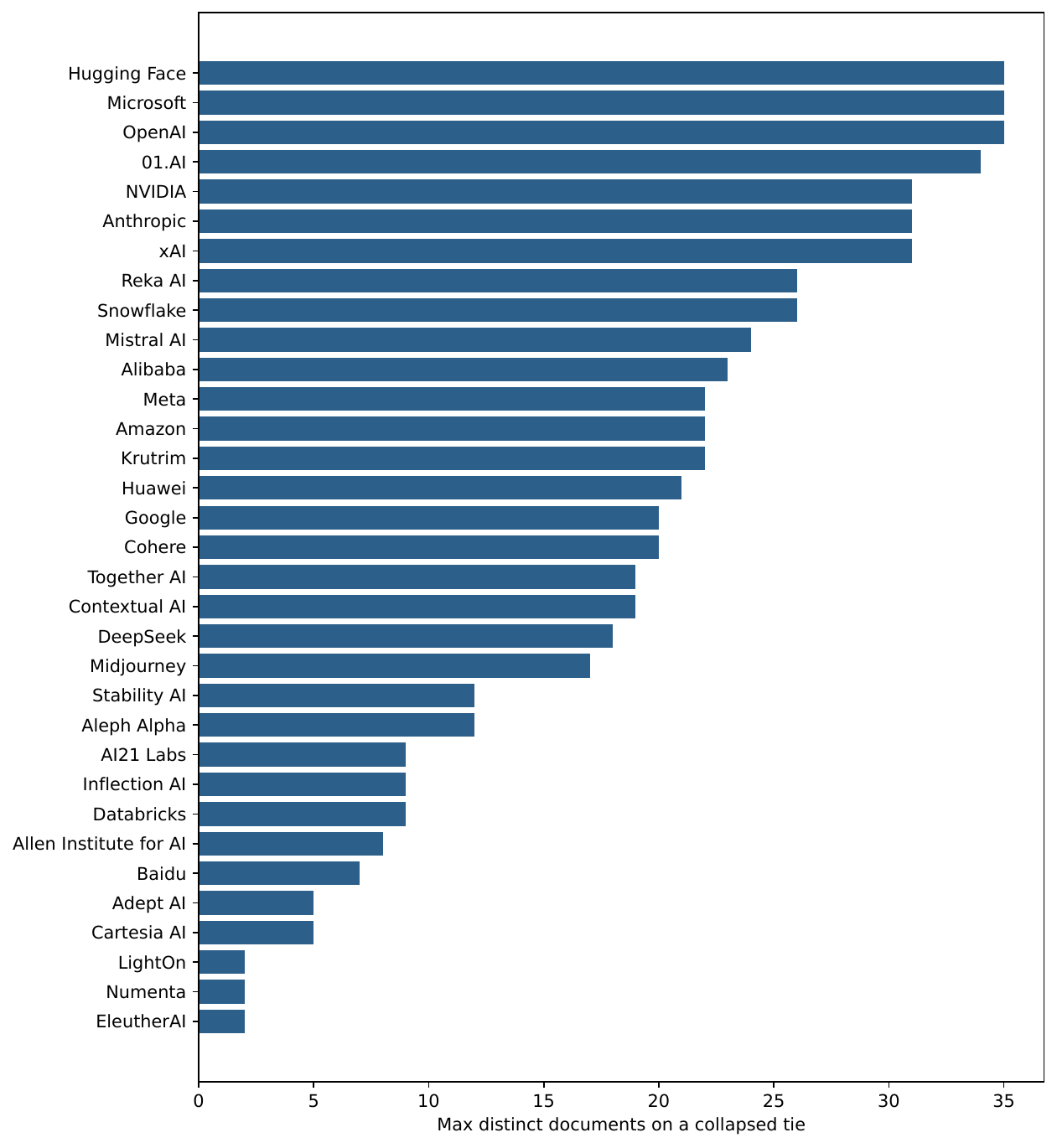}
\caption{Maximum number of distinct documents on a single collapsed tie, by seed organization.
A tie shares endpoints, relationship type, directionality, and calendar month, and is counted for an organization when that organization is an endpoint.}
\label{fig:docs_by_seed_max}
\end{figure}

\Cref{fig:docs_by_seed_max} reports the largest number of distinct documents on any one collapsed tie for each seed organization.
A tie is included for an organization when that organization is an endpoint.
OpenAI, Microsoft, and Hugging Face reach 35 documents
(OpenAI and Microsoft on a shared April 2026 partnership tie; OpenAI and Hugging Face on a shared July 2026 incident tie).
Anthropic, NVIDIA, and xAI reach 31.

\begin{figure}[h!]
\centering
\includegraphics[width=\linewidth]{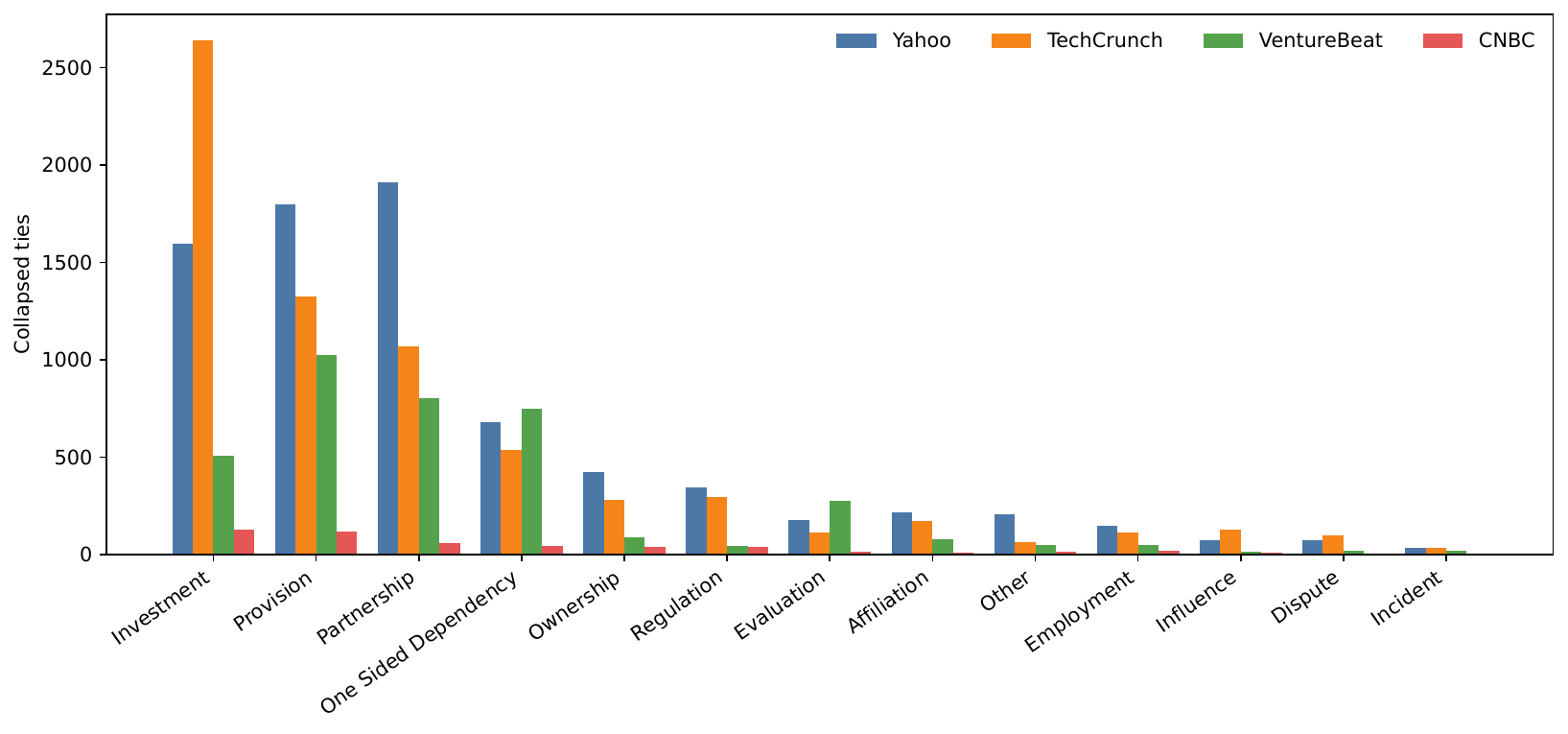}
\caption{Monthly collapsed ties by relationship type for four outlets.
Each tie is counted once, under the publisher that contributes the most of its extracted rows.
Yahoo includes Yahoo Finance and its regional sites.}
\label{fig:publisher_by_reltype}
\end{figure}

\Cref{fig:publisher_by_reltype} breaks the same monthly ties out by relationship type for the four outlets that carry the most of them.
Each tie is assigned to the publisher that accounts for the most of its extracted rows.
Yahoo's regional finance sites are counted together.
Yahoo is the largest source of partnership, provision, and ownership ties.
TechCrunch is the largest source of investment ties, with about 2{,}600 such ties against about 1{,}600 from Yahoo.
CNBC appears in every major type, at much lower volume.

\clearpage
\section{Dataset Documentation}
\label{app:dataset}

The public Hugging Face release (\texttt{sarahcen/ai-supply-chain}) is a parquet table of AI-related relationship rows.
\Cref{tab:hf_dataset_fields} gives an abbreviated overview in the main text; \Cref{tab:hf_dataset_fields_full} lists every published column with full categorical value sets.
The release includes \texttt{ai\_relevance}, \texttt{relationship\_status}, and \texttt{ai\_stack\_layers}.
Directed rows are stored as \texttt{ACTOR\_1\_TO\_ACTOR\_2} or \texttt{ACTOR\_2\_TO\_ACTOR\_1}.
The analysis graphs rewrite both labels to \texttt{DIRECTIONAL} and, for the second label, swap the endpoints so actor~1 is the source.
Published rows keep only \texttt{AI\_RELATIONSHIP} and \texttt{INVOLVES\_AI\_ACTOR}, and drop rows whose relationship type, status, or directionality is \texttt{CONTRADICTORY\_INFORMATION}.

{\small
\begin{longtable}{@{}
  >{\raggedright\arraybackslash\ttfamily\footnotesize}p{0.36\linewidth}@{\hspace{1.2em}}
  >{\raggedright\arraybackslash\footnotesize}p{0.22\linewidth}@{\hspace{1.2em}}
  >{\raggedright\arraybackslash\footnotesize}p{0.32\linewidth}@{}}
\caption{Full field schema for the publicly released Hugging Face dataset}
\label{tab:hf_dataset_fields_full}\\
\toprule
\multicolumn{1}{l}{\textbf{Field}} & \textbf{Description} & \textbf{Possible values} \\
\midrule
\endfirsthead
\caption[]{Full field schema for the publicly released Hugging Face dataset (continued)}\\
\toprule
\multicolumn{1}{l}{\textbf{Field}} & \textbf{Description} & \textbf{Possible values} \\
\midrule
\endhead
\midrule
\multicolumn{3}{r}{\emph{Continued on next page}} \\
\endfoot
\bottomrule
\endlastfoot
link & Source document URL & Text (URL) \\
article\allowbreak\_title & Title associated with the source page & Text \\
document\allowbreak\_date & Parsed document/article date, when available & Date, or missing \\
actor\allowbreak\_1\allowbreak\_name & Name of actor~1 & Text \\
actor\allowbreak\_1\allowbreak\_type & Entity type of actor~1 & \texttt{COMPANY}, \texttt{NONPROFIT\_OR\_CIVIL\_SOCIETY}, \texttt{ACADEMIC\_INSTITUTION}, \texttt{COALITION\_OR\_CONSORTIUM}, \texttt{PERSON}, \texttt{GOVERNMENT\_BODY}, \texttt{GEOPOLITICAL\_ENTITY\_OR\_COUNTRY}, \texttt{INTERGOVERNMENTAL\_OR\_SUPRANATIONAL}, \texttt{OTHER\_ACTOR}, \texttt{NOT\_AN\_ACTOR}, \texttt{UNKNOWN\_OR\_UNCLEAR} \\
actor\allowbreak\_2\allowbreak\_name & Name of actor~2 & Text \\
actor\allowbreak\_2\allowbreak\_type & Entity type of actor~2 & Same enum as \texttt{actor\_1\_type} \\
relationship\allowbreak\_type & Primary kind of link & \texttt{INVESTMENT}, \texttt{PROVISION}, \texttt{OWNERSHIP}, \texttt{EVALUATION}, \texttt{PARTNERSHIP}, \texttt{ONE\_SIDED\_DEPENDENCY}, \texttt{EMPLOYMENT}, \texttt{AFFILIATION}, \texttt{REGULATION}, \texttt{INFLUENCE}, \texttt{COMPETITION}, \texttt{DISPUTE}, \texttt{INCIDENT}, \texttt{PUBLIC\_STATEMENT}, \texttt{NO\_RELATIONSHIP}, \texttt{OTHER}, \texttt{UNCLEAR} \\
relationship\allowbreak\_status & Lifecycle status of the relationship & \texttt{ESTABLISHED}, \texttt{PROPOSED}, \texttt{ENDED}, \texttt{NOT\_APPLICABLE}, \texttt{UNCLEAR} \\
ai\allowbreak\_stack\allowbreak\_layers & Relationship-level AI stack layers (1--2) & List from \texttt{CAPITAL}, \texttt{HARDWARE}, \texttt{INFRASTRUCTURE}, \texttt{DATA}, \texttt{RESEARCH}, \texttt{MODELS}, \texttt{PLATFORMS}, \texttt{APPLICATIONS}, \texttt{ADOPTION}, \texttt{OPERATIONS}, \texttt{SAFETY}, \texttt{SECURITY}, \texttt{GOVERNANCE}, \texttt{TALENT}; \texttt{NOT\_AI\_STACK} and \texttt{UNKNOWN\_OR\_UNCLEAR} appear alone \\
directionality & Source and target of the tie & \texttt{ACTOR\_1\_TO\_ACTOR\_2}, \texttt{ACTOR\_2\_TO\_ACTOR\_1}, \texttt{SYMMETRIC}, \texttt{UNCLEAR} \\
relationship\allowbreak\_description & One-sentence description of the relationship & Text \\
evidence & Short supporting excerpt from the source & Text \\
evidence\allowbreak\_strength & How directly the source supports the extraction & \texttt{EXPLICIT}, \texttt{STRONGLY\_IMPLIED}, \texttt{IMPLIED} \\
ai\allowbreak\_relevance & Whether the tie is AI-related (published rows keep only these two values; extraction also produces \texttt{NOT\_AI\_RELATED} and \texttt{UNCLEAR}) & \texttt{AI\_RELATIONSHIP}, \texttt{INVOLVES\_AI\_ACTOR} \\
ai\allowbreak\_relationship\allowbreak\_categories & AI topical tags ($\le$3) & Full AI-tag taxonomy (117 labels): \texttt{AI\_SEMICONDUCTORS}, \texttt{AI\_COMPUTE}, \texttt{AI\_HARDWARE}, \texttt{AI\_NETWORKING}, \texttt{AI\_STORAGE}, \texttt{AI\_INFRASTRUCTURE}, \texttt{AI\_DATA\_CENTERS}, \texttt{AI\_REAL\_ESTATE}, \texttt{AI\_COOLING}, \texttt{AI\_CLOUD}, \texttt{AI\_ENERGY}, \texttt{AI\_ENVIRONMENTAL\_IMPACT}, \texttt{AI\_DATA}, \texttt{AI\_DATA\_ANNOTATION}, \texttt{AI\_SYNTHETIC\_DATA}, \texttt{AI\_MODELS}, \texttt{AI\_MODEL\_TRAINING}, \texttt{AI\_MODEL\_ADAPTATION}, \texttt{AI\_INFERENCE}, \texttt{AI\_SOFTWARE}, \texttt{AI\_DEVELOPER\_TOOLS}, \texttt{AI\_DEPLOYMENT}, \texttt{AI\_OPERATIONS}, \texttt{AI\_APPLICATIONS}, \texttt{AI\_AGENTS}, \texttt{AI\_COMPANIONS}, \texttt{AI\_INTEROPERABILITY}, \texttt{AI\_ROBOTICS}, \texttt{AI\_AUTONOMOUS\_SYSTEMS}, \texttt{AI\_SIMULATION}, \texttt{AI\_CODING}, \texttt{AI\_INFORMATION\_RETRIEVAL}, \texttt{AI\_RECOMMENDATION}, \texttt{AI\_CONTENT\_GENERATION}, \texttt{AI\_SYNTHETIC\_MEDIA}, \texttt{AI\_AUDIO}, \texttt{AI\_IMAGE}, \texttt{AI\_VIDEO}, \texttt{AI\_MULTIMODAL}, \texttt{AI\_EDGE} \\
\textit{(continued)} & \textit{(continued)} & \texttt{AI\_FOR\_HEALTHCARE}, \texttt{AI\_FOR\_LIFE\_SCIENCES}, \texttt{AI\_FOR\_FINANCIAL\_SERVICES}, \texttt{AI\_FOR\_EDUCATION}, \texttt{AI\_FOR\_GOVERNMENT}, \texttt{AI\_FOR\_MARKETING}, \texttt{AI\_FOR\_MANUFACTURING}, \texttt{AI\_FOR\_TRANSPORTATION}, \texttt{AI\_FOR\_LOGISTICS}, \texttt{AI\_FOR\_AGRICULTURE}, \texttt{AI\_FOR\_COMMERCE}, \texttt{AI\_FOR\_MEDIA}, \texttt{AI\_FOR\_ENTERTAINMENT}, \texttt{AI\_FOR\_LEGAL\_SERVICES}, \texttt{AI\_FOR\_ENERGY}, \texttt{AI\_FOR\_SCIENCE}, \texttt{AI\_RESEARCH}, \texttt{AI\_EVALUATION}, \texttt{AI\_INTERPRETABILITY\_AND\_EXPLAINABILITY}, \texttt{AI\_STANDARDS}, \texttt{AI\_ASSURANCE}, \texttt{AI\_TALENT}, \texttt{AI\_LABOR}, \texttt{AI\_FUNDING}, \texttt{AI\_GRANTS}, \texttt{AI\_PHILANTHROPY}, \texttt{AI\_OWNERSHIP}, \texttt{AI\_ACQUISITIONS}, \texttt{AI\_MERGERS}, \texttt{AI\_ORGANIZATIONAL\_CONTROL}, \texttt{AI\_GOVERNANCE}, \texttt{AI\_LICENSING}, \texttt{AI\_DISTRIBUTION}, \texttt{AI\_INTEGRATION}, \texttt{AI\_ADOPTION}, \texttt{AI\_PARTNERSHIP}, \texttt{AI\_MARKET\_ACCESS}, \texttt{AI\_COMPETITION}, \texttt{AI\_INSURANCE}, \texttt{AI\_LIABILITY} \\
\textit{(continued)} & \textit{(continued)} & \texttt{AI\_OPEN\_SOURCE}, \texttt{AI\_OPEN\_WEIGHTS}, \texttt{AI\_TECHNOLOGY\_TRANSFER}, \texttt{AI\_ACCESSIBILITY}, \texttt{AI\_INCLUSION}, \texttt{AI\_POLICY}, \texttt{AI\_REGULATION}, \texttt{AI\_ENFORCEMENT}, \texttt{AI\_ADVOCACY}, \texttt{AI\_TRADE\_RESTRICTIONS}, \texttt{AI\_DIPLOMACY}, \texttt{AI\_SOVEREIGNTY}, \texttt{AI\_MARKET\_POWER}, \texttt{AI\_INTELLECTUAL\_PROPERTY}, \texttt{AI\_PRIVACY}, \texttt{AI\_DATA\_GOVERNANCE}, \texttt{AI\_CONTENT\_PROVENANCE}, \texttt{AI\_CONTENT\_MODERATION}, \texttt{AI\_HUMAN\_RIGHTS}, \texttt{AI\_SOCIAL\_IMPACT}, \texttt{AI\_SURVEILLANCE}, \texttt{AI\_BIOMETRICS}, \texttt{AI\_IDENTITY}, \texttt{AI\_ACCESS\_CONTROL}, \texttt{AI\_INFORMATION\_INTEGRITY}, \texttt{AI\_FRAUD}, \texttt{AI\_CONSUMER\_PROTECTION}, \texttt{AI\_CHILD\_SAFETY}, \texttt{AI\_ALIGNMENT}, \texttt{AI\_SAFETY}, \texttt{AI\_SECURITY}, \texttt{AI\_INCIDENT\_RESPONSE}, \texttt{AI\_DEFENSE}, \texttt{AI\_NATIONAL\_SECURITY}, \texttt{AI\_RISKS}, \texttt{AI\_HARMS}, \texttt{AI\_DISPUTES} \\
relationship\allowbreak\_pass & Independent extraction pass index & Integer (1-based; currently 1 or 2) \\
\end{longtable}
}
\FloatBarrier
\clearpage

\end{document}